\documentclass[a4paper,notitlepage]{report}   	%
\renewcommand{\thechapter}{{\textbf{\arabic{chapter}}}}
\renewcommand{\thesection}{\thechapter:$\,$\arabic{section}}

\renewcommand{\appendix}{
	\section*{Appendices}
	\setcounter{section}{0}
	\renewcommand{\thesection}{\thechapter:$\,$\Alph{section}}
}
\newcommand{\appendixend}{
	\setcounter{section}{0}
	\renewcommand{\thesection}{\thechapter:$\,$\arabic{section}}
}
\usepackage{amsmath}
\usepackage{amssymb}
\newcommand{\bmf}{} %
\usepackage{xr}
\usepackage{refcount}

\usepackage{graphicx}
\usepackage{mathrsfs}
\usepackage{bm}
\usepackage[hyperref]{xcolor} %
\definecolor{darkgreen}{rgb}{0,0.4,0} %
\definecolor{darkred}{rgb}{0.55,0,0} %
\definecolor{navy}{rgb}{0,0,0.55} %
\usepackage[hypertexnames=false,breaklinks=true,colorlinks=true,citecolor=darkgreen,linkcolor=black,urlcolor=navy]{hyperref} %
\usepackage{amsthm}
\usepackage{lscape}
\theoremstyle{definition} %

\newcommand{\tbf}[1]{\textbf{#1}}

\newcommand{\G}{\mc{G}}
\newcommand{\Z}{\mc{Z}}
\newcommand{\ke}[2]{k^{(e)}_{#1}(\mc{E}_{#2})}

\newcommand{\E}[1]{\textrm{E}#1}
\newcommand{\prm}[1]{\protect{$#1$}}

\newcommand{\sol}[1]{}
\newcommand{\bmh}{{\bm{\mathrm{H}}}}%

\newcommand{\ta}{{\tilde{a}}}
\newcommand{\tb}{{\tilde{b}}}
\newcommand{\tc}{{\tilde{c}}}
\newcommand{\td}{{\tilde{d}}}
\newcommand{\tW}{{\tilde{W}}}

\newcommand{\ulambda}{\lambda} %
\newcommand{\mrm}[1]{\mathrm{#1}}

\newcommand{\mbb}[1]{\mathbb{#1}}
\newcommand{\mc}[1]{\mathcal{#1}}

\newcommand{\mscr}[1]{\mathscr{#1}}
\newcommand{\bgrid}{\left(\begin{array}{rrr}}
\newcommand{\egrid}{\end{array}\right)}
\newcommand{\bgridt}{\left(\begin{array}{rr}}
\newcommand{\egridt}{\end{array}\right)}
\newcommand{\bgridtt}{\left[\begin{array}{rr}}
\newcommand{\egridtt}{\end{array}\right]}
\newcommand{\eref}[1]{(\ref{#1})}
\newcommand{\Eref}[1]{Eq.~(\ref{#1})}

\newcommand{\Erefr}[2]{Eqs.~(\ref{#1}--\ref{#2})}

\newcommand{\erefr}[2]{(\ref{#1}--\ref{#2})}

\newcommand{\Erefs}[2]{Eqs.~(\ref{#1}) and~(\ref{#2})}
\newcommand{\Ereft}[1]{Eqs.~(\ref{#1})}
\newcommand{\Eqrefs}[2]{Equations~(\ref{#1}) and~(\ref{#2})}

\newcommand{\erefs}[2]{(\ref{#1}) and~(\ref{#2})}

\newcommand{\cref}[1]{Chapter~\ref{#1}}
\newcommand{\Cref}[1]{Chapter~\ref{#1}}
\newcommand{\crefs}[2]{Chapters~\ref{#1} and~\ref{#2}}
\newcommand{\crefr}[2]{Chapters~\ref{#1}--\ref{#2}}

\newcommand{\Pref}[2]{\ref{#2}}
\newcommand{\Peref}[2]{(\ref{#2})}
\newcommand{\PEref}[2]{Eq.~(\ref{#2})}

\newcommand{\PErefr}[3]{Eqs.~(\ref{#2}--\ref{#3})}

\newcommand{\Perefr}[3]{(\ref{#2}--\ref{#3})}

\newcommand{\PErefs}[3]{Eqs.~(\ref{#2}) and~(\ref{#3})}
\newcommand{\PEreft}[2]{Eqs.~(\ref{#2})}

\newcommand{\fref}[1]{Fig.~\ref{#1}}
\newcommand{\Fref}[1]{Figure~\ref{#1}}
\newcommand{\frefs}[2]{Figs.~\ref{#1} and~\ref{#2}}
\newcommand{\Freft}[1]{Figures~\ref{#1}}
\newcommand{\freft}[1]{Figs.~\ref{#1}}
\newcommand{\sref}[1]{Sec.~\ref{#1}}
\newcommand{\Psref}[2]{\sref{#2}}
\newcommand{\Psrefs}[3]{\srefs{#2}{#3}}
\newcommand{\Sref}[1]{Section~\ref{#1}}
\newcommand{\srefs}[2]{Secs.~\ref{#1} and~\ref{#2}}

\newcommand{\Srefs}[2]{Sections~\ref{#1} and~\ref{#2}}
\newcommand{\srefr}[2]{Secs.~\ref{#1}--\ref{#2}}
\newcommand{\aref}[1]{Appendix~\ref{#1}}

\newcommand{\tref}[1]{Table~\ref{#1}}
\newcommand{\bsm}{\bar\sigma^\mu}
\newcommand{\bsmm}{\bar\sigma_\mu}
\newcommand{\bsn}{\bar\sigma^\nu}

\newcommand{\nn}{\nonumber}
\newcommand{\rmd}{\mathrm{d}}
\newcommand{\rmi}{\mathrm{i}}%
\newcommand{\la}{\langle}
\newcommand{\ra}{\rangle}

\newcommand{\bdag}{{(\dagger)}}

\newcommand{\rcite}[1]{Ref.~\cite{#1}}

\newcommand{\rcites}[2]{Refs.~\cite{#1} and~\cite{#2}}
\newcommand{\rcitess}[1]{Refs.~\cite{#1}}
\newcommand{\p}[1]{\phantom{#1}}
\newcommand{\pt}[1]{\protect{#1}}

\newcommand{\triplet}[3]{\left(\begin{aligned}&#1\\&#2\\&#3\end{aligned}\right)}
\newcommand{\ctriplet}[3]{\left(\begin{array}{c}#1\\#2\\#3\end{array}\right)}

\newcommand{\fg}{\mrm{fg}}
\newcommand{\bgfield}[1]{[#1]_\mrm{bg}}
\newcommand{\fgfield}[1]{[#1]_\fg}
\newcommand{\OO}[1]{\mrm{O}\left(#1\right)}
\newcommand{\OOO}[1]{\mrm{O}\left[#1\right]}
\newcommand{\OOOO}[1]{\mrm{O}\left\{#1\right\}}
\newcommand{\OOOOx}[1]{\mrm{O}\Bigg\{#1\Bigg\}}
\newcommand{\eV}{\mrm{eV}}

\newcommand{\MeV}{\mrm{M}\eV}
\newcommand{\GeV}{\mrm{G}\eV}
\newcommand{\TeV}{\mrm{T}\eV}
\newcommand{\SU}[1]{\mrm{SU}(#1)}
\newcommand{\su}[1]{\mrm{su}(#1)}

\newcommand{\U}[1]{\mrm{U}(#1)}

\newcommand{\MSbar}{\overline{\mathrm{MS}}}
\newcommand{\pref}[1]{\protect{\ref{#1}}}
\newcommand{\peref}[1]{\protect{(\ref{#1})}}

\newcommand{\pfref}[1]{\protect{Fig.~\ref{#1}}}
\newcommand{\pfreft}[1]{\protect{Figs.~\ref{#1}}}
\newcommand{\ptref}[1]{\protect{Table~\ref{#1}}}
\newcommand{\psref}[1]{\protect{Sec.~\ref{#1}}}
\newcommand{\paref}[1]{\protect{Appendix~\ref{#1}}}
\newcommand{\pcref}[1]{\protect{Chapter~\ref{#1}}}

\newcommand{\GL}[2]{\mrm{GL}(#1,\mbb{#2})}

\newcommand{\gl}[2]{\mrm{gl}(#1,\mbb{#2})}
\newcommand{\SL}[2]{\mrm{SL}(#1,\mbb{#2})}

\newcommand{\GLTR}{\GL{3}{R}}%
\newcommand{\GLNR}{\GL{9}{R}}%
\newcommand{\gltr}{\mrm{gl}(3,\mbb{R})}
\newcommand{\SLTC}{\mrm{SL}(2,\mbb{C})}

\renewcommand{\bar}[1]{\overline{#1}}

\newcommand{\preon}{\psi}

\newcommand{\thf}{\theta_f}

\newcommand{\Kf}{\mc{K}}
\newcommand{\ket}[1]{|#1\ra}

\newcommand{\K}{K_\ell}
\newcommand{\Ke}{K_e(\theta_e)}
\newcommand{\Kee}[1]{\left[K_e(\theta_e)\right]^{#1}}
\newcommand{\Keij}[1]{\left[\Ke\right]_{#1}}
\newcommand{\ILO}[1]{\mrm{O}(#1)}
\newcommand{\ILOO}[1]{\mrm{O}[#1]}
\newcommand{\ILOOO}[1]{\mrm{O}\{#1\}}
\newcommand{\ILOOOO}[1]{\mrm{O}\bm{(}#1\bm{)}}

\newcommand{\bmmf}{\tilde{f}}
\renewcommand{\bmf}[1]{\bm{\mrm{f}}\!\left(#1\right)}
\newcommand{\bmfcdot}{\bm{\mrm{f}}(\cdot)}

\newcommand{\N}{{\mathfrak{n}}}%
\newcommand{\RM}{{{\mbb{R}^{1,3}}}}

\newcommand{\Cw}[1]{{{\mbb{C}^{\wedge #1}}}}

\newcommand{\vp}{\varphi}

\newcommand{\tagref}[1]{{\tag{\ref{#1}}}}
\newcommand{\Ptagref}[2]{{\tag{\ref{#2}}}}%

\newcommand{\taue}{e}

\newcommand{\ff}{\mrm{f}}
\newcommand{\Gversion}[1]{#1}
\newcommand{\quietGversion}[1]{}
\newcommand{\onlyinsummary}[1]{}

\newcommand{\mcst}{\big[m_c^*(\mc{E}_{\nu_i})\big]}
\newcommand{\BK}{B^+\!\rightarrow\!K^+}

\newcommand{\completed}[1]{}
\newcommand{\chapeight}[1]{}
\newcommand{\notinsummary}[1]{#1}
\newcommand{\levelone}{\subsection}
\newcommand{\leveltwo}{\subsubsection}
\newcommand{\chap}[1]{}
\newcommand{\notchap}[1]{#1}
\newcommand{\standalone}[1]{}
\newcommand{\notstandalone}[1]{#1}
\newcommand{\paper}{chapter}
\usepackage{geometry}
\usepackage[numbers,square,sort&compress]{natbib}

\begin{document}

\title{A Classical Analogue to the Standard Model\\and General Relativity}
\author{Chapters 4--11\\~\\R. N. C. Pfeifer}

\date{01 January 2024}

\maketitle
\thispagestyle{empty}
\newpage

{\small
\newgeometry{left=1.5cm,right=1.5cm,top=2cm,bottom=3cm}
\!\!\!\!\!\!\!\!\!\!The relationships determined to date between fundamental constants in %
CASMIR are:
\begin{align}
\frac{m_{e_i}^2}{m_e^2}=\,&\left.\frac
{\left[k^{(e)}_i(\mc{E}_{e_i})\right]^4\left[1+\Delta_e(m_{e_i},\mc{E}_{e_i})\right]}
{\left[k^{(e)}_1(\mc{E}_{e_i})\right]^4\left[1+\Delta_e(m_e,\mc{E}_{e_i})\right]}\left[1+\mc{O}_e(m_{e_i},\mc{E}_{e_i})\right]\quad\right|\quad e_i\in\{e,\mu,\tau\}~\mrm{for}~i\in\{1,2,3\}~\mrm{respectively}\nn
\\
\frac{m_\tW^2}{m_e^2}=\,&18{N_0}^{4}\left(1+\frac{2}{N_0}\right)^{-4}\left(1+\frac{1}{N_0}\right)^{-4}\frac
{\Biggl[1+\left(64+\frac{3}{2\pi}-f_Z\right)\frac{\alpha}{2\pi}\Biggr]\left\{1+\frac{51}{18\left[k^{(e)}_{1}(\mc{E}_e)\,{N_0}\right]^4}\right\}%
}
{\left[1+\Delta_{e}(m_e,\mc{E}_e)\right]%
}\left[1+\mc{O}_b+\mc{O}_e(m_e,\mc{E}_e)\right]\nn
\\
\frac{m_\tW^2}{m_Z^2}=\,&\frac
{3\left[1+\left(64+\frac{3}{2\pi}-f_Z\right)\frac{\alpha}{2\pi}\right]\left\{1+\frac{51}{18\left[k^{(e)}_{1}(\mc{E}_e)\,{N_0}\right]^4}\right\}}
{4\left[1+\left(\frac{401}{12}+\frac{3}{2\pi}\right)\frac{\alpha}{2\pi}\right]
\left\{1+\frac{55}{18\left[k^{(e)}_{1}(\mc{E}_e)\,{N_0}\right]^4}\right\}
}\left(1+\mc{O}_b\right)
\nn
\\
\frac{m_\tW^2}{m_{\bmh}^2}=\,&\frac
{9\left[1+\left(64+\frac{3}{2\pi}-f_Z\right)\frac{\alpha}{2\pi}\right]\left\{1+\frac{51}{18\left[k^{(e)}_{1}(\mc{E}_e)\,{N_0}\right]^4}\right\}}
{20\left\{\left(1-\frac{2}{3N_0}+\frac{1}{3{N_0}^2}\right)\left[1+\frac{30\alpha}{9\pi}\left(1+\frac{1}{3N_0}\right)\right]+\frac{1}{2\pi}\left[1+\frac{30\alpha}{\pi}\left(1-\frac{1}{3N_0}\right)\right]\right\}
\left\{1+\frac{39}{18\left[k^{(e)}_{1}(\mc{E}_e)\,{N_0}\right]^4}\right\}
}
\left(1+\mc{O}_b\right)
\nn
\Gversion{%
\\G_N=\,&
\frac{2\alpha^4hc\left[1+\Delta_e(m_e,\mc{E}_e)\right]^\frac{1}{2}
\left(1+\frac{7\alpha}{2\pi}+\frac{229\alpha}{12\pi N_0}\right)}{\pi\left[k^{(e)}_1(\mc{E}_e)\right]^4m_e^2\bm{\left(}1+\frac{\alpha}{2\pi}+%
\left\{\frac{197}{144}+\frac{\pi^2[1-6\,\mrm{ln}(2)]}{12}+\frac{3\,\zeta(3)}{4}\right\}\frac{\alpha^2}{\pi^2}\bm{\right)}^8\left\{1+\frac{281}{36\left[k^{(e)}_{1}(\mc{E}_e)\,{N_0}\right]^4}\right\}
}\nn
\\&\times 
\frac{{N_0}^6}{\{2(N_0+4)^4(N_0+3)^4+[8(N_0+3)^4+6(N_0+2)^4](N_0+1)^4\}(N_0+2)^5(N_0+\tfrac{5}{4})^3(N_0+1)^8}
\nn
\\&\times \nn\left[1+\mc{O}_b+\mc{O}_e(m_e,\mc{E}_e)+\OO{\frac{\alpha^2 m_e^2}{\pi^2 m_\mu^2}}\right]
}\end{align}\\
where $m_\tW$ is the effective $W$~boson mass appearing in first order loop corrections to the lepton gyromagnetic ratio in the low-energy regime, noting that in CASMIR this value differs from the mass of the free $W$~boson, $m_{W}$.
The above expressions yield six independent relationships. Taking $\alpha$, $m_e$, and $m_\mu$ as input, these may be solved for $m_\tW$, $m_Z$, $m_\bmh$, $m_\tau$, $G_N$, and $N_0$ (which is an internal parameter of the CASMIR model). %
Additional symbols appearing in the above expressions are defined as follows:
\begin{align}
\Delta_{e}(m_{e_i},\mc{E}):=\,&\nn
\frac{8\alpha}{3N_0(3\alpha+2\pi)}\left\{1+\frac{(10\pi+180\alpha)m_{e_i}^2}{3\pi\left[m_c^*(\mc{E})\right]^2}+\frac{(5-4f_Z)\alpha m_{e_i}^2}{4\pi m_\tW^2}\right\}%
+\frac{90\alpha m_{e_i}^2}{\pi\left[m_c^*(\mc{E})\right]^2}+\frac{(5-4f_Z)\alpha m_{e_i}^2}{2\pi m_\tW^2}
\\
&+\frac{5m_{e_i}^2}{[m_c^*(\mc{E})]^2}\left\{1+\frac{90\alpha m_{e_i}^2}{\pi\left[m_c^*(\mc{E})\right]^2}+\frac{(25-12f_Z)\alpha m_{e_i}^2}{6\pi m_\tW^2}\right\}+\frac{40m_{e_i}^2}{3m_{\bmh}^2\left[k^{(e)}_{1}(\mc{E}_e)\,{N_0}\right]^4}
+\mc{O}_e(m_{e_i},\mc{E})\nn
\\
\nn\theta_e(\mc{E}) :=\,& -\frac{3\pi}{4}\bm{\Biggl(}1-\frac{4\sqrt{%
\delta_e\{r[\Delta_\taue(m_\tau,\mc{E})-\Delta_\taue(m_\mu,\mc{E})]\}}}{3\pi}\bm{\Biggr)}
\bm{\left(}1+\frac{4}{3\pi}\sqrt{\delta_e\left\{r\left[\frac{1+\Delta_\taue(m_e,\mc{E})}{1+\Delta_\taue(0,\mc{E})}-1\right]\right\}}\bm{\right)}
\nn
\\
&\!\!\!\!\!\!\!\!\!\!\!\!\!\!\!\!\!\!\!\!\!\!\!\!\!\!\!\!\!\!\!\!
\begin{aligned}
\begin{aligned}
f_Z&:=\frac{1}{3}\left(4-24\frac{m_\tW^2}{m_Z^2}+16\frac{m_\tW^4}{m_Z^4}\right)\nn
\\
\left[m_c^*(\mc{E})\right]^2&:=m_c^2\left(1-\frac{27}{10}\frac{\mc{E}^2}{m_c^2c^4}\right)\nn
\end{aligned} &\quad\quad\!\!
\begin{aligned}
\delta_e(n) &:= \sqrt{1+\frac{\pi^2n}{8}\left(1+\frac{\pi^2n}{32}\right)+\ILO{n^3}}-1\nn
\qquad\quad\!\! r(n) := n\cdot \sqrt{1-\frac{n}{9}\;}\\
k^{(\ell)}_n(\mc{E}) &:= 1+\sqrt{2}\cos{\left[\theta_\ell(\mc{E})-\frac{2\pi(n-1)}{3}\right]}\qquad\qquad\!\!~\mc{E}_\ell := m_\ell c^2
\end{aligned}
\end{aligned}\\
&\!\!\!\!\!\!\!\!\!\!\!\!\!\!\!\!
m_c^2:=m_\tW^2\left\{\frac
{1+\frac{131}{18\left[k^{(e)}_{1}(\mc{E}_e)\,{N_0}\right]^4}}
{1+\frac{51}{18\left[k^{(e)}_{1}(\mc{E}_e)\,{N_0}\right]^4}}\right\}
\left(1+\mc{O}_b\right)\qquad
\mc{O}_b:=\OOOO{\frac{\alpha}{\pi}\left[k^{(e)}_1(\mc{E}_e)\,{N_0}\right]^{-4}}+\OO{\frac{\alpha^2}{\pi^2}}\nn
\end{align}
\begin{align}
\begin{split}
\mc{O}_e(m_{e_i},\mc{E})
:=\,&\OO{\frac{\alpha}{\pi{N_0}^2}}+\OO{\frac{\alpha^2}{\pi^2{N_0}}}+\OOO{\frac{\alpha m_{e_i}^4}{\pi N_0[m_c^*(\mc{E})]^4}}+\OOOO{\frac{\alpha^2 m_{e_i}^2}{\pi^2\left[m_c^*(\mc{E})\right]^2}}
+\OOOO{\frac{\alpha m_{e_i}^2}{\pi m_\bmh^2\left[k^{(e)}_1(\mc{E}_e)\,{N_0}\right]^4}}
\\&
+\OOOO{\frac{m_{e_i}^6}{\left[m_c^*(\mc{E})\right]^6}}
+\OOOO{\frac{m_{e_i}^2}{m_\bmh^2\left[k^{(e)}_1(\mc{E}_e)\right]^4{N_0}^5}}
+\OOOO{\frac{m_{e_i}^4}{m_\bmh^4\left[k^{(e)}_1(\mc{E}_e)\,{N_0}\right]^4}}\nn.
\end{split}\nn%
\end{align}
The mass of the free $W$~boson is then denoted $m_{W}$, and this is accompanied by eight coloured counterparts $W^\tc$ having masses denoted $m_{{W^\tc}}$. When modelled in CASMIR, ATLAS measures the value of $m_{W}$ while CDF~II measures the root mean square mass of a mixture of $W^\tc$ and $W$ bosons with ratio 8:1 (up to small energy-dependent corrections not evaluated here).
There also exist coloured counterparts to the $Z$ boson, denoted $Z^{\tc}$, which have masses $m_{{Z^\tc}}$ and are not directly1 detectable by current measurements. 
The values of these parameters are given by
\begin{align}
\begin{aligned}
m_{W^{\tc}}^2&:=m_\tW^2\left\{\frac
{1+\frac{2491}{288\left[k^{(e)}_{1}(\mc{E}_e)\,{N_0}\right]^4}}
{1+\frac{51}{18\left[k^{(e)}_{1}(\mc{E}_e)\,{N_0}\right]^4}}\right\}
\left(1+\mc{O}_b\right)\\
m_W^2&:=m_\tW^2\left\{\frac
{1+\frac{19}{18\left[k^{(e)}_{1}(\mc{E}_e)\,{N_0}\right]^4}}
{1+\frac{51}{18\left[k^{(e)}_{1}(\mc{E}_e)\,{N_0}\right]^4}}\right\}
\left(1+\mc{O}_b\right)
\end{aligned}
\qquad\quad
\begin{aligned}
m_{Z^{\tc}}^2&:=m_Z^2\left\{\frac
{1+\frac{3979}{1152\left[k^{(e)}_{1}(\mc{E}_e)\,{N_0}\right]^4}}
{1+\frac{55}{18\left[k^{(e)}_{1}(\mc{E}_e)\,{N_0}\right]^4}}\right\}
\left(1+\mc{O}_b\right)\\
[m_W^2]&{}_\mrm{CDF~II}:=\frac{m_W^2+8m_{W^{\tc}}^2}{9}.
\end{aligned}\nn
\end{align}
Software solving all of the above equations numerically may be found at \href{https://www.academia.edu/65931512}{https://www.academia.edu/65931512}

~

Regarding confidence intervals in \tref{tab:overview}, it would be easy to assume that unevaulated higher-order terms such as those in $\mc{O}_e$ and $\mc{O}_b$ attract coefficients of $\ILO{1}$ or less. However, in CASMIR situations frequently arise where $\ILO{10}$ degenerate channels may reinforce one another (e.g.~equivalent loop correction diagrams due to eight species of coloured $W$ bosons and one colourless $W$ boson). The confidence intervals of \tref{tab:overview} are therefore evaluated for coefficients in the range $\pm10$. While this is a reasonable precaution in the absence of any exploration of these higher-order terms, it may potentially overestimate the uncertainty associated with the CASMIR values of the calculated parameters, artificially reducing tension. As a check on the values calculated, an upper bound on tension between experiment and the CASMIR expressions to present order may be obtained by taking the CASMIR uncertainty to zero. 
On the rare occasion \notinsummary{within this monograph }that a discernable impact occurs, this is %
no larger than $0.3\,\sigma$, %
and is remarked on in the associated text and/or table caption. For \tref{tab:overview} the tensions for $m_W$~(ATLAS), $m_W$~(CDF~II), $m_Z$~(LEP), $m_Z$~(CDF~II), $m_\bmh$, and $m_\tau$ remain unchanged to the precision shown, while that associated with $G_N$ is now bounded from above by $0.3\,\sigma$ and that associated with $a_\mu$ is bounded from above by $0.5\,\sigma$. %
This change has minimal consequence for interpretation of the results presented.

\onlyinsummary{For derivation, see Refs.~\cite{pfeifer2022CASM1,pfeifer2022CASM2,pfeifer2022CASM3,pfeifer2022CASM4,pfeifer2022CASM5}.

~

~

R. N. C. Pfeifer, Dunedin, Otago, New Zealand. 17th October 2022}

\restoregeometry
}

\tableofcontents

\setcounter{chapter}{3}

\chapter{Calculation of fermion masses in the \protect{$\mathbb{C}^{\wedge 18}$ model}\label{ch:fermion}}

\begin{abstract}
The $\Cw{18}$ model \standalone{is an analogue model capable of emulating the complete particle spectrum of the Standard Model, including interactions,}\notstandalone{reproduces the particle spectrum and interactions of the Standard Model} using only free scalar fields on a manifold with anticommuting co-ordinates. Solitonic excitations in a pseudovacuum state behave as coloured preons, forming triplets which behave as emergent fermions and pairs which behave as emergent bosons. This chapter demonstrates how the emergent fermions acquire mass through interaction with the pseudovacuum, with particle generations arising from diagonalisation of colour interactions within the triplet. 
\end{abstract}

\section{Introduction}

Analogue models are one of the most powerful concepts in modern physics, and at their most abstract, they may be summarised by the observation that two systems described by the same mathematical systems will exhibit equivalent behaviours.\notchap{ This equivalence may be exact, as in the application of Onsager's exact solution to the 1-D quantum Ising chain \cite{onsager1944,suzuki1976}, or appoximate and valid only in appropriate regimes, with examples including the effective field theories of quasiparticles \cite{srivastava1990,drazin1989,cooper1956,bardeen1957}.} These analogies are bidirectional, such that the microscopic model emulates the behaviour of quasiparticles, and the quasiparticle model provides an effective description of the microscopic system in the appropriate regime.
\notchap{Copious examples of analogue systems abound in nature, in models, and in the laboratory \cite{maynard2001,dragoman2004,lewenstein2007,visser2002,liberati2009,barcelo2011,unruh1981,garay2000,garay2001,lahav2010,gordon1923,leonhardt1999,leonhardt2000,jacobson1998,reznik2000,schutzhold2005,schutzhold2002}.}

One may argue that the value of an analogue model is determined by its utility in performing calculations or in granting qualitative insight into the behaviour of a physical system. While the mapping of the $\Cw{18}$ model to the low-energy regime of a quantum field theory resembling the Standard Model in \cref{ch:SM} provides qualitative insight into the behaviour of the $\Cw{18}$ model, of far greater interest is whether the $\Cw{18}$ model can enable the calculation of observable quantities in the Standard Model, especially if it exhibits relationships which the quantum field theory does not. Such calculations are indeed possible, but to perform them it is first necessary to elucidate the mass mechanisms of the $\Cw{18}$ model. Thus the fermionic mass mechanism is presented to tree level in the present chapter, and the bosonic mass mechanism is presented in \cref{ch:boson}. Of particular note in the present chapter is the mechanism giving rise to particle generations, which is found to have resemblance to earlier work by \citeauthor{koide2000} \cite{koide2000}. Koide's work is remarkable for proposing an approximate relationship between the masses of the different generations of leptons \cite{koide1983}, and this similarity in structure %
yields particle generations from the mass matrix on $\Cw{18}$.

\section{Conventions}

This chapter follows the same conventions as \crefr{ch:simplest}{ch:SM}. %
\notchap{Units are chosen such that $c=1,~h=1$.
When equations and lemmas from \crefr{ch:simplest}{ch:SM} are referenced, they take the forms (\textbf{1}.1), (\textbf{2}.1), %
etc.}%
Symmetry factors arising from Feynman diagrams are typically denoted $n_\mrm{sym}$ whereas those arising from the Fundamental Scalar Fields (FSFs) are typically denoted~$n_\vp$.

\notchap{Regarding terminology around Feynman diagrams and symmetry factors:
\begin{itemize}
\item Where there exist multiple ways to connect up sources, vertices, and sinks to obtain equivalent diagrams up to interchange of non-distinguishable co-ordinates, the same term is obtained from the generator $\Z$ in multiple different ways and thus the diagram acquires a multiplicative factor. This is referred to in the present volume as a \emph{symmetry factor}.
\item Where integration over the parameters of a diagram (for example, over source/sink co-ordinates) 
yields the same diagram multiple times up to interchange of labels on these parameters, 
this represents a double- (or multiple-)counting of physical processes. It is then necessary to eliminate this multiple-counting by dividing by the appropriate symmetry factor. This is referred to in the present volume as \emph{diagrammatic redundancy} or \emph{double- (multiple-)counting.}
\end{itemize}
}

\section{Fermion masses\label{sec:fermionmasses}}

\subsection{General considerations\label{sec:generalconsider}}

\begin{figure}
\includegraphics[width=\linewidth]{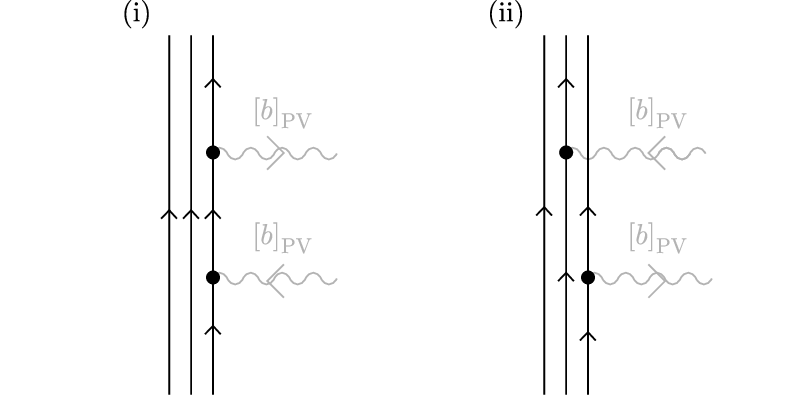}
\caption{Interactions between preons in an arbitrary foreground fermion and %
a pseudovacuum (PV) vector boson field $b$. Each of the bosons may interact with any of the three preons, and in any order, with two example configurations shown here. Arrows are present on the boson lines for interactions with complex vector bosons such as \protect{$W^\bdag_\mu$}.\label{fig:leptonmassterm}}
\end{figure}

For any species of fermion, mass terms may arise from interactions between its constituent preons and the vector or complex scalar boson fields of the pseudovacuum, which correspond to specific gradients of the fundamental scalar fields. In order for their expectation values~(\Pref{I}{eq:Exy},~\Pref{I}{eq:<hh>}) to be nonvanishing, the boson fields in such an interaction must appear as conjugate pairs. The relevant interactions between a fermion and a vector boson are shown in \fref{fig:leptonmassterm}; couplings to the scalar boson are analogous but much weaker (\sref{sec:scalbosint} and \fref{fig:scalarspread}), so are ignored.

The preon lines in these diagrams correspond, in theory, to the whole preon field, and in principle any particle (foreground or background/pseudovacuum) may acquire mass from this mechanism. However, the normalisation of the generating functional adopted in \sref{sec:normWrtBgFields} factors out interactions involving background particles only, with the result that this interaction only need be evaluated when a foreground field is present. Preons obey the Pauli exclusion principle, and field excitations satisfying the long-range correlation properties of a foreground field excitation cannot also be pseudovacuum excitations, and thus in effect only foreground preons participate in mass interactions.

Let a foreground fermion propagate over a distance or time large compared to $\mc{L}_0$ in the isotropy frame of the pseudovacuum, with %
the fermion source and sink in \fref{fig:leptonmassterm} separated by distance or time large compared to $\mc{L}_0$. %
If these diagrams are to be non-vanishing then the pseudovacuum fields at the two boson sources must be correlated. They must therefore constitute a conjugate or self-conjugate pair, and it is convenient to adopt the approximation of \PEref{I}{eq:window} 
such that they are
separated by a distance of $\OO{\mc{L}_0}$ or less. 
The interactions of the boson pair may
either be both with the same preon, as in \fref{fig:leptonmassterm}(i), or with different preons as in \fref{fig:leptonmassterm}(ii). The preons are bound into a triplet by the exchange of gluons over length and timescales $\mc{L}_\preon$, which is anticipated to %
be small compared with
the characteristic length of the %
pseudovacuum, $\mc{L}_0$. Given this, it follows that both \fref{fig:leptonmassterm}(i) and~(ii) contribute to the lepton mass, with %
preon-preon %
interactions within the fermion (not shown) distributing any transferred momentum evenly among the three component preons.

Considering that a bosonic background field has units of $L^{-1}$, a non-vanishing conjugate pair of interactions with the pseudovacuum may in general be expected to yield terms contributing to the square of the particle mass. 
The tree-level couplings with the pseudovacuum may be viewed as analogous to the mass vertices, and any higher-order corrections to these diagrams will either yield corrections intrinsic to the mass vertex, or contributions to the diagrams of the proper self-energy (PSE). In the present chapter diagrams are only evaluated to tree level and thus no examples of these corrections are shown, though they are discussed at length in \cref{ch:detail}. %
For the moment, it suffices to note that if a 1-particle-irreducible (1-PI) foreground field correction could equally well be applied either to a pseudovacuum interaction mass vertex or to a conventional mass vertex arising from a Lagrangian term such as $m_e\bar e_Re_L$ in the Standard Model, it belongs in the PSE and not in the definition of the mass vertex itself. Terms not meeting this criterion are instead accounted part of the expression for the mass vertex, and in consequence, the structures of the PSE terms are identical in the $\Cw{18}$ model and the Standard Model.

Now note that, in general, the two interaction vertices illustrated in either diagram of \fref{fig:leptonmassterm} may appear anywhere in the series expansion of the fermion propagator, and since the fermion line between these two interactions is foreground, there will typically be other 1-PI vertices intercalated in between. %
To allow for this, 
let the propagator linking the two vertices be the dressed propagator, incorporating additional mass vertices equivalent to those presently being computed. To represent this, treat the propagator as that of a massive fermion, and solve by requiring consistency. %

\begin{figure}
\includegraphics[width=\linewidth]{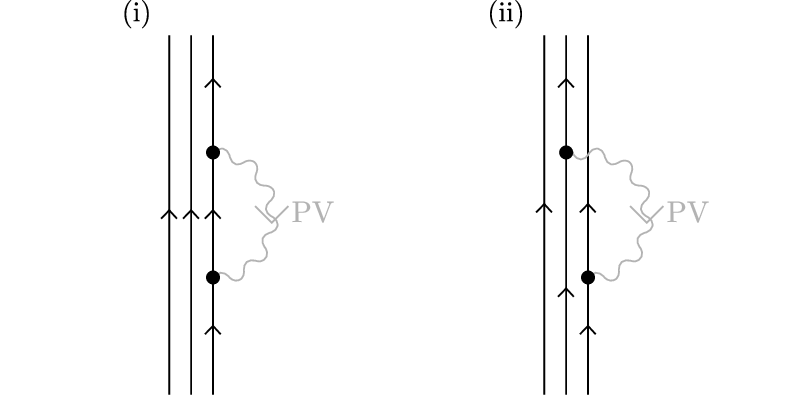}
\caption{Interactions with the pseudovacuum (PV) may be thought of as the mean-field term in a perturbation expansion of a loop correction to the particle propagator.  These diagrams reproduce \pfref{fig:leptonmassterm} with the pseudovacuum interactions redrawn as loops. The boson propagator in grey is replaced by the mean field expansion of the pseudovacuum. %
\label{fig:QLloop}}
\end{figure}

Returning to \fref{fig:leptonmassterm}, note that the two vertices %
(whose locations will be denoted $x$ and $y$) are separated on all axes by less than $\OO{\mc{L}_0}$ in the isotropy frame of the pseudovacuum, and momentum of the foreground fermion and the pseudovacuum need not be separately conserved over this interval. Indeed, although the pseudovacuum field is evaluated as a mean field value at points $x$ and $y$, the fermion field in the loop must propagate across the intervening distance, and can surrender momentum to the pseudovacuum at $x$ and recover it at $y$, or vice versa.
This ability for a propagating particle to surrender momentum to and recover momentum from the pseudovacuum is responsible for the 4-momentum fluctuations described in \Psref{I}{sec:4momflucs}, and is directly equivalent to the emission and absorption of a boson in a loop correction (\fref{fig:QLloop}). 
Provided the magnitude of the 4-momentum transferred to the pseudovacuum loop is sufficiently small compared with $\mc{E}_\Omega$, evaluation of these diagrams is essentially unchanged and continues to be dominated by the pseudovacuum mean field term. Since there are no background fields in the Standard Model, these interactions are associated to the mass vertices, and not the PSE, even when written in the form of a background field loop. Meanwhile, by construction, the PSE terms of the two models remain largely equivalent.

Discussion of PSE terms leads naturally to considerations of renormalisation. The $\Cw{18}$ model emulates renormalisation through the existence of a real UV cutoff $\Lambda:=\frac{1}{2}\mc{E}_\Omega$, with this emulation being good in a regime where the error introduced on taking $\Lambda\rightarrow\infty$ is negligible.\footnote{The factor of \prm{\frac{1}{2}} is discussed in \psref{sec:meaningofEOmega} and \pcref{ch:CDF2}%
.} However, all physical predictions of QFTs are independent of renormalisation methods used, and thus the $\Cw{18}$ model equivalently emulates a QFT subject to either UV cutoff or $\MSbar$ renormalisation. To achieve a practical realisation of $\MSbar$ renormalisation, the portions of all PSE diagrams which are finite in the limit $\Lambda\rightarrow\infty$ may be absorbed into the definition of the mass vertex such that the bare mass vertex factor for a particle corresponds to the observable mass of that particle. %
Further to this, however, note that there are also loop corrections to the mass interactions which have no counterpart in the Standard Model. These diagrams necessarily contribute to the bare mass vertex, and a number of them are evaluated in \cref{ch:detail}. Given the ability to change renormalisation scheme, it is convenient to evaluate these higher-order corrections to the bare mass vertices in the UV cutoff scheme for which \fref{fig:leptonmassterm} maps to a pair of simple vertices. Having evaluated these corrections to whatever order is desired, and reduced them to numerical multipliers on the vertices of \fref{fig:leptonmassterm}, the PSE diagrams may then be reintroduced, being corrections to these vertices which are common to $\Cw{18}$ and the Standard Model. A further choice of emulated renormalisation scheme may then be made, determining whether the numerical multipliers from the PSE terms are absorbed into a further redefinition of the bare mass vertices. Next, recognise that with both the PSE terms and the obligate bare mass vertex terms reducing to numerical multipliers on \fref{fig:leptonmassterm}, the order of this process may be inverted. Thus the net effect of the PSE terms may be evaluated first, and reduced to a numerical multiplier of the bare vertices, with the higher-order non-PSE terms then correcting bare vertices corresponding to the PSE rescaling of \fref{fig:leptonmassterm}. This approach is frequently the most convenient, and permits the correction of \fref{fig:leptonmassterm} by higher-order non-PSE terms to be identified with the physical mass of the participating fermion. A similar treatment applies to bosons, which are discussed in \crefr{ch:boson}{ch:detail}.

It is also worth noting that when a calculation yields a diagram which corresponds to a PSE loop on a massive propagator which does not enclose an explicit mass vertex, regardless of the renormalisation scheme being emulated that diagram may be discarded as a redundant representation of terms already incorporated into the mass vertices and loop corrections which are implicitly associated with the massive propagator.

Finally, to determine which pseudovacuum bosons make a contribution to fermion mass, recognise that the preons making up a fermion in $\bm{\Psi}^{ag\alpha}$~\Peref{III}{eq:collectivecompositefermions} carry charges with respect to both $\SU{3}_A$ and $\SU{3}_C$, and thus interact with vector bosons from both the electroweak and the gluon sector (counting $N_\mu$ as a ninth gluon).
Contributions to particle mass %
may then be separated according to sector of origin, and by the requirement that the species in \PEref{I}{eq:Exy} be conjugate, any pairwise term such as that shown in \fref{fig:leptonmassterm} receives contributions only from one sector. Further, since the propagators in \fref{fig:leptonmassterm} are taken to be dressed propagators in which only one pair of pseudovacuum interactions has been made explicit, there is no need to explicitly evaluate terms of higher order in the pseudovacuum fields. %

\subsection{Leptons\label{sec:leptons}}

\subsubsection{\prm{A}-sector interactions\label{sec:Asector}}

In evaluating \fref{fig:leptonmassterm}, initially treat particle generation as an \emph{ad hoc} concept, and specialise to a single generation of leptons.
For now also ignore the contribution of coloured and scalar pseudovacuum bosons (these may be written as small corrections to the photon term, and will be reintroduced in \cref{ch:detail}), so that for any given generation $g$
the amplitudes associated with \fref{fig:leptonmassterm} may simply be written down in terms of $\Psi$, the vector of composite leptons $\Psi^{ag\alpha}$. %
Since the interacting bosons are of type $a^\ta_\mu$, the colour of the interacting preon remains unchanged. %
Each boson may interact with any of the three components, %
giving nine %
distinct diagrams, but the corresponding factor of nine is cancelled by factors from \PEref{III}{eq:compositeleptonspre} and this cancellation is incorporated into %
the resulting boson/fermion coupling in \PEref{III}{eq:LPsibm} and \Psref{III}{sec:EWint_numerical}. %
Taking into account these factors,
the total amplitude is therefore
\begin{equation}
\begin{split}
{n_\vp n_\mrm{sym}f^2}%
\!\!\iint\!\!\rmd^4x\,\rmd^4y\,&\fgfield{\bar\Psi(x)\lambda^A_\ta\bsm\Psi(x)}{a^\ta_\mu(x)}\\
\times&\fgfield{\bar\Psi(y)\lambda^A_\tb\bsn\Psi(y)}{a^\tb_\nu(y)}
\end{split}\label{eq:lepmass1}
\end{equation}
where $a^\ta_\mu(x)$ and $a^\tb_\nu(y)$ represent the total boson fields, i.e.~not just background or foreground, at $x$ and $y$ respectively. Factor~$n_\mrm{sym}$ is a diagrammatic symmetry factor %
comprising
\begin{itemize}
\item the number of choices for which vertex the fermion source should be connected to (2 if the boson species is self-dual, 1 if not), 
\item the number of different ways to connect the photon sources to the interaction vertices (2),
\item and a diagrammatic redundancy factor of $\frac{1}{2}$ as the photon exchange factor is redundant, leading to double counting, %
\end{itemize}
and factor~$n_\vp$ contains symmetry factors of $\ILO{N_0}$ arising from FSF exchange similar to that described in \sref{sec:interactions}.

Over length scales of order $\mc{L}_0$ or greater, %
expression~\eref{eq:lepmass1} is dominated by boson contributions from the pseudovacuum, and thus it will ultimately be helpful to take a form of mean field approximation and 
replace the fields $a^\ta_\mu(x)$ with $\bgfield{a^\ta_\mu(x)}$. 
However, 
with $x$ and $y$ being separated on all axes by less than $\OO{\mc{L}_0}$ in the isotropy frame of the pseudovacuum, 
first note that the background fields may be rewritten as a loop, as per \sref{sec:generalconsider} and \fref{fig:QLloop}.
Provided the magnitude of the 4-momentum transferred to the pseudovacuum loop is sufficiently small compared with $\mc{E}_0$, evaluation of these diagrams continues to be unambiguously dominated by the pseudovacuum mean field term.

Furthermore, note that both limbs of the loop are potentially massive. The fermion remains a foreground particle and thus may undertake an arbitrary excursion \emph{en route} between $x$ and $y$---it need not remain within the region of characteristic dimension $\OO{\mc{L}_0}$---and may engage in additional mass interactions while doing so. If it has surrendered momentum to the boson field, it does so with a 4-momentum deficit. %
The surrendered momentum is carried in the indicated boson channel, is necessarily also foreground in character, and the ends of the loop are separated by up to \protect{$\mc{L}_0$} in the isotropy frame of the pseudovacuum. This allows the boson excitation associated with the surrendered 4-momentum to also engage in %
additional mass interactions not explicitly shown. The excitation of the boson channel corresponding to the surrendered momentum therefore acquires %
mass interactions appropriate to the boson species.

\begin{figure}
\includegraphics[width=\linewidth]{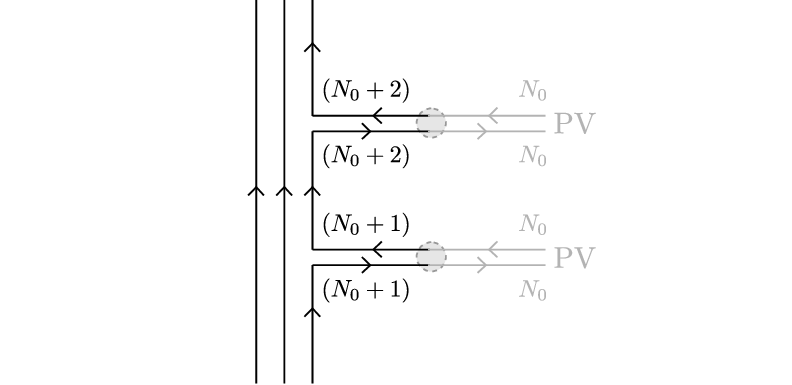}
\caption{Symmetry factors arising from the exchange of fundamental scalar fields (FSFs). One factor attaches to each field operator appearing on a vertex.
The net FSF symmetry factor for this diagram is \prm{(N_0+2)^2(N_0+1)^2{N_0}^4}. [Each vertex also attracts a factor of~\prm{\frac{1}{3}} from the normalisation of \protect{\PEref{III}{eq:generalfermion}} but this is offset as each photon may couple to any of the three preons.]\label{fig:evalNsyms}}
\end{figure}

For a lepton $\ell$ the factor associated with such a loop, accounting for the transient fermion momentum surfeit or deficit, %
may be written
\begin{equation}
n_\vp n_\mrm{sym}\frac{ \Xi f^2}{4\pi}\,\bmf{\frac{m^2_\ell}{m^2_{a^\ta}}}\label{eq:mloopfactor}
\end{equation}
where 
\begin{equation}
\bmf{\frac{m_\ell^2}{m_{b}^2}}\longrightarrow\left\{\begin{aligned}1~~~~\quad&\textrm{if }m_{b}^2=0\\
\OO{\frac{m_\ell^2}{m_{b}^2}}\quad&\textrm{if }m_{b}^2\gg m_\ell^2.\end{aligned}\right.
\end{equation}
$\Xi$ is a structure factor equal to~2 for the photon, $10/3$ for the $W^\bdag$ boson, etc. \cite{peskin1995}, and $n_\vp$ and $n_\mrm{sym}$ are symmetry factors. %
Anticipating a mean field substitution for the boson fields %
yields
\begin{equation}
\begin{split}
n_\vp\,{n_\mrm{sym}f^2}%
\!\!\iint\!\!\rmd^4x\,\rmd^4y\,&\fgfield{\bar\Psi(x)\lambda^A_\ta\bsm\Psi(x)}\bgfield{a^\ta_\mu(x)}\\
\times&\fgfield{\bar\Psi(y)\lambda^A_\tb\bsn\Psi(y)}\bgfield{a^\tb_\nu(y)},
\end{split}\label{eq:lepmass1QL}
\end{equation}
which is exactly equal to \Eref{eq:lepmass1} in the absence of foreground bosons. To incorporate the loop factors~\eref{eq:mloopfactor}, recognise that the vertex couplings and symmetry factors are already present in \Eref{eq:lepmass1QL}, and that for consistency the factor of $(4\pi)^{-1}$ (if present) and the structure factor must be absorbed into the mean field value associated with the pseudovacuum.\footnote{In practice the factor of $(4\pi)^{-1}$ vanishes when the loop field is background, as is discussed in \paref{apdx:massloops}. %
However, this detail is unimportant to the present calculation.} 
Introducing the remaining factor $\bmfcdot$ %
yields
\begin{equation}
\begin{split}
{n_\vp n_\mrm{sym}f^2}\,\bmf{\frac{m^2_\ell}{m^2_{a^\ta}}}\!\iint\!\!\rmd^4x\,\rmd^4y\,&\fgfield{\bar\Psi(x)\lambda^A_\ta\bsm\Psi(x)}\bgfield{a^\ta_\mu(x)}\\
\times&\fgfield{\bar\Psi(y)\lambda^A_\tb\bsn\Psi(y)}\bgfield{a^\tb_\nu(y)}.
\end{split}\label{eq:lepmass2}
\end{equation}
In the absence of external foreground bosons, by gauge choices~\Peref{III}{eq:U1gauge} and~\Perefr{III}{eq:bga12gauge}{eq:bga8gauge} this expression~\eref{eq:lepmass2} may be non-vanishing only for $\ta=\tb=3$ on the $A$~sector, corresponding to interactions with the pseudovacuum %
photon field $\bgfield{A_\mu}$. (Note that contributions from the $N_\mu$~boson are prohibited by gauge even though $\la\bgfield{N^\mu N_\mu}\ra$ may be nonvanishing.) The electrons and quarks may acquire mass as a result of this interaction (for the quarks this will be a bare mass), but the neutrino does not as it does not couple to this species of boson. %
To proceed with a definite example, %
specialise to the electron fields to obtain
\begin{equation}
\begin{split}
n_\vp{f^2}\!\!\iint\!\!\rmd^4x\,\rmd^4y\!\!\!\!\!\!\sum_{e\in\{e_L,\bar e_R\}}\prod_{z\in\{x,y\}}&%
\left\{\fgfield{\bar e(z)\bsm e(z)}\bgfield{A_\mu(z)}\right\}.
\end{split}\label{eq:unreducedeLmassint}
\end{equation}
Using spinor and sigma matrix identities this may be rewritten
\begin{equation}
\begin{split}
\!\!\!-\frac{n_\vp f^2}{2}\!\!\iint\!\!\rmd^4x\,\rmd^4y%
\!\!\!\!\!\!\sum_{e\in\{e_L,\bar e_R\}}\!\!\!\!\!
&\fgfield{\bar e(x)\bar e(y)\, e(x)e(y)}\bgfield{A^\mu(x) A_\mu(y)}.\label{eq:reducedeLmassint}
\end{split}
\end{equation}
Provided the two vertices are sufficiently close, i.e.~$\delta_{\mc{L}_0}^{(4)}(x^\mu-y^\mu)$ is non-vanishing in the rest frame of the pseudovacuum, this admits the per-excitation mean field %
substitution 
\begin{equation}
\bgfield{A^\mu(x)A_\mu(y)}\rightarrow -%
{\omega_0}^2\label{eq:HHsub}
\end{equation}
where $\omega_0$ is the mean energy per FSF gradient, as in \PEref{I}{eq:N0terms}. In contrast with \PEref{I}{eq:Exy}, but in keeping with the approach of \sref{sec:interactions}, symmetry factors from FSF exchange are evaluated separately and represented by $n_\vp$. %
These factors are shown in \fref{fig:evalNsyms}.

To appreciate the origin of these factors, first recall that the fermion is foreground, and thus an extra FSF is recruited for both the inbound and outbound legs of each vertex. This reflects that after the first interaction the intermediate fermion segment may propagate and interact outside of the locally correlated area before returning, and regardless of whether it does so or not, the foreground nature of the correlations it carries---which are in addition to the correlations of the background field---necessitate the introduction of additional FSFs. (At its simplest this may be a single additional FSF capable of hosting both a holomorphic and an antiholomorphic derivative with foreground correlations, which may bridge between the lower and the upper vertices.) Note that since a nonvanishing correlator requires that vertex co-ordinates $x$ and $y$ lie within the same correlation region, all additional FSFs are accessible to both vertices. Evaluating the FSF symmetry factors for the preon lines within the fermion (and recalling that factors associated with the noninteracting lines are subsumed into the fermion definition as per \Psrefs{III}{sec:complepE}{sec:photonint}), the presence of these additional correlated FSFs yields factors of \prm{(N_0+2)} twice and \prm{(N_0+1)} twice. This may be compared with the factors on the emission vertex in \pfref{fig:fermionsymfactors}. 

Next, consider the preons in the background photons, which must come from the background fields. Considering the upper vertex first, the holomorphic and antiholomorphic preons at the upper vertex each attract factors of \prm{N_0}. Recalling that preons in photons may carry labels \prm{a=1} or \prm{a=2}, if the preons of the lower vertex do not match those of the upper vertex then they also attract a factor of \prm{N_0} apiece. However, if they do match then each attracts a factor of \prm{N_0-1} from the background fields, plus a further factor of one as the inbound preon on the upper vertex may also be understood as an outbound antipreon, and the inbound antipreon as an outbound preon, giving one additional source apiece, replacing the effective free background preons which were consumed and once again returning a total FSF symmetry factor of ${N_0}^2$ for the background fields associated with the vertex. 
The net FSF symmetry factor for this diagram is thus 
\begin{align}
\begin{split}
n_\vp&=(N_0+2)^2(N_0+1)^2{N_0}^4\\
&={N_0}^8\left[1+6{N_0}^{-1}+13{N_0}^{-2}+\OO{{N_0}^{-3}}\right]\label{eq:nvp_pre}
\end{split}\\
&=:{N_0}^8S_{6,13},\label{eq:IV:defS613}
\end{align}
where $S_{6,13}$ is a convenient notation for this power series in ${N_0}^{-1}$ introduced in \PEref{III}{eq:III:defS613}.

By the choice of gauge described in \sref{sec:GL18Cgauge}, \Eref{eq:lepmass2} is the only non-vanishing colourless boson correlator for the pseudovacuum at $x$ and $y$: 
The $a^{12\bdag}$ bosons have been eliminated, the $G^\bdag$ pseudovacuum correlator always vanishes, the $N$ boson has in this context been restricted to the colour sector, and the pseudovacuum correlators for $W^\bdag$ and $Z$ boson fields vanish in the absence of a foreground particle of the same type.

Defining 
\begin{equation}
m_e:=\sqrt{\frac{n_\vp}{2}}f\omega_0%
,\label{eq:earlyelecmass}
\end{equation}
\Eref{eq:reducedeLmassint} becomes
\begin{equation}
\begin{split}
\iint\!\!\rmd^4x\,\rmd^4y\!\!\!\!\!\!\sum_{e\in\{e_L,\bar e_R\}}\!\!\!\!\!%
&\fgfield{\bar e(x)\bar e(y)}\,m_e\, \fgfield{e(x)e(y)}\,m_e.
\end{split}\label{eq:actualelecmass}
\end{equation}
When working at energy scales $\mc{E}\ll\mc{E}_0$ or over probe scales $\mc{L}_p\gg\mc{L}_0$ %
this will be indistinguishable from
\begin{equation}
\begin{split}
\iint\!\!\rmd^4x\,\rmd^4y\,&\fgfield{\bar e_L(x)e_R(x)}\,m_e\, \fgfield{\bar e_R(y)e_L(y)}\,m_e\label{eq:effectiveelecmass}
\end{split}
\end{equation}
due to the expression being non-vanishing only when co-ordinates $x$ and $y$ satisfy the window function constraint
\begin{equation}
\delta_{\mc{L}_0}^{(4)}(x^\mu-y^\mu)\not=0\label{eq:windowconstraint}
\end{equation}
arising from \PEref{I}{eq:window}, and due to the intermediate lepton state between the two mass vertices not being accessible at probe scales large compared to $\mc{L}_0$. Note that symmetry factors arising from different ways to connect the $\Cw{18}$ mass diagram have already been accounted for in constructing \Eref{eq:lepmass1}, and there are no further factors arising %
from \Eref{eq:effectiveelecmass}, so identifying \Ereft{eq:reducedeLmassint}, \eref{eq:actualelecmass}, and~\eref{eq:effectiveelecmass} with one another corresponds directly to equating the electron mass $m_e$ in %
\Eref{eq:actualelecmass} with the leading order expression given in \Eref{eq:earlyelecmass}.

Expression~\eref{eq:effectiveelecmass} is exactly the Standard Model term for electron mass, applied twice to the left-handed electron and written in terms of Weyl spinors:
\begin{equation}
\begin{split}
\iint\!\!\rmd^4x\,\rmd^4y\,&\bar e_L(x)e_R(x)m_e\, \bar e_R(y)e_L(y)m_e.
\end{split}
\end{equation}
In both instances the electron spinors pass through an intermediate state which is not observed, in which their helicities reverse. In the present model this is only an approximation introduced on going from \Eref{eq:actualelecmass} to \Eref{eq:effectiveelecmass}, whereas in the Standard Model observation of this state is prohibited by imposing conservation of spin as a superselection criterion. Note that 
\begin{itemize}
\item[(i)] a similar argument may be followed for the right-handed electron, and
\item[(ii)] in the present model it is the window constraint~\eref{eq:windowconstraint} which requires close proximity of the interaction vertices, making the details of the mass mechanism undetectable at probe scales $\mc{L}_p\gg\mc{L}_0$ in the rest frame of the pseudovacuum. 
\end{itemize}
Taking all of this into account, including the discussion of probe scales in \Psref{I}{sec:ProbeOmegaScale}, interaction with the background photon field $\bgfield{A_\mu}$ is anticipated to give rise to electron mass terms structurally equivalent to those of the Standard Model in current experimentally accessed %
regimes.

\subsubsection{\prm{C}-sector interactions\label{sec:Csector}}

\begin{figure}
\includegraphics[width=\linewidth]{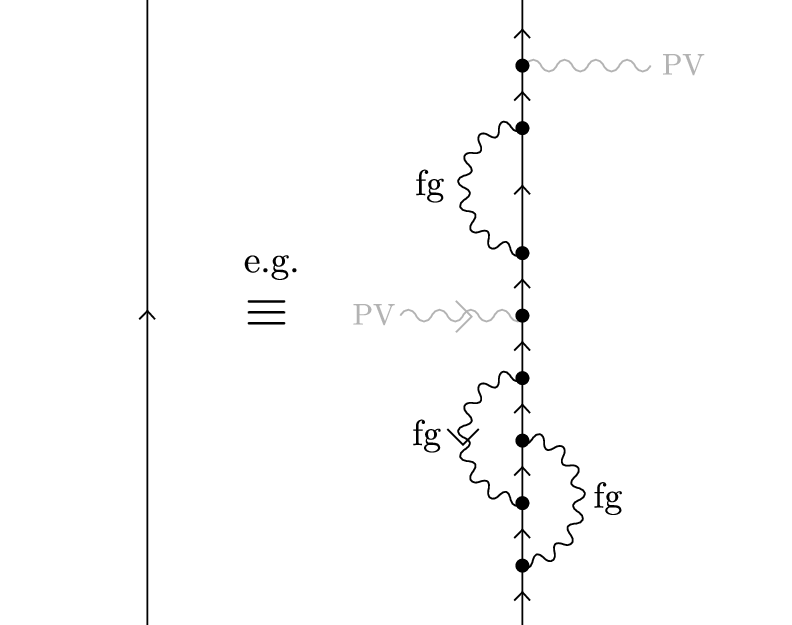}
\caption{To represent the preon propagator as a single fermion line, even over length scales of \protect{$\OO{\mc{L}_0}$}, is a major oversimplification neglecting the multiple interactions taking place between the preon, its neighbours, and the background field which may potentially change preon colour. %
\label{fig:interactingpreon}}
\end{figure}

Now consider the role of the colour charge on interacting preons.
First, note that the preon propagator lines in \fref{fig:leptonmassterm} have been oversimplified, ignoring both unpaired interactions with pseudovacuum photons, and a multitude of colour-mediated interactions between a preon and (i)~its neighbours and (ii)~the coloured bosons of the background field (\fref{fig:interactingpreon}). 

Thus far, only interactions with pairs of colourless bosons separated by at most $\mc{L}_0$ have been considered. However, by definition of the pseudovacuum, interactions with unpaired bosons of any sort (or paired and separated by more than $\mc{L}_0$) must leave the pseudovacuum unchanged over length and timescales large compared with $\mc{L}_0$, and thus unpaired bosons may impart no net mass or momentum to the propagating preon triplet. 

The only remaining interactions to consider are thus paired coloured bosons engaging in the interactions shown in \fref{fig:QLloop}, and colour transfer to and from the pseudovacuum fields (which approaches negligible over scales large compared with $\mc{L}_0$, and is guaranteed to vanish by a process described below). Since the gluons are seen in \cref{ch:boson} to have masses on order of $m_W$ at this length/energy scale, %
\Eref{eq:lepmass2} implies that their direct contribution to particle mass may reasonably be ignored at leading order as it is small compared with that arising from the photon. However, each interaction between a preon and a coloured boson may change the colour of that preon, and this effect is distinct from that direct mass contribution and must still be accounted for. There are nine gluon-type bosons (three diagonal, six off-diagonal), and %
all are of equal mass. Although over the totality of the preon propagator they all appear in conjugate pairs, these may be overlapping and intercalated. Once again this is represented by allowing the fermion propagator between the two vertices to be massive.

Now note that the decomposition of $\GL{9}{R}$ to yield $\SU{3}_A$ and $\SU{3}_C$ in \PEref{III}{eq:GL9Rdecomp} is only strictly valid in the continuum limit, and that any interaction with a $\GLNR$ boson corresponds to an interaction with one boson from each of the $A$ and $C$ sectors (which may in either case be associated with the trivial representation).
Thus for each application of the (totality of the) $\SU{3}_A$ sector's contribution to fermion mass, there is %
also one application of the (totality of the) $\SU{3}_C$ sector's contribution. [This argument is %
reviewed in more detail in \aref{apdx:gaugeSU9} on examining the extension of the gauge choices of \sref{sec:gaugechoice} to $\SU{9}$ and its consequences for the construction of $A$-sector bosons.]
Only the photon makes a meaningful contribution to the $A$~sector's mass interaction and %
this is accompanied by application of the entire $C$~sector.

To evaluate the $\SU{3}_C$ sector's action on a preon triplet,
introduce the vector notation
\begin{equation}
\bm{\psi}^{a\alpha}=\triplet{\psi^{ar\alpha}}{\psi^{ag\alpha}}{\psi^{ab\alpha}}\label{eq:psitriplet}
\end{equation}
for an interacting preon, and recognise that on the colour sector the $N$ boson acts as an additional, ninth gluon associated with the representation matrix %
\begin{equation}
\lambda^C_9=\frac{1}{\sqrt{3}}\bgrid 1&0&0\\0&1&0\\0&0&1 \egrid.
\end{equation}
Triplet~\eref{eq:psitriplet} is therefore acted on by both the trivial and non-trivial (dimension-8) matrix representations of $\su{3}_C$, collectively $\{\lambda^C_i|i\in 1,\ldots,9\}$.
This symmetry is %
unbroken in the low-energy limit ($\mc{E}_p\ll\mc{E}_0$). %

Now recognise that for a colour-neutral triplet of three preons at $\bm{\{}x_i|i\in\{1,2,3\}\bm{\}}$, a basis on the colour sector for a preon triplet is given by the six configurations of \tref{tab:preoncolourconfigs}. 
\begin{table}
\begin{center}
\begin{tabular}{cc}
(i)&~\\
~&~\\
~&~\\
~&~
\end{tabular}
\begin{tabular}{c|c|c}
$c_1$&$c_2$&$c_3$\\\hline
$r$&$g$&$b$\\
$g$&$b$&$r$\\
$b$&$r$&$g$
\end{tabular}\qquad\qquad
\begin{tabular}{cc}
(ii)&~\\
~&~\\
~&~\\
~&~
\end{tabular}
\begin{tabular}{c|c|c}
$c_1$&$c_2$&$c_3$\\\hline
$r$&$b$&$g$\\
$b$&$g$&$r$\\
$g$&$r$&$b$
\end{tabular}
\end{center}
\caption{Colour configurations forming a basis on the colour sector for a colour-neutral preon triplet. Configurations~(ii) are obtained by any pairwise column exchange of configurations~(i).\label{tab:preoncolourconfigs}}
\end{table}%
However, in the low-energy limit the position of individual preons is not precisely determined, and must be integrated over a region of characteristic length at least $\mc{L}_\preon$. Where two preons have identical $A$-charges, exchange of colour is equivalent to exchange of position, and is therefore seen to be a double-counting. Since all fermions not eliminated by gauge contain at least one pair of preons with identical $A$-charge, in the low-energy limit it therefore suffices to consider only \tref{tab:preoncolourconfigs}(i).
Putting preon~1 into a superposition of colour states then corresponds to a superposition of cyclic spatial rearrangements of the members of the triplet, with colours $c_1=r$, $c_1=g$, and $c_1=b$ corresponding to colour assignments \emph{with respect to preon spatial co-ordinates} of $rgb$, $gbr$, and $brg$ respectively. It is then incidentally clear that for leptons at least, colour superselection is %
not violated in the low-energy regime, with any superposition of colour states on the constituent preons mapping into a superposition of preon spatial configurations.\footnote{The same argument cannot be applied directly to the preons in quarks due to inhomogeneity of $A$-charge, but may be applied in a more general form: Any superposition of states involving different colour configurations on one or more preons must have equal total charge across the whole state, and will admit interpretation as the result of colour-redistributing procedures and/or co-ordinate transformations---which for leptons may be reinterpreted as particle position exchange, with due care and attention to sign.}

Next, consider the action of the bosons of both representations of $\su{3}_C$ on a single preon in a fermion triplet. On average this preon will interact with background bosons of all such species with equal frequency, and each boson will act the associated representation matrix on the preon's colour state~\eref{eq:psitriplet}. Bosons in the non-trivial representation which have off-diagonal elements
change the colour of the preon on which they act, breaking colour neutrality of the preon triplet.

On average any such colour change is reverted over a length scale of $\ILO{\mc{L}_0}$. However, %
over any finite timescale there are only a finite number of such interactions, and measurement of the net colour charge of the preon triplet will interrupt some number of vertex pairs. 
As the number of interactions thus interrupted is finite, there is no guarantee that the net colour shift will vanish and hence
colour measurements may be inconsistent even over macroscopic length and time scales. Since the macroscopic theory must be expressible without reference to the presence of the pseudovacuum, it is necessary to restore colour neutrality. Let a preon triplet with vanishing net foreground gluon interaction over macroscopic length scales therefore \emph{define} colour neutrality, and---as foreshadowed in \sref{sec:EWint_Wintdetail}---for each pair of interactions with conjugate bosons (\fref{fig:QLloop}) perform a local change of co-ordinates on a patch of space--time enveloping the non-interacting preons such that the triplet as a whole continues at all times to be colour-neutral. As this choice of co-ordinate frame affects only the $C$~sector, it is always compatible with the earlier choice of co-ordinate frame introduced in \sref{sec:EWint_Wintdetail} which acts purely on the $A$~sector to ensure that the bosons of $\SU{3}_A$ to act as a representation of $\SU{2}\otimes\U{1}\subset\SU{3}_A$ on the left-handed leptons. %

As an aside, note that although measurements performed on a fermion may interrupt conjugate pairs of interactions with off-diagonal background gluons, %
these colour-unpaired interactions must nevertheless still occur in even numbers on any given subsystem. This is necessary in order that the background gluon correlator~\Peref{III}{eq:<cc>} which arises from these interactions does not vanish. This is automatically satisfied in the current scenario, as the colour interactions under consideration occur pairwise as per \fref{fig:QLloop}.

Returning to the effects on preon colour, for any given off-diagonal colour mapping, for example from $g$ to $r$, %
contributions are received from two representation matrices which must be weighted with equal magnitude by non-violation of $\SU{3}_C$ symmetry. 
Associating
\begin{equation}
r\equiv 1,\quad g\equiv 2,\quad b\equiv 3,
\end{equation}
transformation of a preon from green to red is enacted by %
the representation matrices %
$\lambda^C_1$ and $\lambda^C_2$, and these may be combined according to 
\begin{equation}
\frac{1}{\sqrt{2}}(\pm\lambda^C_1\pm\lambda^C_2). 
\end{equation}
A preon being transformed from green to red therefore attracts a factor of
\begin{equation}
A = \pm\frac{1\pm\rmi}{2}
\end{equation}
with the net transformation corresponding to matrix $A\, e_{rg}$.
Likewise, one being transformed from red to green attracts the conjugate transformation $A^* \, e_{gr}$. Other off-diagonal coefficients follow from colour cycle invariance.
These coefficients may be collected into a matrix $K$, 
\begin{equation}
K=\left(\begin{array}{lll} \,1&A&A^*\\A^*&\,1&A\\A&A^*&\,1\end{array}\right),\quad A=\pm\frac{1\pm\rmi}{2},\label{eq:IV:Kmatrix}
\end{equation}
with the diagonal elements being~1 as couplings which leave colour unchanged do not induce any co-ordinate transformation.
For any given gluon acting at a lower vertex in \fref{fig:QLloop}, colour cycle invariance implies this is associated with the action of $K^{i+1,j+1}e_{i+1,j+1}$ and $K^{i+2,j+2}e_{i+2,j+2}$ on the other two preons of the triplet to maintain colour neutrality, with this action corresponding to the application of the colour-neutrality-maintaining co-ordinate transformation. The upper vertex is then acted on by the conjugate gluon and attracts the conjugate factors
\begin{align}
K^{j+1,i+1}e_{j+1,i+1}=\left(K^{i+1,j+1}e_{i+1,j+1}\right)^\dagger\\
K^{j+2,i+2}e_{j+2,i+2}=\left(K^{i+2,j+2}e_{i+2,j+2}\right)^\dagger.
\end{align}

\begin{figure}
\includegraphics[width=\linewidth]{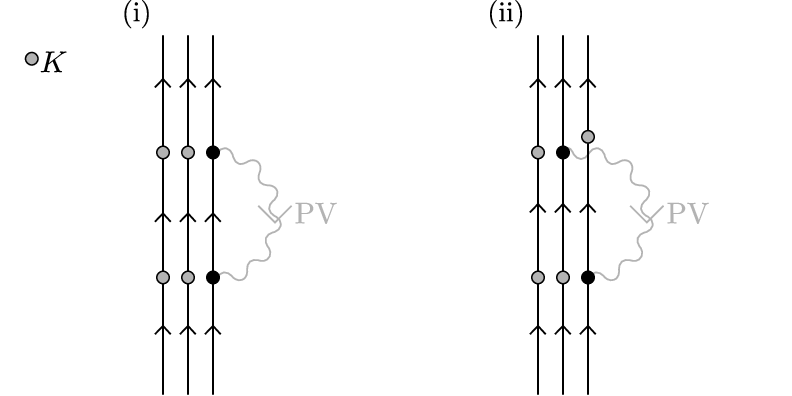}
\caption{The gluon contributions to fermion mass, represented as a single \pt{$\GLTR$}-valued boson loop. The \pt{$K$}-matrices act on all preons not coupling to the boson, and this is simultaneous with the boson interaction in the isotropy frame of the pseudovacuum. The \pt{$K$} matrix is slightly offset in diagram~(ii) for clarity.\label{fig:Kmatrices}}
\end{figure}

Collectively the nine gluon-type bosons may be summed to yield a single $\gltr_C$-valued boson 
\begin{equation}
c^{ij}_\mu e_{ij}:=c^\tc_\mu\ulambda^C_\tc, 
\end{equation}
with concommittant summing of the co-ordinate transformations yielding a copy of matrix $K$ acting on the colour space of each non-interacting preon simultaneous with the boson interaction (as determined in the isotropy frame of the pseudovacuum), as shown in \fref{fig:Kmatrices}. 
To collect these effects together, define an operator
\begin{equation}
\begin{split}
\hat\Kf_\mu =~&c^{ij}_\mu e_{ij}\otimes K\otimes K\\
&+K\otimes c^{ij}_\mu e_{ij}\otimes K\\
&+K\otimes K\otimes c^{ij}_\mu e_{ij}.
\end{split}\label{eq:defMCK}
\end{equation}
As each preon of the propagating fermion triplet undergoes a near-arbitrarily large number of such interactions %
over the course of any macroscopic propagator, it follows that the fermion triplet as a whole must be an eigenstate of pairwise applications of operator $\bgfield{\hat\Kf_\mu}$ (constructed using bosons $\bgfield{c^{ij}_\mu}$). 
For leptons, it further follows from colour-cycle invariance that each individual preon must be an eigenstate of matrix $K=K^{ij}e_{ij}$, and that that eigenstate must be the same for each member of the triplet. %
For leptons, it is therefore convenient to define an operator $\hat K^{(\ell)}$ which acts on a lepton $\Psi^{ag\alpha}$ to yield $k_g\Psi^{ag\alpha}$ (no sum over $g$) where $k_g|_{g\in\{1,2,3\}}$ is an eigenvalue of matrix $K$. Note that $g$ is a label, not an index, so %
the choice of upper or lower position is not meaningful. %

To complete the determination of matrix $K$, note that
while the sign on $\rmi$ is %
free to be chosen by convention, the overall sign on $A$ is fixed by %
requiring that a full cyclic permutation of colours ($r\rightarrow g\rightarrow b\rightarrow r$ and similar), which is in $K^3$, %
must leave the sign of an eigenstate of $K$ unchanged in accordance with the colour cycle invariance of the unbroken $\GLTR_C$ symmetry. The eigenvalues of $K$ must therefore be non-negative, %
which requires that
\begin{equation}
A=-\frac{1\pm\rmi}{2}. \label{eq:defA}
\end{equation}
It is useful to write the mixing matrix $K$ as a function of an angle $\thf$. In the notation of \rcite{koide2000}:
\begin{align}
K &= a_f E - b_f S(\thf)\label{eq:IV:KfES}\\
E&=\frac{1}{\sqrt{3}}\,\mbb{I}_3\\
S(\thf)&=\frac{1}{\sqrt{6}}\left(\begin{array}{ccc} 0 & e^{\rmi\thf} & e^{-\rmi\thf}\\
e^{-\rmi\thf} & 0 & e^{\rmi\thf}\\
e^{\rmi\thf} & e^{-\rmi\thf} & 0 \end{array}\right)\\
a_f&=b_f=\sqrt{3}\qquad\thf=\frac{\pi}{4}.
\end{align}

Now recognise that when operator $\hat{\mc{K}}_\mu$ acts on a preon, implicitly representing a gluon/preon coupling, this occurs at an energy scale $\mc{E}_0\ll\mc{E}_\preon$ and consequently this interaction is necessarily a gluon/fermion coupling. All preons in the fermion are involved in the vertex, and thus all preons attract FSF symmetry factors when they connect to a $K$ matrix as in \fref{fig:Kmatrices}. In order to maintain normalisation of the nonparticipatory preons, these factors are accompanied by compensatory factors of ${N_0}^{-2}$ per $K$~matrix such that the norm of a free preon propagating in the presence of the pseudovacuum is maintained. Each preon in the triplet is of unique colour, and thus the symmetry factor $n_\vp$ is corrected to
\begin{align}
\begin{split}
n_\vp&=(N_0+2)^6(N_0+1)^6{N_0}^4\cdot {N_0}^{-8}\\
&={N_0}^8(1+18{N_0}^{-1}+147{N_0}^{-2}+\ldots)\label{eq:newnvp}
\end{split}\\
&=:{N_0}^8S_{18,147}\label{eq:defS18147}
\end{align}
where the name $S_{18,147}$ has been assigned to this power series in ${N_0}^{-1}$.

Next, return to considering the mass interaction as a whole. The pseudovacuum photon interaction is summed over all choices of preon coupling, and since there are no foreground $W$ or $Z$ bosons in \fref{fig:leptonmassterm}, this sum corresponds to the totality of the $A$-sector acting on all preons. 
This interaction is accompanied by the action of the totality of the $C$~sector on all preons (including the preon(s) being acted on by the photon). The pseudovacuum photon interaction is capable of coupling any of the three preons to any of the three preons, while each gluon can couple one specific preon to one specific preon; as noted before, the direct contribution to particle mass arising from interactions between a fermion and the pseudovacuum gluon fields is small compared to the contribution from the photon field, and it is therefore convenient to treat each gluon contribution as a multiplicative correction to the corresponding photon contribution. %
As per \Eref{eq:defMCK}, whenever gluons act on a preon, the corresponding co-ordinate transform from~$K$ acts on the other two preons of the triplet, and thus each photon (plus implicit gluon correction) is accompanied by a pair of $K$-matrices
(\fref{fig:Kmatrices}). %
 
Writing the gluon contribution as a multiplicative correction %
of order $[1+\mrm{O}(m_\ell^2/m_c^2)]$, it thus suffices to approximate each lepton/pseudovacuum interaction vertex by
\begin{equation}
\left.\frac{f}{\sqrt{2}}\,\bm{\bar\psi}^{\dot a}\!\hat K^{(\ell)\dagger}\bsm \hat K^{(\ell)}\bm{\psi}^{a} \bgfield{a^\ta_\mu}\,(\lambda^A_\ta)_{\dot aa}\left[1+\OO{\frac{m_\ell^2}{m_c^2}}\right]\right|_{\ta=3}.
\end{equation}
Denoting the eigenvalues of matrix $K$ as $k%
_g$, $g\in\{1,2,3\}$, the resulting leading-order expressions for the electron, muon, and tau masses are thus
\begin{equation}
\begin{split}
m_{e_g}&:={%
\sqrt{\frac{n_\vp}{2}}k_g^2f\omega_0}\label{eq:leptonmasses}\\
m_{e_1}\equiv m_e,&\quad m_{e_2}\equiv m_\mu,\quad m_{e_3}\equiv m_\tau,
\end{split}
\end{equation}
justifying the earlier introduction of %
index $g$ in \PEref{III}{eq:compositeleptonspre}.

On computing the %
lepton masses as 
per \Eref{eq:leptonmasses}
above, $m_\mu$ and $m_\tau$ are non-zero but $m_e$ is found to vanish. However, a non-vanishing $\tau$ mass and non-vanishing gluon, $W$, and $Z$ boson masses (see \cref{ch:boson}) %
imply the existence of non-vanishing electroweak loop corrections to \freft{fig:QLloop}(i)-(ii). 
Taking into account the evaluation of the weak interaction strength as per \sref{sec:weakint}, the lowest-order correction is of $\OO{m_\tau^2/m_W^2}$. Noting that 
\begin{equation}
1+\frac{m_\tau^2}{m_W^2}=\left(1+\rmi \frac{m_\tau}{m_W}\right)\left(1-\rmi \frac{m_\tau}{m_W}\right)
\end{equation}
it is reasonable to speculate that this generation-dependent correction may be approximated by a correction to $\thf$ of order $\OO{m_\tau/m_W}$. 
The dependency of $K$ on preon species is not fully explored in the present chapter, but for now it suffices to note that higher-order diagrams do indeed give rise to corrections to $\thf$ which are
dependent on fermion mass, and which result in a nonvanishing mass for the electron in \cref{ch:detail}.

With angle $\thf$ depending on fermion mass, and fermion mass in turn depending on preon charge, it follows that matrix $K$ will in general depend on the species of preon to which it is applied. In the quarks, the dependency of the eigenvalues of $K$ on preon type then selectively favours or disfavours interactions with the unique preon compared with the others, perhaps yielding an effective overall colour charge in addition to the dipole effect discussed in \srefs{sec:complep1}{sec:quarksgluons}, again corresponding to the colour (or anti-colour) of the unique preon (or its inverse). %

With regard to the construction presented in this Section, and in particular with 
the reduction of electroweak corrections to a shift in $\thf$, it is worth noting several important points:
\begin{enumerate}
\item %
Mass matrices derived from~\Eref{eq:IV:KfES} are known not to be capable of reproducing the observed lepton masses precisely~\cite{koide2000}.
\item However, the generation-dependent electroweak loop corrections to \fref{fig:QLloop} also represents a deviation from form~\eref{eq:IV:KfES}.
\item The best fit to observed values of the lepton masses obtainable using form~\eref{eq:IV:KfES} is given by setting \cite{koide2000}
\begin{align}
\frac{b_f}{a_f}&=\left[3\frac{m_e+m_\mu+m_\tau}{\left(\sqrt{m_e}+\sqrt{m_\mu}+\sqrt{m_\tau}\right)^2}-1\right]^\frac{1}{2}\\
\thf&=\tan^{-1}\left(\sqrt{3}\frac{\sqrt{m_\tau}-\sqrt{m_\mu}}{\sqrt{m_\tau}+\sqrt{m_\mu}-2\sqrt{m_e}}\right)
\end{align}
in $K$, yielding
\begin{equation}
\frac{b_f}{a_f}=0.999991(10) \qquad \thf=0.8249679(83)~\mrm{radians}.%
\end{equation}

\item This best-fit value of $\thf$ corresponds to 
\begin{equation}
\frac{\pi}{4}\left[1+2.27856(68)\frac{m_\tau}{m_W}\right]~\mrm{radians}%
\end{equation}
(for $m_W=80.360(16)$ \cite{the-ATLAS-collaboration2023}), %
so the correction to $\thf$ which yields the best possible fit is therefore of similar magnitude to both the %
the gluon contributions to particle mass, and the electroweak loop
corrections required to \fref{fig:QLloop}.
\end{enumerate}
It must be emphasised that the gluon and electroweak connections do \emph{not} in fact take the form of a pure correction to $\thf$, and this is just a convenient estimate allowing confirmation that the electroweak loop correction is of the correct order of magnitdue to yield the observed lepton mass ratios.
These corrections are addressed in detail in \cref{ch:detail}.
[Note that purely electromagnetic loop corrections to \fref{fig:QLloop}(i)-(ii) also apply, but are independent of lepton mass so may be factored out into a generation-independent correction to \Eref{eq:leptonmasses} and thus do not appear in lepton mass ratios.]

Also note that neutrinos acquire no mass through the photon term of this mechanism. However, the colour mixing matrix still admits the same three eigenvectors. Implications for neutrino generations and mixing are explored in \sref{sec:neutrinos}.

Finally, two observations with regards to the local change of co-ordinates which permits lepton triplets to remain colour-neutral. First, note that this change of co-ordinates is not a choice of gauge. It is therefore worth asking whether this transformation may be rewritten as the introduction of additional boson species. However, by construction these bosons are very tightly constrained, appearing only in conjunction with the $K$-matrices, acquiring no 4-momentum at their interaction vertices, and having %
no interactions other than to perform the desired preon colour changes. By the scale of the fermionic mass interaction, their effects vanish when integrated over regions larger than $\ILO{\mc{L}_0}$. To represent these colour transformations as bosons is therefore not productive. (This may be contrasted with the \emph{energy-dependent} co-ordinate transforms of \sref{sec:1storderK}, where the boson description is very useful indeed.)
Second, recognise that unpaired interactions with the background gluon field may also impart random shifts in preon colour. These too may be eliminated by adjusting the local $\SU{3}_C$ co-ordinate transformations, but these random shifts do not yield any systematic contribution to the $K$ matrix and thus over length scales large compared with $\mc{L}_0$ they have no effect on the eigenvalues of operator $\hat\Kf$. This %
observation is necessary to the earlier assertion that all intercalated interactions may be represented by making the intermediate fermions of \fref{fig:QLloop} massive.

\subsection{Quarks and preons\label{sec:quarksandpreons}}

As commented on briefly above, %
a calculation similar to that for leptons %
also applies for quarks. 
However, the varying $A$-charges within the quark triplets imply that colour cycle invariance is only respected across superpositions of colour combinations for the triplet as a whole, and %
is anticipated to result
in a %
quark colour mixing matrix $K_q$ which is $A$-charge-dependent. If the colour of the unique quark is taken to provide a reference point in colour space, then this dependency %
breaks colour cycle invariance, which may have a role in the residual colour charge on the quarks. Further, it is also likely that this $A$-charge dependency will permit calculation of this model's counterpart to the CKM matrix. It is interesting to speculate that this difference between lepton and quark eigenvectors may perhaps facilitate neutrino generation mixing during interactions with baryonic matter, though this idea is not explored further here.

It is also noted that: 
\begin{itemize}
\item Individual foreground preons may acquire mass through interactions with the background fields in the same way as foreground fermions, and that for preons with $A$-charge $a=1$ or $a=2$ this effect will be dominated by interaction with the background photon field, giving these preons non-zero rest masses. It follows that the masses of summed preon pairs $\psi^{1c_1\alpha}\psi^{2c_2\alpha}$ will not in general vanish, in conflict with the choices of gauge on $\SU{3}_C$ discussed in \sref{sec:GL18Cgauge}, with the outcome that species incorporating such pairs have no valid foreground on-shell excitations as noted in \sref{sec:consequences}.
\item $K$-matrices are only associated with %
{all-species} interaction vertices, involving nontrivial representations on both $A$ and $C$ sectors. While the gluon sector is conveniently viewed as a small but nonvanishing correction to the photon sector, %
that $C$-sector interaction nevertheless performs the preon mixing which yields the $K$-matrices. That the associated eigenvalues are also applicable to the $A$-sector contribution is perhaps better understood by recognising that the interactions of \fref{fig:QLloop} are in fact with bosons of $\GLNR$, carrying both $A$ and $C$ charges, and the $C$~charges are associated with both non-trivial and trivial representations of $\SU{3}_C$. This is in contrast with the electromagnetic interaction, for example, which is associated with only the trivial representation on the $C$~sector. It is therefore
crucial to recognise all-species processes as distinct from situations involving only a specific boson. For example:
\begin{itemize}
\item Interaction of a fermion with the background boson fields over length scales of $\ILO{\mc{L}_0}$ is an all-species interaction as per \sref{sec:leptons} above. It acquires $K$-factors.
\item The electroweak interactions of \Psref{III}{sec:interactions} are single-species interactions
with no associated $K$~factors.
\item Further examples will be encountered in \crefs{ch:boson}{ch:gravity}.
\end{itemize}
\end{itemize}

\section{Conclusion}

This chapter has introduced the mass mechanism for fermions in the $\Cw{18}$ model, as illustrated by a tree-level calculation of the lepton masses. An interesting feature of this model is its ability to approximately %
reproduce the Koide relationship between the charged lepton masses, and a preliminary calculation indicates that higher-order corrections to the mass process are of the correct magnitude to potentially improve upon this approximation.

The next chapter presents an equivalent calculation for bosons, following which the evaluation of higher-order corrections to both of these calculations in \cref{ch:detail} yields the first predictive calculations of the $\Cw{18}$ analogue model.

\appendix

\section{Extension of gauge choices to \prm{\SU{9}}\label{apdx:gaugeSU9}}

It is frequently convenient to decompose the local $\SU{9}\oplus\GL{1}{R}$ symmetry of the $\Cw{18}$ model as $[\SU{3}_A\oplus\GL{1}{R}_A]\otimes[\SU{3}_C\oplus\GL{1}{R}_C]$, as is done through much of the present paper.
However, in the corresponding quantum field theory the bosons associated with $\su{9}$ may carry charges in both the~$A$ and the $C$ sector on a single quantum (as per \sref{sec:symC18}) whereas those of $\su{3}_A$ or $\su{3}_C$ are restricted to the $A$~sector or the $C$~sector respectively.
The situation is further complicated by the representations of $\gl{1}{R}_A$ and $\gl{1}{R}_C$ being accidentally degenerate. 

To extend the gauge choices of \sref{sec:gaugechoice} to the bosons of the full $\SU{9}$ symmetry, recognise as follows:
\begin{itemize}
\item A choice of gauge involving a boson $b^\ta_\mu$ in $\su{3}_A$ may be understood as generalising to the nine bosons carrying the same $A$-sector charge in $\su{3}_A\otimes[\su{3}_C\oplus\gl{1}{R}_C]$.
\item A choice of gauge involving a boson $b^\tc_\mu$ in $\su{3}_C$ may be understood as generalising to the nine bosons carrying the same $C$-sector charge in $[\su{3}_A\oplus\gl{1}{R}_A]\otimes\su{3}_C$.
\item Due to accidental degeneracy, the representations associated with $\gl{1}{R}_A$ and $\gl{1}{R}_C$ correspond to a single boson $N_\mu$. The gauge choice eliminating coupling of foreground fermions to $\fgfield{N_\mu}$ may be understood as imposing that bosons with trivial representation ($\lambda_9$) on the $A$-sector do not couple to fermions unless this may be attributed to their $C$-charge, and that bosons with trivial representation on the $C$-sector do not couple to fermions unless this may be attributed to their $A$-charge. Extension of the gauge choice eliminating coupling of foreground fermions to $\bgfield{N_\mu}$ proceeds similarly. The consequence of these two extensions is that bosons which decouple from other bosons on the $A$~sector due to a representation $\lambda^A_9$ on subgroup $\GL{3}{R}_A$ consistently likewise decouple from foreground fermions unless the interaction is mediated by the $C$~sector. Similarly, bosons which decouple from other bosons on the $C$~sector due to a representation $\lambda^C_9$ on subgroup $\GL{3}{R}_C$ consistently likewise decouple from foreground fermions unless the interaction is mediated by the $A$~sector.
\item However, note that gauge choices involving $\gl{1}{R}_A$ and/or $\gl{1}{R}_C$ act at the level of composite fermions as per \Erefr{eq:U1gauge}{eq:GL1RNgauge}. They consequently place no constraints on interactions at the level of individual preons.
\end{itemize}

Now revisit the construction of single-sector bosons, and begin with the bosons of the $C$~sector. Recognise that all leptons, even quarks, are colour-neutral sums over preon triplets as per \sref{sec:catalogueall} (with the apparent quark colour residual arising from a difference in spatial distribution of the constituent preons in conjunction with colour shielding). Each of the nine gluons is only able to interact with one preon out of a colour-neutral triplet, and thus gluon interactions are always gluon/preon interactions, not gluon/fermion interactions. Couplings involving the $\gl{1}{R}_B$ symmetry groups are therefore permitted.
The nine bosons of the $C$~sector are then readily constructed as representations of $\gl{1}{R}_A\otimes\su{3}_C$.
Pure $A$-sector bosons are likewise constructed as representations of $\su{3}_A\otimes\gl{1}{R}_C$

Finally, there also exist bosons carrying nontrivial representations in both sectors. Recognising that the reduction in the number of bosons introduced in \sref{sec:symC18} is valid in the continuum limit, it is only necessary to explicitly address these bosons in low-particle-number contexts. In particular, since the pseudovacuum is represented by the mean field approximation, it is never necessary to consider their contribution to nonzero pseudovacuum expectation values. Instead, contributions to particle masses may be attributed in their entirety to the couplings associated with $A$-sector and $C$-sector charges as if these are exclusively couplings to the pure $A$-sector and $C$-sector background fields. If these hybrid bosons are referred to as coloured $A$-sector bosons, they are of relevance only as foreground excitations. In \sref{sec:ACbosonmassespre} it is suggested that the coloured photon is unlikely to be readily distinguished from the gluon, with
deviation from the Standard Model only being notable in current experiments through the appearance of coloured $W$ and $Z$ bosons as discussed in \crefs{ch:detail}{ch:CDF2}.

\appendixend

\notchap{
\section*{Acknowledgements}
This research was supported in part by the Perimeter Institute for Theoretical Physics.
Research at the Perimeter Institute is supported by the Government of Canada through Industry Canada and by the Province of Ontario through the Ministry of Research and Innovation.
The author thanks the Ontario Ministry of Research and Innovation Early Researcher Awards (ER09-06-073) for financial support.
This project was supported in part through the Macquarie University Research Fellowship scheme.
This research was supported in part by the ARC Centre of Excellence in Engineered Quantum Systems (EQuS), Project No.~CE110001013.
}

\chapter{{Calculation} of boson masses in the \protect{$\mathbb{C}^{\wedge 18}$ model}\label{ch:boson}}

\begin{abstract}
The $\Cw{18}$ model \standalone{is an analogue model capable of emulating the complete particle spectrum of the Standard Model, including interactions,}\chap
{reproduces the particle spectrum and interactions of the Standard Model}\notchap
{reproduces the particle spectrum and interactions of the Standard Model} using only free scalar fields on a manifold with anticommuting co-ordinates. Solitonic excitations in a pseudovacuum state behave as coloured preons, forming triplets which behave as emergent fermions and pairs which behave as emergent bosons. This chapter demonstrates how the emergent bosons acquire mass through interaction with the pseudovacuum. The existence of higher-generation massive bosons is predicted (subject to limitations of the $\Cw{18}$ analogue model), with the lightest being the second-generation $W$ boson at almost $17~\TeV/c^2$. %
\end{abstract}

\section{Introduction}

The $\Cw{18}$ model introduced in \cref{ch:SM} is a classical analogue model \cite{maynard2001,dragoman2004,lewenstein2007} remarkable for the behaviour of its quasiparticle description in the low-energy limit, which qualitatively resembles the quantum field theory of the Standard Model. The value of an analogue model comes from its ability to grant physical or numerical insight into the system being described, prompting the evaluation of quantities in the $\Cw{18}$ model which might be observed, such as particle mass. The tree-level evaluation of fermion masses in \cref{ch:fermion} %
hints at an approximately Koide-like \cite{koide1983,koide2000} relationship between the masses of different generations of leptons.
This chapter continues the calculation of tree-level mass interactions in the $\Cw{18}$ model, this time for the vector and scalar boson fields, with higher-order corrections following in \cref{ch:detail}. The $\Cw{18}$ model is seen to predict the existence of higher generations of massive bosons, though some caution may be required as the masses of these bosons are above the threshold below which the $\Cw{18}$ model behaves as a good analogue to a quantum field theory.

\section{Conventions}

This chapter follows the same conventions as \crefr{ch:simplest}{ch:fermion}. 
\standalone{Units are chosen such that $c=1,~h=1$.
When equations and lemmas from \crefr{ch:simplest}{ch:fermion} are referenced, they take the forms (\textbf{1}.1), (\textbf{2}.1), %
etc.

In particular, regarding terminology around Feynman diagrams and symmetry factors:
\begin{itemize}
\item Where there exist multiple ways to connect up sources, vertices, and sinks to obtain equivalent diagrams up to interchange of non-distinguishable co-ordinates, the same term is obtained from the generator $\Z$ in multiple different ways and thus the diagram acquires a multiplicative factor. This is referred to in the present volume as a \emph{symmetry factor}.
\item Where integration over the parameters of a diagram (for example, over source/sink co-ordinates) 
yields the same diagram multiple times up to interchange of labels on these parameters, 
this represents a double- (or multiple-)counting of physical processes. It is then necessary to eliminate this multiple-counting by dividing by the appropriate symmetry factor. This is referred to in the present volume as \emph{diagrammatic redundancy} or \emph{double- (multiple-)counting.}
\end{itemize}
}

\section{Boson masses\label{sec:V:bosonmasses}}

\subsection{Vector boson masses\label{sec:vecbosonmasses}}

\begin{figure}
\begin{center}
\includegraphics[width=5in]{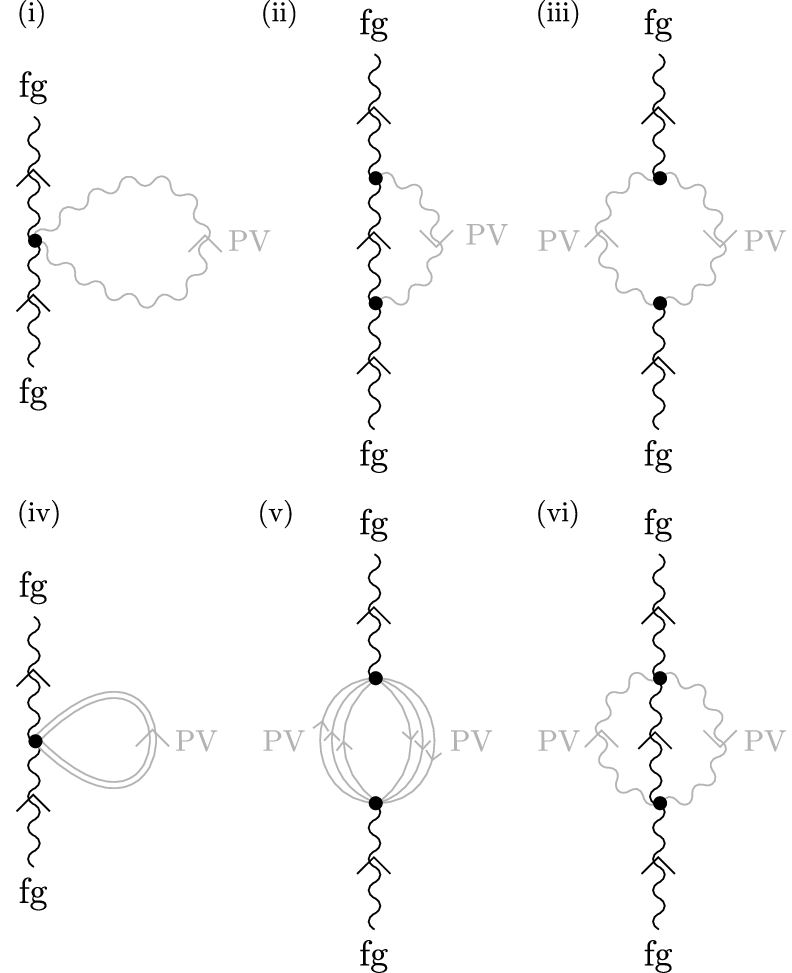}
\end{center}
\caption{Graph types for interactions between a foreground vector boson and the pseudovacuum (PV) vector boson, scalar boson (double line), and fermion fields. (i)~\protect{$\OO{f^2}$} interaction with two PV bosons. (ii)~Two consecutive \protect{$\OO{f}$} interactions with single PV bosons. (iii)~Two disjoint \protect{$\OO{f}$} interactions, each with two PV bosons. (iv)~\protect{$\OO{f^2}$} interaction with two PV scalar bosons. %
This diagram is the scalar counterpart to diagram~(i); scalar counterparts to diagrams~(ii) and~(iii) are also discussed in the text. (v)~Two disjoint $\OO{f}$ interactions, each with six PV preons. %
Further diagrams can be constructed, e.g.~(vi), but reduce to composites of diagrams already discussed. For example, if the central boson in diagram~(vi) is foreground (as shown), this corresponds to two copies of the interaction of diagram~(ii). If the central boson is background, this corresponds to compounding diagrams~(ii) and~(iii).
\label{fig:vecbosonmass}
}
\end{figure}

For vector bosons, multiple candidates for mass terms arise from interactions between the propagating boson and both the background fermion and boson fields as shown in \fref{fig:vecbosonmass}. 
To address each in turn for a general vector boson $V^\bdag_\mu$:

\subsubsection{\pfref{fig:vecbosonmass}(i):\label{sec:vecbosonmass(i)}}
For a vector boson with nonvanishing EM charge the $\OO{f^2}$ four-boson vertex with two background fields yields the background (pseudovacuum) photon interaction term %
\begin{align}
-\frac{n_\vp f^2}{2}\!\iint\!\!\rmd^4x\,\rmd^4y\,&\fgfield{V^\dagger_\mu(x) V^\mu(x)}\bgfield{A_\nu(x) A^\nu(x)}.\label{eq:vecbosonmassi}
\end{align}
The boson may interact with any of the three preons in the fermion pair, but this cancels with factors of $1/\sqrt{3}$ from \PEref{III}{eq:generalfermion} (see also \Psref{III}{sec:EWint_numerical} and \fref{fig:evalNsyms}). %
It vanishes when $V^\mu$ is a photon due to the absence of colour or self-coupling, and for the $Z$ boson as it has no charge. In the $\SU{3}_A$ sector this diagram therefore contributes mass only to the $W^\bdag$ and $G^\bdag$ bosons. By comparison with \PEref{IV}{eq:leptonmasses} its contribution is of %
$\OO{k_1^{-4}m_e^2}$, and is unlikely to exceed $\OO{m_\tau^2}$. (This is confirmed in \cref{ch:detail}.) %
There exists a similar contribution to the gluon mass $m_c$ based on coupling to the background gluon field, which arises instead from the bosons of $\gltr_C$.

Note that in contrast to the two-photon interactions of \fref{fig:leptonmassterm}, there is no foreground propagator separating the two pseudovacuum interaction vertices. This diagram therefore does not
attract
a factor of $\bmfcdot$ \Peref{IV}{eq:mloopfactor} even when the background boson is a gluon.

\subsubsection{\pfref{fig:vecbosonmass}(ii):}
This diagram yields background photon terms of the form
\begin{align}
-&\frac{n_\vp f^2}{4}\!\iint\!\!\rmd^4x\,\rmd^4y\label{eq:vecbosonmassii}\\
\nn&\left\{(\delta^{\mu\rho}\delta^{\nu\sigma}+\delta^{\nu\rho}\delta^{\mu\sigma})\bgfield{A_\mu(x)}\fgfield{V^\dagger_\nu(x) \partial_\rho V'_\sigma(x)}\right.\\
\nn&\p{+\{}+\left.(\delta^{\mu\rho}\delta^{\nu\sigma}+\delta^{\nu\rho}\delta^{\mu\sigma}) \bgfield{\partial_\mu A_\nu(x)}\fgfield{V^\dagger_\rho(x) V'_\sigma(x)}\right\}\\
\nn\times&\left\{(\delta^{\lambda\pi}\delta^{\kappa\tau}+\delta^{\kappa\pi}\delta^{\lambda\tau})\bgfield{A_{\lambda}(y)}\fgfield{V'^\dagger_{\kappa}(y) \partial_{\pi} V_{\tau}(y)}\right.\\
\nn&\p{+\}}+\left.(\delta^{\lambda\pi}\delta^{\kappa\tau}+\delta^{\kappa\pi}\delta^{\lambda\tau}) \bgfield{\partial_{\lambda}A_{\kappa}(y)}\fgfield{V'^\dagger_{\pi}(y) V_{\tau}(y)}\right\}
\end{align}
and similar for pseudovacuum gluons up to $\bmfcdot$. In the electron mass interaction of \PEref{IV}{eq:lepmass2}, spinor identities allowed two spatially separated one-boson interactions to be rewritten as an absolute square. 
In this diagram there is no equivalent identity, so instead the properties of the pseudovacuum may be used to contract the two vertices, replacing
\begin{equation}
\la\bgfield{A_\mu(x)A_\nu(y)}\ra = -\frac{1}{4}\bm{f}(x-y){N_0}^4{\omega_0}^2\eta_{\mu\nu}.\label{eq:bgreduction}
\end{equation}
The resulting contraction of the two vertices yields an expression equivalent to an all-foreground loop diagram weighted by a mass coefficient. %
Comparing with %
\sref{sec:vecbosonmass(i)}, this expression corresponds to one of the diagrams obtained on expanding the $W^\bdag$ propagators with Proper Self Energy terms. It is therefore not necessary to count this diagram separately. The same argument applies if the background vector boson is replaced by a background scalar boson.

\subsubsection{\pfref{fig:vecbosonmass}(iii):}
This diagram is very similar to \fref{fig:vecbosonmass}(ii), and with good reason, as both are obtained from mean field theory expansion of the same original figure. This diagram is the lowest-order in foreground fields, and the foreground momentum is understood to travel from the lower to the upper vertex as perturbations about this mean field term. If the interaction vertices are not collocated, a basis (not necessarily spatial) may always be chosen such that all foreground momentum is transferred in a single limb of the loop, recovering \fref{fig:vecbosonmass}(ii). Alternatively, if the vertices are collocated, the diagram reduces to the duplication of \fref{fig:vecbosonmass}(i). Either way, it is redundant.

\subsubsection{\pfref{fig:vecbosonmass}(iv):\label{sec:vecbosonmassiv}}
This diagram is similar in structure to \fref{fig:vecbosonmass}(i), but with the composite scalar boson in place of a vector boson. Mapping $\partial^\mu\partial_\mu$ to $\partial\partial\bar\partial\bar\partial$ attracts a factor of $-2$, but this is offset by a factor of $-\frac{1}{2}$ on evaluating $\la\bgfield{\bar\psi\bar\psi\psi\psi}\ra$ using \PEref{III}{eq:bghh*}.
Unusually for a scalar boson the vertex factor is augmented by a FSF symmetry multiplier ${N_0}^8[1+\ILO{{N_0}^{-1}}]$---this is not reduced by ${N_0}^2$ as the composite scalar bosons are emitted at a single vertex and are therefore immediately mapped to their local mean field values without first requiring propagation (see discussion in \sref{sec:scalbosint}).
Overall this term again evaluates to
$\OO{k_1^{-4}m_e^2}\lesssim\OO{m_\tau^2}$ so contributes only a relatively small amount to $m_W^2$.

\subsubsection{\pfref{fig:vecbosonmass}(v):\label{sec:Wmass5v}} 
The main contribution to vector boson masses comes from diagrams of the form of \fref{fig:vecbosonmass}(v). This diagram is non-vanishing if the %
preon sources and sinks are all separated by a distance of $\ILO{\mc{L}_0}$ or less, and thus appear collocated to a probe with energy $\mc{E}_p\ll\mc{E}_0$ or timescale $\mc{L}_p\gg\mc{L}_0$ (essentially all probes---see \Psref{I}{sec:ProbeOmegaScale}). That is, the two vertices are effectively (though not strictly) collocated as discussed in the context of \fref{fig:vecbosonmass}(iii), though this time the resulting figure has not previously been accounted for.

Regarding the preon lines in this diagram, %
note that %
\begin{itemize}
\item A composite boson necessarily has energy small compared with $\mc{E}_\preon$ in the isotropy frame of the pseudovacuum. Since $\mc{E}_\preon$ is also anticipated to exceed the energy scale of the pseudovacuum, $\mc{E}_0$, %
the preons making up the composite particles in \fref{fig:vecbosonmass}(v) appear as bound triplets. %
\item In the presence of the background fields, the dominant contribution to the average value of \fref{fig:vecbosonmass}(v) is given by replacing the fields of the loop with the mean-field value for the pseudovacuum contribution. Consequently these fields do not give rise to a loop factor $\bmfcdot$, as there is no foreground propagator between the two interaction vertices in this mean-field term. 
\item Nevertheless, foreground momentum must be transferred between the lower and the upper vertex and this takes place through perturbations about this mean-field value. These may be viewed as equivalent to foreground fields. 
\item As discussed in \sref{sec:Csector}, whenever a fermion engages in an $A$-sector boson interaction (such as the $W$ interaction discussed here), this may be associated with a set of co-occuring $C$-sector interactions, corresponding in this case to couplings to the background gluon fields at the upper and lower vertices of the loop. 
For the background component of the fermion loop the normalisation convention of \Psref{I}{sec:normWrtBgFields} ensures that these interactions make no direct contribution to the $W$~boson mass vertex. However, the freedom to pair non-conjugate gluons as per \PEref{III}{eq:<cc>} indicates that nonvanishing colour perturbations may take place in these interactions, and following \Psref{IV}{sec:Csector} it is convenient to evaluate the impact of these couplings by associating them with the dominant $A$-sector interaction. Since $\SU{3}_C$ symmetry is preserved, these couplings place the background fermion into eigenstates of matrix~$K$ as before.
\item In contrast with \Psref{IV}{sec:Csector}, however, these $C$-sector couplings are optional. In \Psref{IV}{sec:Csector} a sum over all propagation processes comprised terms involving fermion/photon coupling and terms involving fermion/gluon coupling, with each vertex appearing in equal numbers, such that each photon interaction was accompanied by nine gluon interactions and thus by matrices~$K$. In the present context, the sum over all propagation processes includes the $W$~boson in every diagram, but the preons of the loop may interact either as two background fermions, which accrue $K$ matrices, or as three background bosons, which do not accrue $K$ matrices, in an alternative process discussed further in \sref{sec:universalcorrs}. %
(The background preons must always arise from composite background particles, as the background energy scale $\mc{E}_0$ is assumed small compared with $\mc{E}_\preon$.)
Consequently there \emph{may} be $K$ matrices associated with the $W$/preon triplet vertices, and summation over all propagation processes includes diagrams both with and without these matrices present. 
However, only diagrams with $K$ matrices are considered in the present chapter, as the value of the diagram is much reduced when the $K$ matrices are absent. 
\item Given the presence of $K$ matrices, it is convenient to work in the generations basis for fermions. At tree level this corresponds to eigenvectors of the colour mixing matrix $K$ given in \PEref{IV}{eq:IV:KfES}. The vertices may admit triplets from any generation.
\item Symmetry factors are computed as normal, as they arise through the extraction of interaction vertices from the generating functional and are independent of the manner by which the propagators are evaluated (e.g.~Green's function vs.~mean field theory), though care must be taken not to double-count any symmetry factors already present in the mean field expressions. For this reason, as in \PEref{IV}{eq:HHsub}, evaluation in terms of $\omega_0$~\Peref{I}{eq:N0terms} is preferred over %
$\mc{E}_0$~\Peref{I}{eq:E0}.
\end{itemize}

It is convenient to represent preon triplets of any generation by the symbols of their associated fermions. %
The $K$ matrices implicit in the vertices of \fref{fig:vecbosonmass}(v) similarly imply a generation structure for massive bosons where (to tree level) the lightest second-generation boson is a $W$ boson with a first approximation to its mass satisfying
\begin{equation}
m_{W_2}%
\approx \frac{m_{W_1}m_\mu}{m_e} \approx 17~\TeV/c^2. 
\end{equation}
Since only one generation of $W^\bdag$ bosons has yet been observed, for now restrict attention to generation~1 before further discussion in \sref{sec:tooheavy}.

Taking the $W^\bdag$ boson as the first example of a vector boson acquiring mass from \fref{fig:vecbosonmass}(v), and deferring calculation of FSF symmetry factors, the lepton channel is
\begin{align}
-36f^2k_1^4\!\!\iint\!\!\rmd^4x\,\rmd^4y\,&\fgfield{W^\dagger_\mu(x)W_\nu(y)}\label{eq:leptonvecbg1}\\
\times& \bgfield{\bar{\nu}_e(x)\bsm e_L(x)\bar{e}_L(y)\bsn\nu_e(y)}\nn
\end{align}
where the factor of 36 arises as there are $3!$ ways to order the electron propagators on the left (with the exchange of two lines %
corresponding to the exchange of two spinor operators on the upper vertex and two on the lower vertex
for a net factor of $+1$) and $3!$ ways to order the neutrino propagators on the right. 
The quark channel is similarly %
\begin{align}
-36f^2k_1^4\!\!\iint\!\!\rmd^4x\,\rmd^4y\,&\fgfield{W^\dagger_\mu(x)W_\nu(y)}\label{eq:leptonvecbg2}\\
\times& \bgfield{\bar{u}_L(x)\bsm d_L(x)\bar{d}_L(y)\bsn u_L(y)}\nn
\end{align}
where the factors for reordering of identical propagators are now $2!\times 2!$, but each triplet also acquires a factor of 3 corresponding to the three different choices as to the colour of the unique preon, once again giving a net total symmetry factor of 36.

\begin{figure}
\includegraphics[width=\linewidth]{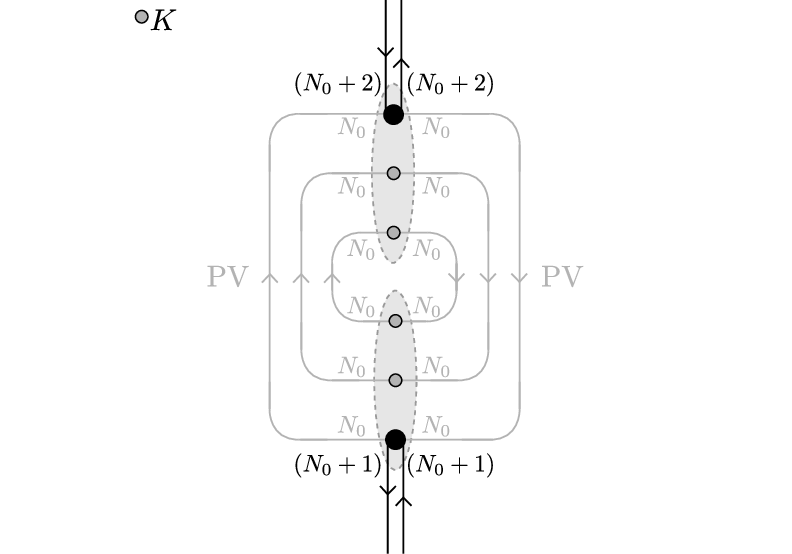}
\caption{Fundamental scalar field symmetry factors in the \prm{W} boson/background fermions mass interaction. One factor attaches to each field operator appearing on a vertex (grey ellipse). Within the fermion-boson interaction, \prm{K}-matrices are shown as grey dots, and preon-boson interactions as black dots. [Each vertex also attracts a factor of~\prm{\frac{1}{3}} from the normalisation of \protect{\PEref{III}{eq:generalfermion}} but this is offset as each boson may couple to any of the three preons---see also \protect{\Psref{III}{sec:EWint_numerical}}.]\label{fig:evalWNsyms}}
\end{figure}

Expanding the fermions and bosons as preons and calculating the FSF symmetry factors proceeds very similarly to \fref{fig:evalNsyms}, only now the foreground preons are in the boson part of the diagram, not the fermion part. These factors are shown in \fref{fig:evalWNsyms}. 
As before, background fields contribute symmetry factors of $N_0$ per preon and thus the overall FSF symmetry factor is
\begin{equation}
\begin{split}
n_{\vp,W}&=(N_0+2)^2(N_0+1)^2{N_0}^{12}\\
&={N_0}^{16}S_{6,13}\label{eq:WFSFsymfactor}
\end{split}
\end{equation}
where 
\begin{equation}
\begin{split}
S_{6,13}:=\,&{N_0}^{-4}(N_0+2)^2(N_0+1)^2\\
\approx\,&(1+6{N_0}^{-1}+13{N_0}^{-2}+\ldots)
\end{split}
\end{equation}
as in \PEref{III}{eq:III:defS613}.
From $n_{\vp,W}$, a factor of ${N_0}^4$ might customarily be absorbed into the definition of the fermion so that the factors associated with the noninteracting preons can be suppressed. However, to eliminate the noninteracting preons entirely it is more convenient to leave the FSF factors explicit and integrate over $\rmd^4y$. 
The fermions therefore take the form of \PEref{III}{eq:generalfermion}. 
The paired background preons thus take the form
\begin{equation}
f^2\Big<\bar\psi^{\dot a\dot c}\bar\psi^{\dot a\dot d}\psi^{ac}\psi^{ad}\Big>_{\dot a=a,\dot c=c,\dot d=d}
\end{equation}
which evaluates as
\begin{align}
\begin{split}
-\frac{f^2}{2}\Big<\big\|\bar\psi^{\dot a\dot c}\bsmm\psi^{ad}\big\|^2\Big>_{\dot a=a}&= -\frac{1}{2}\left<\left\|\vp^{\dot aa\dot cd}_\mu\right\|^2\right>\\
&= \p{-}\frac{1}{2}{N_0}^2{\omega_0}^2
\end{split}
\end{align}
with integral with respect to $y$ of
\begin{equation}
\begin{split}
\int\!\rmd^4y~\left[f^2\Big<\bar\psi^{\dot a\dot c}\bar\psi^{\dot a\dot d}\psi^{ac}\psi^{ad}\Big>_{\dot a=a,\dot c=c,\dot d=d}\right]^2 %
&=\frac{1}{4}.
\end{split}
\end{equation}
The sum of \Erefs{eq:leptonvecbg1}{eq:leptonvecbg2} thus reduces to
\begin{equation}
\begin{split}
&9f^2k_1^4{\omega_0}^2{N_0}^{12}~\!\!\int\!\!\rmd^4x\, W^\dagger_\mu(x) W^\mu(x)~\left[1+\OO{{N_0}^{-1}}\right],
\end{split}
\end{equation}
corresponding to
\begin{equation}
\begin{split}
m_W^2&=9f^2k_1^4{\omega_0}^2{N_0}^{12}\left[1+\OO{{N_0}^{-1}}\right]\\
&=18m_e^2{N_0}^4\left[1+\OO{{N_0}^{-1}}\right]
\end{split}\label{eq:mWfromme}
\end{equation}
at tree level by \PEref{IV}{eq:leptonmasses}. Higher-order corrections to \fref{fig:vecbosonmass}(v) amend this to
\begin{equation}
m_W^2=9f^2k_1^4{\omega_0}^2{N_0}^{12}\left[1+\OO{{N_0}^{-1}}+\OO{\alpha}\right]\label{eq:mWsq}
\end{equation}
and are evaluated in \cref{ch:detail}.

Interestingly, the higher-order corrections to the FSF symmetry factor $n_{\vp,W}$ for the $W$ boson are given by series $S_{6,13}$ which also appears when evaluating electron mass~\Peref{IV}{eq:IV:defS613}. For the electron this factor subsequently attracts further corrections~\Peref{IV}{eq:defS18147}, so the two do not cancel exactly. %
Nevertheless, \Erefr{eq:mWfromme}{eq:mWsq} permit the value of $N_0$, the mean number of fundamental scalar field centres per hypervolume ${\mc{L}_0}^4$, to be evaluated up to corrections of $\ILO{\alpha}$ and $\ILO{{N_0}^{-1}}$ as
\begin{equation}
\begin{split}
N_0&=\sqrt{\frac{m_W}{3\sqrt{2}m_e}}\left[1+\OO{{N_0}^{-1}}+\OO{\alpha}\right]\\&\approx 193.
\end{split}\label{eq:N0value} %
\end{equation}

To now proceed similarly for the photon yields the pre-FSF vertex expression
\begin{equation}
-{36f^2}k_1^4\!\!\iint\!\!\rmd^4x\,\rmd^4y\fgfield{A_\mu(x)A_\nu(y)}\bgfield{X_A^\mu(x)X_A^\nu(y)}\label{eq:photonQLint}
\end{equation}
where $X_A^{\mu}(x)$ is the net vertex interaction on summing over all background fields,
\begin{align}
\!\!\!\!\!X_A^\mu(x) = \biggl\{&\label{eq:XforA}
\bgfield{e_R(x)\bsm \bar{e}_R(x)}-\bgfield{\bar{e}_L(x)\bsm e_L(x)}\!\\
&+\frac{1}{3}\bgfield{d_R(x)\bsm\bar{d}_R(x)}-\frac{1}{3}\bgfield{\bar{d}_L(x)\bsm d_L(x)}\nn\\
&-\frac{2}{3}\bgfield{u_R(x)\bsm \bar{u}_R(x)}+\frac{2}{3}\bgfield{\bar{u}_L(x)\bsm u_L(x)}\biggr\}.\nn
\end{align}
In \Eref{eq:photonQLint}, $f^2$ has been replaced by $f^2/2$ due to the weaker coupling at the photon vertex, but a compensatory factor of 2 is acquired from exchange symmetry of the two identical interaction vertices. %
Any diagrams containing ``tadpole'' fermions (fermion lines whose source and sink are on the same vertex) must vanish, so the per-diagram symmetry factor for the persisting terms is once again $3!\times 3!=36$, and cross terms (where the fermion species interacting with the two vertices differ) disappear.

As per \sref{sec:GL18Cgauge}, gauge choice~\Peref{III}{eq:ma3gauge} is chosen such that \Eref{eq:photonQLint} vanishes---up to a small correction. More precisely it is not \Eref{eq:photonQLint} which is chosen to vanish, but rather the sum of \Eref{eq:photonQLint} (plus FSF factors and higher-order corrections) and the smaller terms arising from \freft{fig:vecbosonmass}(i) and~(iv) (plus FSF factors and higher-order corrections). 
As previously noted in \sref{sec:consequences}, at any given point $x$ the values of each of these terms may be positive or negative and the action of the gauge transform is to selectively eliminate the dominant contributions, bringing the net mass to zero. With \freft{fig:vecbosonmass}(i) and~(iv) having symmetry factors smaller than \fref{fig:vecbosonmass}(v) by a factor of ${N_0}^4$, they pose no obstacle to ensuring the mean vanishing of photon mass to probes of scale $\mc{L}_p\gg\mc{L}_0$ (essentially, all probes---see \Psref{I}{sec:ProbeOmegaScale}). Having established the mechanism by which masslessness of the photon may be arrived at, it is arguably now convenient to rewrite this gauge choice in the simpler form
\begin{equation}
\left\|\fgfield{A_\mu(x)}\right\|^2=0.\label{eq:ma3gaugenew}
\end{equation}

For the $Z$ boson, the corresponding pre-FSF expressions are
\begin{align}
-{72f^2}k_1^4\!\!\iint&\rmd^4x\,\rmd^4y\fgfield{Z_\mu(x)Z_\nu(y)}\bgfield{X_Z^\mu(x)X_Z^\nu(y)}\label{eq:ZQLint}\\
X_Z^\mu(x) = \frac{2}{\sqrt{6}}&\biggl\{
\frac{1}{2}\bgfield{e_R(x)\bsm \bar{e}_R(x)}+\frac{1}{2}\bgfield{\bar{e}_L(x)\bsm e_L(x)}\nn\\
&-\frac{1}{2}\bgfield{d_R(x)\bsm \bar{d}_R(x)}-\frac{1}{2}\bgfield{\bar{d}_L(x)\bsm d_L(x)}\nn\\
&-\bgfield{\bar\nu_e(x)\bsm\nu_e(x)}\biggr\},
\end{align}
where the factor of~72 in \Eref{eq:ZQLint} is made up of 36 from preon exchange symmetries and~2 from interchange of source and sink. %
(The two vertices may also be interchanged, but when source, sink, and both vertices are exchanged, the original diagram is recovered, reflecting that these are equivalent descriptions of a single twofold symmetry.) %
Incidentally, note that the factor of two is also present in the Standard Model, where it multiplies the Lagrangian term $\frac{1}{2}m_Z^2Z^\mu Z_\mu$ to yield a mass vertex $m_Z^2 Z^\mu Z_\mu$.

On evaluating \Eref{eq:ZQLint}, the resulting leading-order expression for the $Z$ boson mass is
\begin{equation}
\begin{split}
m_Z^2&=12f^2k_1^4{\omega_0}^2{N_0}^{12}\left[1+\OO{{N_0}^{-1}}+\OO{\alpha}\right]\\
&\approx\frac{4m_W^2}{3}.\label{eq:Zmass}
\end{split}
\end{equation}

Finally, the nine gluons $c^\tc_\mu$ also acquire a mass $m_c$ through interaction with the pseudovacuum, primarily also through coupling to the fermion fields. 
This is due to coupling of gluons to the colour charges on preons in the pseudovacuum. %
By maximal entropy of the pseudovacuum, and the absence of symmetry-breaking gauge choices on the bosonic sector of the $\SU{3}_C$ symmetry, the resulting %
mass contribution imparted to any gluon %
is equal to that acquired by the $W$ boson from the pseudovacuum fermion fields,
\begin{equation}
m_c^2\approx m_W^2.
\end{equation} 
Interaction with background gluons and scalar bosons then contributes additional terms, corresponding to the other diagrams of \fref{fig:vecbosonmass}, correcting the %
gluon mass by
\begin{equation}
m_c^2\longrightarrow m_c^2\left[1+\OO{\frac{m_\tau^2}{m_W^2}}\right].
\end{equation}

In practice this mass is observed only indirectly and in a limited number of circumstances, as the typical gluon lifetime is of $\ILO{\mc{L}_\preon}$ which is taken to be small compared with both $\mc{L}_0$~(the characteristic scale up to which mass interactions are observed, and above which particles are necessarily massive) and $2\mc{L}_\Omega$~(which is the minimum scale over which mass interactions may be observed, and below which particles are necessarily massless).\footnote{Recognise that \prm{\mc{L}_\Omega} corresponds to the characteristic interval between pseudovacuum scattering events. The shortest time interval \prm{I_t} within which a nonvanishing mass interaction can reliably be detected is thus \prm{2c^{-1}\mc{L}_\Omega}, as this is the minimum interval such that a pseudovacuum interaction located at an arbitrary time \prm{t} within this interval may always be matched with another pseudovacuum interaction at a time \prm{t\pm c^{-1}\mc{L}_\Omega} which is likewise within interval \prm{I_t}.}
Consequently gluons typically appear massless, except as discussed in \aref{apdx:gluonmass}.

\subsubsection{\pfref{fig:vecbosonmass}(vi):\label{sec:Wmass5vi}} 
A small number of additional two-vertex diagrams may be constructed by making use of the $g^2$ vertices, but (as noted in the caption to \fref{fig:vecbosonmass}) these may all be equated to sums over diagrams already eliminated and so add nothing further to the vector boson mass term.

\subsubsection{Exception: The \prm{G}~boson\label{sec:Gbosonmassless}}
At first glance the $G$~boson appears to play a very similar role to the $W$~boson, but instead of coupling the electron neutrino and the left-handed electron (or higher-generation equivalents) it instead couples the electron antineutrino and the right-handed electron, and thus acquires mass from the background fermion and photon fields. However, gauge choice~\Peref{III}{eq:bga45gauge} imposes that the magnitude of the background $G$~field vanishes, and this condition is nonsingular so is valid everywhere. This has an interesting collateral effect: If the $G$~boson has a nonvanishing interaction with the pseudovacuum, then 
there must exist nonvanishing momenutm transfer between the background fields and the $G^\bdag$ fields, and transfer of momentum from the background fields to the $G^\bdag$ fields must at times result in $G^\bdag$ fields which carry background character, in violation of choice of gauge. To avoid contradiction it is therefore necessary that the $G$~boson never actually interact with the background fields, and thus its mass-inducing couplings to the background photon and fermion fields remain unrealised. This may always arise through an appropriate choice of co-ordinate frame on $\GL{18}{C}$ as gauge choice~\eref{eq:bga45gauge} is nonsingular, and the $G$~boson is therefore massless. %

\subsection{Boson generations\label{sec:tooheavy}}

A curious feature of \sref{sec:vecbosonmasses} is the prediction of second- and third-generation $W$ and $Z$ bosons, with the lightest being $W_2$ with a mass of approximately $16.6~\TeV/c^2$. However, as per \Psref{I}{sec:extendfgE} the energy scale at which $\Cw{18}$ ceases to behave as a good analogue to a QFT is
\begin{equation}
\mc{E}_\Omega:={\N}{N_0}(N_0-\tfrac{1}{2})\,\omega_0\quad|\quad\N=9,\label{eq:V:EOmega}
\end{equation}
which is seen in \aref{apdx:accessory} to take a value of approximately $6.2~\TeV$. %
In \cref{ch:CDF2} %
it is discussed that the pseudovacuum energy scale $\mc{E}_\Omega$ corresponds to a particle energy of $E_\Omega=\frac{1}{2}\mc{E}_\Omega$,
which is substantially less than the mass of the $W_2$ boson. In the $\Cw{18}$ model the value of $E_\Omega$ corresponds to a real UV cutoff, with the consequence that virtual $W_2^\bdag$ bosons are frozen out of all processes having energy less than about $13.5~\TeV$ %
in the isotropy frame of the pseudovacuum. Even above this threshold, behaviour is atypical as formation of a virtual $W_2^\bdag$ boson requires the obligate contribution of $13.5~\TeV$ of energy from the foreground particles. %
However, real $W_2^\bdag$ bosons in the $\Cw{18}$ model
will still be generated as consistent collision energies and momenta permit.

For further details of the higher-generation weak bosons and their masses, see \sref{sec:heavyWZH}. %

\subsection{Scalar boson mass\label{sec:scalbosonmass}}

\begin{figure}
\begin{center}
\includegraphics[width=4.4in]{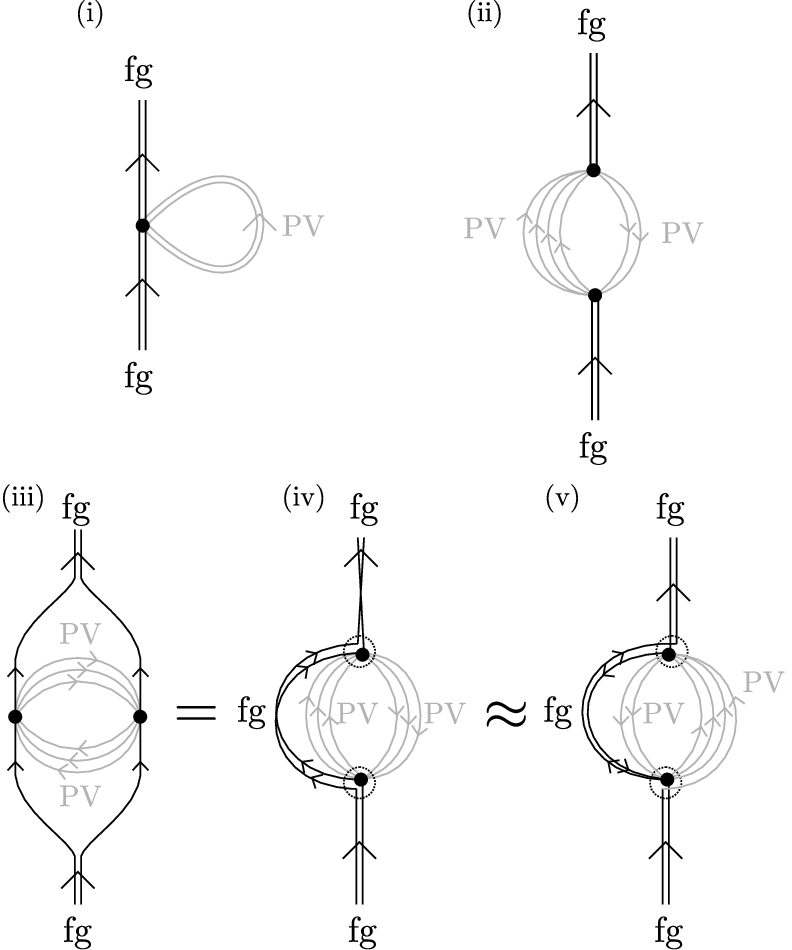}
\end{center}
\caption{Interactions between a foreground scalar boson and (i)~the bosons and (ii)-(iii)~other preon channels of the pseudovacuum. Diagram~(iii) is drawn in a relatively intuitive form, whereas in diagram~(iv), this interaction has been rearranged using diagrammatic isotopy. Dotted circles represent effective net interactions at the upper and lower vertices. In diagram~(v) this identification of effective interaction vertices permits a different choice of fields to be replaced with pseudovacuum mean field terms. Several lines have also been taken over from the left side to the right and vice versa to highlight the similarity to diagram~(ii). Looking inside the effective vertices of diagram~(v), some foreground and pseudovacuum fields are seen to exchange roles; this is permissible due to the quasiparticle nature of the foreground fields, which implies that the foreground momentum is in fact a collective property of all \prm{(N_0+2)} locally correlated FSFs, with the net contribution of \protect{$N_0$} of these being undetectable over length scales \protect{$\mc{L}\gg\mc{L}_0$}. Evaluation of diagram~(v) is therefore equivalent to evaluation of diagram~(iv) up to the presence of a vector boson loop, and a small difference in the FSF symmetry factors of the background fields discussed in the main text. Note also that, as is readily seen from diagram~(iii), the scalar boson preons in diagrams~(iii)-(v) are confined at all times to be separated by no more than \protect{$\mc{L}_\preon$}, and this is entirely compatible with the mass interaction illustrated, as the vertices of this interaction must be separated by no more than \protect{$\mc{L}_0$} which is larger than \prm{\mc{L}_\preon}.\label{fig:scalarbosonmass}}
\end{figure}

The scalar boson is also capable of interacting with the background boson fields and other preon channels of the pseudovacuum including background fermion fields. These interactions are shown in \fref{fig:scalarbosonmass}, and once again the multiple-preon pseudovacuum interactions of diagrams~(ii)-(iii) dominate. 
These contribute a mass on the order of the electroweak scale (consistent with the Higgs boson of the Standard Model) while the pseudovacuum scalar boson interaction of diagram~(i) provides a correction of $\OO{m_\tau^2}$.

To evaluate the scalar boson mass to leading order, first consider 
the preon fields at the interaction vertices in \fref{fig:scalarbosonmass}(ii), and ignore \fref{fig:scalarbosonmass}(iii). In diagram~(ii) the scalar boson $\bmh$ interacts with two preons and four anti-preons, rather than three and three as in \fref{fig:vecbosonmass}(v). However, this may still be understood as an interaction with the background fermion fields. A pair of fermion operators in the background field is typically written $\bgfield{\bar\Psi(x)\Psi(x)}$ for some $x$. However, background fields exhibit correlations over length scales of $\ILO{\mc{L}_0}$ and in practice the preons in the particles associated this expression may be detected at disparate locations within a local correlation region. If the two vertices are at $x$ and $y$, within a common local correlation region, then the preon fields of \fref{fig:scalarbosonmass} may therefore be understood as fermion fields nominally written $\bgfield{\bar\Psi(x)\Psi(x)}$ and $\bgfield{\bar\Psi(y)\Psi(y)}$, but with one preon operator from the former being evaluated at $y$, and one antipreon operator from the latter being evaluated at $x$. The fermion operators continue to represent sources or sinks for fermions at a point (up to confinement scale $\mc{L}_\preon$), but subject to detection at a distance via their correlations.

This effective redistribution, with four preons and two antipreons (or vice versa) at each vertex, does however impact the FSF symmetry factor of diagram~(ii) as the detection operations being performed at the vertices are now grouped as four preon operators and two antipreon operators or vice versa. %
If the preon/antipreon pairs at a vertex are labelled first, each attracts a factor of ${N_0}^2$ and maintains the number of available source/sinks as discussed in \sref{sec:Asector}. However, the final two background operators on each vertex are either two preons or two antipreons. Recognise that evaluation of the background mean fields occurs essentially pointwise at each interaction vertex, and thus for background fields, symmetry factors must be evaluated independently at each vertex. %
In practice, the composite nature of the fermion requires that ``pointwise'' be approximated as ``on a region characterised by length scale $\mc{L}_\preon$'', but for $\mc{L}_\preon$ assumed much smaller than $\mc{L}_0$ the symmetry factors on the vertices effectively continue to be evaluated independently.

The factor for a pair of preons or antipreons then depends on whether or not they are of the same species. As noted in \Psref{IV}{sec:Asector}, such competition for sources is mitigated by the propensity for preons to change charge (on both $A$ and $C$ sectors) as they interact with the background fields, even within a context such as \fref{fig:scalarbosonmass}(ii) where the excursions of the background preons are restricted to a region characterised by dimension $\mc{L}_\preon$. 
If these preons are of identical $A$- and $C$-charges as they are considered for application to a contested source/sink then the resulting factor is $N_0(N_0-1)={N_0}^2(1-{N_0}^{-1})$, whereas if they are of different charges then the factor is ${{N_0}^2}$. 

\begin{figure}
\includegraphics[width=\linewidth]{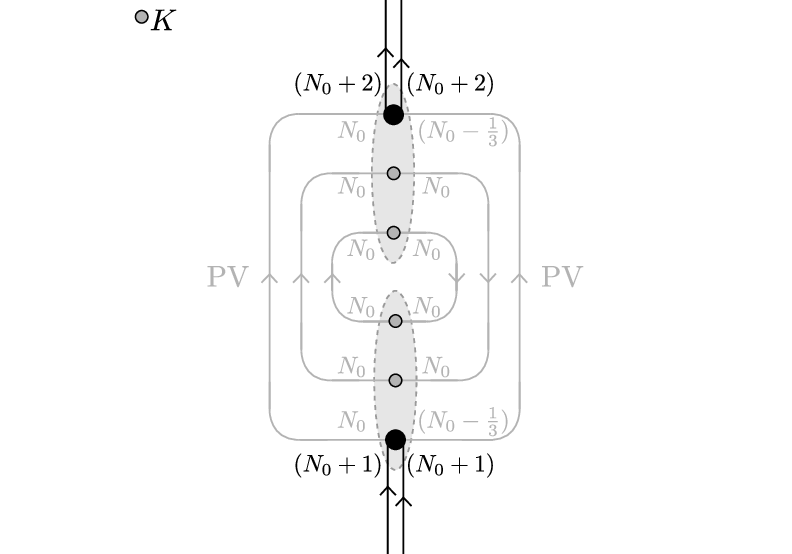}
\caption{Fundamental scalar field symmetry factors in the complex scalar boson/background fermions mass interaction. Note that one orientation arrow is reversed on the pseudovacuum fields compared with \pfref{fig:evalWNsyms}, causing two factors of \prm{N_0} to both be replaced with \prm{(N_0-1)} with one in three probability. [Each vertex also attracts a factor of~\prm{\frac{1}{3}} from the normalisation of \protect{\PEref{III}{eq:generalfermion}} but this is offset as each boson may couple to any of the three preons.] \label{fig:evalHNsyms}}
\end{figure}

As there are nine different charge combinations on each preon, the fourth preon has a $\frac{1}{9}$ chance of matching a given one of the other three preons. However, the %
nominal fourth preon may in fact match \emph{any} of the other three at the vertex, as these three preons are located within $\mc{L}_\psi\ll\mc{L}_0$ of one another, increasing the chance of a contested source/sink to $\frac{1}{3}$ rather than $\frac{1}{9}$. (The other three preons at the vertex are all mutually distinct by construction, and thus always represent three of the nine possibilities.)
The resulting mean factor at a single vertex may therefore be written ${{N_0}^6}[1-(3N_0)^{-1}]$ and the full labelling of FSF symmetry factors accurate to $\ILO{{N_0}^{-1}}$ is shown in \fref{fig:evalHNsyms}. 
Finally, note that if the fourth emitted preon at the lower vertex matches (or does not match) one of the other three, then the fourth absorbed preon at the upper vertex similarly matches (or does not match) one of the other three. The total factor for the two vertices together is consequently the correlated factor ${{N_0}^{12}}[1^2+1^2+(1-{N_0}^{-1})^2]/3$ and not the independent factor ${{N_0}^{12}}\{[1+1+(1-{N_0}^{-1})]/3\}^2$.
The resulting FSF symmetry factor for \fref{fig:scalarbosonmass}(ii) may thus be written
\begin{equation}
\begin{split}
n_{\vp,\bmh}&=n_{\vp,W}\left(1-\tfrac{2}{3N_0}+\tfrac{1}{3{N_0}^{2}}\right)\\
&={N_0}^{12}\left(1+\tfrac{16}{3}{N_0}^{-1}+\tfrac{28}{3}{N_0}^{-2}+\ldots\right).
\end{split}
\label{eq:HFSFsymfactor1}
\end{equation}
Note that this factor applies only to \fref{fig:scalarbosonmass}(ii)---there is no factor of $[1-2/(3N_0)+1/(3{N_0}^2)]$ attached to \fref{fig:scalarbosonmass}(iii) as the number of preon and antipreon fields from the pseudovacuum is equal. The FSF symmetry factor for diagram~(iii) is thus simply given by \Eref{eq:WFSFsymfactor} and is the same as for \fref{fig:vecbosonmass}(v). This must be kept in mind when relating diagram~(iii) to diagram~(ii) as shown in \fref{fig:scalarbosonmass}(iv)-(v).

Proceeding with the evaluation of \fref{fig:scalarbosonmass}(ii),
it is next necessary to count the diagrammatic symmetry factors associated with the different fermions of the background fields. %
To do so, recognise that all preon configurations which may validly be grouped on the same vertex as the $\bmh$ operator may be obtained by iterating over all generation-1 fermion/anti-fermion pairs and considering all ways to replace one preon with an antipreon. For example, briefly ignoring summations on spinor and colour indices, the channel indexed by $e_L\equiv\psi^{2}\psi^{2}\psi^{2}$ has associated vertices resembling
\begin{equation}
\psi^{2}\psi^{2}\bar{\psi}^{2}\bar{\psi}^{2}\bar{\psi}^{2}\bar{\psi}^{2}\bmh
\end{equation}
and hermitian conjugate. As with the $W$ boson, integration over one of the spatial degrees of freedom permits elimination of two preons and two antipreons from each vertex. The resultant expression is nonvanishing only where the preon pair and the antipreon pair remaining on the vertices of \fref{fig:scalarbosonmass} are conjugate. For the lepton channels this is trivially satisfied, whereas for the quark-labelled channels it is satisfied only when the preon which is replaced by an antipreon (or vice versa) is the unique species in the triplet (e.g.~$\psi^{3c\alpha} $ in $u_L$).

\begin{figure}
\includegraphics[width=\linewidth]{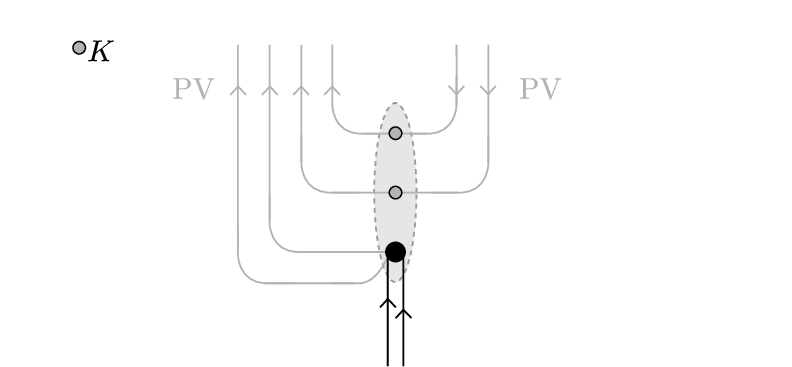}
\caption{Detail of \prm{K} matrices in a $\bmh$/preon triplet interaction vertex, with preons and antipreons grouped as in \pfref{fig:scalarbosonmass}(ii).\label{fig:scalarKmatrices}}
\end{figure}

For the resulting diagrams, the symmetry factors associated with the preon connections may then be evaluated at:
\begin{itemize}
\item 48 for each of the three first-generation lepton channel diagrams ($e_R\bsm\bar e_R$, $\bar e_L\bsm e_L$, $\bar \nu_e\bsm \nu_e$), for a total of 144, comprising $4!\times 2!$ for the reordering of the preon lines and antipreon lines, and 
\item 144 for each of the four first-generation quark channel diagrams ($\bar u_L\bsm u_L$, $u_R\bsm\bar u_R$, $\bar d_L\bsm d_L$, and $d_R\bsm d_R$), for a total of 576, comprising:
\begin{itemize}
\item $2!\times 2!$ for reordering of the lines in the preon summed pair and antipreon summed pair. 
\item $3 \times 3$ for the choice of colour of the unpaired species on each side of the interaction vertices, recalling that the boson coupling vertex is implicitly a sum over photon and gluon terms so may change the colour.
\item Noting that reversing the orientation of the unique antipreon line yields another matched preon pair, $2!$ for exchange of the lines in this pair. 
\item $2!$ for exchanging the two preon pairs---the summed pair, and the pair arising from the unique preon and antipreon on exchange of orientation of the antipreon.
\end{itemize}
Where labels match, an exchange corresponds to the associated symmetry factor. Where they do not match, it is a sum over diagrams correponding to rearrangements of these labels.
\end{itemize}
Then, for additional factors common to both leptons and quarks:
\begin{itemize}
\item Recognise that the scalar boson $\bmh%
=f\psi\psi$ contains an internal $R$-index summation, so is made up of nine terms, which may be thought of as internal channels enumerated by the $R$-index. A single scalar boson excitation will on average excite only one of these channels. The inbound and outbound $R$-channels are required to coincide, with a net outcome that each diagram attracts a factor of $\N^{-1}=\frac{1}{9}$.
\item As described in \sref{sec:scalbosint}, the $\bmh$ and $\bmh^*$ bosons on the inbound and outbound legs of the diagram are represented in the context of a propagator, and so are far field bosons attracting substantially reduced FSF symmetry factors. Expressed relative to the corresponding vector boson diagrams, these lend the diagrams of~\fref{fig:scalarbosonmass} a factor of $4\big[k%
_1{N_0}\big]^{-4}\left[1+\OO{{N_0}^{-1}}\right]$ apiece. %
\item As the background preons correspond to fermion triplets, interaction with the scalar boson may be accompanied by pairs of $K$ matrices as shown in \fref{fig:scalarKmatrices}. Although diagrams may be constructed with and without $K$ matrices, it turns out that only diagrams involving $K$ matrices contribute to the leading terms in scalar boson mass. Diagrams without $K$ matrices are discussed in \sref{sec:universalcorrs}.
\end{itemize}

For vector bosons the Lagrangian yields vertex pairs %
such as
\begin{align}
\begin{split}
-n_{\vp,W}f^2k_1^4\!\!\iint\!\!\rmd^4x\,\rmd^4y\,&\fgfield{W^\dagger_\mu(x)W_\nu(y)}%
\\
\times& \bgfield{\bar{\nu}_e(x)\bsm e_L(x)\bar{e}_L(y)\bsn\nu_e(y)}
\\
=\frac{n_{\vp,W}f^2k_1^4}{2}\!\!\iint\!\!\rmd^4x\,\rmd^4y\,&\fgfield{W^\dagger_\mu(x)W^\mu(y)}\\
\times& \bgfield{\bar{\nu}_e(x)\bar{e}_L(y) e_L(x)\nu_e(y)},
\end{split}
\end{align}
with this example, before diagrammatic symmetry factors, 
yielding $1/72$ of the contribution to the leading-order value of $m_W^2$. If the $W$ boson is replaced by an equivalent hypothetical boson diagonal on $\SU{3}_A$ and taken to couple only to $e_L$,\footnote{This field may be constructed explicitly from \protect{$A_\mu$}, \protect{$Z_\mu$}, and \prm{N_\mu}, as the latter acts trivially on both \protect{$\SU{3}_A$} and \protect{$\SU{3}_C$}, so may be considered to act as a trivial representation of either group.} and vector boson factors are written in terms of spinor derivatives acting on $\varphi$, then the application of spinor identities and integration by parts yields one of the nine internal channels of the scalar boson, not including the preon pair's foreground FSF exchange symmetry factor. %
The equivalent term involving the same components of the background fields therefore imparts mass to the scalar boson as a whole according to
\begin{align}
\frac{n_{\vp,W}f^2k_1^4}{2}\!\!\iint\!\!\rmd^4x\,\rmd^4y\,&\fgfield{W^\dagger_\mu(x)W^\mu(y)}\nn\\
\times& \bgfield{\bar{\nu}_e(x)\bar{e}_L(y) e_L(x)\nu_e(y)}\nn\\
\downarrow\label{eq:Hvert}\\
-\frac{2n_{\vp,\bmh}f^2k_1^4}{9\big[k%
_1{N_0}\big]^4}\left[1+\OO{{N_0}^{-1}}\right]&\iint\!\!\rmd^4x\,\rmd^4y\nn\\\times\fgfield{\bmh^*(x)\bmh(y)}&\,\bgfield{\bar{e}_L(x)\bar{e}_L(y) e_L(x)e_L(y)}.\nn
\end{align}
Summing across channels and symmetry factors for the $W$ boson yields a factor of $36+36=72$ on \Eref{eq:leptonvecbg1}, for a total mass-squared of $m_W^2$. An equivalent summing for \fref{fig:scalarbosonmass}(ii) %
yields $144+576=720$, for an apparent %
mass-squared of 
\begin{equation}
-\frac{40m_W^2}{9\big[k%
_1{N_0}\big]^4}\frac{n_{\vp,\bmh}}{n_{\vp,W}}\left[1+\OO{{N_0}^{-1}}\right].\label{eq:rawHmass}
\end{equation}
This plays a role similar to $m_W^2$ in the Lagrangian, and thus appears in a term with \emph{positive} overall sign (whereas $m_W^2$ appears preceded by a negative sign).
However, %
the propagator term for the complex scalar boson similarly takes the form
\begin{equation}
-2\big[k%
_1{N_0}\big]^{-4}\left[1+\OO{{N_0}^{-1}}\right]\bmh^*\square\bmh\tag{\Pref{III}{eq:Lscalarprop}} %
\end{equation}
with the same higher-order terms in $\big[k%
_1{N_0}\big]^{-1}$ as the mass interaction, and the Lagrangian for the free massive complex scalar boson is consequently
\begin{equation}
\begin{split}
\mscr{L}_{\bmh,m_\bmh}=-&\,\frac{2}{\big[k%
_1{N_0}\big]^4}\left(\bmh^*\square\bmh-\frac{20m_W^2}{9}\frac{n_{\vp,\bmh}}{n_{\vp,W}}\bmh^*\bmh\right)\\&\times\left[1+\OO{{N_0}^{-1}}\right],
\end{split}\label{eq:firstmhsq}
\end{equation}
yielding [at least, for $m_W^2$ and $m_{\bmh}^2$ evaluated to leading order, and ignoring \fref{fig:scalarbosonmass}(iii)]
\begin{equation}
m_{\bmh}^2=\frac{20m_W^2}{9}.
\end{equation}
Recognising that the factor of $-2\big[k%
_1{N_0}\big]^{-4}\left[1+\OO{{N_0}^{-1}}\right]$ in \Eref{eq:firstmhsq} is a common factor to both terms and can be divided out of $\mscr{L}_{\bmh,m_\bmh}$ with no effect on the dynamics of $\bmh$, it is then worthwhile to revisit the difference between FSF symmetry factors for the $W$ and $\bmh$ bosons. \Eqrefs{eq:WFSFsymfactor}{eq:HFSFsymfactor1} differ by a factor of $\left[1-2/(3N_0)+1/(3{N_0}^2)\right]$, correcting only the mass term in $\mscr{L}_{\bmh,m_\bmh}$ to yield
\begin{equation}
m_{\bmh}^2=\frac{20m_W^2}{9}\left(1-\frac{2}{3N_0}+\frac{1}{3{N_0}^2}\right).\label{eq:firstHmass}
\end{equation}
Taking $m_W=80.360(16)~\GeV$~\cite{the-ATLAS-collaboration2023}
and substituting the value of $N_0$ from \Eref{eq:N0value} gives a first approximation to the leading-order value of $m_{\bmh}^2$ of $119.59(2)~\GeV/c^2$ (ignoring the unquantified uncertainty from higher-order terms in $N_0$). %

Now consider \fref{fig:scalarbosonmass}(iii). To evaluate this figure it is useful to employ a technique known as diagrammatic isotopy \cite{kitaev2006,bonderson2007}. When the preons are viewed as intrinsically massless particles, they may be understood as a representation of a conformal field theory (CFT). Their self-consistent exchange relationships imply that this is a braided CFT, and up to the symmetry factors absorbed into the definition of the fermion its vertices also satisfy requirements for unitarity. The modular $S$-matrix ($S_{ab}$) is trivial, ensuring that when restricted to the surface of a torus the CFT is also modular. Provided the number of fermion vertices is held constant, the preon model is consequently a unitary, braided, modular conformal field theory. All particle representations are of dimension~1 and thus Feynman diagrams are automatically normalised in the diagrammatic isotopy convention (up to the factors associated with the definition of fermions). Preon lines may therefore be freely deformed so long as the topology of the diagram remains intact in the sense employed in \rcitess{kitaev2006,bonderson2007}, and the fermion count is left unchanged or appropriate compensatory factors are introduced.

As indicated in the caption to \fref{fig:scalarbosonmass}, diagrammatic isotopy and braiding permit diagram~(iii) to be redrawn as shown in diagram~(v), resembling \fref{fig:scalarbosonmass}(ii) supplemented by an foreground additional loop correction of dimension $L^{-1}$. The value of diagram~(iii) is therefore the value of \fref{fig:scalarbosonmass}(ii) modified by the factor arising from the presence of the loop in diagram~(v)---and also without the reduction in FSF symmetry factors discussed in \fref{fig:evalHNsyms}. The correct FSF symmetry factors are yielded by diagram~(iv) of \fref{fig:scalarbosonmass}, and it is also convenient to evaluate the loop factor on this diagram.

Before doing so, consider diagram~(iii) once more. The constituent preons of the complex scalar boson in diagram~(iii) form a bound pair and thus are separated by a distance of $\ILO{\mc{L}_\preon}$. This distance is smaller than the minimum scale $2\mc{L}_\Omega$ associated with mass interactions (see \srefs{sec:Wmass5v}{sec:chromenv}), and thus when the preons making up the loop corrections in diagrams~(iv) and~(v) traverse this length scale, they do so as massless particles (as the correction is dominated by these short-range contributions). Further, diagram~(iii) represents a mass vertex and thus is implicitly truncated. It is convenient to reintroduce short line segments on the foreground preons to assist in evaluation using diagrammatic isotopy, but these line segments {must} not engage in any further couplings, including mass vertices. For this reason the loop boson in diagrams~(iv)-(v) must be massless.

Since the aim of diagrams~(iii)-(v) is to write \fref{fig:scalarbosonmass}(iii) as a correction to \fref{fig:scalarbosonmass}(ii), and in the latter the vertex factors are the same as those associated with the composite vertices in \fref{fig:scalarbosonmass}(v) but arise entirely from the eight-preon vertices marked by a solid dot $\bullet$, these vertex factors must be recovered after integration out of the loop correction in \fref{fig:scalarbosonmass}(iv). The presence of the loop correction does not affect the reduction of the pseudovacuum lines to a numerical coefficient, and evaluation of this loop correction is therefore performed independently of that reduction, with vertex factors of~$1$. Given a specific labelling of the inbound and outbound preons, the foreground loop is one of nine terms arising from a complex scalar boson loop. The boson mass is zero as discussed above, and with no sum over nine different terms the scalar boson loop correction is thus equivalent to a photon loop up to the following relative factors: 
\begin{itemize}
\item Coupling coefficients yield $1$, not $\alpha$, for a relative factor of~$\alpha^{-1}$.
\item The asymmetry of background preons in diagram~(v) (four preons and two antipreons at once vertex, and the opposite at the other) is taken to be associated with the $\bmh$/preon vertices, leaving the $\bmh$/vector boson loop symmetric under exchange of source and sink vertices, like the photon. Relative symmetry factor of~1.
\end{itemize}
The loop therefore yields a net factor of $1/(2\pi)$. Recalling that also, in contrast with \fref{fig:scalarbosonmass}(ii), diagram~(iii) does not attract a factor of $\left[1-2/(3N_0)+1/(3{N_0}^2)\right]$, \Eref{eq:firstHmass}
is therefore corrected to 
\begin{equation}
m_{\bmh}^2=\frac{20m_W^2}{9}\left[\left(1-\frac{2}{3N_0}+\frac{1}{3{N_0}^2}\right)+\frac{1}{2\pi}\right]\label{eq:secondHmass}
\end{equation}
giving the leading-order value
\begin{equation}
m_{\bmh}=128.78(3)~\GeV/c^2 \label{eq:mH} %
\end{equation}
(again ignoring the unquantified uncertainty from higher-order terms in $N_0$).
This will %
be %
further improved upon by higher-order corrections performed in \cref{ch:detail}.

\subsection{Summary of boson masses\label{sec:bosonmasssummary}}

\subsubsection{Photon}
The photon, $A$, 
attracts no mass to all orders by construction of gauge choice~\eref{eq:ma3gaugenew}.

\subsubsection{$W$ and $Z$ bosons\label{sec:Wbosonmass}}
The $W$ and $Z$ bosons %
attracts the majority of their mass from \fref{fig:vecbosonmass}(v). In contrast to the photon, this is nonvanishing for the $Z$ boson due to symmetry breaking in the quark sector (\sref{sec:GL18Cgauge}). 
The resulting leading-order expressions are
\begin{align}
m_W^2&=9f^2k_1^4{N_0}^{12}{\omega_0}^2\tag{\ref{eq:mWsq}a}\\
m_Z^2&=12f^2k_1^4{N_0}^{12}{\omega_0}^2\tag{\ref{eq:Zmass}a}
\end{align}
yielding
\begin{equation}
\sin^2\theta_W=1-m^2_W/m^2_Z=0.25.
\end{equation}
This only very crudely approximates the observed figure of $0.22290(30)$~\cite{tiesinga2018}, %
but higher-order corrections are calculated in \cref{ch:detail} and yield %
better agreement with observation.

\subsubsection{Gluons and \prm{N}~boson\label{sec:gluonsAndNmass}}
In principle the gluons in $\Cw{18}$ (including the $N$~boson, which predominantly behaves like a ninth gluon) participate in a mass interaction, acquiring mass through similar mechanisms to the $W$ boson. To leading order their masses satisfy $m_c^2\sim m_W^2$. However, it is questionable whether this mass can be observed for any but the $N$~boson, as the other eight gluons are confined on an energy scale $\mc{E}_\preon$ which is large compared with the scale of the mass interactions, $\mc{E}_0$ (see \sref{sec:chromenv}).

\subsubsection{\prm{G}~boson}
At the present stage of exposition of the $\Cw{18}$ model (with flat emergent spacetime) the $G$ boson is present in the model and is a massless particle. In principle it could acquire mass through the same mechanisms as the $W$ boson, but the choice of gauge which zeros the magnitude of background $G^\bdag$ fields also necessarily prevents $G^\bdag$~bosons from acquiring mass through coupling to the background fields. %
The $G^\bdag$~bosons therefore appear to constitute a novel right-handed weak interaction, but are subsequently eliminated in \sref{sec:Rwnf}.

\subsubsection{Scalar boson}
The scalar boson $\bmh$ is a massive particle, with mass given %
approximately %
by
\begin{align}
m_{\bmh}^2&=\frac{20m_W^2}{9}\left[\left(1-\frac{2}{3N_0}+\frac{1}{3{N_0}^2}\right)+\frac{1}{2\pi}\right]\tag{\ref{eq:secondHmass}}\\
N_0&=\sqrt{\frac{m_W}{3\sqrt{2}m_e}}\left[1+\OO{{N_0}^{-1}}+\OO{\alpha}\right].
\tag{\ref{eq:N0value}}
\end{align}

\subsubsection{Extension to the bosons of $\GLNR$\label{sec:ACbosonmassespre}}

As noted in \sref{sec:symC18}, separability of $A$-sector and $C$-sector charges is an approximation only valid in the %
high-particle-number regime. However, on extending gauge choices~\erefs{eq:bga45gauge}{eq:ma3gaugenew} to the full $\GLNR$ symmetry as described in \aref{apdx:gaugeSU9}, all colourations of the photon and $G$~boson are made massless and with current experimental capabilities it is not practical to distinguish between a massless gluon having the trivial $A$-sector representation matrix $\lambda^A_9$ and a massless coloured photon having $A$-sector representation matrix $\lambda^A_3$. 

For the $W$ and $Z$ bosons, there exist chromatic counterparts which have $A$~sector representations equivalent to the usual $W$ or $Z$ bosons while also carrying nontrivial representations on the $C$~sector. In \cref{ch:detail} it is seen that these species %
have masses distinct from those of the usual $W$ and $Z$ bosons (which are associated with the trivial representation $\lambda^C_9$ on the $C$~sector). Detection of these coloured species is discussed in \cref{ch:CDF2}. There is substantial evidence that detection of these species
has already taken place.

\section{Conclusion}

This chapter has introduced the mass mechanisms for bosons in the $\Cw{18}$ model, and illustrated this with a tree-level calculation of the leading terms in $W$, $Z$, and complex scalar boson ($\bmh$) masses. An interesting consequence of these calculations is the ability to relate the ratios of the lepton and electroweak boson scales to the number of fundamental scalar fields within a locally correlated region of the pseudovacuum. When calculated to tree level, the electroweak mass ratios $m_W^2/m_Z^2$ and $m_W^2/m_\bmh^2$ are in crude qualitative agreement with those of the Standard Model.

In the $\Cw{18}$ model the electroweak bosons are predicted to exhibit generations, with the lightest second-generation boson being a heavy $W$ boson with mass of just under $17~\TeV/c^2$. However, this energy scale exceeds the limit at which the $\Cw{18}$ model behaves as an analogue to a quantum field theory, and the second-generation $W$ boson is \emph{less} able to appear as a virtual particle in the $\Cw{18}$ model than would be anticipated from quantum field theory. With divergence of the $\Cw{18}$ model and the Standard Model, caution must be exhibited in extrapolating to this regime.

\notchap{
\section*{Acknowledgements}
This research was supported in part by the Perimeter Institute for Theoretical Physics.
Research at the Perimeter Institute is supported by the Government of Canada through Industry Canada and by the Province of Ontario through the Ministry of Research and Innovation.
The author thanks the Ontario Ministry of Research and Innovation Early Researcher Awards (ER09-06-073) for financial support.
This project was supported in part through the Macquarie University Research Fellowship scheme.
This research was supported in part by the ARC Centre of Excellence in Engineered Quantum Systems (EQuS), Project No.~CE110001013.
}

\chapter{{Particle} generations and masses on \protect{$\mathbb{C}^{\wedge18}$}\label{ch:detail}}

\begin{abstract}
The $\Cw{18}$ analogue model contains counterparts to the full particle spectrum and interactions of the Standard Model, but has only three tunable parameters. %
Predictive relationships may therefore be obtained between its counterparts to many constants of the Standard Model. In this chapter, the model values for the masses of the tau, the $W$ and $Z$ bosons, and a Higgs-like scalar boson are calculated as functions of $\alpha$, $m_e$, and $m_\mu$, with no free fitting parameters. They are shown to be $1776.867413(43)~\MeV/c^2$, $80.3587(22)~\GeV/c^2$, $91.1877(35)~\GeV/c^2$, and $125.1261(48)~\GeV/c^2$ respectively. All are within $0.2\,\sigma$ or better of the corresponding observed values of $1776.86(12)~\MeV/c^2$, $80.360(16)~\GeV/c^2$, $91.1876(21)~\GeV/c^2$, and $125.11(11)~\GeV/c^2$ respectively. These results are suggestive of the existence of a unifying relationship between lepton generations and the electroweak mass scale, which in the $\Cw{18}$ model arises from preon interactions mediated by the strong force.
\end{abstract} 

\section{Introduction}

Introduced in \crefr{ch:simplest}{ch:boson}, the $\Cw{18}$ model is a classical analogue model \cite{maynard2001,dragoman2004,lewenstein2007} which comprises free scalar fields (Fundamental Scalar Fields, FSFs) on a manifold with 18 anticommuting complex co-ordinates, denoted $\Cw{18}$. 
When the FSFs are in a highly disordered and hence (from a coarse-grained perspective) a highly homogeneous state, the product of these fields on a $\mbb{R}^{1,3}$ submanifold admits a description as a pseudovacuum state on which exist soliton waves. These soliton waves, in turn,
behave as interacting quasiparticles governed by the low-energy regime of a quantum field theory with a gaugeable local symmetry. 
The emergent quasiparticles behave as coloured fermionic preons, and they in turn condense into fermions, quarks, and a scalar boson, with choices of gauge yielding a particle spectrum closely resembling that of the Standard Model, supplemented only by
\begin{itemize}
\item a ninth gluon, with mass on the electroweak scale and negligible interactions other than gravity (\cref{ch:gravity}),
\item higher generations of electroweak bosons, with the lightest being $W_2$ at $16.61320(46)~\TeV$, and %
\item a weakly interacting massless complex vector boson $G$ which is eliminated on extending the model to curved space--times (\sref{sec:Rwnf}). 
\end{itemize}
Anticipating %
the weakness of the residual effects demonstrated in
\cref{ch:gravity}, the species $G^\bdag$ may be ignored in the current chapter without introducing significant error.

Previously, \crefs{ch:fermion}{ch:boson} %
have introduced the boson and lepton mass interactions of the $\Cw{18}$ model at tree level, in which the emergent composite bosons and leptons acquire mass through coupling to the high-entropy pseudovacuum. 
At tree level, the mass ratios $m_W/m_Z$ and $m_W/m_\bmh$ of the electroweak bosons have been shown to be in rough qualitative agreement with the Standard Model. A similar calculation at tree level for leptons is not possible due to vanishing of the tree-level electron mass, but order-of-magnitude estimates for higher-order corrections suggest that the electron gains a small mass once the 1-loop electroweak corrections are taken into account, and that the model is then not conspicuously inconsistent with the observed mass ratios $m_\mu/m_e$ and $m_\tau/m_e$. In the present chapter the boson and lepton mass calculations are performed to higher order, and relationships between the sectors are identified which permit the three input parameters of the $\Cw{18}$ model to be taken as $\alpha$, $m_e$, and $m_\mu$. The calculated values of $m_W$, $m_Z$, $m_\bmh$, and $m_\tau$ are then seen to be in exceptional agreement with observation.

\section{Conventions}

This chapter follows the same conventions as \crefr{ch:simplest}{ch:boson}. 
\standalone{Units are chosen such that $c=1,~h=1$.
When equations and lemmas from \crefr{ch:simplest}{ch:boson} are referenced, they take the forms (\textbf{1}.1), (\textbf{2}.1), %
etc.}

In this Chapter, it is generally assumed that any particle under study is at rest or near-rest with respect to the isotropy frame of the pseudovacuum. 

When referring to uncertainty in results, experimental uncertainties are denoted $\sigma_\mrm{exp}$, and uncertainties in theoretical calculations are denoted $\sigma_\mrm{th}$.

\standalone{Regarding terminology around Feynman diagrams and symmetry factors:
\begin{itemize}
\item Where there exist multiple ways to connect up sources, vertices, and sinks to obtain equivalent diagrams up to interchange of non-distinguishable co-ordinates, the same term is obtained from the generator $\Z$ in multiple different ways and thus the diagram acquires a multiplicative factor. This is referred to in the present volume as a \emph{symmetry factor}.
\item Where integration over the parameters of a diagram (for example, over source/sink co-ordinates) 
yields the same diagram multiple times up to interchange of labels on these parameters, 
this represents a double- (or multiple-)counting of physical processes. It is then necessary to eliminate this multiple-counting by dividing by the appropriate symmetry factor. This is referred to in the present volume as \emph{diagrammatic redundancy} or \emph{double- (multiple-)counting.}
\end{itemize}
}

\section{The pseudovacuum}

The FSFs are defined on manifold $\Cw{18}$ and denoted $\vp_q$. Defining a submanifold $M\subset\Cw{18}$ such that $M\cong\RM$, the fields obtained on mapping the FSFs to $\RM$ are denoted $\vp_q(x)$. Since $\RM$ (but not $M$) may support fields of arbitrary power in $x$, the product field $\vp(x)$ is defined on $\RM$ as
\begin{equation}
\varphi(x) := \prod_q\varphi_q(x)\tag{\Pref{I}{eq:defvp}}.
\end{equation}
The observed quasiparticles are then constructed from the gradients of the product field with respect to translations of the $\RM$ submanifold within $\Cw{18}$.
The pseudovacuum is characterised by an energy scale $\mc{E}_0$, a mean expectation value
\begin{equation}
f^{-1}:=\la\vp\ra,
\end{equation}
and a particle number $N_0$ being the number of FSFs having a point of inflection within a hypervolume 
\begin{equation}
{\mc{L}_0}^4:={\mc{E}_0}^{-4}. 
\end{equation}
Each FSF has at most one point of inflection on $M$.
Introducing an energy per FSF, 
\begin{equation}
\omega_0:=\frac{\mc{E}_0}{N_0},
\end{equation}
the macroscopic properties of the pseudovacuum may be parameterised in terms of ($f$,\,$N_0$,\,$\omega_0$) of which $f$ and $N_0$ are unitless.

On $\RM$, the pseudovacuum may conveniently be described in terms of its non-vanishing expectation values. In %
the notation of \crefr{ch:simplest}{ch:boson} the non-vanishing bosonic components are
\begin{align}
\la\bgfield{A^\mu(x)A_\mu(y)}\ra &= -\bm{f}(x-y){\mc{E}_0}^2\label{eq:AQL}\\
\la\bgfield{c^{\tc\mu}(x)c^\td_\mu(y)}\ra&=-\bm{f}(x-y){\mc{E}_0}^2\label{eq:cQL}
\end{align}
where $\bm{f}(x)$ is a Gaussian satisfying
\begin{equation}
\sol{c^{-3}}\!\!\int\!\rmd^4x%
\,{\mc{E}_0}^4\,\bm{f}(x-y)=1\quad\forall\quad y.\tag{\Pref{I}{eq:xynorm}}
\end{equation}
As noted in \Psref{III}{sec:consequences}, the $\SU{3}\oplus\GL{1}{R}$ invariance of the $C$~sector at the preon scale implies that the background fields in \Eref{eq:cQL} need not form a conjugate pair to have nonvanishing pseudovacuum expectation values.

The leptons are made up of three differently-coloured preons, generically
\begin{align}
\Psi^{ag\alpha}(x) \propto &\left(\varepsilon^{\alpha\beta}\varepsilon^{\gamma\delta}-\varepsilon^{\alpha\gamma}\varepsilon^{\beta\delta}+\varepsilon^{\alpha\delta}\varepsilon^{\beta\gamma}\right)\label{eq:compositeleptons}
\\\nn&\times
\mc{C}^g_{c_1c_2c_3}\psi^{ac_1}_\beta(x_1)\psi^{ac_2}_\gamma(x_2)\psi^{ac_3}_\delta(x_3)
\end{align}
following \PEref{III}{eq:compositeleptonspre}.
In this, $g$ is the generation index, and the coefficients $\mc{C}^g_{c_1c_2c_3}$ are constrained by requiring that particle $\Psi^{a\alpha}$ be both colourless and an eigenstate of the mass-generating interaction with the pseudovacuum.
The preon expectation values satisfy
\begin{align}
\begin{split}
\la\bgfield{\bar\psi^{\dot{m}}(x)&\,\bar\psi^{\dot{n}}(y)\,\psi^m(x)\,\psi(y)^n}\ra\\
&=\la\bgfield{\bar\psi^{\dot{m}}(x)\,\bar\psi^{\dot{n}}(x)\,\psi^m(y)\,\psi(y)^n}\ra\\
&=\frac{1}{2f^2}\,\delta_{\dot mn}\delta_{\dot nm}\,\bm{f}(x-y)\,{\mc{E}_0}^2
\end{split}\label{eq:xycorr1}
\end{align}
where $m$ (or $\dot m$) and $n$ (or $\dot n$) range from 1 to 9 and enumerate pairs of index values $a~(\mrm{or}~\dot a)\in\{1,2,3\}$, $c~(\mrm{or}~\dot c)\in\{r,g,b\}$.
The scalar boson is a sum over nine terms, 
\begin{equation}
\bmh:=f\psi^m\psi_m,
\end{equation}
and thus satisfies
\begin{equation}
\la\bgfield{\bmh(x)\bmh^*(y)}\ra=\frac{9}{2}\,\bm{f}(x-y)\,{\mc{E}_0}^2.\label{eq:HQL}
\end{equation}

The field $A_\mu$ corresponds to the photon, and $c^\tc_\mu$ corresponds to gluons associated with the Gell-Mann basis of $\SU{3}_C$ plus an additional diagonal species, as enumerated in \PErefr{II}{eq:Cbasis1}{eq:Cbasis2}. The construction of the pseudovacuum introduces a preferred rest frame, but this is essentially undetectable at energy scales small compared with 
\begin{equation}
\frac{1}{2}\mc{E}_\Omega:=\frac{\N}{2}{N_0}(N_0-\tfrac{1}{2})\omega_0\quad|\quad\N=9,\Ptagref{V}{eq:V:EOmega}
\end{equation}
which is seen to be approximately $3.1~\TeV$~\eref{eq:EOmegavalue}. %
To leading order the value of $N_0$ is
\begin{equation}
\begin{split}
N_0&=\sqrt{\frac{m_W}{3\sqrt{2}m_e}}\left[1+\OO{{N_0}^{-1}}+\OO{\alpha}\right]\\&\approx 193.
\end{split}\Ptagref{V}{eq:N0value} %
\end{equation}

\section{Boson Mass Interactions\label{sec:VI:bosonmasses}}

\subsection{$W$ mass\label{sec:Wmass}}

In \Psref{V}{sec:vecbosonmasses}, a first-order expression for the $W$ boson mass was obtained in terms of the free parameters of the $\Cw{18}$ model, $f$, $N_0$, and $\omega_0$:
\begin{equation}
\begin{split}
m_W^2&=9f^2{k_1}^4{\omega_0}^2{N_0}^{12}\left[1+\OO{{N_0}^{-1}}\right]\\
&=18m_e^2{N_0}^4\left[1+\OO{{N_0}^{-1}}\right].
\end{split}\tag{\Pref{V}{eq:mWfromme}}
\end{equation}
To obtain the high-precision numerical results presented in the present chapter, it is necessary to evaluate some higher-order corrections to this expression. However, before proceeding with these corrections there is another, more important question---which $W$?

As noted in \aref{apdx:gaugeSU9}, bosons in the model on $\Cw{18}$ all carry representations on both the $A$~and $C$~sectors. The conventional $W$ boson carries representation $\frac{1}{\sqrt{2}}(\lambda^A_6-\rmi\lambda^A_7)$ on the $A$~sector and the trivial representation $\lambda^C_9$ on the $C$~sector. However, there are also coloured $W$~bosons having nontrivial representations on the $C$~sector and these attract additional mass interactions beyond those of the colourless $W$ boson. The mass of the colourless $W$ boson is considered first, with calculation of the mass of coloured $W$ bosons being deferred to \sref{sec:colouredWmass}.

\subsubsection{Boson loops---overview\label{sec:Wmassbosonloops}}

\begin{figure}[p]
\begin{center}
\includegraphics[width=4.25in]{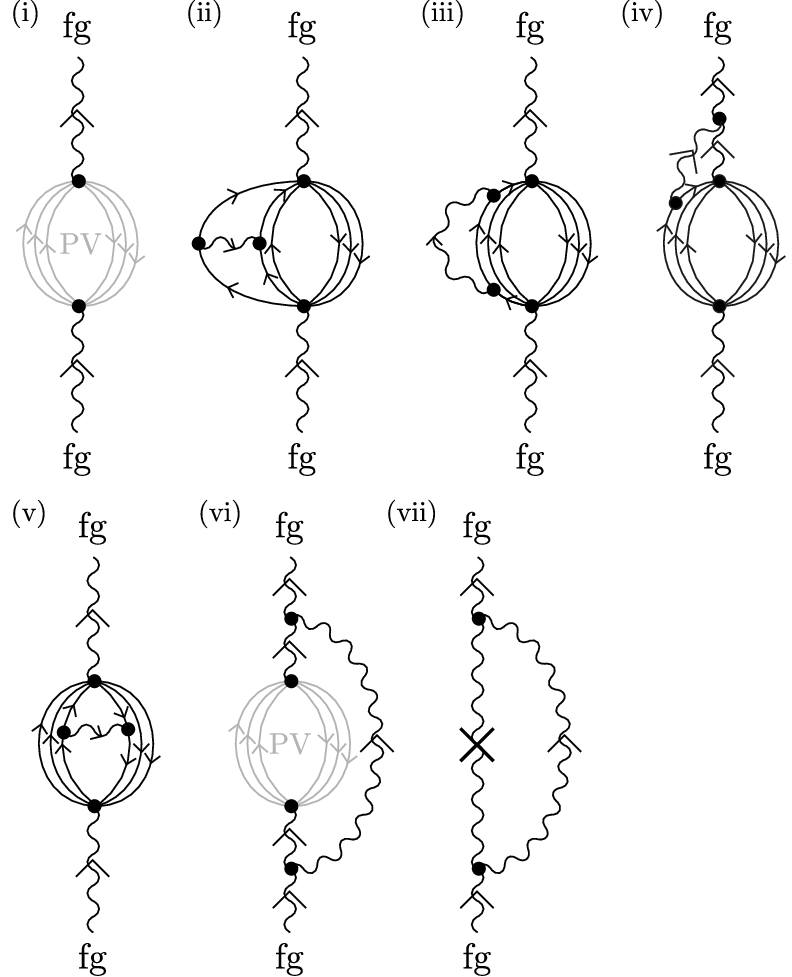}
\end{center}
\caption{(i)~Leading-order contribution to the \protect{$W$} boson mass. External legs are truncated. 
(ii)-(v)~Candidate structures for vector boson loop corrections to diagram~(i). Note that loop diagrams may (ii)-(v)~modify the \protect{$W$} boson mass-squared interaction, or (vii)~merely involve it. They must continue to include six background preons, but on diagrams~(ii)-(v) there are now more than six preon/antipreon lines so there are choices as to which are expanded using the mean background field approximation. These choices are not shown here, so it is not specified %
which preons are foreground and which are background in these diagrams. In diagrams~(i) and (vi), PV indicates pseudovacuum. To consider in turn the structures shown:
The loop in diagram~(ii) may be contracted onto either the upper or lower vertex, and thus this constitutes a loop correction to diagram~(i). The loop in diagram~(iii) does not contract onto a vertex and thus constitutes part of the preon propagator. It does not contribute to loop corrections, and is already implicit in the mean-field substitution of diagram~(i). Diagram~(iv) looks like a correction to diagram~(i) but is more properly understood as involving rather than contributing to the mass vertex. %
Diagram~(v) looks like it could yield corrections to diagram~(i) but on expanding in terms of preons as per \pfref{fig:Wmassbosoncorrsv}(i) it may be shown to contain implicit preon tadpoles and therefore vanishes. %
Diagram~(vi) involves but does not contribute to the vertex, and its Standard Model counterpart is shown in diagram~(vii). All diagrams are discussed further in the text.\label{fig:Wmassbosoncorrsnew}}
\end{figure}

The leading contribution to the mass of the $W$ boson arises from interactions between the $W$ boson and the fermion components of the pseudovacuum. This interaction is corrected by numerous boson loops, 
with the candidate structures being illustrated in \fref{fig:Wmassbosoncorrsnew}. 
Note that these structures may include loop corrections taking place over length scales small compared with the preon binding scale $\mc{L}_\preon$. On these scales, bosons are understood as preon/antipreon or preon/preon pairs; however, the pair acts with the same net ineraction strength and representation in $\GLNR$ as the corresponding boson, and thus it is convenient to retain the notation of bosons at these scales as a shorthand for the associated pairs. %

With regards to \fref{fig:Wmassbosoncorrsnew}, many of these diagrams may be rapidly dismissed. To contribute a loop correction to \fref{fig:Wmassbosoncorrsnew}(i), a diagram must contain a loop which may be contracted down onto one of the vertices of \fref{fig:Wmassbosoncorrsnew}(i). Diagram~(ii) meets this criterion, but diagram~(iii) does not as the loop contracts down onto an arbitrary point on the preon line. Diagram~(iii) is therefore a correction to the preon propagator, so is already accounted for in the mean-field expansion of the pseudovacuum in diagram~(i). Diagram~(iv) also looks like it could contain a loop which could contract onto the upper vertex, but is better understood as a diagram \emph{involving} the mass vertex rather than a \emph{correction to} the mass vertex. This distinction is discussed further in the context of diagram~(vi). %

\begin{figure}
\includegraphics[width=\linewidth]{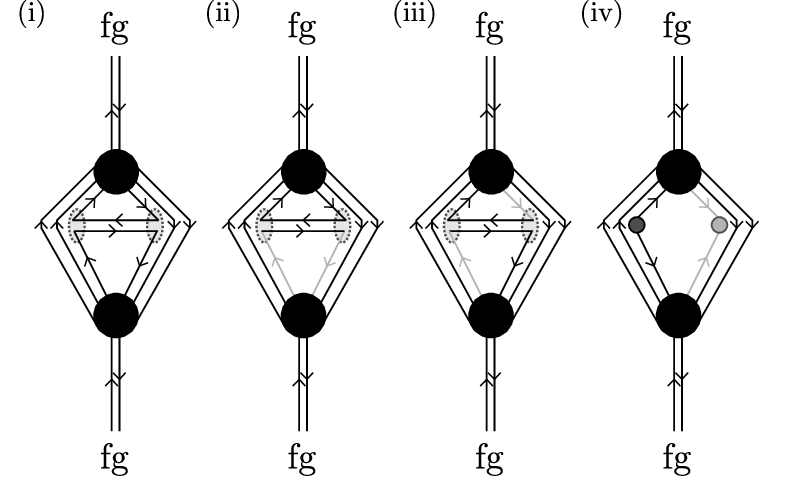}
\caption{(i)~Preon expansion of \pfref{fig:Wmassbosoncorrsnew}(v). (ii)~When both lower or both upper preon lines connecting to the loop boson are background, on averaging over many interactions the foreground field lines behave as a tadpole diagram. (iii)~When one upper and one lower preon line are background, perturbations about the mean background value cannot transport foreground momentum as this would once again result in an upper or lower tadpole for the momentum transported. This implies that the momenta associated with the background preons at their vertices with the loop boson are not independent, permitting diagram~(iii) to be redrawn as diagram~(iv). The grey dots are orientation-reversing (and perhaps colour-changing) vertices. The colour transforms they perform are correlated, but they do not exchange momenta. They may be absorbed into the lower vertex, effectively reducing the number of background field lines by one.\label{fig:Wmassbosoncorrsv}}
\end{figure}

Diagram~(v) requires a little more attention as it contains a loop which can be contracted onto the upper or lower vertex. It is expanded in terms of preons in \fref{fig:Wmassbosoncorrsv}(i). 
Two of the preons connecting with the loop boson must be expanded about the pseudovacuum term, and where these are both on the same vertex as in, for example, \fref{fig:Wmassbosoncorrsv}(ii), the average over many such interactions reduces to constant factors. The foreground momenta coupling to the loop boson then constitute a tadpole diagram which vanishes. When the pseudovacuum terms are on different vertices as in \fref{fig:Wmassbosoncorrsv}(iii), recognise that the mean background field term is only the first in a perturbative series describing fluctuations in the pseudovacuum. Some foreground momentum may therefore be transferred along pseudovacuum lines in the form of these fluctuations, but this transfer again constructs a tadpole so such contributions to foreground momentum transport necessarily vanish. The foreground and background momenta therefore propagate independently, equivalent to \fref{fig:Wmassbosoncorrsv}(iv). As the two vertices where the background field contacts the loop boson are not capable of independently borrowing or lending momentum to the foreground fields, the effective number of interface vertices between the foreground and background fields is reduced by two. This diagram therefore does not mediate a coupling between the $W$~boson and the background fermion fields (which require six preon and six antipreon operators), and thus is not part of the mass series for $W$.

Finally, diagram~(vi) again does not contribute as it requires non-truncated external legs. In fact, diagrams~(iv) and~(vi) are better understood as processes \emph{involving} rather than \emph{correcting} diagram~(i). 
This is readily seen for diagram~(vi) when the leading-order interaction of \fref{fig:Wmassbosoncorrsnew}(i) is associated with a simple mass vertex $m^2_W W^\dagger W$. In both the $\Cw{18}$ model and the Standard Model, expansion of the propagator yields higher-order corrections to the mass vertex as shown in diagram~(vii). In the Standard Model a renormalisation scheme such as $\MSbar$ may be chosen to eliminate these terms, such that the coefficient on the simple mass vertex corresponds to the true mass of the $W$ boson. This also holds for the $\Cw{18}$ model in the regime in which it is a good analogue to the QFT. The correction given in diagram~(vi) is therefore a Proper Self Energy (PSE) term extraneous to the simple mass vertex in both the Standard Model and the $\Cw{18}$ model. For diagram~(iv) the situation is similar: For a gluon, the net colourlessness of the $W$~boson causes this diagram to vanish, whereas for $A$-sector bosons, on summing over all possible locations for the lower vertex to couple to a preon or antipreon, conservation of charges and momenta ensures that equivalency is recovered to a situation where a loop boson couples two separate locations in the propagator of the $W$ boson.

\begin{figure}
\begin{center}
\includegraphics[width=5.25in]{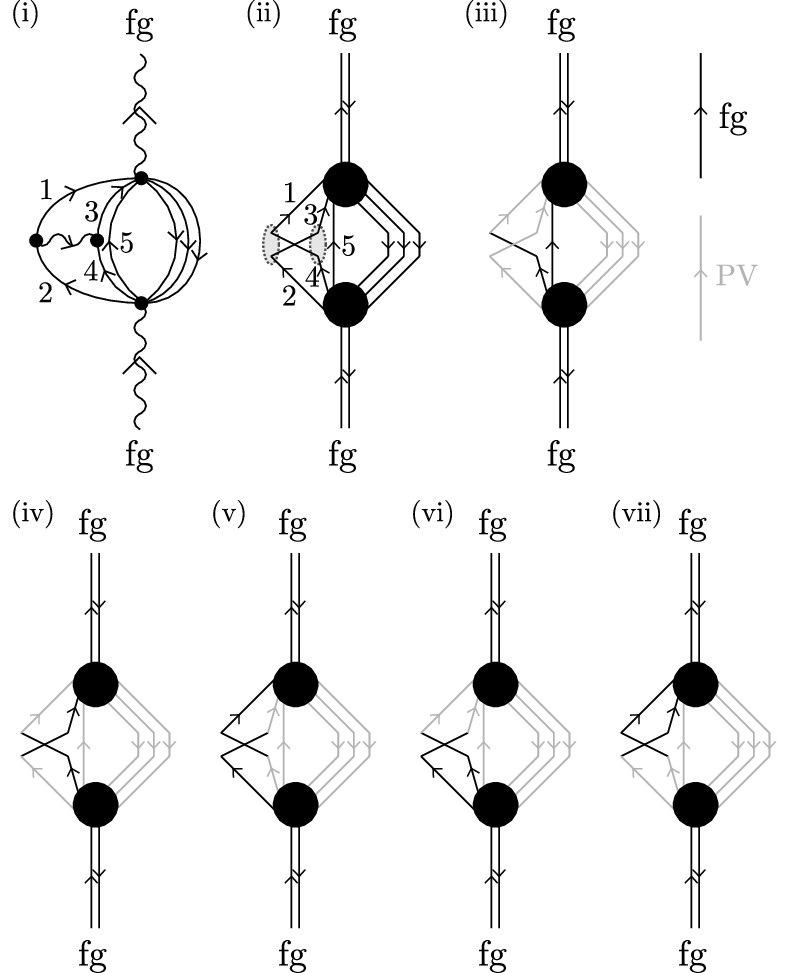}
\end{center}
\caption{(i)~Taking the specific diagram shown in \pfref{fig:Wmassbosoncorrsnew}(ii), label the preon lines as shown. (ii)~Preon form of diagram~(i). %
(iii)~Option~1 in \ptref{tab:listofterms} has an effective foreground connection between upper and lower vertices, discussed further in the text, and thus only five background preon lines. It is not a valid \prm{W}~boson mass term. Options~2, 4, 6, 7, and~8 may be discarded similarly. Options~3 and~10 are shown in diagrams~(iv) and~(v) and are seen to be %
equivalent up to two braids and a cycling in colour labels, so only one need be retained. The other valid diagrams are options~5 and~9, shown in diagrams~(vi) and (vii) respectively. In diagrams~(iii)-(vii), foreground preons are drawn in black and pseudovacuum preons in grey.\label{fig:Wmassbosoncorrsdetail}}
\end{figure}

Having eliminated the other candidate diagrams, the only corrections remaining to be evaluated are those having the form of \fref{fig:Wmassbosoncorrsnew}(ii). Since these diagrams constitute couplings to background fermions, they must contain couplings between the $W$ boson, three background preon lines, and three background antipreon lines. Numbering the preon lines as per \fref{fig:Wmassbosoncorrsdetail}(i) there are ten candidate terms in the mean background field expansion, listed in \tref{tab:listofterms}.
\begin{table}
\caption{List of labellings of preon lines in \pfref{fig:Wmassbosoncorrsdetail}(i). Labellings which connect the upper and lower vertices with a foreground preon are described as ``connected'' and may be discounted.\label{tab:listofterms}}
\begin{center}
\begin{tabular}{c|c|c|c}
Option&Background&Foreground&Status\\\hline\hline
1&1,2,3&4,5&Connected\\ %
2&1,2,4&3,5&Connected\\ %
3&1,2,5&3,4&Valid\\ %
4&1,3,4&2,5&Connected\\ %
5&1,3,5&2,4&Valid\\ %
6&1,4,5&2,3&Connected\\ %
7&2,3,4&1,5&Connected\\ %
8&2,3,5&1,4&Connected\\ %
9&2,4,5&1,3&Valid\\ %
10&3,4,5&1,2&Equivalent to option 3  %
\end{tabular}
\end{center} 
\end{table}

Examining option~1, in this diagram only one foreground preon (line~4) connects with the loop boson. The background fields proceeding from the loop boson to the upper vertex (lines~1 and~3) may transport foreground momentum as fluctuations around the background mean field state, but the parameter space of this connection corresponds to two preons, and is therefore over-large for transmitting the momentum of a single foreground preon. Performing a unitary mixing of these two background preons, there exists a (generally nonlocal) basis in which fluctuations about the background mean field value are transmitted along only one of the two upper left background preon lines. Also, fluctuations are on average not transmitted along the lower background field line (line~2) as this creates a tadpole as discussed for \fref{fig:Wmassbosoncorrsv}(ii). Consequently, as with \fref{fig:Wmassbosoncorrsv}(iii), the lower background preon (line~2) and the uninvolved degrees of freedom obtained from lines~1 and~3 may be mapped to a single, connected background field extending from the lower to the upper vertex as was done in \fref{fig:Wmassbosoncorrsv}(iv) (though this time no reversal of orientation is required). This diagram therefore also has the incorrect number of independent background preon operators to mediate an interaction between the $W$~boson and background fermion fields. The same argument applies to options~2, 4, and~7, and a similar argument separating the foreground and background degrees of freedom applies to options~6 and~8.

Further, option~10 is %
equivalent to option~3 up to a cycling in colour labels. Other diagrams having the same structure as \fref{fig:Wmassbosoncorrsnew}(ii) correspond to precisely these cyclings of the colour labels, and are included in a summation over diagrams below, so option~10 is redundant. Therefore only options~3, 5, and~9 remain. It is worth noting that interactions between preons within the loop boson of option~3 [\fref{fig:Wmassbosoncorrsdetail}(iv)] may also in theory cause construction of foreground momentum loops with both termini connected to the upper or lower vertex---however, these tadpoles vanish and thus the only admissible mixing processes are those which preserve the diagrammatic form shown. These mixing processes may consequently be ignored.

Having identified the valid choices as to which preons participate in the mean background field expansion,
now recognise that \fref{fig:Wmassbosoncorrsnew}(ii) is one of six diagrams having equivalent structures. On the left of the figure there are three choices of pairs of preons which may be linked by a boson, and on the right side of the figure there are three choices of pairs of antipreons which may similarly be pairwise linked to yield equivalent factors. It suffices to evaluate one of these diagrams, and multiply by six.

In evaluating these corrections, note that the preons which are evaluated using the pseudovacuum mean field values no longer have all field operators on the same vertices. For these preons to remain correlated with their counterparts, such that the diagram's contribution to $m_W^2$ does not vanish, the pseudovacuum sources and sinks must continue to be within $\ILO{\mc{L}_0}$ of one another, and the loop correction must not introduce any correlations with particles outside the local region (both spatial and temporal) within which the pseudovacuum is self-correlated. %
This region has dimensions of order $\mc{L}_0$, and is termed the autocorrelation region, or local correlation region. This is similar to the treatment of the scalar boson mass diagram having four preon and two antipreon lines [\fref{fig:scalarbosonmass}(ii)]. %

Further recognise that in \fref{fig:Wmassbosoncorrsnew}(ii) the loop spans between two components of a composite fermion. These are necessarily separated, on average, by a distance of $\ILO{\mc{L}_\preon}$ in the rest frame of the fermion, which is %
the maximum separation of the preon components of a fermion triplet. This length scale is smaller than $2\mc{L}_\Omega$, which is the minimum length scale for mass interactions, and thus the loop boson on-shell propagation across this distance is as a massless particle. 
It may also engage in lengthier excursions characterised by scales such as $\OO{\mc{L}_0}$, on which it will acquire a nonzero mass, but over a net distance of $\OO{\mc{L}_\preon}$ from one preon to another, such excursions will inevitably be far off shell and may be disregarded.
As noted in \Psref{V}{sec:vecbosonmasses}, the underlying vertex diagram is restricted to a region characterised by $\mc{L}_\preon$ and thus the added vertices on any loop corrections must also lie within such a region. By the same argument as for \fref{fig:Wmassbosoncorrsnew}(ii), the particles participating in these loops are likewise massless. (This argument also applies to corrections arising from %
the complex scalar boson.)

Having thus established 
\begin{itemize} %
\item the diagrams which need to be evaluated to compute vector boson loop corrections to \fref{fig:Wmassbosoncorrsnew}(i), and %
\item that all bosons in these loops are effectively massless over the length scales involved, 
\end{itemize}
these loop corrections may be evaluated as follows.

\subsubsection{Gluon loops\label{sec:Wmassgluoncorr}}

It is convenient to work in the $e^C_{ij}$ basis of $\gl{3}{R}_C$, noting the caveat of %
\Psref{III}{sec:consequences} regarding counting of FSF symmetry factors. 
As noted above, there are six topologically distinct diagrams having the general form of \fref{fig:Wmassbosoncorrsnew}(ii) where a gluon propagates either from one preon to another, or from one antipreon to another. Each of these six diagrams admits multiple colour labellings.
By $\SU{3}_C$ symmetry, all six choices as to which preons engage in gluon exchange, and all valid choices of colour labels, make equal contributions to $m_W^2$. Interaction with the pseudovacuum is with randomly chosen background fermion fields compatible with a given diagram, granting colour neutrality to the preon triplet and to the antipreon triplet, and the colour symmetry of the pseudovacuum implies a summing over possible colour configurations. 
Since the colours of the three members of the triplet are unique, the factor associated with this sum may be equivalently represented either by explicitly summing over all colour labellings, or by suppressing these labellings and instead counting the exchange symmetry of the preon lines. (This works even when the $A$-charges are not homogeneous, as a sum is then also made over the position of the mismatched $A$-charge.)

Given one specific choice of diagram with one specific colour labelling, the gluon vertices are each associated with a factor of $f$, and FSF symmetries yield a factor of ${N_0}^3\left[1+\frac{33}{8}{N_0}^{-1}+\ILO{{N_0}^{-2}}\right]$ with counting being the same as for the photon \Peref{III}{eq:EMsymmetrised} (recalling that for the purpose of counting FSF symmetry factors, the $W$ boson and gluons must be written in terms of their real hermitian components). Comparing with \PEref{III}{eq:f(alpha)} the product of the gluon vertex factors is therefore equal to\footnote{Corrections of \prm{\ILO{\alpha^n}} are implicitly of \prm{\ILOO{(\alpha/\pi)^n}}, with factors of \prm{\pi} frequently being omitted in Chs.~\pref{ch:detail} and~\pref{ch:gravity} for brevity.} 
\begin{equation}
2\alpha[1+\ILO{\alpha}].\label{eq:gluonvertexinalpha}
\end{equation}
Further, taking an approach similar to \Psref{III}{sec:EWint_numerical}, consider a specific one of the six topologically distinct diagrams [e.g.~gluon links preon~1 to preon~2, as shown in \fref{fig:Wmassbosoncorrsnew}(ii)] but let this diagram be averaged over all colour labellings. For conciseness, include FSF symmetry factors in the vertex coefficients ($f\rightarrow\bmmf_A$). Choosing one gluon vertex to nominally be the emission vertex, it attracts factors as follows:
\begin{itemize}
\item One (spatial) choice of which preon to interact with.
\item Three choices of colour on that preon prior to interaction.
\item Interaction strength $\bmmf_A$.
\end{itemize} %
Similarly, for the nominal absorption vertex:
\begin{itemize}
\item One (spatial) choice of which preon to interact with.
\item Two choices of colour on that preon prior to interaction (which must differ to the preon at the emission vertex prior to interaction).
\item Interaction strength $\bmmf_A$.
\end{itemize} %
For the gluon:
\begin{itemize}
\item There are nine different species of gluons.
\item However, only one in three is compatible with the preon colour at the emission vertex.
\item Similarly, only one in three of those remaining is compatible with the preon colour at the absorption vertex.
\end{itemize}
Finally, to go from a sum to an average over colour labellings on the preon lines, introduce a further factor of $\frac{1}{6}$. After evaluating all colour labellings in this way, the net product of vertex factors for a given position of the loop is $\bmmf_A^2=2\alpha[1+\ILO{\alpha}]$.

\begin{figure}
\includegraphics[width=\linewidth]{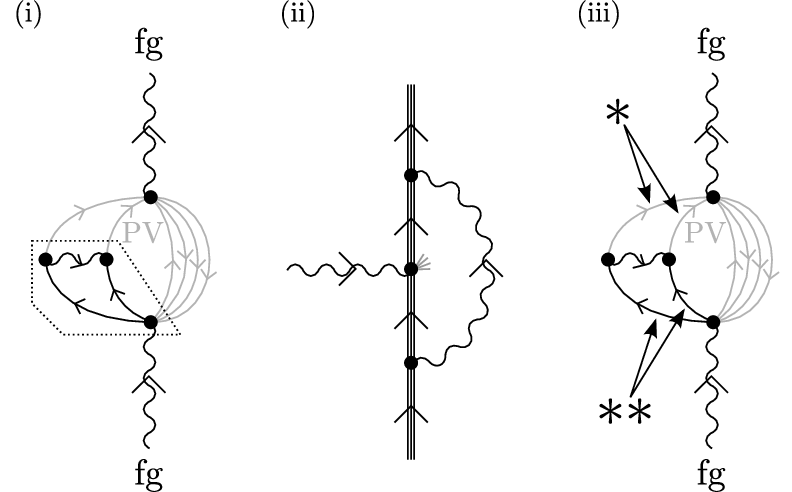}
\caption{Applying diagrammatic isotopy permits (i)~gluon exchange to be rewritten as (ii)~a gluon-mediated loop correction to \prm{W}~boson emission via either of the processes illustrated in \pfref{fig:Wmassbosoncorrsloopfactor}. [The other four preons at the vertex are uninvolved, and are shown as truncated stubs in diagram~(ii).] This gives numerical equivalence between the vertex loop correction of diagram~(ii) and the correction generated by the indicated region of diagram~(i), up to a sign due to the crossing in \freft{fig:Wmassbosoncorrsloopfactor}(ii) and~(vi) and a factor of~2 for the symmetry which interchanges the two different ways this reduction can be achieved.
This symmetry is illustrated in diagram~(iii), where simultaneous exchange of the two preon lines marked $*$ and the two preon lines marked $**$ leaves the figure unchanged up to a permutation of colour indices, and interchanges \pfreft{fig:Wmassbosoncorrsloopfactor}(i) and~(v).
\label{fig:subdiagram}}
\end{figure}

\begin{figure}
\begin{center}
\includegraphics[width=5.4in]{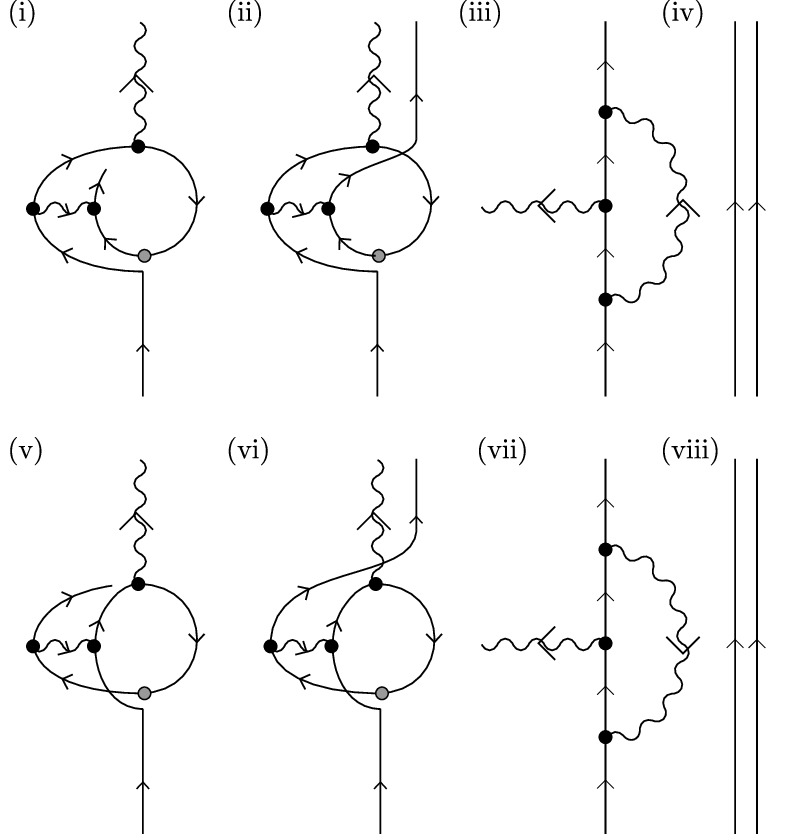}
\end{center}
\caption{(i)~A subset of the preon lines in %
\pfref{fig:Wmassbosoncorrsdetail}(i)-(ii), chosen for consistency of colour, construct an off-diagonal boson loop correction to \prm{W}~boson emission at the upper vertex. The grey dot denotes a change in $A$-sector charge with no consequence for the colour interaction. If the free preon is taken to the far field as shown in diagram~(ii), this is then equivalent to the loop correction shown in diagram~(iii) up to a crossing, which yields a factor of $-1$. Adding further preons~(iv) to diagram~(iii) permits the emitting particle to be converted to a fermion. Internal symmetrisation within the fermion then allows the loop boson to couple to any member of the triplet at each vertex as discussed in \psref{sec:leptonloops}, but the resulting factors of~3 cancel with factors of \prm{1/\sqrt{3}} in the definitions of the fermions in \protect{\Psref{III}{sec:complep1}}. %
An alternate and equivalent construction, which may also be obtained from \pfref{fig:Wmassbosoncorrsdetail}(i)-(ii), is shown in diagrams~(v)-(viii).
Note that the other preons from \pfref{fig:Wmassbosoncorrsdetail}(i)-(ii) may also optionally be reintroduced into diagrams~(i)-(ii) or~(v)-(vi), where they comprise one free preon with arrow oriented downwards, and either two more free preons (one up, one down) or a closed loop which is eliminated by vacuum normalisation. Their reintroduction introduces no further factors into the multiplicative constant associated with the loop correction.
\label{fig:Wmassbosoncorrsloopfactor}}
\end{figure}

Next, recognise that each interaction between a preon and a gluon takes place in the presence of two other preons, with the three preons arising from a background fermion. This is a single-gluon-species interaction so $K$ matrices are not required (\Psref{IV}{sec:quarksandpreons}). For purposes of evaluation it is useful next to consider the choice of diagram and pseudovacuum expansion shown in \fref{fig:subdiagram}(i) and map the indicated subdiagram into a loop diagram involving fermions. 
To do so, recognise that a subset of the preon lines in \fref{fig:Wmassbosoncorrsdetail}(i)-(ii) may be identified with a boson loop correction to $W^\bdag$ emission, up to an additional crossing. A suitable subset is shown in \fref{fig:Wmassbosoncorrsloopfactor}(i)-(ii), with an alternative choice being shown in \fref{fig:Wmassbosoncorrsloopfactor}(v)-(vi). 
An isotopy transformation on either set yields diagram~(iii) up to a sign corresponding to the preon crossing. It is then convenient to introduce a further two preon lines which have the same $A$-charge as the interacting preon, and with colours chosen to yield overall colour neutrality. If these are balanced by a numerical factor of $\mc{N}^{-1}=(N_0+1)^{-2}$ then this leaves the value of the diagram unchanged. The resulting diagram has definite colour on each preon, and a specific pair of preons of definite colour engage in the gluon interaction. However, if the diagram is averaged over %
the locations of the preons,\footnote{This averaging takes place with respect to the preons of \pfref{fig:Wmassbosoncorrsloopfactor}(iv), which are a numerical convenience---it has no bearing on the sum over the six different loop positions available in \pfref{fig:Wmassbosoncorrsnew}(ii) or the three different pseudovacuum expansions apiece, and the calculation thus far is still only for a single one of these six diagrams and three expansions.} %
and these are assumed subject to random redistribution as they propagate, for independence of arrangement at upper and lower vertices, then this gives three configurations at each loop vertex, each accompanied by a factor of $1/3$. This has no effect on the vertex factors (see \Psref{III}{sec:EWint_numerical_f/g}), but completes the mapping of the preon diagram in \fref{fig:subdiagram}(i) to the fermion diagram of \fref{fig:subdiagram}(ii). The loops in the two diagrams are numerically equivalent up to a factor of~$-1$ for the crossing in \fref{fig:Wmassbosoncorrsdetail}(ii) or~(vi), and a factor of~2 for the symmetry present in \fref{fig:subdiagram}(i) [and made explicit in \fref{fig:subdiagram}(iii)] which gives rise to the two choices of \freft{fig:Wmassbosoncorrsdetail}(i) and~(v).

It is convenient to have a systematic means of presenting the factors associated with the many loop diagrams encountered in this chapter. Therefore let the overall correction to the original vertex be written in terms of the following factors:
\begin{itemize}
\item A combined vertex and structural factor,
\item A mass-related factor, and
\item A common factor of $(4\pi)^{-1}$ written separately for clarity.
\end{itemize}
For reference, in this notation the factors associated with the electromagnetic correction to lepton magnetic moment at the one-loop level would be
\begin{equation}
2\alpha\left[1+\OO{\alpha}\right]\cdot \bmf{\frac{m_\ell^2}{m_A^2}}\cdot \frac{1}{4\pi}\label{eq:Aloopcorr}
\end{equation}
and those for the one-$W$-loop correction to lepton magnetic moment would be \cite{peskin1995}
\begin{equation}
-\frac{10\alpha}{3}\left[1+\OO{\alpha}\right]\cdot\bmf{\frac{m_\ell^2}{m_W^2}}\cdot\frac{1}{4\pi}.\label{eq:Wloopcorr}
\end{equation}
Where the loop boson is a foreground field, the mass dependency $\bmfcdot$ evaluates as
\begin{equation}
\bmf{\frac{m_\ell^2}{m_b^2}}\longrightarrow\left\{\begin{aligned}1~~~~\quad&\textrm{if }m_b^2=0\\
\frac{m_\ell^2}{4\pi m_b^2}\quad&\textrm{if }m_b^2\gg m_\ell^2.\end{aligned}\right.
\end{equation}
(Note that this expression attracts a modification for background fields, discussed in \aref{apdx:massloops}.) %

From the above breakdown of factors associated with vector boson loop corrections to vector boson emission, it is readily seen that the off-diagonal gluon attracts a structure factor of $-10\alpha/3$, 
supplemented by a factor of~2 from the two ways which \fref{fig:subdiagram}(i) may be mapped to \fref{fig:subdiagram}(ii),
and a factor of $-1$ for the crossing in \freft{fig:Wmassbosoncorrsloopfactor}(ii) and~(vi) respectively. The mass factor vanishes because the gluon is massless over scale $\mc{L}_\preon$.
Finally, multiplying by three for the three different pseudovacuum expansions and six for the six numerically equivalent diagrams yields
\begin{equation}
3\cdot6\cdot\frac{20\alpha}{3}\left[1+\OO{\alpha}\right]\cdot 1\cdot\frac{1}{4\pi}=\frac{60\alpha}{2\pi}\left[1+\OO{\alpha}\right].\label{eq:Wgluonloop}
\end{equation}
These loop corrections apply identically to the diagrams in which the $W$~boson interacts with background lepton fields (``$W$/lepton diagrams'') and the diagrams in which the $W$~boson interacts with background quark fields (``$W$/quark diagrams'') as the colour structure of both diagrams is the same, and the gluons are agnostic of $A$-charge.

\subsubsection{Photon, $W$, and $Z$ boson loops\label{sec:WcorrsAWZ}}

Calculation of the contribution of gluon loops is relatively straightforward because the $W$~boson is colour-agnostic, having trivial representation on the $C$~sector. However, more caution is required on the $A$~sector, as the $W$~boson is not agnostic of $A$-charges.

For the photon, proceed as with the gluons by enumerating all the interactions at the preon level. Begin with the $W$/lepton diagrams.
In contrast with the preceding calculation there are only three such interactions, as the photon may only couple to the charged preons in the $e_L$ side of the loop. However, there are still three pseudovacuum expansions per diagram. Next, as per \Psref{III}{sec:EWint_numerical}, note that lepton couplings to photons arise in equal measure from all three preons. \emph{Within a fermion}, the coupling of a single preon to a photon carries an effective coefficient of $\sqrt{\alpha}/3$ due to the normalisation factor in \PEref{III}{eq:generalfermion}.
This may be contrasted with gluon couplings which are always mediated by the single preon carrying the appropriate charge to participate in the interaction---but for which there is then a %
sum over three different positions for that colour of preon.
For the photon the net EM vertex factor per preon is therefore $\sqrt{\alpha}/3$, in contrast with $\sqrt{2\alpha}[1+\ILO{\alpha}]$ for gluons. Equivalently, emission by a fermion of a specific gluon is a single-preon process, whereas emission by a fermion of a photon is a collective process.

\begin{figure}
\includegraphics[width=\linewidth]{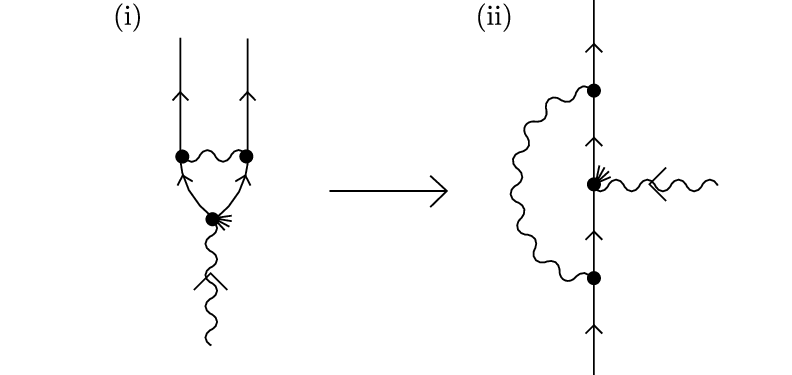}
\caption{Mapping of (i)~a photon-mediated preon-preon interaction in \pfref{fig:Wmassbosoncorrsnew}(ii) to (ii)~an emission vertex interaction exploits colour agnosticism, and involves both time reversal and charge and parity inversion for no net change of sign. The other four preons at the vertex are uninvolved, and are shown as truncated stubs.\label{fig:photonmapping}}
\end{figure}

For the rest of the structure factor, 
the diagram this time is a simple photon loop correction to a boson emission vertex, for a factor of $1/(2\pi)$, and has no hidden crossings. Mapping to a standard photon loop correction to an emission vertex (\fref{fig:photonmapping}) involves reversing the time orientation of one of the preon limbs, but also reversal of charge and parity at its source and sink operators, for a sign of $+1$. 
Both diagrams have equivalent symmetry factors from vertex exchange. The boson is diagonal, so the total structure factor is~$2\alpha/9$. %
The net correction is therefore
\begin{equation}
3\cdot3\cdot \frac{2\alpha}{9}\cdot1\cdot\frac{1}{4\pi}=\frac{\alpha}{2\pi}.\label{eq:Wphotonloop}
\end{equation}

Next, the photon loop corrections to the $W$/quark diagrams of \PEref{V}{eq:leptonvecbg2}. For these it is helpful to compare the photon/preon interactions involved with those encountered in the photon loop correction to the $W$/lepton vertex just completed. There are numerous ways to count the interactions; one of the simplest is as follows: First, recognise that to host a diagram having the form of \fref{fig:Wmassbosoncorrsnew}(ii), a group of three preons or antipreons must include two charged constituents. This restricts such diagrams to the up quark limb, and for a given configuration of preons there is only one such pair, in contrast with three choices in the $W$/lepton diagram. However, even while remaining colour-agnostic, there are now three different configurations of preons available, corresponding to the choice of which preon is uncharged. Overall, the photon loop may therefore once again be in any of three different positions
(one for each of the three configurations). %
The remainder of the calculation proceeds identically, and the electromagnetic loop factor for the $W$/quark mass vertex is %
therefore equivalent to that for the $W$/lepton mass vertex. %

For $Z$ boson loop corrections, a little more caution is needed. When computing photon loop corrections, all photon emission processes are additive. However, when emitting the bosons for $Z$~loops, some destructive interference takes place. Recognise that there are two vertices which may contribute to the sign of $Z$~emission---the coupling between the $W$ boson and the background preon fields, which gives rise to different preon and antipreon triplets with varying weights, and the coupling between these triplets and the emitted $Z$~field which goes on to form the loop correction. When looking for cancellations it is convenient to treat the preon triplet collectively as a fermion, and the same with the antipreon triplet. However, once cancellations have been identified, it is then necessary to return to a preon picture to evaluate the surviving loops.
In the present context the $W$~boson couples to $e_L\bar\nu_e$ and $d_L\bar{u}_L$ with equal vertex weight, and these constituents then couple to the $Z$~field with the strengths shown in \tref{tab:WZweights}.
\begin{table}
\caption{When a \prm{W}~boson couples to the background fields, it does so to multiple particle species. This table summarises the weights of the \prm{W}/preon couplings (vertex weight), the coefficients of these species' couplings to the \prm{Z}~boson field, and the relative contributions of each choice of fermion species to diagram~\pfref{fig:Wmassbosoncorrsnew}(i) as a whole (loop weight). FSF factors are omitted for brevity%
. Species \prm{e_L} and \prm{d_L} are seen to occur with equal vertex weights, and have opposite coupling coefficients to the \prm{Z}~field, indicating that their contributions to \prm{Z}~emission cancel. There is no residual $Z$~emission on the preon lines, only the antipreon lines.\label{tab:WZweights}}
\begin{center}
\begin{tabular}{cccc}
\hline\hline\multicolumn{4}{c}{Preon lines}\\\hline
Species & ~Vertex weight~ & $Z$ coefficient & Loop weight\\\hline
$e_L$&$f$&$\frac{2f}{\sqrt{6}}\cdot\frac{1}{2}$&$\frac{1}{2}$\\
$d_L$&$f$&$\frac{2f}{\sqrt{6}}\cdot-\frac{1}{2}$&$\frac{1}{2}$\\\hline\hline
\end{tabular}\\~\\~\\
\begin{tabular}{cccc}
\hline\hline\multicolumn{4}{c}{Antipreon lines}\\\hline
Species & ~Vertex weight~ & $Z$ coefficient & Loop weight\\\hline
$\bar\nu_e$&$f$&$\frac{2f}{\sqrt{6}}\cdot-1$&$\frac{1}{2}$\\
$\bar u_L$&$f$&$0$&$\frac{1}{2}$\\\hline\hline
\end{tabular}
\end{center}
\end{table}%
Comparison of coefficients reveals that the different $Z$~emission processes on the preon lines cancel, as do the couplings of the individual preons within $u_L$, and only emission from the antineutrino sector of the antipreon line persists.

The $Z$~loops on the antineutrino preon triplet may then be mapped to an effective overall vertex correction exactly as was done for the photon loops.
To evaluate this loop correction by reduction to a previously solved problem, recognise that the relative strengths of the one-photon and one-$Z$~boson loop corrections to lepton magnetic moment are
\begin{equation}
2\alpha\cdot1\cdot\frac{1}{4\pi}\quad\textrm{and}\quad-f_Z\alpha\cdot\bmf{\frac{m_\ell^2}{m_W^2}}\cdot\frac{1}{4\pi}
\end{equation}
where $f_Z$ is the $Z$~boson loop structural factor
\begin{align}
\begin{split}
f_Z&:=\frac{1}{3}\left[\left(4\sin^2\theta_W-1\right)^2-5\right]\\
&=\frac{1}{3}\left(4-24\frac{m_W^2}{m_Z^2}+16\frac{m_W^4}{m_Z^4}\right)\label{eq:fZ}
\end{split}\\
\sin^2\theta_W&=1-\frac{m_W^2}{m_Z^2}.
\end{align}
In the present context
\begin{enumerate}
\item the $W$ and $Z$ bosons are effectively massless, eliminating $\bmfcdot$, and
\item the average vertex factors are $-1$ (for neutrinos) instead of $1/2$ (for electrons), for a relative factor of~4.
\end{enumerate}
The one-$Z$-loop correction term arising from the $e_L\bar e_L$ base diagram after eliminating cancelled terms may thus obtained by taking the one-photon-loop correction to the lepton magnetic moment, multiplying by the above two factors, and then further multiplying by the ratio of the structure coefficients for the one-$Z$-loop and one-photon-loop corrections to the lepton magnetic moment. This gives
\begin{equation}
\frac{\alpha}{2\pi}\cdot4\cdot-\frac{f_Z\alpha}{2\alpha}=-\frac{f_Z\alpha}{\pi}%
\end{equation}
which is positive as $f_Z<0$. %

This is the correction term arising from the $e_L\bar e_L$ loop after eliminating cancelled terms, whereas the correction term arising from the $d_L\bar u_L$ loop after eliminating cancellations is zero. The relative contributions of each type of loop to the overall mass process are given by squaring the vertex weights and normalising so that they sum to~1, to obtain the loop weights. In this instance these correspond to equal contributions, and the average correction factor is thus
\begin{equation}
-\frac{f_Z\alpha}{\pi}\cdot\frac{1}{2}+0\cdot\frac{1}{2} = -\frac{f_Z\alpha}{2\pi}.
\end{equation}

Finally, for $W$ boson loop corrections, 
recognise that for fermions, $W^\bdag$~emission is a species-changing activity accompanied by a change of co-ordinates on the $A$~sector (see \Psref{III}{sec:EWint_Wintdetail}). Collectively the preon limb may convert from $e_L$ to $\nu_e$ or from $d_L$ to $u_L$ with emission of $W$, followed shortly by reabsorption of the same, with conjugate processes on the antipreon limb.

\begin{figure}
\includegraphics[width=\linewidth]{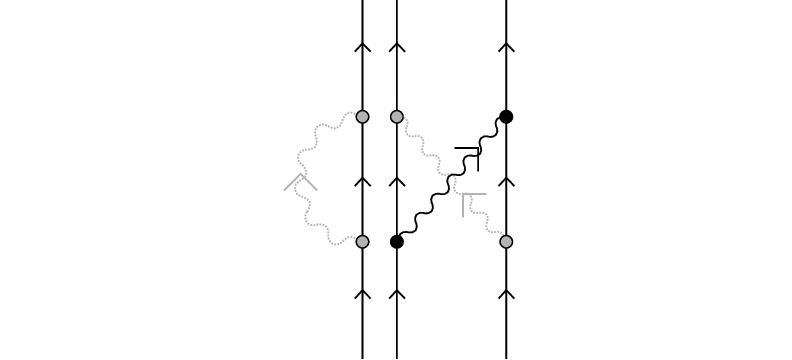}
\caption{Prototype of the boson exchange process involved in a $W$~loop correction to \pfref{fig:Wmassbosoncorrsnew}(i), with co-ordinate transformations represented as pale dotted bosons carrying off-diagonal $A$-sector representations but zero momentum.\label{fig:Wproto}}
\end{figure}

However, now consider the accompanying co-ordinate changes which maintain the fermion as a valid species in accordance with the gauge choices of \Psref{III}{sec:GL18Cgauge} and their implications for supported fermions (\Psref{III}{sec:catalogue}). These changes are conjugate, and may be represented as exchange of a boson in the $e^A_{ij}$ basis with zero momentum but appropriate $A$-charges. The $W$~boson must be emitted and absorbed on different preons, and thus the prototype for such diagrams is as per \fref{fig:Wproto}.
This diagram is the same for leptons and quarks:
\begin{itemize}
\item For leptons, the interaction may be with any preon, but diagrams yielding preon PSE terms do not contribute to the vertex correction.
\item For quarks, the interaction must be with the unique preon, but that preon may do so from one of three different positions:
\begin{itemize} 
\item As noted in \Psref{IV}{sec:Csector}, each $A$-sector interaction is accompanied by one $C$-sector interaction. %
Over the course of a loop correction involving two preons, labelled~1 and~2, and not involving preon~3, the two associated $C$-sector interactions may realise any two of the following with equal likelihood:
\begin{itemize} %
\item preon~1 may interact with preon~3,
\item preon~2 may interact with preon~3, and
\item preon~1 may interact with preon~2.
\end{itemize}
\item Any of these interactions may freely transfer any amount of the foreground momentum being carried as fluctuations by the preons involved, and any colour rearrangement over the three preons may be realised. In a context where position is integrated over, this arbitrary rearrangement of all properties except $A$-charge is equivalent to an arbitrary repositioning of the unique $A$-charge between the two vertices.
Thus the position of the interacting preon is independent at source and sink.
\end{itemize}
\end{itemize}
Now recognise that as per \aref{apdx:massloops}, the massless foreground $W$~correction reduces to an interaction at a spatial point. Similarly, the co-ordinate change boson connecting the participating preons is a $W$~boson exchange which is restricted to an interaction at a single point in momentum space. The resulting correction factors are equivalent under fourier transform, implying equivalent boson emission strengths, and the bosons have opposed orientations. The result is net zero $W$~field at the point of emission, indicating that the contribution from the $W$~loop must vanish.

\subsubsection{Scalar boson loops\label{sec:Wmass_scalbosloop}}

\begin{figure}[p]
\begin{center}
\includegraphics[width=4in]{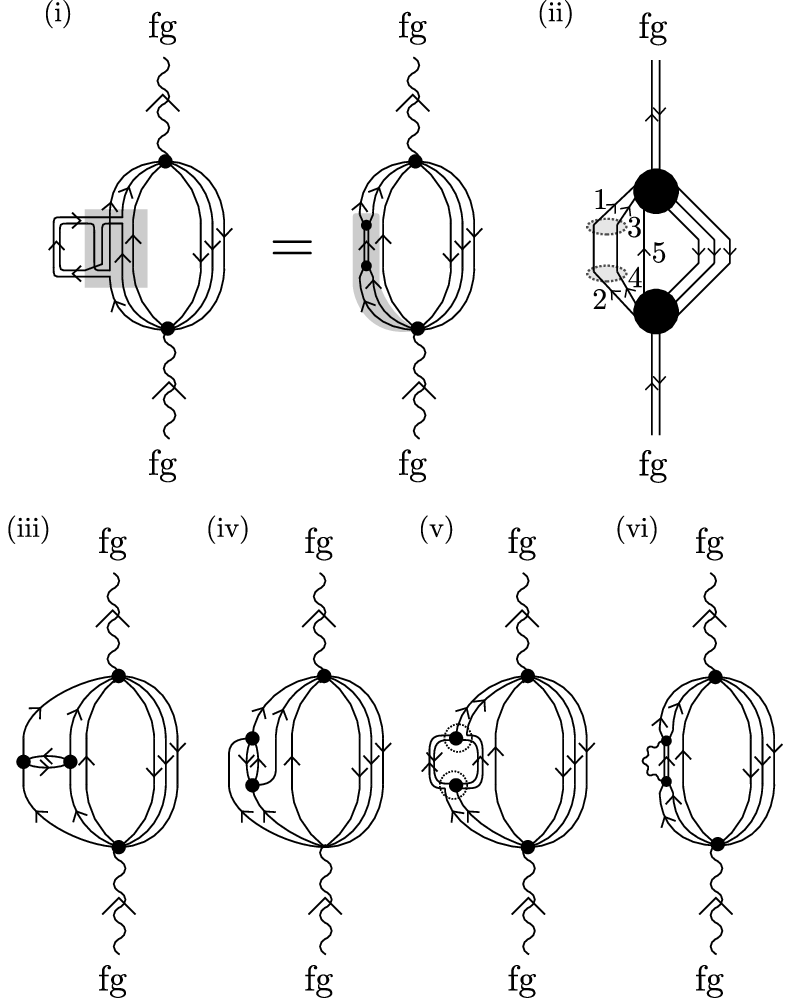}
\end{center}
\caption{Structures of loop corrections to \prm{W}/fermion vertices involving scalar bosons. Diagram~(i) shows the simplest scalar boson loop correction---in the first form, the shaded area corresponds to scalar boson emission as per \pfref{fig:scalarspread}. In the second form this is simplified by diagrammatic isotopy. The shaded region of the second form comprises the correction loop, which may be contracted into a numerical correction to the lower vertex (or into the upper vertex if preferred, though this represents two ways to evaluate the same diagram, not a twofold degeneracy). 
Within the scalar boson, the preons may be crossed or uncrossed, though only uncrossed is shown. Colour inconsistency within the scalar boson is addressed in the main text, by reduction to \pfref{fig:Wmassscalarcorrs_loop}. Diagram~(ii) shows the preon expansion for the uncrossed configuration, and labels the preons which may participate in the pseudovacuum expansion. The different labellings are enumerated in \ptref{tab:listofterms2}. Diagrams~(iii)-(vi) construct a further scalar boson loop correction: Begin with \pfref{fig:Wmassbosoncorrsnew}(ii) and first (iii)~redraw the vector boson as a pair of preons. Then (iv)~use diagrammatic isotopy to rotate the resulting loop. (v)~Group the preons as shown, and introduce effective vertices at the top and bottom of the loop which encompass both the existing vertices (factor of \prm{f} apiece) and the preon redistribution operations (factor of \prm{1} apiece). The end result is diagram~(vi), which is equivalent to diagram~(i) supplemented by a further massless vector boson loop with effective vertex factors of~1. This must be considered in addition to \pfref{fig:Wmassbosoncorrsnew}(ii) as the mass shell of the resultant foreground scalar boson is distinct from that of the vector boson in \pfref{fig:Wmassbosoncorrsnew}(ii) so it is treated as a separate effective excitation.\label{fig:Wmassscalarcorrs}}
\end{figure}

For scalar bosons the overall approach is similar to that taken for vector bosons, though the diagrams are different, being shown in \fref{fig:Wmassscalarcorrs}.
The preons within a scalar boson may be crossed or uncrossed. 

\begin{figure}
\includegraphics[width=\linewidth]{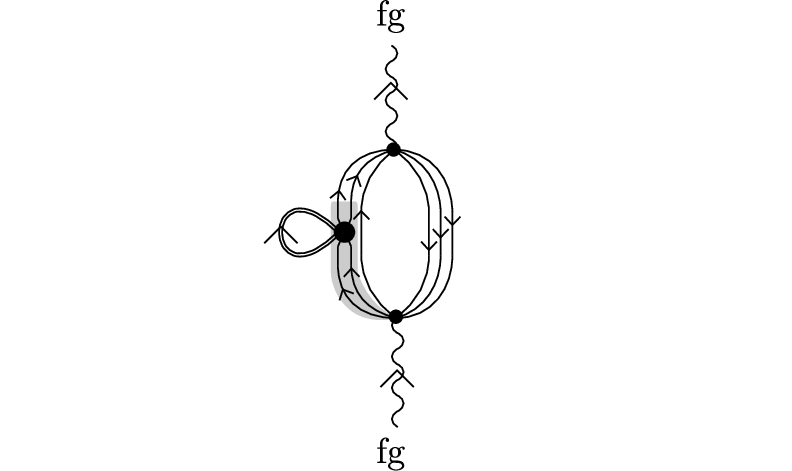}
\caption{The contribution of \pfref{fig:Wmassscalarcorrs}(i) to \prm{W}~boson mass is dominated by contributions made when the two interaction vertices coincide, as shown here. However, given the underlying structure of \pfref{fig:Wmassscalarcorrs}(i) this should not be mistaken for a loop correction to the propagator. As a loop correction to \pfref{fig:Wmassbosoncorrsnew}(i), the resulting factor is still absorbed into either the upper or the lower vertex, with the latter being indicated by the shading. 
\label{fig:Wmassscalarcorrs_loop}}
\end{figure}

Considering first \fref{fig:Wmassscalarcorrs}(i), for a given set of preons in the fermion fields this diagram yields one of nine terms in the sum making up the scalar boson. 
Recalling the factor of $\ILO{{N_0}^{-2}}$ attracted by any scalar boson propagating to the far field, this term is dominated by the contribution when the two vertices coincide as shown in \fref{fig:Wmassscalarcorrs_loop}(i). 
As per \Psref{III}{sec:scalbosint}, terms where the source and sink are separate then correct this diagram by at most $\ILO{{N_0}^{-4}}$. Further, now consider colour consistency and recognise that at either of the loop vertices in \fref{fig:Wmassscalarcorrs}(i), both preons in the fermion must be of different colour while both preons in the scalar boson must be the same. This can only be reconciled if the two vertices are %
collocated, and thus may be rewritten as a single colour-conserving vertex. 
Any diagrams where the termini of the scalar boson loop do not coincide---written as $\ILO{{N_0}^{-4}}$ above---must therefore vanish. (This collocation is relaxed very slightly on noting that colour may be borrowed from the background field, but only over $\mc{L}_\preon$, assumed to be the smallest scale of the model. Similar considerations are applicable to all fermion/$\bmh$ interactions of the sort shown in \fref{fig:scalarspread}.)
Note that the factor associated with the effective colour-conserving vertex is determined by its construction from two disparate vertices (and integrating out their accessory degrees of freedom)---there is no need to assign an additional factor of two due to on-vertex symmetries.

It is now useful to compare this scenario with the vector boson loop correction of \pfref{fig:Wmassbosoncorrsnew}(ii):
\begin{itemize}
\item For the vector boson loop, the contribution is determined by the contribution when the loop vertices coincide, as discussed in the \aref{apdx:massloops}.
\item For the scalar boson loop, the contribution is determined by the contribution when the loop vertices coincide, as discussed for \fref{fig:Wmassscalarcorrs_loop} above.
\item For the vector boson loop, this is equivalent to contracting the boson loop into either the upper or the lower vertex.
\item For the scalar boson loop, writing the correction as a numerical multiplier on the original diagram is likewise equivalent to contracting the boson loop into either the upper or the lower vertex.
\item In both cases, the choice of contracting into the upper or lower vertex is a descriptive one. Although it represents a symmetry in the calculation such that there are two equivalent ways of \emph{evaluating} the associated Feynman diagrams, these do not correspond to different diagrams, and hence there is no factor of~2 associated with this.
\end{itemize}

To evaluate \fref{fig:Wmassscalarcorrs}(i), first recognise that with the preons crossed, the resulting preon diagram has the same preon structure as \fref{fig:Wmassbosoncorrsdetail}(ii)---the colour sector is summed (corresponding to the sum over terms in the scalar boson) rather than averaged, but the net result is the same up to an overall multiplier. Further, both of these diagrams are dominated by contributions in which the separation of the boson vertices approaches zero.
This diagram will consequently evaluate equivalent to the photon loop of \fref{fig:Wmassbosoncorrsdetail}(i) up to an overall multiplier to be determined explicitly now. Whereas the photon loop accrues an overall factor of $2\cdot(4\pi)^{-1}$, for the scalar boson 
take note of the following:
\begin{itemize}
\item For coupling coefficients, 
\begin{itemize}
\item the raw vertex coefficient is $f$ in place of $f/\sqrt{2}$,
\item Each preon/scalar boson coupling should be understood in the context of a fermion/scalar boson coupling and thus attracts a factor of $\frac{1}{3}$ per vertex from the definition of the fermion~\Peref{III}{eq:generalfermion}. As with the photon, the total net coupling is averaged across the three preons by this factor.
\end{itemize}
The coupling coefficient is consequently $(2\alpha/9)[1+\ILO{\alpha}]$. 
\item Also contributing to the structure factor, 
\begin{itemize}
\item there is no symmetry under exchange of boson source/sink (factor~1), and 
\item unlike the photon the complex scalar boson is not written in terms of $\sigma^\mu\sigma_\mu$ (factor~$-\frac{1}{2}$).
\end{itemize}
The structure factor, incorporating the vertex coefficient, is consequently~$-(\alpha/9)\left[1+\ILO{\alpha}\right]$ compared with the photon's~$2\alpha/9$.
\item Regarding the pseudovacuum expansion and crossed and uncrossed preon lines,
\begin{itemize}
\item the crossed preon configuration admits three pseudovacuum expansions, directly equivalent to those of \fref{fig:Wmassbosoncorrsnew}(ii), while
\item the uncrossed preon configuration admits four pseudovacuum expansions, elaborated below. These have one less crossing and are therefore of opposite sign.
\end{itemize}
The net factor arising from pseudovacuum expansions and preon crossings in combination is therefore~$-1$.
\item For other factors:
\begin{itemize}
\item Three choices of preon pairs in the ascending limb, three choices in the descending limb, yield six diagrams.
\item The above evaluates the contribution of one of nine terms in the scalar boson sum. All terms contribute, as excitation of any one leads to excitation of all by unbroken $\SU{3}_C\oplus\GL{1}{R}_N$ symmetry at the preon scale, for a further factor of~9.
\end{itemize}
\end{itemize}
To evaluate pseudovacuum expansions for \fref{fig:Wmassscalarcorrs}(i) with the preon lines in the scalar boson uncrossed, label as per \fref{fig:Wmassscalarcorrs}(ii) and expand as per \tref{tab:listofterms2}.
\begin{table}
\caption{List of labellings of preon lines in \pfref{fig:Wmassscalarcorrs}(ii). For brevity, labellings in which preon~5 is foreground have been omitted as these are always ``connected'' so never contribute to the mass correction.\label{tab:listofterms2}}
\begin{center}
\begin{tabular}{c|c|c|c}
Option&Background&Foreground&Status\\\hline\hline
1&1,2,5&3,4&Connected\\ %
2&1,3,5&2,4&Valid\\ %
3&1,4,5&2,3&Valid\\ %
4&2,3,5&1,4&Valid\\ %
5&2,4,5&1,3&Valid\\ %
6&3,4,5&1,2&Connected %
\end{tabular}
\end{center} 
\end{table}
Following the same arguments employed to identify connected diagrams in \sref{sec:Wmassbosonloops} above, only one option out of each pair (1,3) and (4,6) need be retained, and this should be the one in which there is no foreground connection between upper and lower vertices.

It is also worth noting the caution required around the sign of scalar boson contributions to pseudovacuum expansions. In identifying the crossed preon configuration with the preon expansion of \fref{fig:Wmassbosoncorrsnew}(ii), these diagrams are seen to have equivalent sign, and the uncrossed configurations to have the opposite sign. However, without making the pseudovacuum expansion explicit, it would have been easy to assume that all scalar boson loop terms had a minus sign relative to the vector boson terms due to the presence of an additional fermion loop in \fref{fig:Wmassbosoncorrsnew}(i). It is always advisable to perform explicit preon expansions to check signs of terms in diagrams involving both scalar bosons and loops.

Putting together the above, the net factor associated with \fref{fig:Wmassscalarcorrs}(i) [with terms ordered as per \Eref{eq:Wgluonloop}, and the extra factor from the nine terms in the scalar boson at the end] is 
\begin{equation}
\begin{split}
&-1\cdot 6\cdot -\frac{\alpha}{9}[1+\ILO{\alpha}]\cdot 1\cdot\frac{1}{4\pi}\cdot 9=\frac{3\alpha}{2\pi}[1+\ILO{\alpha}].
\end{split}\label{eq:Wscalbosloop}
\end{equation}

Next, consider \fref{fig:Wmassscalarcorrs}(vi). This diagram is obtained by transforming the photon loop correction of \fref{fig:Wmassbosoncorrsnew}(ii), but must be considered a separate figure as it involves the scalar boson, which propagates as a distinct entity under the Lagrangian on $\RM$. The pair of loops may be decomposed into two independent multiplicative factors, one corresponding to a scalar boson and directly equivalent to \fref{fig:Wmassscalarcorrs}(i), and another arising from the massless vector boson with interaction vertices of value~1. Note that as with \fref{fig:Wmassscalarcorrs}(i) reducing to \fref{fig:Wmassscalarcorrs_loop}, the vertices of the scalar boson in \fref{fig:Wmassscalarcorrs}(vi) must likewise be brought to coincide.
The resulting scenario is very similar to that encountered in \Psref{V}{sec:scalbosonmass}---compared with a photon loop, relative factors are as follows:
\begin{itemize}
\item The boson loop contains only the appropriate preons, so there is no factor of $1/\sqrt{2}$ associated with the vertex. The coupling would consequently be $2\alpha[1+\ILO{\alpha}]$ in place of $\alpha$. However, this is then replaced with~1 as there are no factors of $f$ associated with the vector boson loop. Net relative factor: $\alpha^{-1}$.
\item No extra symmetry factor for on-vertex interchange of boson ends where compatible, as the effective vertex has been synthesised from two subvertices and not drawn down from the generator~$\Z$.
\end{itemize}
The vector boson loop on diagram~(iv) therefore contributes an additional multiplier of
$2\cdot({4\pi})^{-1}$
for a total factor from \fref{fig:Wmassscalarcorrs}(vi) of
\begin{equation}
\frac{3\alpha}{2\pi}\cdot\frac{1}{2\pi}.
\end{equation}
Combining with \fref{fig:Wmassscalarcorrs}(i) yields a net scalar boson loop correction factor of
\begin{equation}
\frac{3\alpha}{2\pi}\left(1+\frac{1}{2\pi}\right).
\end{equation}

\subsubsection{Net effect of all boson loops\label{sec:netallbosonloopsvecbmass}}
The net effect of the boson loop corrections is therefore to amend the $W$ boson mass equation to
\begin{align}
\label{eq:Wmasswithloops}
m_W^2=\,&9f^2{k_1}^4{\omega_0}^2{N_0}^{12}S_{6,13}\\
&\times\left[1+\left(64+\frac{3}{2\pi}-f_Z\right)\frac{\alpha}{2\pi}+\OOO{(k_1N_0)^{-4}}+\OO{\alpha^2}\right]
\nn
\end{align}
\begin{align}
\begin{split}
S_{6,13}&:={N_0}^{-4}(N_0+2)^2(N_0+1)^2\\&=1+6{N_0}^{-1}+13{N_0}^{-2}+\ldots
\end{split}\label{eq:defS613}\\
f_Z&=\frac{1}{3}\left(4-24\frac{m_W^2}{m_Z^2}+16\frac{m_W^4}{m_Z^4}\right)\label{eq:fZshort_preupdate}
\end{align}
where $S_{6,13}$ was previously introduced in \PEreft{III}{eq:III:defS613}, \Peref{IV}{eq:IV:defS613} and~\Peref{V}{eq:WFSFsymfactor}, and
the next-most-relevant members of the correction series are those due to the coupling of the $W$ boson to the background photon and scalar fields [which are smaller by a factor of~${\big(k_1{N_0}\big)^{-4}}$], %
the second-order electromagnetic corrections [with leading contribution being of $\ILO{\alpha^2}$], and the various factors of $[1+\ILO{\alpha}]$ in the above calculations which correct terms already containing $\alpha$ and thus are also of $\ILO{\alpha^2}$. 

\subsubsection{Species dependence of \prm{k_1}\label{sec:speciesdependenceofk}}

Regarding the factor of $k_1$ in the above: 
\begin{itemize}
\item When particle masses are evaluated beyond tree level, the mass matrices incorporate higher-order corrections which are themselves dependent on the particle masses (as seen, for example, in the $f_Z$ dependence of $m_W^2$). Similar corrections will also apply in fermion mass calculations.
\item It immediately follows that the mass eigenvalues $k_i$ are thus dependent on particle species. 
\end{itemize}
For fermion masses:
\begin{itemize}
\item At a fermion mass vertex, the only participating fermion is the foreground fermion, and thus the eigenvalues are calculated with respect to the foreground fermion.
\item In fermion mass calculations, therefore replace $k_i$ with $k^{(\ff)}_i$ where $\ff$ is a representative member of the fermion family. For example, $k^{(e)}_i$ is mass eigenvalue~$i$ for the family $\{e,\mu,\tau\}$.
\end{itemize}
For boson masses:
\begin{itemize}
\item At a boson mass vertex, the boson interacts with the background fields of the pseudovacuum, which may be viewed as containing a multiplicity of fermions (and bosons).
\item Attention may be restricted to the fields appearing within a region of dimension $\ILO{\mc{L}_0}$. By construction of the pseudovacuum, this collection is (on average) colour-neutral and uncharged.
\item Colour interactions within this region emulate the exchange of gluons between $\ILO{N_0}$ preons. In lieu of the three-preon $K$-matrix of \sref{sec:Csector}, the background field theoretically attracts a $K$-matrix acting on all $\ILO{N_0}$ preons simultaneously.
\item However, it is almost universally possible to identify colour-neutral triplets of identical $A$-charge within this region. Although such triplets are not necessarily as tightly bound as the foreground triplets of \sref{sec:fermionmasses}, they are nevertheless within a common correlation region and thus their exchange of coloured bosons is non-negligible and also generates a reduced $K$-matrix for the triplet.
\item By randomness (and thus cancellation) of the interactions outside the triplet, on average such triplets are necessarily also eigenstates of this reduced $K$-matrix, $\K$, with eigenvalues in $\{k^{(\ell)}_i\}$, $\ell\in\{e,\nu\}$.
\item Next, recognise that although \fref{fig:Wmassbosoncorrsnew}(i) is an interaction with the collection of background fields as a whole, the normalisation of \sref{sec:normWrtBgFields} eliminates all background-field-related terms except for the fields at the interaction vertices, and associated with these, up to four instances of eigenvalues to be denoted $k^{(\ell)}_i$.
\item However, the colour interactions generating these $\K$-matrix eigenvalues are not constrained to act on the same preons as the $A$-sector interactions of \fref{fig:Wmassbosoncorrsnew}(i). Further, the preon triplets interact with both $C$-sector and $A$-sector bosons from the background field and although only the effect on the $\K$~matrix is retained, the dominant interaction types are interactions with the pseudovacuum photon and gluon fields, which are equivalent to the lepton mass interactions described in \sref{sec:lepmassint}. It is therefore these interactions which determine the values of the eigenvalues $k^{(\ell)}_i$, and these are identical to those computed when determining the lepton masses in \sref{sec:thetacorr}.
Recognising that factors of $\ILO{N_0}$ are accounted for elsewhere, the outcome is a sum over all admissible choices of eigenvalues $k^{(\ell)}_i$.
\item It is seen in \sref{sec:bosonkmatrixspeciesdependency} that contributions from the electron sector render the neutrino sector negligible, and thus these eigenvalues take the form $k^{(e)}_i$.
\item All fermions acted on by a single $\K$-matrix acquire the same generation label~$i$.
\item Particle generations within the pseudovacuum are only consistent if both $\K$-matrices involved in the evaluation of \fref{fig:Wmassbosoncorrsnew}(i) (one introduced in association with the upper vertex and one introduced in association with the lower vertex) yield eigenvalues of the same generation.
\item Thus there exist three sets of mass eigenvalues for \fref{fig:Wmassbosoncorrsnew}(i), corresponding to $\K$-matrix eigenvalues $k^{(e)}_i$, $i\in\{1,2,3\}$.
\end{itemize}
In \sref{sec:thetacorr} it is subsequently also seen that the $\K$-matrix eigenvalues $k^{(\ff)}_i$ exhibit dependency on an energy scale. Identification of this energy scale is addressed in \sref{sec:bosonkmatrixenergydependency}, in which it is seen that the eigenvalues appearing in the boson mass vertices are always evaluated at energy scale $\mc{E}_{e_i}$, $e_i\in\{e,\mu,\tau\}$, where the generation of the lepton label $e_i$ matches the generation of the massive boson. For the present, this energy dependency of $k^{(\ff)}_i(\mc{E})$ is left implicit until it is demonstrated in \sref{sec:thetacorr}. %

Later evaluation of $k_1^{(e)}%
$~\eref{eq:ke1} shows the term $\mrm{O}{\big[\big(k_1N_0\big)^{-4}\big]}$ in \Eref{eq:Wmasswithloops}, which is now understood to be $\mrm{O}\big\{\big[k_1^{(e)}%
N_0\big]^{-4}\big\}$, to be between $\alpha/(4\pi)$ and $\alpha^2/(4\pi)^2$ in magnitude. This correction is evaluated shortly. %

\subsubsection{Background photon and scalar interactions\label{sec:WmassQLphotonandscalar}}

\begin{figure}
\includegraphics[width=\linewidth]{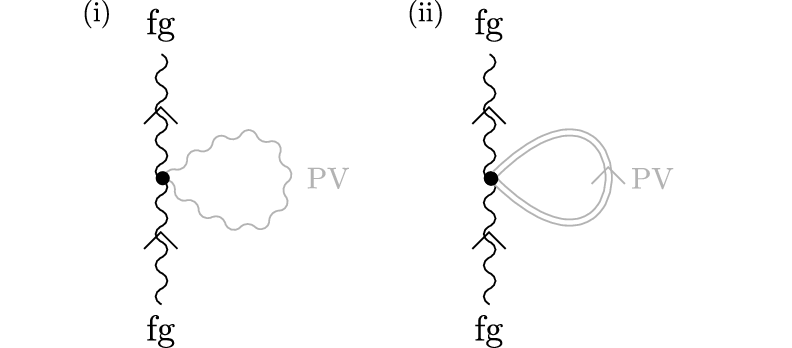}
\caption{(i)~Coupling between $W$ boson and background photon field. (ii)~Coupling between $W$ boson and background scalar field. For the scalar boson there is no need to separately consider crossed and uncrossed configurations or FSF exchange in the absence of foreground fields, as all such symmetry factors are incorporated into the mean-squared background field value.\label{fig:Wphotons}}
\end{figure}

Another potentially relevant correction is the $W$/background photon coupling, \fref{fig:Wphotons}(i).
At tree level this readily evaluates to
\begin{equation}
\frac{f^2}{2}{N_0}^8S_{6,13}\label{eq:WAfactor}
\end{equation}
where a symmetry factor of 2 arises from the presence of two identical photon operators on the interaction vertex, but is cancelled by the diagrammatic %
equivalence of exchanging the connected photon lines/sources/sinks. This
gives a net expression %
\begin{align}
\nn m_W^2=\,&9f^2\left[k_1^{(e)}\right]^4{\omega_0}^2{N_0}^{12}S_{6,13}\\
&\times\Biggl\{1+\left(64+\frac{3}{2\pi}-f_Z\right)\frac{\alpha}{2\pi}&\\
&\qquad+\frac{1}{18\left[k_1^{(e)}{N_0}\right]^{4}}\left[1+\OO{\alpha}\right]+\ldots+\OO{\alpha^2}\Biggr\}.\nn
\end{align}

The composite pseudovacuum scalar boson field yields a similar contribution, represented by ``\ldots'' in the above. In comparison with the photon term:
\begin{itemize}
\item In contrast with $A^\mu A_\mu$, the complex scalar boson operators $\bmh$ and $\bmh^*$ are no longer interchangeable. However, each vertex may be either an $\bmh$ vertex or an $\bmh^*$ vertex. The resulting factors of~2 and~$\frac{1}{2}$ cancel,
for no net change in on-vertex symmetry.
\item Because scalar boson source and sink are located on the same vertex, the loop may be contracted to a point and as discussed in \Psref{III}{sec:scalbosint} there is therefore no factor of~$2\big[k^{(e)}_1{N_0}\big]^{-4}$ for taking the scalar bosons to the far field.
\item The vertex coupling factor is obtained from the Lagrangian. It includes evaluation of $(\rmi/2)^2\Upsilon^\mu\bar\Upsilon_\mu$, and recognition that the pair may be constructed either as $\bmh\bmh^*$ or $\bmh^*\bmh$. It evaluates as $-2f^2$. This is compared with $f^2/2$ for the photon, for a relative factor of~$-4$. (It is %
interesting to compare this with the approach taken for the scalar boson loops of \sref{sec:Wmass_scalbosloop}, in which loops were evaluated by reduction to a previously solved problem rather than direct construction from the Lagrangian. %
The minus sign seen here corresponds to the one obtained in \sref{sec:Wmass_scalbosloop} from sigma matrices, which was included in the loop structure factor.)
\item Comparing \Erefs{eq:AQL}{eq:HQL}, the mean field value for $\bgfield{\bmh\bmh^*}$ 
attracts a factor of $-9/2$ relative to that for $\bgfield{A^\mu A_\mu}$. 
\end{itemize}
The scalar boson term is consequently larger than the photon term by a factor of
\begin{equation}
1\cdot -4\cdot -\frac{9}{2}=18.
\end{equation}
Incorporating the background scalar boson contribution to $m_W^2$ therefore yields 
\begin{align}
\nn m_W^2=9&f^2\left[k_1^{(e)}\right]^4{\omega_0}^2{N_0}^{12}S_{6,13}\\
&\times\Biggl\{1+\left(64+\frac{3}{2\pi}-f_Z\right)\frac{\alpha}{2\pi}&\\
&\qquad~~+\frac{19}{18\left[k_1^{(e)}{N_0}\right]^4}\left[1+\OO{\alpha}\right]+\OO{\alpha^2}\Biggr\}.\nn
\end{align}

There is no equivalent gluon coupling as the $W$~boson is colourless. Any attempts to construct a vector coupling to the $C$~sector background vanishes on summation over emission coefficients, similar to the cancellation seen while evaluating the $Z$~coupling in \sref{sec:WcorrsAWZ} above.

\subsubsection{Universality of loop corrections\label{sec:universalcorrs}}

Up to now, the photon in \fref{fig:Wphotons} has been treated as a fundamental particle. However, in principle any occurrence of a fundamental boson may be re-expressed as a pair of preons using the identity
\begin{equation}
\varphi^{\dot a\dot cac}_\mu=\bar\partial^{\dot a\dot c}\bsmm\partial^{ac}\varphi\approx f\bar\partial^{\dot a\dot c}\varphi\bsmm\partial^{ac}\varphi=f\bar\psi^{\dot a\dot c}\bsmm\psi^{ac}
\end{equation}
derived from \PErefs{I}{eq:defvarphimu}{eq:vphipreonsub}.
Such pairs are bound by the colour interaction with a characteristic separation of $\ILO{\mc{L}_\preon}$, but this is also the preon separation observed at the vertices of the $W$/lepton diagrams such as \fref{fig:Wmassbosoncorrsnew}(i). 

With this in mind, consider again the fermion-mediated boson mass vertex of \fref{fig:Wmassbosoncorrsnew}(i). In \Psref{V}{sec:vecbosonmasses} the interaction between the vector boson and the preon fields was taken to involve all three preons, collected as a composite fermion, leading to the introduction of $K$ matrices on two of the preon lines at each vertex. Although integration reduced two preon lines and two antipreon lines to a numerical factor using \Eref{eq:xycorr1}, these lines still interacted with the foreground field, permitting them to acquire FSF symmetry factors (which may be understood as corresponding to a choice over which preon lines the $K$ matrices are applied to).

An alternative approach to the evaluation of \fref{fig:Wmassbosoncorrsnew}(i) is to interpret the six preons as three vector bosons. This construction has no $K$~matrices, and corresponds to the diagrams without $K$~matrices discussed in \Psref{V}{sec:Wmass5v}. After integrating out four of the lines, the remaining lines are collected as emission and absorption of vector bosons from the background field. This requires that each boson constructed involves a preon from one vertex and an antipreon from the other vertex. (Choosing both from the same vertex just yields a propagator of the foreground boson over the intervening space.)
In contrast with the background fermion diagrams, since there are no $K$ matrices associated with boson-boson interactions, %
the normalisation convention of \Psref{I}{sec:normWrtBgFields} requires that the eliminated lines yield a net factor of~1.

If the residual background bosons are conjugate, then this constitutes a reduction of \fref{fig:Wmassbosoncorrsnew}(i) to a diagram having the general form of \fref{fig:Wphotons}(i). However, the species involved may be unexpected, and this may represent an additional channel of interaction between the foreground boson and the background boson fields which must also be taken into account.
For the $W$ boson this contribution vanishes, however, as the resulting background bosons are associated with orthogonal representation matrices on the $A$~sector ($e^A_{22}$ and $e^A_{33}$) and therefore have a vanishing mean product on the pseudovacuum.

Now consider diagrams having the form of \fref{fig:Wmassbosoncorrsnew}(ii) which contain loop corrections. The preons persisting after integration no longer necessarily arise from the same vertices, but this is unimportant as they are still within $\mc{L}_\preon$ of one another and can therefore once again be considered to make up composite vector bosons, again yielding a coupling to the boson sector of the pseudovacuum. (Again these vanish for the $W$~boson.) 

Crucially, however, the converse process may be applied to any boson loop diagram such as \fref{fig:Wphotons}(i), recruiting an additional two bosons (two preon lines and two antipreon lines) from the pseudovacuum. Choice of gauge ensuring that this recruitment yields triplets consistent with fermions. The recruited preons permit reconstruction of a diagram having the form of \fref{fig:Wmassbosoncorrsnew}(i), without $K$ matrices.
Such a reconstructed six-preon-line diagram then admits all of the loop corrections identified for \fref{fig:Wmassbosoncorrsnew} above.
This mapping therefore identifies a set of nontrivial higher-order correction to the boson loop diagram of \fref{fig:Wphotons}(i) which can \emph{only} be obtained by mapping the interacting bosons into preon constituents and recruiting additional preons from the pseudovacuum to make up preon vertices before integrating them out again. [Note that this recruitment (i)~is always possible due to the homogeneity of the pseudovacuum and the large number of background particles within autocorrelation length \prm{\mc{L}_0}, and (ii)~is obligatory as it provides additional channels for interaction between a propagating boson and the background fields.] %
Further recalling that the scalar boson loop in \sref{sec:Wmass_scalbosloop} was evaluated by mapping to a vector boson loop, %
equivalent corrections also apply to the scalar boson loop diagram of \fref{fig:Wphotons}(ii).

The net outcome is that every term in \Eref{eq:Wmasswithloops} has a counterpart on the boson loops of \fref{fig:Wphotons}. %
The $W$ boson mass %
may then concisely be written as
\begin{align}
\begin{split}
m_W^2=&\,\,9f^2\left[k_1^{(e)}\right]^4{\omega_0}^2{N_0}^{12}S_{6,13}\\
&\times\left[1+\left(64+\frac{3}{2\pi}-f_Z\right)\frac{\alpha}{2\pi}+\OO{\alpha^2}\right]
\label{eq:Wwithk1a}\\
&\times\left\{1+\frac{19}{18\left[k^{(e)}_{1}{N_0}\right]^{4}}\left[1+\OO{\alpha}\right]\right\}.
\end{split}
\end{align}
Later evaluation of $k_1^{(e)}$~\eref{eq:ke1} shows the term in $\big[k_1^{(e)}N_0\big]^{-4}$ to be between $\alpha/(4\pi)$ and $\alpha^2/(4\pi^2)$ in magnitude, and the factor $[1+\ILO{\alpha}]$ on this term therefore introduces corrections which are small compared with $\ILO{\alpha^2}$. 
However, pending demonstration of this, \Eref{eq:Wwithk1a} may be written
\begin{align}
\nn
m_W^2=\,&9f^2\left[k_1^{(e)}\right]^4{\omega_0}^2{N_0}^{12}S_{6,13}\Biggl[1+\left(64+\frac{3}{2\pi}-f_Z\right)\frac{\alpha}{2\pi}\Biggr]\\
&\begin{aligned}
&\times\left\{1+\frac{19}{18\left[k^{(e)}_{1}{N_0}\right]^4}\right\}\\
&\times\bm{\left(}1+\OOOO{\alpha\left[k^{(e)}_1{N_0}\right]^{-4}}+\OO{\alpha^2}\bm{\right)}.\end{aligned}
\label{eq:Wwithk2}
\end{align}

This universality of the preon-level loop corrections to all interactions---boson and fermion---and the involvement of both boson and fermion terms in the calculation of boson masses---even where, as for the $W$~boson, these disappear---has the important implication that corrections to the Weinberg angle~$\theta_W$ are likewise universally applied, regardless of whether $\theta_W$ is expressed in terms of boson masses or interaction strengths. Thus the mass ratio expression for $\theta_W$ in the definition of $f_Z$~\eref{eq:fZ} is consistent with the interaction ratio expression $\tan\theta_W=g'_\mrm{eff}/g_\mrm{eff}$ once all corrections to interaction strengths with no counterparts in the Standard Model have been taken into account.

This completes calculation of the standard (colourless) $W$~boson mass to the level of precision employed in this paper.

\subsection{$Z$ mass\label{sec:Zmass}}

Higher order corrections to the $Z$ boson mass are also required, and their calculation is similar to that for the $W$ boson. 

\subsubsection{Boson loops}

\paragraph[Gluon loops]{Gluon loops}

The calculation is analogous to that performed for the $W$ boson. However, introduction of an off-diagonal gluon coupling rearranges colour and so eliminates the end-to-end symmetry of the $Z$ boson mass-squared interaction, resulting in the loss of a symmetry factor of 2 relative to the original diagram. 
There are no diagonal gluon couplings consistent with \fref{fig:Wmassbosoncorrsnew}(ii), so this reduction applies to all gluon loop corrections.
The net gluon contribution is therefore
\begin{equation}
\frac{30\alpha}{2\pi}.
\end{equation}
An alternative explanation of this reduction in factors is given by noting that the original background fields are summed over connection to the source and sink in two different orientations (corresponding to the symmetry factor of two). When the same exchange is applied to a diagram with a loop correction in the preon limb, this maps to a loop correction in the antipreon limb. Counting all six positions for loop corrections is therefore double counting, and again the contribution must be reduced by a factor of two.

\paragraph[Photon, \prm{W}, and \prm{Z} boson loops]{Photon, $W$, and $Z$ boson loops:}

These calculations follow the approach described in \sref{sec:WcorrsAWZ}. The photon and $W$~boson are readily dismissed:
\begin{itemize}
\item The $Z$~boson has no electromagnetic charge, so on constructing a photon-loop coupling table all terms are immediately seen to vanish.
\item The $W$~boson loop correction vanishes by the same argument as \sref{sec:WcorrsAWZ} above.
\end{itemize}

For $Z$ loops, however, the coupling table does not sum to zero and so is computed here. Consider the relative contributions of the different fermions which contribute to the leading order diagram [analogous to \fref{fig:Wmassbosoncorrsnew}(i)]. At tree level the $Z$ boson couples only to electrons, neutrinos, and the down quarks. The relative contribution of each line to the $Z$~field of the loop boson is the product of the vertex weights and the $Z$~coefficients, given for the preon limb in \tref{tab:ZcorrZ}.
\begin{table}%
\caption{List of channels contributing to the preon limb of the leading order \pt{$Z$}~boson mass diagram, the \prm{Z}/preon couplings (vertex weights), coefficients of coupling to the (loop) \prm{Z}~boson field, and the relative contributions of each choice of fermion species to the leading-order mass as a whole (loop weight). Terms for the antipreon limb are identical up to a sign on both vertex weight and \prm{Z} coefficient representing the selection of the \prm{CP}-conjugate at the boson/fermion vertex, and the opposite sign of interaction of this conjugate with the \prm{Z}~boson.
\label{tab:ZcorrZ}}
~\\
\begin{center}
\begin{tabular}{cccc}
\hline\hline\multicolumn{4}{c}{Preon lines}\\\hline
Species & ~Vertex weight~ & $Z$ coefficient & Loop weight\\\hline
$e_L$		&$\frac{2f}{\sqrt{6}}\cdot\frac{1}{2}$	&$\frac{2f}{\sqrt{6}}\cdot\frac{1}{2}$
&$\frac{1}{8}$\\
$\bar e_R$	&$\frac{2f}{\sqrt{6}}\cdot\frac{1}{2}$	&$\frac{2f}{\sqrt{6}}\cdot\frac{1}{2}$
&$\frac{1}{8}$\\
$\nu_e$		&$\frac{2f}{\sqrt{6}}\cdot-1$			&$\frac{2f}{\sqrt{6}}\cdot-1$
&$\frac{1}{2}$\\
$d_L$		&$\frac{2f}{\sqrt{6}}\cdot-\frac{1}{2}$	&$-\frac{2f}{\sqrt{6}}\cdot-\frac{1}{2}$
&$\frac{1}{8}$\\
$\bar d_R$	&$\frac{2f}{\sqrt{6}}\cdot-\frac{1}{2}$	&$-\frac{2f}{\sqrt{6}}\cdot-\frac{1}{2}$
&$\frac{1}{8}$\\\hline\hline
\end{tabular}
\end{center}
\end{table}%
All such terms are of identical sign so no cancellation takes place. 

\begin{figure}
\includegraphics[width=\linewidth]{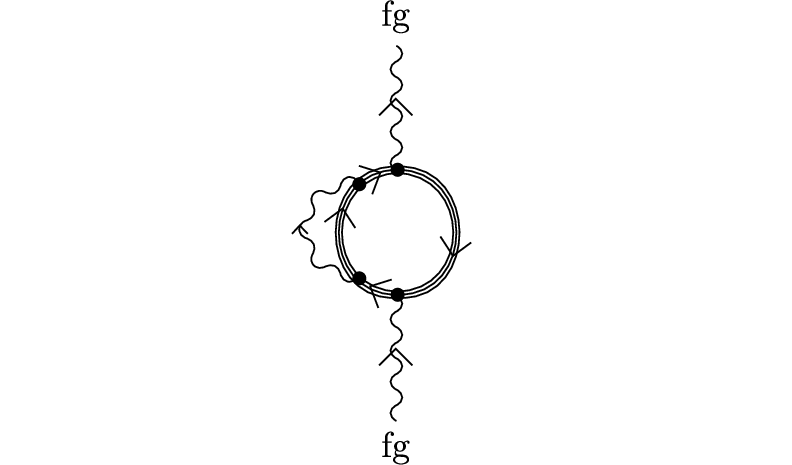}
\caption{A boson loop coupling a fermion to itself as shown may be expanded to yield preon proper self-energy terms and three preon-preon couplings between members of the fermion triplet.\label{fig:Zfermiloop}}
\end{figure}

Each choice of fermions in the loop then acquires corrections from all uncancelled terms (i.e.~all terms, in this context). These must be evaluated at the preon level, but a shortcut is possible on recognising that: 
\begin{itemize}
\item In the leptons, \begin{itemize}
\item the three preons are identical, and after incorporating the fermion normalisation, each contributes an effective $Z$~coupling which is $\frac{1}{3}$ that of the corresponding fermion as per \Psref{III}{sec:EWint_numerical}. 
\item The sum over three loop positions and three pseudovacuum expansions per loop cancels with resulting the factor of $\frac{1}{9}$.
\end{itemize}
\item In the quarks, \begin{itemize}
\item as discussed in \sref{sec:WcorrsAWZ}, each all-$A$-sector interaction is accompanied by one all-$C$-sector interaction. Collectively these interactions have the effect of arbitrarily rearranging the preons between the two vertices, so the sum over position of the interacting preon proceeds independently at source and sink. 
\item If source and sink are time-ordered this yields six loop diagrams; they are not time-ordered, so this reduces to three.
\item Again, each diagram has three different pseudovacuum expansions.
\item Again this cancels a factor of~$\frac{1}{9}$ from fermion normalisation.
\end{itemize}
\end{itemize}
The net result is that for both leptons and quarks, the numerical factor associated with all preon/$Z$/preon loops within a given fermion is precisely the same as the factor associated with a $Z$~loop coupled at each end to that fermion, as shown in \fref{fig:Zfermiloop}, if one ignores that such a term would normally be absorbed in to the fermion PSE. 
Indeed, expanding each vertex in terms of the three possible preon couplings yields nine diagrams, of which three couple a preon to itself so are not loop corrections to the vertex, but the other six are precisely those diagrams which provide the loop corrections. (More generally, \fref{fig:Zfermiloop} is used to describe both a PSE correction to the fermion propagator \emph{and} the loop corrections to a boson/fermion vertex, being normalised away in the former, and yielding a numerical factor in the latter.)

In light of the above, the mean $Z$~loop coefficient on the preon limb, after summing over preon diagrams and pseudovacuum expansions, may be calculated as the square of the $Z$/fermion~coefficient (the vertex factor in \tref{tab:ZcorrZ}) multiplied by the loop weight. This evaluates as ${5f^2}/{12}$. The calculation is the same on the antipreon limb, so the mean value is the same across all species and diagrams. It will be corrected to $\ILO{\alpha}$ by higher-order terms not explicitly evaluated.

Compare this with the photon loop correction to the $W$~boson mass, where the equivalent mean factor across all species and diagrams is $f^2/2$, corrected by higher-order terms corresponding to $(1+a_e)$. For the photon, the loop correction factor is $\alpha/(2\pi)$. On na\"\i{}eve inspection the $Z$~boson loop correction to $Z$~mass has twice as many loop diagrams as the photon correction to $W$~mass (as the former includes both preon and antipreon terms, but the latter acquires loops only on the preon limb). Recalling that the $Z$~boson is also effectively massless in this context, this would yield a net factor of
\begin{equation}
2\cdot\frac{5f^2}{12}\cdot\left(\frac{f^2}{2}\right)^{-1}\cdot\frac{\alpha}{2\pi}\cdot\frac{1+\OO{\alpha}}{1+a_e}=\frac{5}{3}\cdot\frac{\alpha}{2\pi}\left[1+\OO{\alpha}\right].\label{eq:ZZpre}
\end{equation}
However, this calculation overlooks additional symmetries present in the $Z$~boson diagrams:
\begin{itemize}
\item For the photon loop corrections to the $W$~mass: 
\begin{itemize}
\item With the exception of interchange of the loop vertices, which is accounted for in the structure factor, there are no other vertex interchange symmetries.
\end{itemize}
\item For the $Z$~loop corrections to the $Z$~mass: 
\begin{itemize}
\item There are four identical $Z$/fermion vertices, which admit~24 permutations. One such permutation suffices to generate all preon diagrams associated with a given position of the fermion vertices, so this appears to be a 24-fold overcounting.
\item However, symmetry under exchange of the two inner loop vertices is also present in the $W$/fermion reference diagram and its associated factor of $\alpha/(2\pi)$, so factor this out and divide the count by two.
\item Similarly, symmetry under exchange of the two external loop vertices is included in the leading-order diagram of \fref{fig:Wmassbosoncorrsnew}(i) which these figures correct, so again divide the overcount by a factor of two.
\item Now consider a nonvanishing loop correction diagram. 
\begin{itemize}
\item To yield a nonvanishing correction the inner loop must have the form shown in \fref{fig:Wmassbosoncorrsnew}(ii), and not that of \fref{fig:Wmassbosoncorrsnew}(iii) which is instead a preon PSE diagram. However, the positions of the outer loop vertices are not similarly constrained.
\item All remaining vertex exchange operations swap at least one vertex from the inner loop with a vertex from the outer loop. 
\item There is therefore a one-in-three chance that the diagram resulting from a vertex exchange will have vanishing contribution to the loop correction. %
\end{itemize}
This reduces the count by a factor of $\frac{2}{3}$, for a final net overcount by a factor of~4.
\end{itemize}
\end{itemize}
To compensate for this overcount on mapping to fermions, introduce a complementary factor of~$\frac{1}{4}$. The factor associated with the $Z$~loop corrections to $Z$~boson mass therefore becomes
\begin{equation}
\frac{1}{4}\cdot2\cdot\frac{5f^2}{12}\cdot\left(\frac{f^2}{2}\right)^{-1}\cdot\frac{\alpha}{2\pi}\cdot\frac{1+\OO{\alpha}}{1+a_e}=\frac{5}{12}\cdot\frac{\alpha}{2\pi}\left[1+\OO{\alpha}\right].
\end{equation}

\paragraph[Scalar boson loop]{Scalar boson loop:}

The scalar boson loop calculation is identical to that for the $W$ boson, yielding a correction of
\begin{equation}
\frac{3\alpha}{2\pi}\left(1+\frac{1}{2\pi}\right).
\end{equation}

\paragraph[Net effect of all boson loops]{Net effect of all boson loops:}
The net effect of the boson loop corrections is therefore to amend the $Z$ boson mass equation to
\begin{align}
m_Z^2=\,&12f^2\left[k_1^{(e)}\right]^4{\omega_0}^2{N_0}^{12}S_{6,13}\\
&\times\left[1+\left(\frac{401}{12}+\frac{3}{2\pi}\right)\frac{\alpha}{2\pi}+\OO{{N_0}^{-4}}+\OO{\alpha^2}\right]\nn
\end{align}
where the next-most-relevant corrections are those due to the coupling of the $Z$ boson to the pseudovacuum scalar field, and the second-order electromagnetic corrections.

\subsubsection{Background photon, gluon, and scalar interactions\label{eq:ZmassQLphotonandscalar}}

\paragraph[Direct coupling]{Direct coupling:\label{sec:Zdirect}}
As the $Z$ boson is uncharged and colourless, it acquires no mass through direct coupling to the background photon or gluon fields. However, it still interacts with the background scalar boson field. Following a similar calculation to \sref{sec:WmassQLphotonandscalar} and separating the series as per \sref{sec:universalcorrs} yields
\begin{align}
\nn
m_Z^2=\,&12f^2\left[k_1^{(e)}\right]^4{\omega_0}^2{N_0}^{12}S_{6,13}\Biggl[1+\left(\frac{401}{12}+\frac{3}{2\pi}\right)\frac{\alpha}{2\pi}
\Biggr]\\
&\begin{aligned}
&\times\left\{1+\frac{1}{\left[k^{(e)}_{1}{N_0}\right]^4}\right\}\\
&\times\bm{\left(}1+\OOOO{\alpha\left[k^{(e)}_1{N_0}\right]^{-4}}+\OO{\alpha^2}\bm{\right)}.
\end{aligned}
\end{align}

\paragraph[Indirect (universality) coupling]{Indirect (universality) coupling:\label{sec:Zuniversalcplg}}
Rather surprisingly, however, there does exist a mechanism whereby the $Z$ boson may acquire mass from the background photon and gluon fields. Begin with the photon field, and consider again the mechanism behind the universal applicability of boson mass-squared vertex loop corrections described in \sref{sec:universalcorrs}. When the $Z$ boson is interacting with the preons of the pseudovacuum, this attracts the obvious leading-order term associated with the fermion interpretation of \fref{fig:Wmassbosoncorrsnew}(i). However, for the $Z$ boson it is also possible to construct nonvanishing diagrams from the three-boson interpretation discussed in \sref{sec:universalcorrs}. Since the $A$-sector representation matrix associated with the $Z$ boson is diagonal, the background bosons constructed from the residual preons are identical and self-conjugate, generating a diagram having the form of \fref{fig:Wphotons}(ii). 

A basis of diagonal bosons may be chosen consisting of the photon, $Z$~boson, and $N$~boson from which
it is possible to construct any given composite vector boson with representation matrix $e_{ii}$. %
However, given the gauge choices of \Psref{III}{sec:GL18Cgauge}, any mass arising from this sector must always be attributed to the contributions of the background photon field.
Consequently it is only necessary to consider triplets involving charged preons.

First, consider the lepton channels.
When the preons are electron-type preons ($a\in\{1,2\}$) there are three choices of charged preon and three choices of charged antipreon, and freedom to choose which preons to integrate out gives nine ways to make a composite vector boson whose $a$-charges indicate it relates to the photon. When the preons are neutrino-type preons ($a=3$), these have no overlap with the photon and so can be ignored. The lepton sector thus offers a total of 18 channels (nine from $e_L$ and nine from $\bar e_R$).

Next, consider the quark channels.
Again, only diagonal contributions from electron-type preons are nonvanishing and thus each down quark contributes only one channel. The up quark does not couple to the $Z$ boson.

For these twenty channels (nine each from the $e_L$ and $\bar e_R$ channels and two from $d_L$), now determine the weight of each channel when compared with the $WW^\dagger AA$ vertex of \fref{fig:Wphotons}(i)%
:
\begin{itemize}
\item On the electron diagrams:
\begin{itemize}
\item The background preon line retained at the upper vertex may be any one of three available.
\item To construct a photon, the preon line at the lower vertex must be of matching colour. However, this may be in any of the three positions for a further factor of three (either from a sum over positions of an explicit label, or if colour labelling is suppressed, then from freedom to choose among three identical preons). %
\item After integrating out the eliminated preons, the upper and lower vertex act as a single composite vertex which exhibits two photon couplings. In scenarios where the preons \emph{become} photons, as opposed to \emph{emit} photons, there is consequently no cancellation between photon interactions arising from the $e_L$ and $\bar e_R$ sectors of the pseudovacuum.
\end{itemize}
Preon configurations consistent with background fermions $e_L$ and $\bar e_R$ therefore contribute nine figures apiece.
\item On the down quark diagrams:
\begin{itemize}
\item The upper preon must be the unique preon. Its colour is whatever it is.
\item The lower preon must likewise be the unique preon. Its colour matches.
\end{itemize}
Preon configurations consistent with background quarks $d_L$ and $\bar d_R$ therefore contribute one figure apiece.
\item In each of these twenty figures, there are three choices for colour of the retained preons, for a factor of~3.
\item Preon vertex factors yield $f^2/6\equiv(\alpha/3)\left[1+\ILO{\alpha}\right]$.
\item Symmetry under interchange of the $Z$~boson source and sink is also present in the diagram which this figure corrects, so must be factored out---multiply by~$\frac{1}{2}$.
\end{itemize}
Compared with the factor of $\frac{1}{2}f^2{N_0}^8S_{6,13}$ for the $W/A$ coupling~\eref{eq:WAfactor}, %
this yields $\frac{5}{2}f^2{N_0}^8S_{6,13}$ for a net correction weight of
\begin{equation}
\frac{5}{18\left[k^{(e)}_1{N_0}\right]^4}\left[1+\OO{\alpha}\right],
\end{equation}
increasing the 
$Z$~boson mass to
\begin{align}
\nn
m_Z^2=\,&12f^2\left[k_1^{(e)}\right]^4{\omega_0}^2{N_0}^{12}S_{6,13}\Biggl[1+\left(\frac{401}{12}+\frac{3}{2\pi}\right)\frac{\alpha}{2\pi}
\Biggr]\\
&\times\left\{1+\frac{23}{18\left[k^{(e)}_{1}{N_0}\right]^4}\right\}%
\label{eq:Zwithk2pre}\\
&\times\bm{\left(}1+\OOOO{\alpha\left[k^{(e)}_1{N_0}\right]^{-4}}+\OO{\alpha^2}\bm{\right)}.\nn
\end{align}

There is also an indirect coupling to the gluon fields of the pseudovacuum. For this coupling, only the colour charges on the background preon lines of \fref{fig:Wmassbosoncorrsnew} are relevant, and unlike in the $A$~sector, \sref{sec:consequences} allows that these charges need not form dual pairs. In principle there may therefore exist nine distinct choices for the boson at the upper vertex which is not integrated out, and nine at the lower. However, this space of bosons includes the one-dimensional subspace associated with $\lambda^C_9$, corresponding to $N_\mu$. This species does not couple to other bosons due to the vanishing of all commutators involving $\lambda^A_9\otimes\lambda^C_9=\mbb{I}_9$, reducing the number of available boson species at each vertex to eight. There are thus $8\cdot 8=64$ channels compared with 20 for the photon sector. Further, the interaction vertices correspond to gluon couplings of strength $f$ rather than photon couplings of strength $\frac{f}{\sqrt{2}}$ for a further relative factor of~$2$.
This yields a net contribution from the gluon sector of
\begin{equation}
\frac{5\cdot8\cdot8\cdot2}{18\cdot20\left[k^{(e)}_1{N_0}\right]^4}\left[1+\OO{\alpha}\right]=\frac{32}{18\left[k^{(e)}_1{N_0}\right]^4}\left[1+\OO{\alpha}\right],\label{eq:Zbosonccterm}
\end{equation}
increasing the $Z$~boson mass to
\begin{align}
\nn
m_Z^2=\,&12f^2\left[k_1^{(e)}\right]^4{\omega_0}^2{N_0}^{12}S_{6,13}\Biggl[1+\left(\frac{401}{12}+\frac{3}{2\pi}\right)\frac{\alpha}{2\pi}
\Biggr]\\
&\times\left\{1+\frac{55}{18\left[k^{(e)}_{1}{N_0}\right]^4}\right\}%
\label{eq:Zwithk2}\\
&\times\bm{\left(}1+\OOOO{\alpha\left[k^{(e)}_1{N_0}\right]^{-4}}+\OO{\alpha^2}\bm{\right)}.\nn
\end{align}

This completes calculation of the standard (colourless) $Z$~boson mass to the level of precision employed in this paper. 

\subsection{Weak mixing angle\label{sec:weakmix}}

If the weak mixing angle is defined in terms of $W$ and $Z$ boson mass, the above results for $m_W^2$ and $m_Z^2$ imply a weak mixing angle
\begin{align}
\sin^2\theta_W=1-&\frac{m_W^2}{m_Z^2}\label{eq:weakmix}\\
\nn=1-&\frac{3\left[1%
+\left(64+\frac{3}{2\pi}-f_Z\right)\frac{\alpha}{2\pi}\right]%
\left\{1+\frac{19}{18\left[k^{(e)}_{1}\right]^4{N_0}^4}\right\}}
{4\left[1%
+\left(\frac{401}{12}+\frac{3}{2\pi}\right)\frac{\alpha}{2\pi}\right]%
\left\{1+\frac{23}{18\left[k^{(e)}_{1}\right]^4{N_0}^4}\right\}}\\
&\times\bm{\left(}1+\OOOO{\alpha\left[k^{(e)}_1{N_0}\right]^{-4}}+\OO{\alpha^2}\bm{\right)}
\end{align}
where $f_Z$ in turn depends on $\sin^2\theta_W$ \eref{eq:fZ} and it is necessary to solve for consistency.
It is worth noting that the corrections described above for the $W$ and $Z$ boson mass diagrams also apply to foreground fermion/weak boson interaction vertices. 
For the $Z$ boson the magnitude of the corrections show some variation between interactions with different species, with (for example) the electrons attracting EM loop corrections which are not present for the neutrino. For foreground weak sector lepton/boson interactions these vertex corrections mirror those observed in the Standard Model, and the precision of the CASMIR reproduction of the weak boson mass ratio therefore implies a comparable precision of concordance with the all-order Standard Model weak interaction vertices.

\subsection{Coloured boson masses\label{sec:colouredbosonmasses}}

\subsubsection{Gluon and \prm{N}~boson masses\label{sec:gluonmasses}}

Recognising that the emergent local $\GLTR$ symmetry of the colour sector is unbroken at the preon scale, it is reasonable to work in the $e^C_{ij}$ basis. As noted in \sref{sec:consequences}, in this basis the fields $c^{ij}_\mu$ and $c^{kl}_\mu$ need not be dual to one another for the expectation value $\la\bgfield{c^{ij\mu}c^{kl}_\mu}\ra$ to be nonvanishing. It follows that the fields $c^{ij}_\mu$ are not a basis of mass eigenstates. However, it is convenient to work in this basis %
because $\GLTR$ symmetry ensures that the total mass in the gluon sector is divided evenly across these nine components. The nominal gluon mass in the $e^C_{ij}$ basis is denoted $m_c$.

To determine $m_c$ it suffices to calculate the mass of one off-diagonal gluon [though remaining aware that FSF symmetry factors must be computed using a basis of $\SU{3}_C\oplus\GL{1}{R}_N$ as per \sref{sec:consequences}]. Evaluation of the gluon mass is therefore similar to evaluation of $W$~boson mass, and indeed the lepton/fermion contribution of the leading order diagram and the preon-to-preon gluon loop corrections to this diagram proceed equivalently. Where the $W$ and gluon mass calculations differ is in the contribution from interactions with the pseudovacuum boson fields. For the $W$~boson this contribution arises from the pseudovacuum photon and scalar boson fields. For the gluons, there are couplings to the pseudovacuum gluon, photon, and scalar boson fields.

\begin{figure}
\includegraphics[width=\linewidth]{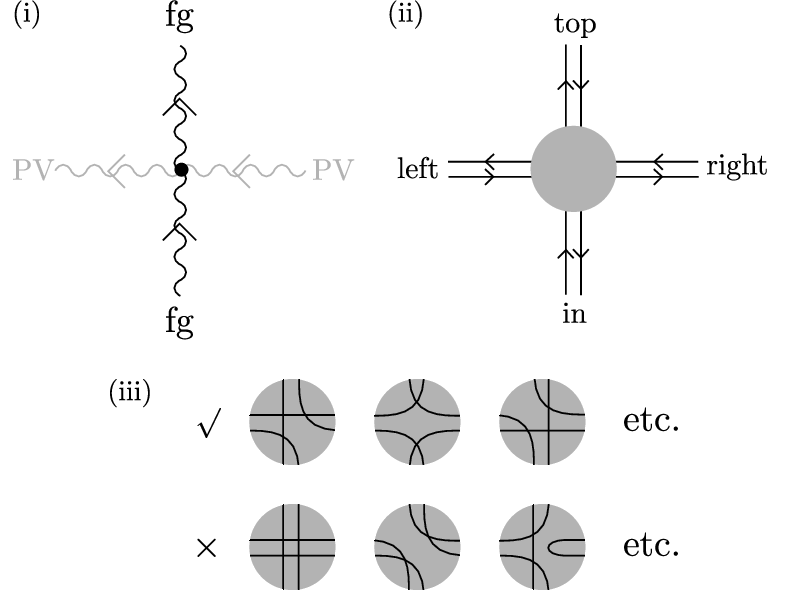}
\caption{(i)~As the gluon interaction with the background gluon field need not conserve colour on a per-interaction basis, foreground gluons may interact with a pair of background gluons having non-complementary charges. (ii)~All valid nontrivial vertices having the form of diagram~(i) correspond to colour-consistent preon interconnections of the participating gluons. Labelling of gluons (``in'', ``left'', ``right'', and ``top'') is for reference in the main text (iii)~The valid nontrivial vertices are those which are not disconnected (\prm{\surd}). In each such diagram the preon lines of any given gluon are connected to two other gluons. Excluded diagrams (\prm{\times}) are disconnected diagrams, and comprise
those in which the gluons approach closely without engaging in preon exchange, or in which a preon line bends back on itself.\label{fig:gluonQLgluoninteraction}}
\end{figure}

To evaluate the gluon contribution to the $\ILO{{N_0}^{-4}}$ term, recognise that the preservation of colour cycle invariance across the entirety of the $\Cw{18}$ analogue model guarantees that all gluons always appear in the context of a superposition of all nine possible species. Interactions with the pseudovacuum in the $e^C_{ij}$ basis need not therefore conserve colour charge on an individual gluon on a term-by-term basis provided colour charge is collectively conserved across the superposition. (Individual gluon colour will, however, be preserved on average over length or time scales sufficiently large compared with $\mc{L}_0$ as the pseudovacuum has net trivial colour.) For mass interactions, the consequence of this is that rather than interacting with a single looped boson as per \fref{fig:Wphotons}(i), a gluon can interact with a pair of different gluons from the pseudovacuum as per \fref{fig:gluonQLgluoninteraction}.

To evaluate the background gluon contribution, note:
\begin{itemize}
\item This interaction has coefficient $f^2$, compared with $f^2/2$ for the photon. 
\item A diagram in which the two background gluons have different field operators on the vertex, $\varphi^{c_1\dot{c}_1}\varphi^{c_2\dot{c}_2}$, receives a factor of $\frac{1}{2}$ relative to the photon term due to loss of vertex symmetry, but a factor of two as these fields may be pulled down from the generator~$\Z$ in either order. A diagram in which the two background gluons have the same field operator attracts neither of these factors.
\item To count valid colourations of \fref{fig:gluonQLgluoninteraction}, note that:
\begin{itemize}
\item the inbound preon in the ``in'' gluon of \fref{fig:gluonQLgluoninteraction}(ii) may be connected to any of ``top'', ``left'', or ``right'' for a factor of~$3$, and
\item the outbound preon in the ``in'' gluon may be connected to either of the other two for a factor of~$2$. It may not be connected to the same gluon as the inbound preon or else this yields a diagram in which two gluons pass close but do not interact.
\item Excluding diagrams in which one gluon corresponds to a wholly disconnected noninteracting preon line bent back on itself, the remaining two line completions then admit only one arrangement, for a factor of~$1$. %
\item There are nine choices of colouration for these two remaining lines, for a factor of~$9$.
\item Note that there is no symmetry factor associated with which gluon line is labelled as foreground. With the ``left'', ``right'', and ``top'' gluons' co-ordinates ranging freely across all space, any such factor~$n$ would be offset by an equivalent $n$-fold multiple-counting.
\item Further, it is in principle not necessary to specify explicitly which preon lines are associated with the foreground momentum in \fref{fig:gluonQLgluoninteraction}(ii) as this is carried diffusely across all local FSFs (\sref{sec:naturefg}). %
\end{itemize}
There are therefore $3\cdot 2\cdot 9=54$ valid diagrams having the form of \fref{fig:gluonQLgluoninteraction}(i).
\end{itemize}
The net contribution to gluon mass from the background gluon field is therefore $2\cdot\frac{1}{2}\cdot 2\cdot 54$ times larger than the contribution to $W$ boson mass from the background photon field, %
\begin{equation}
\frac{108}{18\left[k^{(e)}_{1}N_0\right]^4}\left[1+\OO{\alpha}\right].
\end{equation} 

Next, consider the background photon contribution to gluon mass via the indirect (universalilty) coupling mechanism of \sref{sec:universalcorrs}. Evaluation of this contribution proceeds identically to that for the $Z$~boson up to the following considerations:
\begin{itemize}
\item The gluon is associated with $A$-sector representation $\lambda^A_9$, having coefficients $\frac{1}{\sqrt{3}}$ in the $(\bar e_L,e_L)$ and $(e_R,\bar e_R)$ positions, compared with $\frac{1}{\sqrt{6}}$ for the $Z$~boson representation~$\lambda^A_8$. This yields a relative factor of~2.
\item The $Z$~boson mass interaction attracts a factor of~$2$ for the two ways the vertex may be connected to source and sink. For the prototypical off-diagonal gluon under consideration, $c_{ij}|_{i\not=j}$, this factor is absent for a relative factor of~$\frac{1}{2}$.
\end{itemize}
Consequently, as with the $Z$~boson, the gluon coupling to the background photon field yields a contribution
\begin{equation}
\frac{5}{18\left[k^{(e)}_{1}N_0\right]^4}.
\end{equation} 

The scalar contribution is unchanged from both the $W$ and $Z$ boson calculations at
\begin{equation}
\frac{18}{18\left[k^{(e)}_{1}N_0\right]^4}.
\end{equation} 

To the same order as used in \Eref{eq:Wwithk2} above, the gluon mass is therefore given by
\begin{align}
m_c^2=\,&9f^2\left[k_1^{(e)}\right]^4{\omega_0}^2{N_0}^{12}S_{6,13}\left[1+\left(64+\frac{3}{2\pi}-f_Z\right)\frac{\alpha}{2\pi}\right]\nn\\
&\begin{aligned}
&\times\left\{1+\frac{131}{18\left[k^{(e)}_{1}N_0\right]^4}\right\}%
\\
&\times\bm{\left(}1+\OOOO{\alpha\left[k^{(e)}_1{N_0}\right]^{-4}}+\OO{\alpha^2}\bm{\right)}.
\end{aligned}
\label{eq:gluonbaremass}
\end{align}

As noted in \cref{ch:boson}, this mass %
is unobservable under normal circumstances as the gluon is confined on a length scale which is small compared with both the characteristic scale of the mass interaction, $\mc{L}_\preon\ll\mc{L}_0$, and the minimum scale of the mass interaction, $\mc{L}_\preon<2\mc{L}_\Omega$ (see \sref{sec:Wmass5v} and \cref{ch:CDF2}). However, as discussed briefly in \aref{apdx:gluonmass}, mass interactions themselves are an exception to this rule and contain gluon loops dominated by intermediate-scale contributions $2\mc{L}_\Omega<\mc{L}<\mc{L}_0$ 
for which the gluon demonstrates an effective mass $m_c^2$. (This is further corrected to an energy-dependent value denoted $[m_c^*(\mc{E})]^2$ in \sref{sec:gluonscalarmassdeficit}.) %

Finally, although the $e^C_{ij}$ basis is convenient when working with nine gluons collectively, the other basis frequently used is that of $\lambda^C_\tc$ where $\tc\in\{1,\ldots,8\}$ is an 8-dimensional representation of $\SU{3}$ and $\tc=9$ corresponds to the identity matrix and yields the neutral vector boson $N_\mu$. This neutral boson has no direct colour coupling because its representation $\lambda^C_9$ commutes with all the other representation matrices. 
The indirect $C$-sector coupling for $\lambda^C_9$ is the same as that of the $Z$~boson, attracting a factor of
\begin{equation}
\frac{32}{18\left[k^{(e)}_1{N_0}\right]^4}\left[1+\OO{\alpha}\right],\label{eq:Nbosonccterm}
\end{equation}
for a total $N$~boson mass of
\begin{align}
m_N^2=\,&9f^2\left[k_1^{(e)}\right]^4{\omega_0}^2{N_0}^{12}S_{6,13}\left[1+\left(64+\frac{3}{2\pi}-f_Z\right)\frac{\alpha}{2\pi}\right]\nn\\
&\begin{aligned}
&\times\left\{1+\frac{55}{18\left[k^{(e)}_{1}N_0\right]^4}\right\}%
\\
&\times\bm{\left(}1+\OOOO{\alpha\left[k^{(e)}_1{N_0}\right]^{-4}}+\OO{\alpha^2}\bm{\right)}.
\end{aligned}
\label{eq:Nmass}
\end{align}

This completes calculation of %
gluon and $N$~boson masses to the level of precision employed in this paper. 

~

Next, consider the bosons in $\GLNR$ which carry charges on both the $A$ and $C$ sectors. These bosons may be considered coloured counterparts to the $A$-sector bosons already discussed.

\subsubsection{Coloured~$W$ boson masses\label{sec:colouredWmass}}

The coloured $W$ bosons %
are necessarily massive. As previously, the background fields exist in the large particle number regime and it is largely possible to treat them as being made up of pure $A$-sector and $C$-sector bosons (\aref{apdx:gaugeSU9}). On the $A$~sector, a coloured $W$ boson exhibits the same interactions with the background photons as its uncoloured counterpart by virtue of its nonvanishing electromagnetic charge. 
On the $C$~sector, the colourless $W$~boson exhibits no interaction with the gluons of the pseudovacuum and thus the interactions between the $C$-sector charges of a coloured $W$ boson are to first approximation independent of the $A$-sector character of the boson. To the extent that this is true, the coloured $W$ boson therefore exhibits the same interactions on the $C$~sector as a gluon. 

To move beyond this first approximation, consider again the diagrams of \fref{fig:gluonQLgluoninteraction}(ii) in which a background gluon merely passed near the foreground boson and did not interact through colour exchange. For the coloured $W$ boson, in the absence of colour exchange there may still be an interaction with these background fields which are nominally gluons, but in reality also possess a foreground $A$-sector charge. The separable sector regime implicitly corresponds to summation over all $A$-sector charges which these bosons may carry, and by gauge only that associated with $\lambda^A_3$ (the photon) makes a nonvanishing contribution to the mass interaction. These particular previously-eliminated diagrams are, in effect, restored to the colour sector sum, up to the following corrections:
\begin{itemize}
\item The interaction strength $f^2$ becomes ${f^2}/{2}$ due to a $\lambda^A_3$-mediated coupling replacing the $e^C_{ij}$-mediated gluon coupling, for a relative factor of $\frac{1}{2}$.
\item Without loss of generality, suppose the ``in'' boson connects to the ``left'' boson. Interchange of the two ends of the other boson (``right'' and ``top'') is then a symmetry of the diagram family, and yields a double counting. (Where the two colours in the boson match, integration of both co-ordinates over all spacetime yields a twofold redundancy. Where they do not match, integration over spatial position and summing over labelling yields the equivalent twofold redundancy.) Elimination of this double-counting introduces a relative factor of $\frac{1}{2}$.
\end{itemize}
The diagrams in which a preon loops back on itself within a single gluon continue to be excluded.

As a result of these corrections there are 27 recovered figures, each weighted by $\frac{1}{2}\cdot\frac{1}{2}$ relative to the 54~figures of \sref{sec:gluonmasses}, for a net adjustment to the background gluon contribution
\begin{equation}
\frac{108}{18\left[k^{(e)}_{1}N_0\right]^4} = \frac{2\cdot\frac{1}{2}\cdot2\cdot54}{18\left[k^{(e)}_{1}N_0\right]^4}\longrightarrow\frac{2\cdot\frac{1}{2}\cdot2\cdot\left(54+27\cdot\frac{1}{2}\cdot\frac{1}{2}\right)}{18\left[k^{(e)}_{1}N_0\right]^4} = \frac{243}{36\left[k^{(e)}_{1}N_0\right]^4}.
\end{equation}
The photon and scalar boson contributions are unchanged from before, and in the $e^C_{ij}$ basis, the coloured $W$ bosons therefore have a mean effective mass
\begin{align}
m_{W^{\dot cc}}^2=\,&9f^2\left[k_1^{(e)}\right]^4{\omega_0}^2{N_0}^{12}S_{6,13}\left[1+\left(64+\frac{3}{2\pi}-f_Z\right)\frac{\alpha}{2\pi}\right]\nn\\
&\begin{aligned}
&\times\left\{1+\frac{281}{36\left[k^{(e)}_{1}N_0\right]^4}\right\}%
\\
&\times\bm{\left(}1+\OOOO{\alpha\left[k^{(e)}_1{N_0}\right]^{-4}}+\OO{\alpha^2}\bm{\right)}.
\end{aligned}
\label{eq:Wccbaremass}
\end{align}
Recalling that the $e^C_{ij}$ basis is not a mass eigenbasis, recognise that for the $W$~boson family this superposition includes the colourless $W$ boson (associated with $\lambda^C_9$), which does \emph{not} acquire a colour coupling through the indirect (universality) coupling mechanism as the available $A$-sector charges on \fref{fig:Wmassbosoncorrsnew}(i) are not compatible with this mechanism. In contrast with the gluons, the $W$~sector therefore breaks $\GLTR_C$ symmetry and two subgroups of $W$~boson may be identified: The colourless $W$ boson, associated with $\GL{1}{R}_C$ and having a mass given by \Eref{eq:Wwithk2}, and eight coloured $W$ bosons with unbroken $\SU{3}_C$ symmetry having mass
\begin{equation}
m_{W^\tc}^2=m_W^2 + \frac{9}{8}\left(m_{W^{\dot cc}}^2-m_W^2\right).\label{eq:Wcmass}
\end{equation}

\subsubsection{Coloured~$Z$ boson masses}

As with the coloured $W$ boson, begin by calculating $m_{Z^{\dot cc}}$ in the $e^C_{ij}$ basis. The calculation initially proceeds as for the colourless $Z$ boson. In particular, note that couplings have already been described between the colourless $Z$ boson and the gluon sector of the pseudovacuum. The mechanism of this coupling is independent of the specific colours of the participating preons, exhibiting $\GLTR$ symmetry, and thus proceeds identically for $m_{Z^{\dot cc}}$ as for $m_Z$. As with the gluon and coloured $W$ boson, some diagrams are excluded on account of vanishing colour-sector interactions; for the $Z$ boson, specifically those associated with colour sector representations $\lambda^9_C$. For the colourless $Z$ boson, operating in the sector-separable regime, this reduces the number of interaction channels from $9\cdot 9=81$ to $8\cdot 8=64$.

However, as with the coloured $W$ boson, the coloured $Z$ boson does not respect separability of the $A$~and $C$ sectors. The background colour excitations must therefore be acknowledged to carry $A$-sector charges associated with representation $\lambda^A_3$, and not $\lambda^A_9$ as was assumed in the mass calculation for the colourless $Z$ boson. Much as the colourless $Z$ boson is able to acquire mass from the coloured species of the pseudovacuum on account of being made up of preons, despite the boson as a whole carrying no net charge with respect to the $C$~sector, the coloured $Z$ boson is similarly able to acquire mass from interaction with the $A$-sector charge of the excluded gluon-like components of the pseudovacuum despite carrying no net electromagnetic charge. 

It turns out that it is necessary and sufficient to consider additional interactions \emph{only} on those vector components of the pseudovacuum which do not interact with the colourless $Z$ boson:
\begin{itemize}
\item Regarding necessity: To observe that these interactions exist and have not already been accounted for in the $A$-sector contributions to coloured $Z$ mass, begin by recognising that the assumption of separability of the pseudovacuum into $A$~and $C$ sectors approximates the vector boson component of the pseudovacuum by species interacting purely in the $A$~sector and species interacting purely in the $C$~sector. To the extent that the foreground boson acquiring mass also behaves as a pure $A$- or $C$-sector boson, mass is acquired from these background species according to the pure $A$-sector or pure $C$-sector interactions, whether direct or indirect as per \sref{sec:universalcorrs}. However, in an appropriate basis there exist $C$-sector components of the separated pseudovacuum which carry representation $\lambda^C_9$, which necessarily does not interact with the foreground boson. On the other hand, in contexts where separability does not hold (for example when the foreground boson carries both $A$- and $C$-sector charges), it must be acknowledged that all vacuum particles in fact carry $A$-sector representation $\lambda^A_3$. So long as the foreground particle interacts nonvanishingly with pseudovacuum components carrying representation $\lambda^A_3$, it must therefore interact nonvanishingly with \emph{all} elements of the pseudovacuum which have nonvanishing vacuum expectation value. A coloured particle of $Z$~boson character, having indirect interactions with representation $\lambda^A_3$ via the universality coupling of \srefs{sec:universalcorrs}{sec:Zuniversalcplg}, is therefore an example of a species which interacts with all elements of the vacuum having nonzero vacuum expectation values (VEVs). As the $\lambda^C_9$ gluons in the separated vacuum have nonzero VEVs, they must represent excitations which interact with the coloured $Z$ boson but which are not faithfully represented by the separated pseudovacuum. This interaction must take place on the $A$~sector, and therefore this interaction deficit may be represented by additional particles carrying representations $\lambda^A_3\otimes\lambda^C_9$ replacing the excluded neutral gluon ($\lambda^A_9\otimes\lambda^C_9$) interactions of \sref{sec:Zuniversalcplg}. Note that this is not a statement that all (or, indeed, any) particles carrying these exact charges were neglected by the separation process---rather, that introducing additional particles with this character and associated with the specified interactions suffices to remedy this specific deficit arising from assuming separability of the pseudovacuum.
\item Regarding sufficiency, i.e.~to confirm that a secondary sector ($A$~sector for a gluon or $C$~sector for a photon) need only be considered for those components of the $A/C$-separated pseudovacuum which would otherwise be neglected, rather than enhancing the interactions of all sectors: If a species interacts with the colourless $Z$ boson via a sector $A$ or~$C$, then by the large particle limit of the vacuum and the pure $A$-sector character of the colourless $Z$ boson, this interaction is separable, and is accounted for in the pure-photon and pure-gluon treatment of \sref{sec:Zmass}. Thus the only nonseparable interactions are ones which do not take place with the colourless $Z$ boson.
Any vector mass interactions which are nonseparable must therefore involve species which have nonvanishing vacuum expectation value but do not contribute to the mass of the colourless $Z$ boson. The only candidate species are those identified as gluons but carrying representation $\lambda^C_9$.
\end{itemize}

To take into account these additional interactions, reintroduce the seventeen excluded interactions. These diagrams are weighted by additional numerical factors arising from differences in their structure and associated particle representations compared with the original sixty-four interactions of \sref{sec:Zuniversalcplg}. %
Regarding these associated numerical factors:
\begin{itemize}
\item There are $81-64=17$ gluon interaction diagrams which were excluded from the colourless $Z$ mass due to the appearance of representation $\lambda^C_9$, compared to 64 which were accepted. There are therefore $\frac{17}{64}$ times as many recovered diagrams as there are original gluon diagrams.
\item Particle interactions necessarily involve preon exchange, and the gluon interactions for the colourless $Z$ boson consequently involve exchange of one colour charge with the pseudovacuum. (Not both, as exchanging both preons would correspond to an exchange of position not an interaction, and not zero, as this is also not an interaction.) The colour charge trajectories must therefore resemble the upper row of \fref{fig:gluonQLgluoninteraction}(iii), with the nominal foreground input and output sharing exactly one preon line. However, the colourless $Z$ mass vertex (and any $\GLTR$-symmetric transformation of it) must conserve colour on the foreground boson, implying that the colour of one of the background preons is fixed to match the foreground preon it replaces. In contrast, for processes involving $\lambda^C_9$, neither background preon is constrained to match a foreground preon as no colour exchange takes place. Therefore the seventeen recovered diagrams admit three times as many colourings each than do the original sixty-four.
\item The recovered interactions are associated with $\lambda^A_3$ rather than $e^C_{ij}$ and therefore they attract an interaction coefficient which is smaller by a factor of $\frac{1}{2}$.
\item As discussed for the coloured $W$~boson, the interplay of summing over colours and integrating over position yields a double-counting, therefore multiply by a further factor of~$\frac{1}{2}$.
\end{itemize}
Compared with the colourless $Z$ boson, the background gluon contribution is therefore adjusted as
\begin{equation}
\frac{32}{18\left[k^{(e)}_{1}N_0\right]^4}\longrightarrow\frac{32+32\cdot\frac{17}{64}\cdot3\cdot\frac{1}{2}\cdot\frac{1}{2}}{18\left[k^{(e)}_{1}N_0\right]^4} = \frac{307}{144\left[k^{(e)}_{1}N_0\right]^4}.
\label{eq:Zcadjustment}
\end{equation}
This results in a mean effective mass in the $e^C_{ij}$ basis 
\stepcounter{equation} %
\begin{align}
m_{Z^{\dot cc}}^2=\,&12f^2\left[k_1^{(e)}\right]^4{\omega_0}^2{N_0}^{12}S_{6,13}\Biggl[1+\left(\frac{401}{12}+\frac{3}{2\pi}\right)\frac{\alpha}{2\pi}
\Biggr]\nn\\
&\times\left\{1+\frac{491}{144\left[k^{(e)}_{1}{N_0}\right]^4}\right\}%
\label{eq:Zccbaremass}\\
&\times\bm{\left(}1+\OOOO{\alpha\left[k^{(e)}_1{N_0}\right]^{-4}}+\OO{\alpha^2}\bm{\right)}.\nn
\end{align}
Again $\GLTR_C$ symmetry is broken, yielding %
a mass for the $\SU{3}_C$-symmetric coloured $Z$ boson subgroup
\begin{equation}
m_{Z^\tc}^2=m_Z^2 + \frac{9}{8}\left(m_{Z^{\dot cc}}^2-m_Z^2\right).\label{eq:Zcmass}
\end{equation}

\subsubsection{$W$~bosons in loop corrections\label{sec:tW}}

Having established that there are multiple different bosons carrying the $A$-sector charges associated with the $W$ and $Z$ bosons, it is now necessary to revisit the weak loop corrections to the $W$ and $Z$ boson masses and ask, what sort of $W$ or $Z$ boson(s) participate in these loops?
This question may be better phrased by recognising that any colour-agnostic boson loop consists of an %
average over preon/antipreon pairs, and asking what mass-generating interactions these preon pairs may participate in while generating the loop correction.

\begin{figure}
\includegraphics[width=\linewidth]{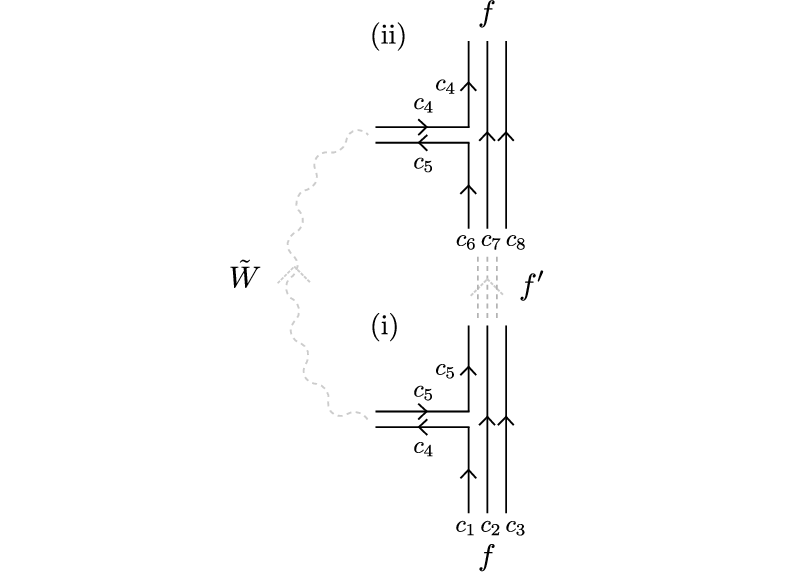}
\caption{In order for the \prm{C}-sector pseudovacuum interaction to give mass to a loop \prm{W} boson arising from a fermion $f$ with preon vertices as shown: (i)~At the source, colour \prm{c_1} must match colour \prm{c_4} (probability \prm{\frac{1}{3}}). Colour \prm{c_5} is then fixed by the other colour in the loop boson. (ii)~At the sink, colour \prm{c_5} must match colour \prm{c_6}, where colours \prm{c_6}, \prm{c_7}, and \prm{c_8} are a permutation of \prm{c_5}, \prm{c_2}, and \prm{c_3}. For \prm{c_4\not=c_5}, the probability of a match is \prm{\frac{2}{3}}. Note that signs associated with preon reordering are incorporated into the propagator~\prm{f'} between diagrams~(i) and~(ii) and there are consequently no minus signs associated with the mass interaction which is restricted to the two fully illustrated vertices.\label{fig:Wloopcolourmassterm}}
\end{figure}

As a prototype, it is convenient to consider a $W$~boson loop which is emitted and absorbed by a foreground fermion~$f$. A preon schematic for this process is shown in \fref{fig:Wloopcolourmassterm}, with some colours labelled for reference.
Since a $W$~loop is an $A$-sector loop correction, its couplings to fermions are implicitly a sum over the three possible preon couplings at each vertex, weighted by a factor of~$\frac{1}{3}$. Extension of the following discussion to include loops which couple to bosons is straightforward,
as couplings to colourless bosons are numerically equivalent, being an average over the three possible colour labels of the participating preon in that boson. %

Now, without reference to source or sink, consider couplings between a coloured $W$ boson in the $e^C_{ij}$ basis and the background gluon fields. For reference the $W$~boson couplings to the background photon field contribute a factor
\begin{equation}
\left\{1+\frac{1}{18\left[k^{(e)}_{1}{N_0}\right]^4}\right\}\left[1+\OO{\alpha}\right]
\end{equation}
to $W$~boson mass. The vertex factor associated with $W$/gluon coupling is larger by a factor of~2, and noting that the participating gluons need not be dual to one another (\sref{sec:consequences}) means there are 81 such diagrams. However, this most general interaction yields an output superposition for which, when decomposed in the $\SU{3}_C\otimes\GLTR_C$ basis, one ninth is associated with representation $\lambda^C_9$ and therefore corresponds to a colourless $W$ boson. Since the coloured and colourless $W$ bosons are different mass eigenstates, the interaction taking a coloured $W$ boson into a colourless $W$ boson does not correspond to a valid mass vertex,\footnote{Such interactions still happen, though they are not mass vertices, and by the net colour neutrality of the pseudovacuum their overall effect on \prm{W}~boson colour must vanish. Further, if the pseudovacuum transiently surrenders colour to transform a colourless \prm{W}~boson into a chromatic \prm{W} boson (or vice versa) then the associated increase (decrease) in boson mass is accompanied by a colour deficit (surfeit) in the pseudovacuum and a hole with effective negative (positive) mass. The collective mass of these excitations remains unchanged, and by the large-scale homogeneity of the pseudovacuum they may be treated as if they co-propagate, with any deviation from doing so being subsumed into pseudovacuum fluctuations. Thus the effective mass of any given boson, chromatic or otherwise, is unaffected by such transient excursions of colour charge.} and thus the number of channels (i.e.~effective number of diagrams) is reduced by a factor of $\frac{8}{9}$, for a total of 72 diagrams and a factor
\begin{equation}
\left\{1+\frac{144}{18\left[k^{(e)}_{1}{N_0}\right]^4}\right\}\left[1+\OO{\alpha}\right].
\end{equation}

Collectively, these 72 figures may be viewed as the action of a $\GLTR_C$-symmetric $\gltr_C$-valued boson field at a mass vertex. This $\gltr_C$-valued boson, in turn, acts on an excitation which is itself necessarily $\GLTR_C$-symmetric, and thus although this interaction may be colour-changing on individual figures when labelled in the $e^C_{ij}$ basis, collectively the distribution of the excitation across the different colourations of fields remains unchanged. This is most simply realised by associating the factor of ${144}/\{18[k^{(e)}_{1}{N_0}]^4\}$ with the $C$-sector mass vertex contribution but leaving the colour of the participating preons unchanged.

Finally, a given term in the $\GLTR_C$-symmetric mass vertex only contributes to the mass of the loop $W$ boson if its colours match those at the source and sink vertices. Exploit $\GLTR_C$ symmetry to treat inbound and outbound colourations on the loop as identical as described above. Colour consistency at the source vertex then requires that the outbound preon in the $W$-type boson match the colour of the inbound preon in the emitting particle, with probability~$\frac{1}{3}$. The other colour at this vertex is then determined by the colour of the inbound preon in the $W$-type boson. This is illustrated with a fermion as the source particle in \fref{fig:Wloopcolourmassterm}(i).

For definiteness, assume that the $W$-type boson carries an off-diagonal colour charge $e^C_{ij}|_{i\not=j}$, and recognise that terms with $i=j$ will yield equivalent contributions by $\GLTR_C$ symmetry. When a fermion emits a boson which is off-diagonal in colour, the fermion now contains two preons of the same colour. When the source is an $A$-sector boson, two terms in the superposition of colour combinations which make up the boson are now identical.

At the sink vertex for the $W$-type boson, again the preon line in the $W$-type boson which is outbound from the vertex must match the preon from the absorbing particle which is inbound, this time shown in \fref{fig:Wloopcolourmassterm}(ii). If the sink vertex is on a fermion, it may be on any of the three preons. If it is on a boson, it acts on each term in the superposition. The colour of the outbound preon line in the $W$-type boson at the sink vertex is now a match for two of the three preons in the fermion, or two thirds of the sum over colour terms in the boson, for a factor of~$\frac{2}{3}$. These factors of $\frac{1}{3}$ and $\frac{2}{3}$ incorporate the averaging over colour and coupling in the $e^C_{ij}$ basis referenced above. There are no constraints on arrangement of the colours of the outbound preon, and any factors of~$-1$ associated with preon reordering are subsumed into the propagator between the two vertices of the mass interaction---these take place within any free fermion propagator regardless of the presence of mass vertices, whereas the mass interaction comprises processes additional to the propagation of the massless fermion.

Evaluating $144\cdot\frac{1}{3}\cdot\frac{2}{3}=32$, the net contribution to the loop $W$ mass from couplings to the $C$-sector bosons of the pseudovacuum is thus
\begin{equation}
\left\{1+\frac{32}{18\left[k^{(e)}_{1}{N_0}\right]^4}\right\}\left[1+\OO{\alpha}\right],\label{eq:Wloopccterm}
\end{equation}
and the $W$~boson participating in the loop, henceforth denoted $\tW$, is seen to have a mass satisfying %
\begin{align}
\nn
m_\tW^2=\,&9f^2\left[k_1^{(e)}\right]^4{\omega_0}^2{N_0}^{12}S_{6,13}\Biggl[1+\left(64+\frac{3}{2\pi}-f_Z\right)\frac{\alpha}{2\pi}\Biggr]\\
&\begin{aligned}
&\times\left\{1+\frac{51}{18\left[k^{(e)}_{1}{N_0}\right]^4}\right\}\\
&\times\bm{\left(}1+\OOOO{\alpha\left[k^{(e)}_1{N_0}\right]^{-4}}+\OO{\alpha^2}\bm{\right)}.\end{aligned}
\label{eq:mtW}
\end{align}
distinct from all three previous values $m_W$~\eref{eq:Wwithk2}, $m_{W^{\dot cc}}$~\eref{eq:Wccbaremass}, and $m_{W^\tc}$~\eref{eq:Wcmass}.

Consideration for $Z$~bosons in electroweak loop corrections proceeds similarly, likewise yielding factor~\eref{eq:Wloopccterm} arising from couplings to the pseudovacuum gluon fields. However, for the $Z$~boson this term is exactly the same as the colour coupling of the standard $Z$~boson~\eref{eq:Zbosonccterm} and thus the loop $Z$ boson mass is simply $m_Z$.

In hindsight, all loop $W$ bosons in \srefr{sec:Wmass}{sec:Zmass} are also necessarily $\tW$~bosons. The expressions for boson masses in \srefr{sec:Wmass}{sec:Zmass} are unaffected by this change save for the substitution $W\longrightarrow \tW$ in $f_Z$ which becomes
\begin{equation}
f_Z=\frac{1}{3}\left(4-24\frac{m_\tW^2}{m_Z^2}+16\frac{m_\tW^4}{m_Z^4}\right)\label{eq:fZ2}.
\end{equation}

\subsubsection{Coloured photons and their role in mass interactions\label{sec:colouredAmass}}

Now consider the bosons which carry representation~$\lambda^A_3$ on the $A$~sector and a nontrivial representation~$\lambda^C_\tc|_{\tc\not=9}$ on the $C$~sector. On extending the gauge choices of \sref{sec:gaugechoice} to $\GLNR$ as per \aref{apdx:gaugeSU9} all species carrying the photon representation~$\lambda^A_3$ are massless by gauge. With all preons carrying both coloured and electromagnetic charge, the massless coloured photons (gluphotons) behave similarly to massless gluons. 

It is first helpful to compare these particles with the true gluons, which have trivial representation on the $A$~sector. The gluons $c^{ij}_\mu$ are theoretically massive over length scales large compared with $2\mc{L}_\Omega$, but they are almost universally confined to length scales small compared with $\mc{L}_\preon$, which is argued in \sref{sec:chromenv} to satisfy $\mc{L}_\preon=\mc{L}_\Omega$. Thus both the true gluons and the gluphotons are massless on the characteristic scale of colour interactions. 

It is next useful to consider the contribution of the gluphotons to $A$-sector interactions. By \aref{apdx:ACseparability} the $A$~and $C$ sectors are separable in the large-particle-number limit, and consequently the existence of gluphotons may augment $A$-sector interactions where couplings to these bosons exist.

First consider lepton-photon couplings e.g.~$A_\mu \bar{e}_L\bsm e_L$, and recognise that if the colourless photon $A_\mu$ is replaced by a coloured gluphoton, it must change the colour of one of the photons in the fermion. However, the fermion triplet is constrained to remain colour-neutral by choice of co-ordinate frame on $\Cw{18}$ (\sref{sec:Csector}) and thus fermions in isolation cannot couple to gluphotons.\footnote{This is circumvented when the fermion is not isolated, but instead is accompanied by a species which can act as a recipient for colour charge, as with the chromatic \prm{Z}~loop of \psref{sec:vecbosexchg}.}
For $W$~boson couplings, on the other hand, the preon in the $W$~boson which couples to the photon exists in a superposition over colour. Further, each entry in this superposition may couple to any of three gluphotons in the $e^C_{ij}$ basis. If the $W$~boson itself is a sum over possible colour labellings, e.g.~a loop $W$ boson as discussed in \sref{sec:tW}, then the sum over all gluphotons (subject to the ability to construct a colour-consistent completion of the diagram) may leave the construction of that $W$~boson unchanged. By $A$-$C$ separability in the large particle number limit, it is not necessary to explicitly assess how the $C$-sector bosonic charge associated with $\{\lambda^C_\tc\}_{\tc<9}$ remains confined to infer that the $A$-sector bosonic charge $(\lambda^A_3)$ may escape to the far field, and thus this coupling augments the effective electromagnetic 4-vector potential associated with such colour-sum $W$ bosons.

In contrast, $W$~bosons which are known to be colourless (for example by their decay products, e.g.~$W\rightarrow e_L+\nu_e$) do not have the freedom to undergo mixing of $C$-sector representations $\lambda^C_\tc$ and thus their electromagnetic 4-vector potential is not augmented by gluphoton contributions to the $A$~sector.

The only known experiment which might have directly detected coloured $W$ bosons is CDF~II (\sref{sec:weakmasses}). However, even if the CDF~II measurement of $W$~boson mass incorporates contributions from coloured $W$ bosons, %
I am not aware of any reconstruction of their trajectories and thus CDF~II provides no window into whether the chromatic $W$~boson apparent electromagnetic charge attracts a correction through coupling to gluphotons.

The only known experiment which measures the electromagnetic coupling of a loop $W$ boson is the Muon~$g-2$ experiment at Fermilab. The value of the muon gyromagnetic anomaly obtained from this experiment is consistent with the existence of coloured $A$, $W$, and $Z$ counterparts and loop~$W$/gluphoton coupling, with tension~$0.2\,\sigma$ (upper bound~$0.5\,\sigma$). This is discussed in \sref{sec:muong2}. 

Finally, gluphotons provide a signficant correction to the strength of the CASMIR analogue of the gravitational interaction. The strength of this gravitational analogue is also consistent with experiment---tension~$<0.1\,\sigma$ (upper bound~$0.3\,\sigma$). See \cref{ch:gravity} for the analogue interaction and its strength, specifically \sref{sec:pairdecay} for the gluphoton correction term.

\subsection{Scalar boson mass\label{sec:scalbosmass}}

\begin{figure}
\includegraphics[width=\linewidth]{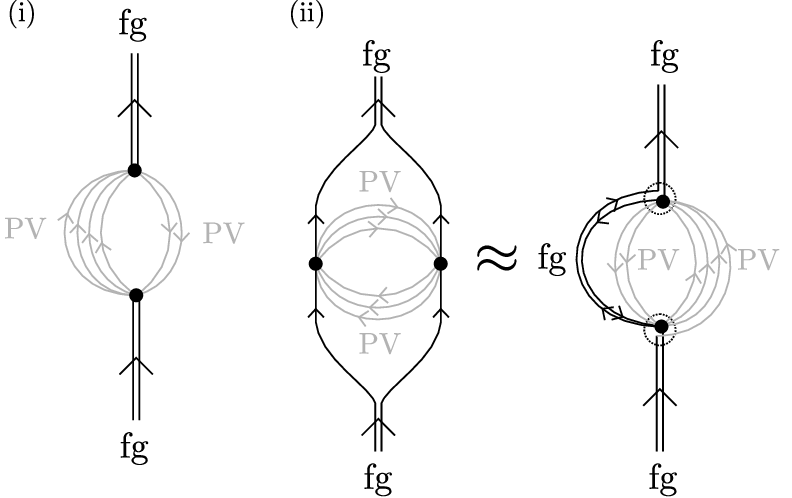}
\caption{(i)~Leading diagram contributing to scalar boson mass. (ii)~First correction.\label{fig:scalbosoncoremassfig}}
\end{figure}

As with the $W$ and $Z$ bosons and the gluon, some higher-order corrections to the scalar boson mass are computed here. 
Recall from \cref{ch:boson} that there are two leading terms in the scalar boson mass interaction, as shown in \fref{fig:scalbosoncoremassfig}(i) and~(ii), and that these carry relative weights of $[1-2/(3N_0)+1/(3{N_0}^2)]$ and $1/(2\pi)$ respectively.

\subsubsection{Boson loops\label{sec:mHvecbosonloops}}

\paragraph[Gluon loops]{Gluon loops:\label{sec:mHgluonbosonloops}}
Begin with evaluation of gluon loop corrections to \fref{fig:scalbosoncoremassfig}(ii). Recognise that the vertex configuration is identical to that of the $Z$~boson, up to some $A$-sector labelling. Evaluation of loop corrections proceeds equivalently, including a factor of $\frac{1}{2}$ because interchangeability of boson/pseudovacuum vertices exchanging the left and right limbs of the diagram. The resulting factor is %
\begin{equation}
\frac{30\alpha}{2\pi}.
\end{equation}

For \fref{fig:scalbosoncoremassfig}(i), recognise that as discussed in \Psref{V}{sec:scalbosonmass} the pseudovacuum elements being interacted with are still fermions, even though the correlation of background fields over ranges of $\ILO{\mc{L}_0}$ permit some of these interactions to take place remotely, creating the appearance of a redistribution of the fermion components. It is the scalar bosons which interact at a distance, as the components of each fermion continue to be bound by colour interactions, and so are separated by at most $\mc{L}_\preon$.\footnote{To the extent that the background fields may be represented as particles within their local correlation region, the emergent Lagrangian favours configurations consistent with the same processes as observed in the foreground fields, and the constituents of background fermions are consequently still represented as bound species. It is only on leaving the local correlation region that this illusion breaks down due to non-normalisation of the background fields, but can be ignored as this is also the regime in which these uncorrelated fields are cancelled out both among themselves and by other fields also in their far field regime. Further, the normalisation of \protect{\Psref{I}{sec:normWrtBgFields}} prevents background field binding interactions from having any effect on numerical results.}

Since the loop corrections take place over distances of $\ILO{\mc{L}_\preon}$, at which the loop bosons are massless, these continue to see the preons as being grouped 3+3 rather than 4+2. However, the gluon correction factor \emph{is} reduced somewhat from $30\alpha/(2\pi)$ because the colour agnosticism of \fref{fig:scalbosoncoremassfig}(ii), in which any preon can carry any colour label, is reduced. Two preons in the group of four must necessarily be the same colour as the foreground preons, %
and as seen in \fref{fig:scalarKmatrices}, looking inside the scalar boson/background preons interaction vertex reveals which two preons this is. Since the specific colour of the pair is fixed for a given term in the sum making up the scalar boson, this eliminates two free colour labellings
compared with \fref{fig:scalbosoncoremassfig}(ii). (Or equivalently, only one colour labelling in nine is compatible with the interaction diagram under evaluation.) Diagram~(i) therefore only attracts a correction of %
\begin{equation}
\frac{30\alpha}{9\cdot2\pi}.
\end{equation}

\paragraph[Photon, \prm{W}, and \prm{Z} boson loops]{Photon, $W$, and $Z$ boson loops:}
Since the scalar boson field couples identically with  $\bar e_R$, $e_L$, and $\nu_e$, the preon limb of the loop couples to the photon and $Z$~boson fields as per the trace of their representation matrices, which vanish. The contribution of the $W$~boson loop also vanishes as per \sref{sec:WcorrsAWZ} above.

\paragraph[Scalar boson loops]{Scalar boson loops:\label{sec:scalbosonloopspara}}
This time, begin with \fref{fig:scalbosoncoremassfig}(i).
As in \frefs{fig:Wmassscalarcorrs}{fig:Wmassscalarcorrs_loop}, scalar boson loops may be constructed on any pair of preons within either the preon limb or the antipreon limb of the fermion loop. Let the scalar boson loop corrections having the form of \fref{fig:Wmassscalarcorrs}(i) and \fref{fig:Wmassscalarcorrs}(iii)-(vi) be termed scalar boson loops of the first and second kind respectively.
Also let the leading-order scalar boson mass diagrams of \freft{fig:scalbosoncoremassfig}(i)-(ii) be described as scalar boson mass term~1 and~2. Let the approximate form of term~2 given in \fref{fig:scalbosoncoremassfig}(iii) be referred to as term~2'.
\begin{enumerate}
\item When a loop of the first kind is applied to term~1, this yields a correction to the mass vertex term obtained from \fref{fig:scalbosoncoremassfig}(i).\label{item1} Let this correction be referred to as ``item~1''.
\item When a loop of the first kind is applied to term~2, this yields a correction to the mass vertex term obtained from \fref{fig:scalbosoncoremassfig}(ii).\label{item2} Let this correction be referred to as ``item~2''.
\item When a loop of the second kind is applied to term~1, the result is equivalent to a loop of the first kind applied to term~2'. In particular, note that through the use of preon decomposition, Fierz identities, and $F$~moves \cite{kitaev2006,bonderson2007} the accessory vector boson's vertices (associated with factors of~1) may be taken out to the vertices of \fref{fig:scalbosoncoremassfig}(i) permitting transformation from the form of term~2' to that of term~2. There are nine possibilities for the accessory boson, fully offsetting the factor of $\frac{1}{9}$ from loss of colour agnosticism. %
This diagram is therefore largely equivalent to item~\ref{item2} (above),
though not quite all terms are duplicated due to the factor of $[1-2/(3N_0)+1/(3{N_0}^2)]$ on term~1 from the regrouping of preon lines. Thus this diagram multiplies item~\ref{item2} by $2[1-2/(3N_0)+1/(3{N_0}^2)]$.
\item When a loop of the second kind is applied to term~2, if the sense of the rotation applied to the loop gluon is chosen counter to that applied in going from term~2 to term~2' then it follows immediately that the induced vector bosons cancel (on appropriate summing across colour labels on both internal and external lines) to yield a loop of the first kind applied to term~1, up to a factor of $%
[1-2/(3N_0)+1/(3{N_0}^2)]^{-1}$ which arises because term~2 has a higher FSF symmetry factor that term~1. Alternatively, if the rotation is chosen such that the rotations do not directly cancel then the same result still follows, essentially from the same isotopy properties as Fig.~16 of \rcite{kitaev2006}. Summation of the loop of the first kind applied to term~1 and the loop of the second kind applied to term~2 is equivalent to multiplying the loop of the first kind applied to term~1 by a factor of $2[1-2/(3N_0)+1/(3{N_0}^2)]^{-1}$.\label{item:scalbosonloopspara_kind2term2}
\end{enumerate}
It therefore suffices to consider only loops of the first kind applied to both diagrams and then multiply by the requisite factors. These loops act on preons in the form of background fermions, and thus it is simplest to evaluate first for \fref{fig:scalbosoncoremassfig}(ii) and then apply the same factor to \fref{fig:scalbosoncoremassfig}(i) modulo a factor of~$\frac{1}{9}$ for colour knowledge on preon lines as before.

Next, note that the vertices between scalar bosons and fermions, and the vertices between scalar bosons and preons, are all interchangeable (because, by construction, a coupling with a fermion may always be written as the average over the couplings of the three constitutent preons).
Not all resulting diagrams yield colour consistency when the scalar boson vertices are combined as per \fref{fig:Wmassscalarcorrs_loop}, but for those which do, the resulting symmetry factors are in 1:1 correspondence with the induced overcounting. To obtain the vertex correction factor, it is either necessary to exploit that all diagrams make equal contributions, and offset the average per-diagram overcounting by dividing by the effective vertex interchange symmetry factor, or more simply just to recognise that the resulting cancellation makes it possible to ignore the symmetry factor altogether.

Now proceed to evaluate the scalar boson loop correction to \fref{fig:scalbosoncoremassfig}(ii). Apply a loop correction of the first kind as per \fref{fig:Wmassscalarcorrs}(i) and combine the vertices as per \fref{fig:Wmassscalarcorrs_loop}. Now recognise that at the cost of a braiding factor of $-1$, diagrammatic isotopy permits reversal of the inner line of the loop. The resulting vector loop is summed over all $A$- and $C$-charges. The nine-element $C$-sector admits a basis with diagonal elements $e^C_{ii}$, each of which is averaged over $A$-charge, so the sum over terms arising from the scalar boson diagram may be rewritten as a sum over gluons.

Evaluating this rewriting:
\begin{itemize}
\item Braid factor:~$-1$
\item Synthesis of $\sigma^\mu\sigma_\mu$:~$-\frac{1}{2}$
\item One $\bmh\bmh^*$ loop yields~81 vector boson terms:
\begin{equation}
\bar\partial\bar\partial\vp\partial\partial\vp\rightarrow-\frac{1}{2}\sum_{\dot a,\dot c,a,c}\bar\partial^{\dot a\dot c}\bsmm\partial^{ac}\vp\bar\partial^{\dot a\dot c}\bsm\partial^{ac}\vp.
\end{equation}
\item However, the scalar boson acts on a single diagram corresponding to an average over colour labelling. A given boson acts on a specific colour labelling, of which there are nine. Thus there is a one in nine chance of any generated gluon being compatible in any given term:~$\frac{1}{9}$. 
\item The gluon correction in \sref{sec:mHgluonbosonloops} was evaluated on preons-in-fermions. To raise the current calculation to the fermion level introduces a factor of~$\frac{1}{9}$ as each preon/boson vertex is acknowledged to occur only one third of the time in any given fermion vertex, by normalisation of the fermion field~\Peref{III}{eq:generalfermion}.
\item Despite the appearance of \fref{fig:Wmassscalarcorrs_loop}, the scalar boson loop does not attract a factor of two for on-vertex symmetry as the underlying construction is as per~\fref{fig:Wmassscalarcorrs}(i) and coincidence of the vertices only occurs through emergent constraints. It is not a natural single vertex pulled down as one piece from the generator~$\Z$.
\item Note that neither the scalar boson in \fref{fig:Wmassscalarcorrs}(i) nor the off-diagonal gluon in \fref{fig:Wmassbosoncorrsnew}(ii) have end-to-end symmetry so there is no need to consider factors of~2 associated with vertex exchange.
\end{itemize}
Each scalar boson loop correction of the first type therefore maps (numerically) to a gluon loop correction on the same preon pair. It remains a separate correction, as the foreground boson making the loop exists relative to a distinct mass shell solution.\footnote{It is interesting that in the \prm{\Cw{18}} model, all on-shell solutions apply to the same underlying preon fields. Thus there exists one set of fields with multiple on-shell solutions, rather than multiple fields with one mass shell apiece. For simplicity all mass shells are treated as independent, which introduces negligible error for sufficiently distinct masses, but is quite readily apparent when different rearrangements of the same preons must be treated as loops constructed on independent foreground fields, as seen here.} 
The numerical coefficient associated with this mapping is
\begin{equation}
-1\cdot -\frac{1}{2}\cdot 81\cdot \frac{1}{9}\cdot \frac{1}{9} = \frac{1}{2}.
\end{equation}
Then there is the factor of $2[1-2/(3N_0)+1/(3{N_0}^2)]$ from also considering loops of the second kind applied to \fref{fig:scalbosoncoremassfig}(i). This yields
\begin{equation}
\frac{30\alpha}{2\pi}\left(1-\frac{2}{3N_0}+\frac{1}{3{N_0}^2}\right).
\end{equation}
Proceeding similarly for loops of the first kind on \fref{fig:scalbosoncoremassfig}(i) and loops of the second kind on \fref{fig:scalbosoncoremassfig}(ii) yields
\begin{equation}
\frac{30\alpha}{9\cdot 2\pi}\left(1-\frac{2}{3N_0}+\frac{1}{3{N_0}^2}\right)^{-1}.
\end{equation}

\paragraph[Net effect of all boson loops]{Net effect of all boson loops:}
Incorporating all of the above corrections arising from the gluon loops and from the scalar boson loops yields a net mass
\begin{equation}
\begin{split}
m_\bmh^2=\,&20f^2\left[k_1^{(e)}\right]^4{\omega_0}^2{N_0}^{12}S_{6,13}\\
&\begin{aligned}
\times\Bigg\{&\left(1-\frac{2}{3N_0}+\frac{1}{3{N_0}^2}\right)\left[1+\frac{30\alpha}{9\pi}\left(1+\frac{1}{3N_0}\right)\right]%
+\frac{1}{2\pi}\left[1+\frac{30\alpha}{\pi}\left(1-\frac{1}{3N_0}\right)\right]\Bigg\}
\end{aligned}\\
&\times\left[1+\OO{{N_0}^{-4}}+\OO{\alpha{N_0}^{-2}}+\OO{\alpha^2}\right].
\end{split}
\end{equation}
By \PEref{V}{eq:N0value} the term in $\OO{\alpha{N_0}^{-2}}$ is small compared with $\OO{\alpha^2}$ and so may be dropped. %

\subsubsection{Background photon, gluon, and scalar interactions}
\paragraph[Direct coupling]{Direct coupling:}
All direct boson/boson couplings require nonvanishing matrix commutators on evaluating the boson term of Lagrangian~\Peref{III}{eq:Lfg1}. However, the scalar boson is associated with the identity matrix on all sectors, and thus any such term must vanish. There are therefore no direct couplings between the complex scalar boson and the background vector or scalar boson fields.

\paragraph[Indirect (universality) coupling]{Indirect (universality) coupling:\label{sec:mHindirect}}
Evaluation of this contribution proceeds as in \sref{sec:Zuniversalcplg}. For the photon sector the contributions from $e_L$ and $\bar e_R$ have weight $3\times 3$, those from $u_L$ and $\bar u_R$ have weight $2\times 2$, and those from $d_L$ and $\bar d_R$ have weight $1\times 1$, for a total count of twenty-eight channels. As before, each channel is associated with a relative factor of $\bigl[72k_1^{(e)}N_0\bigr]^{-4}$ for a contribution of
\begin{equation}
\frac{7}{18\left[k^{(e)}_{1}{N_0}\right]^4}.
\end{equation}
Evaluation of the gluon sector is directly equivalent to that for the $Z$ boson, yielding a contribution
\begin{equation}
\frac{32}{18\left[k^{(e)}_{1}{N_0}\right]^4}.
\end{equation}
The net associated factor is thus
\begin{equation}
\left\{1+\frac{39}{18\left[k^{(e)}_{1}{N_0}\right]^4}\right\}\left[1+\OO{\alpha}\right],
\end{equation}
and together these terms correct the complex scalar boson mass to
\begin{equation}
\begin{split}
m_\bmh^2=\,&20f^2\left[k_1^{(e)}\right]^4{\omega_0}^2{N_0}^{12}S_{6,13}\\
&\begin{aligned}
\times\Bigg\{&\left(1-\frac{2}{3N_0}+\frac{1}{3{N_0}^2}\right)\left[1+\frac{30\alpha}{9\pi}\left(1+\frac{1}{3N_0}\right)\right]%
+\frac{1}{2\pi}\left[1+\frac{30\alpha}{\pi}\left(1-\frac{1}{3N_0}\right)\right]\Bigg\}
\end{aligned}\\
&\times\left\{1+\frac{39}{18\left[k^{(e)}_{1}{N_0}\right]^4}\right\}%
\bm{\left(}1+\OOOO{\alpha\left[k^{(e)}_1{N_0}\right]^{-4}}+\OO{\alpha^2}\bm{\right)}.
\end{split}\label{eq:Hwithk2}
\end{equation}
This completes calculation of $\bmh$~boson mass to the level of precision employed in this paper.

\subsection{Neutral boson gravitation}

It is worth making a note regarding the universality coupling described in \srefs{sec:universalcorrs}{sec:Zuniversalcplg}, in anticipation of the mechanism for $G^\bdag$~boson elimination in \sref{sec:Rwnf}. Although the $Z$ boson is uncharged, through this indirect coupling process it acquires a means of coupling to the photon \emph{pair} field. In \cref{ch:gravity} it is seen that this gives the $Z$~boson a means of influencing space--time curvature when the target manifold is permitted to be non-flat and the $G^\bdag$ bosons are eliminated from the model. %
The $Z$~boson of the $\Cw{18}$ model on curved space--time thus acquires an emergent gravitational mass (though not necessarily one equal to its inertial mass). %

Regarding the foreground gluons (including the $N$~boson), first recognise that these couple to the colour charges on both the pseudovacuum preon and gluon fields, and application of the universality coupling allows the pseudovacuum preon fields to be reduced to composite vector bosons in 1:1 correspondence with the gluons. On %
integrating down from background fermions to background bosons, this does not reveal any new couplings and thus the full inertial mass interaction of the foreground gluons may be accounted for without appeal to the universality coupling. However: The sum over all possible colour couplings is implicitly a sum over all elements of $\GLNR$, of which the portion contributing to inertial mass implicitly carry $A$-sector charge $\lambda^A_3$. Thus the background
composite vector boson fields may also be rewritten to correspond to the two-photon coupling (carrying the photon $\ta$-charge, and being summed over all possible neutral colour combinations). As observed, this coupling is constructed from the same composite vector boson fields already accounted for, so makes no additional contribution to inertial mass, but as a coupling to the photon pair field it does grant gravitational mass to the nine gluon fields proportional to their coupling to the colour sector of the background fields. 
For the eight bosons of $\SU{3}_C$ this is relatively unimportant as these bosons rarely occupy regimes in which they exhibit an inertial mass. More notably, though, this interaction imparts gravitational mass to the $N$~boson, which behaves as a colourless unconfined neutral gluon with inertial mass on the electroweak scale. This boson is therefore a potential dark matter candidate.

\section{Lepton Mass Interaction\label{sec:lepmassint}}

\subsection{Leading order\label{sec:leptonleadingorder}}

\begin{figure}
\includegraphics[width=\linewidth]{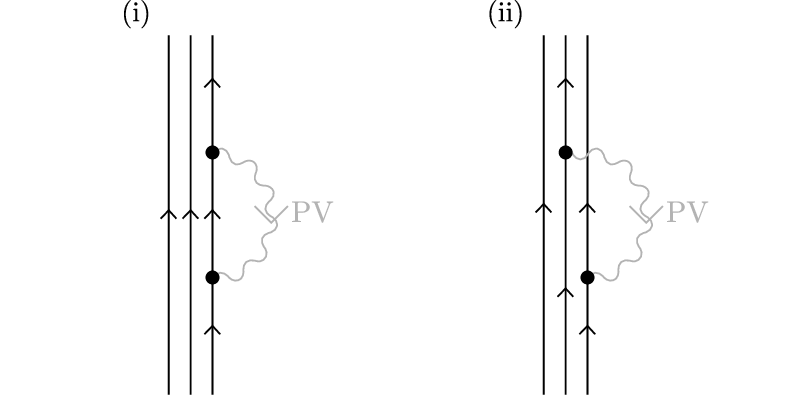} %
\caption{The fundamental interaction giving rise to lepton mass: The triplet of preons scatters twice off the bosonic component of the background quantum liquid. Either (i)~the same, or (ii)~different preons may be involved on each occasion, with the upper and lower vertices independently each connecting to any of the three preons. This results in a total of nine diagrams, three having the form of diagram~(i) and six having the form of diagram~(ii). These nine diagrams are then summed.\label{fig:basicmass}}
\end{figure}

The fundamental interaction giving rise to lepton mass is a double scattering of the preon triplet off the vector boson component of the pseudovacuum (\fref{fig:basicmass}). Both of these diagrams may be considered mean field theory expansions of a loop correction to the fermion propagator in the presence of the pseudovacuum, but it is more convenient to refer to these as ``leading order'' diagrams, and to count the number of (foreground) loop corrections to these leading order diagrams, e.g.~1-loop corrections to the leading order diagram, etc. %
Henceforth such corrections will be termed simply ``1-loop corrections''.

\Fref{fig:basicmass} %
yields a nonvanishing pseudovacuum contribution to fermion mass only when these boson fields are both the photon field or both gluons, as per \Erefr{eq:AQL}{eq:cQL},
and the resulting interactions may each involve any of the three preons making up the lepton. There are also similar diagrams involving scalar boson loops.
By \PEref{I}{eq:window}, for the mass contribution to be nonvanishing, the separation of the two boson source/sinks is at most of order $\mc{L}_0$.

To evaluate the contribution of the pseudovacuum fields to lepton mass, it is helpful to separate the consequences of these boson interactions into two parts as previously discussed in \sref{sec:fermionmasses} and demonstrated in \srefr{sec:Asector}{sec:Csector}. %
First, there is the action of the representation matrices of $\SU{3}_C$ on the preon fields, and second, there is the numerical mass term arising from the mean square value of the pseudovacuum boson field.

\subsubsection{Action on colour sector\label{sec:leptonKmatrix}}

To begin with the action of the colour sector, note that over the course of a propagator of length $\mc{L}\gg\mc{L}_0$, a lepton will engage in a near-arbitrarily large number of interactions with the background fields. Each interaction will apply a $\gltr_C$ %
representation matrix from $\{\lambda_i|i\in 1,\ldots,9\}$ depending on the boson species with which the lepton interacts. In the absence of foreground $W$ or $Z$ bosons, the vector boson sector of the pseudovacuum is made up entirely of photons and gluons. 

Given a preon of colour $c_1$, this may have nonvanishing interaction with the photon or any of three gluons in the elementary basis $e^C_{ij}$. For example, if $c_1=r$ then admissible gluons are $c^{rr}$, $c^{gr}$, and $c^{br}$. Heuristically, their action on the colour space may be represented as
\begin{equation}
c^{rr}\ket{r}\rightarrow \ket{r}\qquad c^{gr}\ket{r}\rightarrow\ket{g}\qquad c^{br}\ket{r}\rightarrow\ket{b}\label{eq:heuristic1}
\end{equation}
where all associated numerical factors have been ignored for illustrative purposes. 

More generally, the family of gluons acts on a vector of preon colours as indicated by
\begin{equation}
\bgrid c^{rr} & c^{rg} & c^{rb} \\
       c^{gr} & c^{gg} & c^{gb} \\
       c^{br} & c^{bg} & c^{bb} \egrid
       \triplet{\ket{r}}{\ket{g}}{\ket{b}}.\label{eq:heuristic2}
\end{equation}
It is worth noting that %
there is no fixed reference point on the colour sector as
the $\SU{3}_C$ symmetry is unbroken, so there exists a freedom of basis %
corresponding to an arbitrary global transformation in $\SU{3}_C$. Any coloured fundamental or composite particle may be put into an arbitrary superposition of colours using such a transformation, though relative colour charges of different particles remain unchanged, as does the magnitude of the overall colour charge of a composite particle.

Recognise now that \fref{fig:basicmass} contains contributions to two mass vertices. Although their contributions to fermion mass are nonvanishing only when they appear pairwise, %
there is no requirement for this pair to be consecutive. It suffices that each vertex be paired with a conjugate vertex separated by distance and time no greater than $\mc{L}_0$ %
in the isotropy frame of the pseudovacuum.
Indeed, these vertices are connected by a foreground fermion propagator which in general also undergoes further interactions with the pseudovacuum, represented by using a massive propagator for this fermion and requiring consistency with the outcome of the mass vertex calculation. %
In general a foreground fermion exhibiting a net propagation over distance or time of $\ILO{\mc{L}_0}$ in the isotropy frame of the pseudovacuum will scatter back and forth multiple times in this process such that the number of unpaired vertices is assumed negligible. %
Furthermore, where unpaired vertices do exist, their net effect vanishes on average over probe length or time scales larger than $\mc{L}_0$. It is therefore reasonable to assume during evaluation that each vertex belongs to a pair. %

Counting vertices arising from \fref{fig:basicmass}, for every photon vertex there is also %
by construction one vertex for %
each of the nine gluon%
s.

Consider now the specific case of interactions between a propagating lepton and the bosons of the pseudovacuum. Each preon may interact either with a photon or with any of the nine gluons, and the action of the photon on the space of preon colours is trivial, so it is convenient to ignore the photon for now and reintroduce it later.

As already noted, paired interactions with the pseudovacuum gluon field conserve the net colour-neutrality of a leptonic preon triplet. However, overlapping and intercalation of multiple interaction pairs implies that this property only holds on average, as any colour measurement will interrupt a finite number of mass interactions and thus summation to yield no net colour charge on the preon triplet cannot be assured. It is desirable that any measurement of lepton colour should be null, not just the average, and thus a local change of co-ordinates on $\SU{3}_C$ must be performed on a co-ordinate patch encompassing the non-interacting preons such that changes in their colours track those of the interacting preon. This change of  
co-ordinates is not part of the choice of gauge on $\SU{3}_C$, and thus is in principle associated with construction of some synthetic boson interactions where it intersects with particle worldlines. As described in \Psref{IV}{sec:fermionmasses}, the vertex factors associated with these interactions arise from the representation matrices of $\SU{3}_C$ given as $\lambda_i$ in \PErefr{II}{eq:Cbasis1}{eq:Cbasis2}.
By construction these bosons are constrained to have no effect beyond the colour shifts associated with the boundary of the patch, and to leading order this effect is parameterless. In the leading order diagram these bosons consequently have no degrees of freedom, carry no momentum, and are associated with a numeric %
factor of~1. Consequently they are not drawn. Only the factors arising from the representation matrices persist, acting on the colour vector of an individual preon as the matrix
\begin{equation}
\K=\bgrid 1&A&A^\dagger\\A^\dagger&1&A\\A&A^\dagger&1\egrid,\quad A=\pm\frac{1\pm\rmi}{2}.
\end{equation}
The sign on $\rmi$ is %
free to be chosen by convention, while the overall sign on $A$ is fixed by noting %
that cyclic permutation of colours, which is in $(\K)^3$, %
is required to leave the sign of an eigenstate of $\K$ unchanged. The eigenvalues of $\K$ must therefore be non-negative, setting 
\begin{equation}
A=-\frac{1\pm\rmi}{2}. 
\end{equation}
Choosing a sign for $\rmi$, the mixing matrix $\K$ may then %
be written
\begin{equation}
\K(\theta_\ell)=\bgrid 1&\frac{e^{\rmi\theta_\ell}}{\sqrt{2}}&\frac{e^{-\rmi\theta_\ell}}{\sqrt{2}}\\
\frac{e^{-\rmi\theta_\ell}}{\sqrt{2}}&1&\frac{e^{\rmi\theta_\ell}}{\sqrt{2}}\\
\frac{e^{\rmi\theta_\ell}}{\sqrt{2}}&\frac{e^{-\rmi\theta_\ell}}{\sqrt{2}}&1\egrid\qquad\theta_\ell=-\frac{3\pi}{4}.\label{eq:VI:Kmatrix}
\end{equation}
As noted in \Psref{IV}{sec:fermionmasses}, this matrix bears a strong resemblance to Koide's $K$ matrix for leptons \cite{koide2000}. The minus sign on Koide's off-diagonal component $S(\theta_{f})$ has been absorbed into the phase $\theta_\ell$, and the free parameters $a_{f}$, $b_{f}$, and $\theta_{f}$ are fixed by the geometry of the model. %

\begin{figure}
\includegraphics[width=\linewidth]{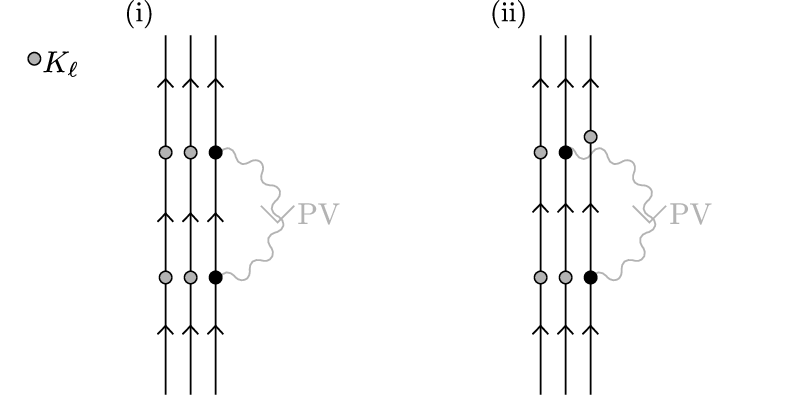}
\caption{When a composite fermion interacts with the gluons from the pseudovacuum, represented as a single \pt{$\gltr$}-valued boson, the colour mixing process represented by matrix \pt{$\K$} acts on %
all preons not coupling to the boson at any given vertex. Diagrams~(i) and~(ii) correspond to \pt{\freft{fig:basicmass}}(i)-(ii) respectively. Once again these are just two representative diagrams from a family of nine, as the upper and lower vertices may independently each be connected to any of the three preons. This results in a total of nine diagrams, three having the form of diagram~(i) and six having the form of diagram~(ii).\label{fig:ActionOfK}}
\end{figure}

Recognising that on average all pseudovacuum gluons act %
with equal frequency, it is convenient to collect these together into a single $\gltr$-valued gluon associated with two applications of matrix $\K$ to the non-interacting preons%
, as shown in \fref{fig:ActionOfK}. %

Now recognise that since the gluons of the $\Cw{18}$ model are massive, and since the fundamental mass interactions take the form of loop diagrams with respect to the lepton propagator (\fref{fig:basicmass}), these will be suppressed by a factor of $\ILO{m_\ell^2/m_c^2}$ relative to the photon contribution. [In practice no single gluon will propagate a distance of $\ILO{\mc{L}_0}$ due to confinement binding it tightly to the emitting lepton, but it is also unnecessary that any single gluon should do so---instead, for gluons the line in \fref{fig:basicmass} represents the propagation of
momentum carried in a distributed fashion within the gluon cloud accompanying the lepton. This propagating momentum carried in the gluon sector then necessarily displays an effective mass shell behaviour consistent with $m_c^2$.]
With each gluon interacting, on average, once for every photon interaction, it is convenient to write the direct contributions of the gluon terms to particle mass as corrections to the larger photon term and to associate copies of the matrix $\K$ with the photon vertices in a manner equivalent to that shown in \fref{fig:ActionOfK} and discussed in \Psref{IV}{sec:Csector}.

Next, consider that the preons on which matrix $\K$ are acting are just two of three preons in a colour-neutral triplet. For leptons, all three $a$-charges are identical and thus over macroscopic scales, where chance fluctuations become negligible, matrix $\K$ will act identically on each member of the triplet. This symmetry is convenient, as it allows the study of an individual preon prior to the reconstruction of the triplet as a whole. %

As per \Eref{eq:compositeleptons}, the preons making up observable leptons are now eigenstates of this matrix $\K$, corresponding to the eigenvectors
\begin{align}
\label{eq:v1}
v_1&=\frac{1}{\sqrt{3}}\triplet{1}{1}{1}\\
v_2&=\frac{1}{\sqrt{3}}\triplet{e^{\frac{\pi\rmi}{3}}}{e^{-\frac{\pi\rmi}{3}}}{-1}\\
v_3&=\frac{1}{\sqrt{3}}\triplet{e^{\frac{2\pi\rmi}{3}}}{e^{-\frac{2\pi\rmi}{3}}}{1}
\label{eq:v3}
\end{align}
which are independent of $\theta_\ell$ and have eigenvalues 
$\bm{\{}k^{(\ell)}_i|i\in\{1,2,3\}\bm{\}}$ given by
\begin{equation}
k^{(\ell)}_n = 1+\sqrt{2}\cos{\left[\theta_\ell-\frac{2\pi(n-1)}{3}\right]}.\label{eq:kell}
\end{equation} 

To reconstruct the lepton as a whole, recognise that for three preons at $\bm{\{}x_i|i\in\{1,2,3\}\bm{\}}$, with corresponding colours $c_i$, and with the preon at $x_1$ having a well-defined color, say $c_1=r$, colour neutrality and colour cycle invariance imply that a choice $c_2=g$, $c_3=b$ is equal up to a sign to the alternative choice $c_2=b$, $c_3=g$ (as this exchange corresponds to spatial exchange of two fermions), and thus it suffices to consider only one such colour assignment (say $c_2=g$, $c_3=b$) along with spatial permutations. Putting preon~1 into a superposition of colour states then corresponds to a superposition of cyclic spatial rearrangements of the members of the triplet, with colours $c_1=r$, $c_1=g$, and $c_1=b$ corresponding to colour assignments \emph{with respect to spatial co-ordinate $x$} of $rgb$, $gbr$, and $brg$ respectively. It follows that for leptons, 
the different spatial configurations of colours on the preon triplet are eigenvectors of a matrix $\K^{(3)}$ with eigenvalues identical to those of $\K$. 

Having established through colour cycle invariance that the matrix $\K$ acts identically on all constituents of a lepton, and through \fref{fig:ActionOfK} that two copies of $\K$ act per pseudovacuum photon interaction, it follows that the effect of matrix $\K$ is to contribute
a factor of $\big[k^{(\ell)}_i\big]^2$ to the mass of a lepton of generation $i$. It might seem problematic that for $\theta_\ell=-3\pi/4$, $k^{(\ell)}_1=0$, but it will be seen in \sref{sec:thetacorr} that $\theta_\ell$ acquires corrections from higher-order diagrams, resulting in $k%
_i>0~\forall~i$, so $k^{(\ell)}_1$ may be assumed real and positive, and this concern may be disregarded.

\subsubsection{Mass from photon and gluon components of the pseudovacuum\label{sec:AcQLcouple}}

The zeroth-order electromagnetic term is readily evaluated by making a mean-field substitution \eref{eq:AQL} for $\bgfield{A^\mu(x)A_\mu(y)}$. 
For a charged lepton $\ell_i$ of generation $i$, this initial approximation may be written
\begin{equation}
m_{\ell_i}^2 = {\frac{f^2}{2}\left[k^{(\ell)}_i\right]^4}{\omega_0}^2{N_0}^8S_{18,147}\left[1+\OO{\alpha}\right]\label{eq:m0}
\end{equation}
as seen in \sref{sec:Csector}, \Eref{eq:leptonmasses}. %
Note that this expression incorporates a symmetry factor of two corresponding to exchange of the two pseudovacuum interactions. 
This may be understood %
by recognising that each term corresponds to a mass vertex and has its external legs truncated independently. These two vertices are then interchangeable for a symmetry factor of~2. 
Alternatively, for any diagram, including ones which do not separate, recognise that the mass-squared is always applied in the context of an untruncated fermion propagator, say from $x$ to $y$. In this context, all fermion connections to the interaction vertices are again untruncated. (Optionally, the full expression for propagation from $x$ to $y$ is then used to infer an equivalent mass term, and the diagram may then be replaced by one in which this mass term is inserted into the propagator twice.) Applying either form of this approach to \fref{fig:basicmass}, the diagram for propagation between two points is again seen to attract a symmetry factor of~2.

Now consider interactions between a foreground fermion and the pseudovacuum gluon fields. As with the photon, these interactions take the form of loop diagrams evaluated in the mean-field regime for the pseudovacuum, and as noted in \Psref{IV}{sec:Asector}, the fermion may transiently surrender momentum to or borrow momentum from the background fields. However, in contrast with the photon loop evaluated to obtain \Eref{eq:m0}, the gluon field is massive, and when a foreground particle transfers momentum to a gluon field, this results in a massive excitation of that gluon field. %
Consequently both limbs of the loop must be considered massive. %
For a general boson $b$ this gives rise to a loop-associated factor of $\bmf{m_{\ell_i}^2/m_b^2}$. This factor goes to~1 for the photon, as it is massless, %
but not for interactions with the background gluon field. It is interesting to compare this situation with the loop corrections to boson mass discussed in \sref{sec:VI:bosonmasses}. In the context of boson mass interactions, the correction loop is constructed on a section of preon line which is \emph{a priori} of dimension $\mc{L}_\preon$ and consequently the loop boson appears massless. In the context of fermion mass interactions, the correction loop is constructed on a section of fermion line which is of nontrivial extent and consequently the loop boson is able to propagate over arbitrary length scales and (for loop bosons other than the photon) must therefore be considered massive.

For both the photon loop and the gluon loop, evaluation of momentum flux around the loop may be taken to yield a factor of
\begin{equation}
\frac{\Xi}{4\pi}\bmf{\frac{m_{\ell_i}^2}{m_b^2}},\quad b\in\{A,c\}
\end{equation}
for some structure factor $\Xi$, where $\bmf{n}$ behaves as described in \aref{apdx:massloops}. For the photon $\Xi=2\alpha$ and the factor $\bmf{\cdot}$ reduces to 1, and the resulting coefficient of $\alpha/(2\pi)$ is absorbed into the pseudovacuum mean-field term by choice of definition. For gluons, dependence on the same energy scale $\mc{E}_0$ indicates that an identical factor of $\alpha/(2\pi)$ is absorbed into the mean field term, while the mass dependence of $\bmf{\cdot}$ reveals that the gluon terms are suppressed by a factor of ${m_{\ell_i}^2}/{m_c^2}$ relative to the photon term. [Although there is some massive transmission of foreground momentum around the gluon loop, the factor arising from $\bmfcdot$ is ${m_{\ell_i}^2}/{m_c^2}$ and not ${m_{\ell_i}^2}/(4\pi{m_c^2})$ as the value of the gluon loop diagram is dominated by the background terms. This is discussed further in \aref{apdx:massloops}.]

To determine the structure factor of the gluon diagram, work in the $e^C_{ij}$ basis, and consider first a specific off-diagonal gluon. As discussed in \Psref{III}{sec:EWint_numerical} and \sref{sec:Wmassgluoncorr} above, mapping to the one-$W$-loop correction to lepton magnetic moment~\eref{eq:Wloopcorr} permits the magnitude of the structure factor to be evaluated as $\frac{10\alpha}{3}\left[1+\ILO{\alpha}\right]$. In comparison with the reference process, however, there is no emitted boson in \fref{fig:ActionOfK}. The sign of the structure factor may then be easily determined by recognising that on mapping \fref{fig:ActionOfK} to a tensor network \cite{penrose1971,pfeifer2010,pfeifer2011} the resulting network is a conjugate %
square and thus the structure factor must be real and positive. For a diagonal gluon $e^C_{ii}$ the calculation is modified as there is only one possible colour, not two, on the target preon (it is the same as the source preon) but this is offset by a factor of two for vertex interchange symmetry so the result is the same.

There is, however, a further correction to the above. In \sref{sec:Wmassgluoncorr}, gluon exchange was restricted by the requirement of colour neutrality on the inbound and outbound triplets. In the present situation the gluon at each vertex represents the $c^{ij}_\mu$ component of the colour mixing operator $\hat\Kf_\mu$~\Peref{IV}{eq:defMCK}, and overall colour neutrality is conserved by the requirement that free fermions be colour-neutral eigenstates of the matrix $K_e(\mc{E})$. Further, since the gluons at the vertices arise from the background fields, by \Eref{eq:cQL} they need not be conjugate to yield a non-vanishing diagram. Momentum transfer through the background field channel may still take place, due to the on-average independent separate conservation of foreground and background momenta, and may be assumed to do so through implicit scattering processes in the background fields. (As noted previously, although the normalisation of \Psref{I}{sec:normWrtBgFields} eliminates contributions of these interactions to numerical results, they may still be considered to occur within the local correlation region.)
This lack of conjugacy corresponds to reopening the loop of \fref{fig:ActionOfK} to recover a diagram more akin to \fref{fig:leptonmassterm}, much as \fref{fig:gluonQLgluoninteraction} is an opening of the loop on \fref{fig:Wphotons}(i), and increases the number of admissible gluon colours by a factor of~3. Overall, the resulting structure factor is
\begin{equation}
\frac{10\alpha}{3}\left[1+\ILO{\alpha}\right]\cdot 3\cdot\bmf{\frac{m_{\ell_i}^2}{m_c^2}}\cdot\frac{1}{4\pi},
\end{equation}
and for background gluon fields %
\begin{equation}
\bmf{\frac{m_{\ell_i}^2}{m_c^2}}\longrightarrow\frac{m_{\ell_i}^2}{m_c^2},
\end{equation}
giving a net relative factor of 
\begin{equation}
\frac{5m_{\ell_i}^2}{m_c^2}\left[1+\ILO{\alpha}\right]. \label{eq:ffgluonrelfac}
\end{equation}
The terms denoted $\ILO{\alpha}$ reflect potential discrepancies in the one-photon-loop corrections to the photon and gluon leading-order diagrams. However, with these corrections not yet having been calculated for either diagram it is convenient to write just the leading global correction to $m_{\ell_i}^2$ as a whole.
Taking both photon and gluon terms into account (but not yet including the scalar boson contribution, denoted \ldots), 
the leading-order expression for %
lepton mass is therefore given by
\begin{equation}
\begin{split}
m_{\ell_i}^2 =\,& {\frac{f^2}{2}\left[k^{(\ell)}_i\right]^4}{\omega_0}^2{N_0}^8S_{18,147}\\
&\times\left\{\frac{q_\ell^2}{e^2}+\frac{5m_{\ell_i}^2}{m_c^2}+\ldots\right\}\left[1+\OO{\alpha}%
\right]
\end{split}
\end{equation}
where $q_\ell$ is the charge of lepton $\ell_i$. %

As an aside, note that for the fermions there is no equivalent to the bosonic universality coupling explored in \sref{sec:Zuniversalcplg}. For the $Z$ boson, this coupling arises as the basic $Z$ mass diagram [equivalent to \fref{fig:Wmassbosoncorrsnew}(i)] intrinsically incorporates six background preon lines, and two co-ordinates to integrate over, permitting reduction to two preon lines when one of these integrals is performed. In contrast, the basic fermion mass diagram (\fref{fig:ActionOfK}) contains no intrinsic mechanism for adding extra preon lines to construct alternative pseudovacuum couplings. Although extra preons may be recruited from the pseudovacuum, consistent normalisation (\Psref{I}{sec:normWrtBgFields}) requires that integrating over the additional co-ordinate thus introduced will inevitably eliminate them again.

\subsubsection{Mass from scalar component of the pseudovacuum\label{sec:HQLcouple}}

\begin{figure}
\includegraphics[width=\linewidth]{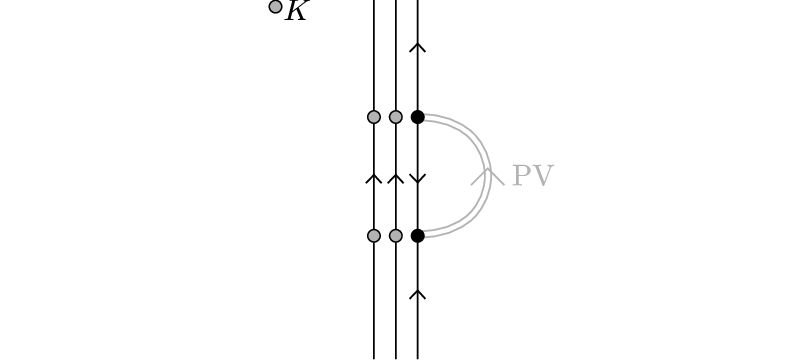}
\caption{Leading-order contribution to fermion mass from the background scalar field. Grey circles represent the action of the \prm{\K}-matrix.\label{fig:f_basicscalmass}}
\end{figure}

Next to be considered is the interaction between the composite lepton and the pseudovacuum complex scalar boson field shown in \fref{fig:f_basicscalmass}. 
Again it is desirable to write this term as a correction to the photon term. As per \Eref{eq:xycorr1} the pseudovacuum expectation value of the complex scalar boson field is nonvanishing, and evaluation of the associated loop factor is most readily performed by determining its weight relative to the photon diagram.

To achieve this, let the $\SU{3}_A$ sector be supplemented by the $N$~boson to obtain the Lie algebra $\gltr_A$ of an effective symmetry group $\GLTR_A$. Adopt a basis $e^A_{ij}$. Decomposing the scalar boson into its terms, the loop associated with each term will now be mapped to an equivalent loop involving a vector boson $e^A_{ij}$. Note, however, that this is just a numerical equivalence used to determine relative loop weights. The actual evaluation of the pseudovacuum term is performed on the complex scalar boson itself, which has trivial $C$-sector and $A$-sector representation and as discussed in \Psref{III}{sec:consequences} this is consequently unaffected by the $\SU{3}_A$ and $\SU{3}_C$ gauge choices presented in \Psref{III}{sec:GL18Cgauge}.

To determine the loop weight, first recognise that each basis element in $e^A_{ij}$ is averaged over colour. The terms in $\bmh\bmh^*$ are summed over $A$-charge and colour, with the result that the terms are equivalent (up to factors to be determined) to a sum over the diagonal bosons $e^A_{ii}$, each multiplied by three to go from an average over colour to a sum.

Now exploit the $\GLTR$-invariance of the $e^A_{ij}$ representation of $\gltr_A$ by noting that the structure factor associated with an diagonal element $e^A_{ii}$ will be the same as that associated with an off-diagonal element $e^A_{ij}|_{i\not=j}$. Therefore begin from the structure factor for the $W$~boson, and multiply by appropriate coefficients to construct the relevant factor for the complex scalar boson. Relevant factors are as follows:
\begin{itemize}
\item Structure factor: $-\frac{10}{3}\left[1+\ILO{\alpha}\right]$ for $W$, 2 for photon. Relative factor:~$-\frac{5}{3}\left[1+\ILO{\alpha}\right]$. %
\item Elimination of sigma matrices from $W$ diagram when mapping to a scalar boson term:~$-2$.
\item The $W$~boson diagram maps to $\frac{1}{9}$ of the $\bmh$ diagram:~$\frac{1}{9}$.
\item There are nine such terms:~9
\item As described in \Psref{III}{sec:scalbosint} the $\bmh$ and $\bmh^*$ vertices are effectively constructed as if in the ``far field'' (in this context, relative to $\mc{L}_\Omega$) and thus their emission from the fermion attracts a factor of $2\big[k^{(e)}_1{N_0}\big]^{-2}\left[1+\ILO{{N_0}^{-1}}\right]$ apiece.
\item The two interaction vertices are within $\ILO{\mc{L}_0}$ of one another and thus attract $\K$-matrices as per \frefs{fig:evalHNsyms}{fig:scalarKmatrices}.
\item As per the caption of \fref{fig:scalarspread}, there are two different ways to assemble the complex scalar boson and conjugate from constituent preons, but these are subsumed into a definition of the complex scalar boson field and thus do not introduce any factors.
\item The complex scalar boson is massive, for a factor of $\bmf{m_{\ell_i}^2/m_\bmh^2}$.
\end{itemize}
On comparing the scalar boson diagram with the vector boson diagram, the net relative factor is thus
\begin{align}
\nn&-\frac{5}{3}\left[1+\ILO{\alpha}\right]\cdot-2\cdot\frac{1}{9}\cdot9\cdot\left[\frac{2}{\big[k^{(e)}_1{N_0}\big]^2}\right]^2\left[1+\ILO{{N_0}^{-1}}\right]%
~\bmf{\frac{m_{\ell_i}^2}{m_\bmh^2}}\\
\nn&=\frac{40m_{\ell_i}^2}{3m_\bmh^2{\left[k^{(e)}_1N_0\right]}^{4}}\left[1+\ILO{{N_0}^{-1}}+\ILO{\alpha}\right]
\end{align}
for a total lepton mass
\begin{align}
m_{\ell_i}^2 =\,& {\frac{f^2}{2}\left[k^{(\ell)}_i\right]^4}{\omega_0}^2{N_0}^8S_{18,147}\label{eq:f_leadingscalar}\\
&\times\Bigg\{\frac{q_\ell^2}{e^2}+\frac{5m_{\ell_i}^2}{m_c^2}\nn
+\frac{40m_{\ell_i}^2}{3m_\bmh^2\left[k^{(e)}_{1}N_0\right]^4}\left[1+\ILO{{N_0}^{-1}}\right]
\Bigg\}\\
&\times\left[1+\ILO{\alpha}\right].\nn
\end{align}

\subsubsection{Gluon and scalar field mass deficits\label{sec:gluonscalarmassdeficit}}

Conservation of energy/momentum implies that the rest mass imparted to the fermion must be compensated by a reduction in the zeroth component of 4-momentum of some of the pseudovacuum fields. Likelihood of contribution from any given pseudovacuum sector will be governed by availability of zeroth-component energy within that sector, i.e.~the rest mass of the associated species, and the strength of coupling to that sector. It therefore follows that this borrowing of rest mass occurs with equal likelihood from each of the nine gluon channels of the pseudovacuum, with much lower likelihood from the scalar boson channel (due to a much weaker coupling), and not at all from the photon channel (due to zero rest mass). For a first approximation, consider only the gluon channel. Borrowing a mass of $m_*^2$ from a background gluon field takes place at the first of the existing gluon/fermion interaction vertices of \fref{fig:basicmass}, and corresponds to deletion of a gluon of mass $m_*$ from the pseudovacuum. %
This hole then propagates as a quasiparticle, and is filled by the conjugate interaction at the second vertex.

More generally, with multiple overlapping pairs of background gluon field interactions occurring along a fermion propagator, there is a consistent propagating hole in the pseudovacuum gluon sector corresponding to an energy deficit of $m_*c^2$, and individual vertices may cause transient fluctuations and may change which specific gluon fields (with respect to some arbitrary choice of colour basis) are involved in propagating this hole, but in general it may be in any of the nine gluon channels at any time. 
This hole is in addition to the effect discussed in \sref{sec:AcQLcouple} where fermions may surrender momentum to or borrow momentum from the pseudovacuum, and thus gives an additional correction factor not yet discussed.

This hole propagates as a quasiparticle accompanying the lepton. Any time that a fermion interacts with the pseudovacuum gluon sector this hole is necessarily also present, and may occupy any of nine channels.

For simplicity, further consider the case when the hole occupies an off-diagonal channel. It would be convenient to write the effect of this hole as a correction to the mass of gluon,
\begin{equation}
m_c^2\longrightarrow m_c^2-km_*^2=:(m_c^*)^2
\end{equation}
for some factor $k$. Recognising that the co-propagating hole's interaction with the fermion is trivial (it is only required to be present), with any local energy/momentum transfer to or from the pseudovacuum being mediated by the fermion/gluon coupling of \fref{fig:basicmass}(i), the hole's interactions attract no structural vertex factor and thus where the direct gluon interactions of \fref{fig:basicmass}(i) acquire a factor of $5/3$ apiece, the hole does not, for an effective relative factor of $3/5$ on the hole's interactions. 
Further, the presence of the hole breaks the time reversal symmetry of a portion of the pseudovacuum and this gives rise to a symmetry factor of $\frac{1}{2}$ relative to the original gluon interaction in which the pseudovacuum was assumed time-reversal-invariant.
Finally, there are nine gluons, and by $\GLTR$ symmetry the mass deficit may propagate via any of them with equal likelihood and equivalent consequence. The multiplicative factor obtained assuming a single channel of propagation for the mass deficit is therefore increased ninefold.
The net outcome is to correct the gluon mass to an effective mass of
\begin{equation}
\begin{split}
(m_c^*)^2 &=m_c^2-\frac{3}{5}\cdot\frac{1}{2}\cdot9\cdot m_*^2\\
&= m_c^2\left(1-\frac{27}{10}\frac{m_*^2}{m_c^2}\right).
\end{split}\label{eq:mc*}
\end{equation}
Note that $m_c^*$ is a function of $m_*$, but for a lepton $\ell_i$ which is on-shell and at (or close to) rest in the isotropy frame of the pseudovacuum this admits the convenient replacement $m_*^2\rightarrow m_{\ell_i}^2$. %
Also note that the gluon mass deficit effect is a whole-field effect, acting on both the foreground and background gluon fields. The corrected gluon mass $m_c^*$ should be used anywhere a particle interacts with a gluon field in the presence of a lepton. 

Similarly, the foreground lepton may also borrow its mass from the scalar boson field. However, coupling between leptons and scalar bosons is weaker than that between leptons and gluons. Much as the gluon deficit corrects $m_c^2$ by $\ILO{m_{\ell_i}^2/m_c^2}$ in expressions for particle rest mass, the scalar mass deficit corrects $m_\bmh^2$ by $\ILO{m_{\ell_i}^2/m_\bmh^2}$. Including this as an unevaluated higher-order term,
the lepton mass equation is amended to
\begin{align}
m_{\ell_i}^2 =\,& {\frac{f^2}{2}\left[k^{(\ell)}_1\right]^4}{\omega_0}^2{N_0}^8S_{18,147}\nn
\\
&\begin{aligned}
\times\Bigg\{&\frac{q_\ell^2}{e^2}+\frac{5m_{\ell_i}^2}{(m_c^*)^2}+\frac{40m_{\ell_i}^2}{3m_\bmh^2\left[k^{(e)}_{1}N_0\right]^4}%
\left[1+\ILO{{N_0}^{-1}}+\OO{\frac{m_{\ell_i}^2}{m_\bmh^2}}\right]
\Bigg\}
\end{aligned}\label{eq:mlbeforefgloops}\\
&\times\left[1+\ILO{\alpha}\right]\nn\\
(m_c^*)^2&=m_c^2\left(1-\frac{27m_{\ell_i}^2}{10m_c^2}\right)\bm{\left(}1+\OOOO{\frac{m_{\ell_i}^2(m_c^*)^2}{m_\bmh^4\big[k^{(e)}_{1}N_0\big]^4}}\bm{\right)}
\end{align}
where the increase in $(m_c^*)^2$ comes from some of the mass deficit being transferred to the scalar boson channel. Noting, however, that this higher-order correction to the $m_{\ell_i}^2/(m_c^*)^2$ term of the r.h.s.{} of \Eref{eq:mlbeforefgloops} is of identical order to the $\ILO{m_{\ell_i}^2/m_\bmh^2}$ correction to the $m_{\ell_i}^2/m_\bmh^2$ term on the r.h.s of \Eref{eq:mlbeforefgloops}, it is convenient to continue to define $m_c^*$ according to \Eref{eq:mc*}, with the error associated with this definition being subsumed into the $\ILO{m_{\ell_i}^2/m_\bmh^2}$ correction to the $m_{\ell_i}^2/m_\bmh^2$ term. In \sref{sec:errors} this term is found to be small. %

\subsection{Foreground loop corrections\label{sec:leptonloops}}

Now consider the effects of foreground loop corrections on the leading-order diagrams of \frefs{fig:basicmass}{fig:f_basicscalmass}. Note that since momentum is continually redistributed among the constituent preons by means of gluon-mediated interactions even over length scales of $\ILO{\mc{L}_\preon}$, a boson need not start and finish its trajectory on the same preon in order to be considered a loop correction to an emission vertex.
Further note that the massive nature of the loop boson does not disrupt the pseudovacuum correlators in these diagrams, as these are brought together through the use of spinor identities at the vertices making these diagrams more robust against interference from intermediate particles propagating outside the autocorrelation region than the boson mass diagrams of \sref{sec:VI:bosonmasses}.

\subsubsection{1-loop EM corrections\label{sec:1loopfgEM}}

\begin{figure}
\begin{center}
\includegraphics[width=5in]{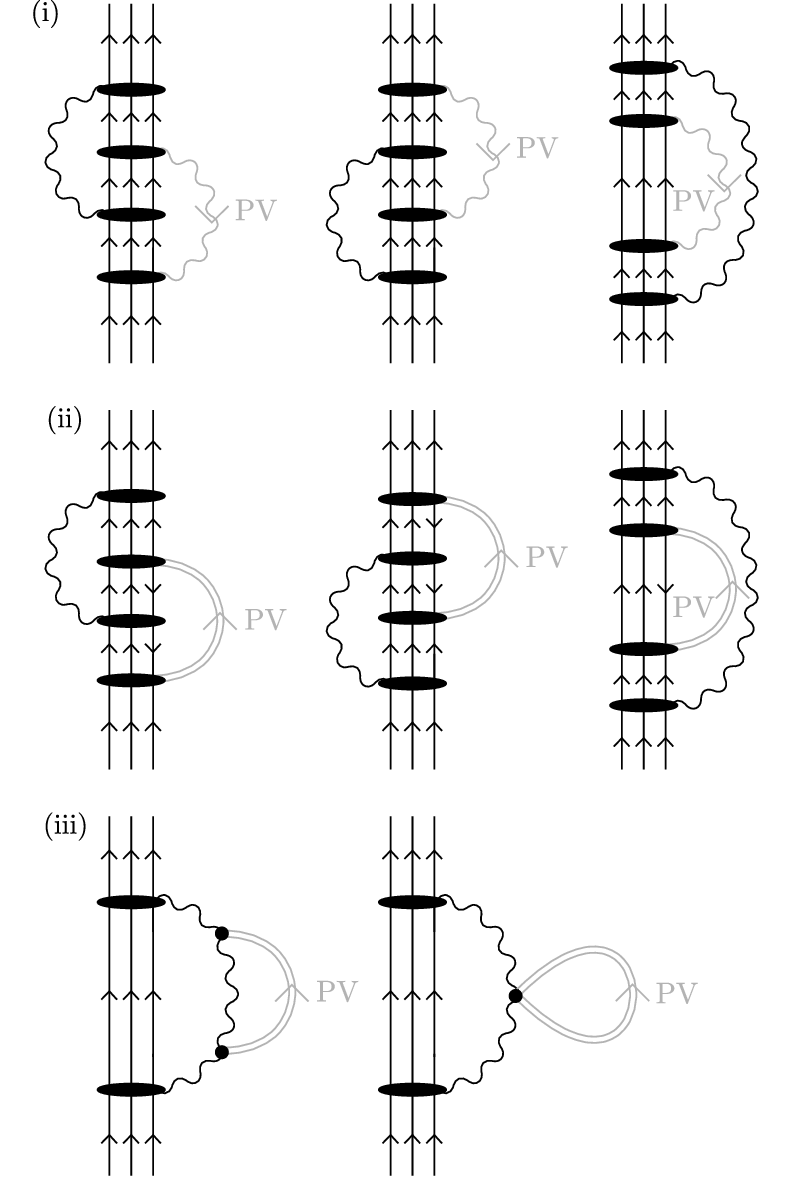} %
\end{center}
\caption{One-foreground-loop EM corrections to (i)~vector boson and (ii)-(iii)~scalar boson lepton mass interactions. The broad oval interaction vertices indicate that the boson may interact with any of the three preons, and all configurations should be summed over.
\label{fig:photonloops}}
\end{figure}

\begin{figure}
\includegraphics[width=\linewidth]{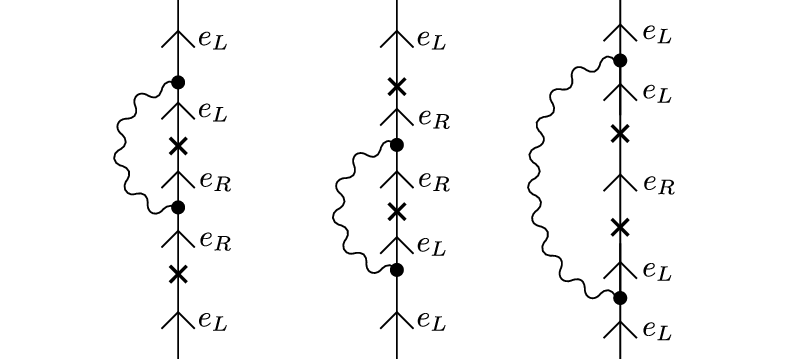} %
\caption{Standard Model one-foreground-loop EM corrections to the electron mass vertex. The example shown is for the left-helicity electron Weyl spinor; equivalent diagrams for the right-helicity spinor exchange \pt{$L$} and \pt{$R$}.
\label{fig:SMphotonloops}}
\end{figure}

The $\ILO{\alpha}$ EM loop corrections to the lepton mass interaction are shown in \fref{fig:photonloops}.
These should be compared with their Standard Model counterparts in \fref{fig:SMphotonloops}.

As in \sref{sec:Wmassbosonloops}, the only corrections which need to be incorporated into the electron mass vertex are those which do not also appear in the Standard Model. Further, as per \Psref{IV}{sec:generalconsider} all PSE terms may be absorbed into the fundamental vertex prior to applying non-PSE corrections.
Begin with the diagrams of \fref{fig:photonloops}(i), which have the Standard Model counterparts shown in \fref{fig:SMphotonloops}. Evaluation of symmetry factors is described in \aref{apdx:symloops} and reveals these to be directly equivalent. The diagrams of \fref{fig:photonloops}(i) therefore correspond to PSE vertex corrections in the Standard Model. Provided the mass vertex after all non-PSE corrections is identified with the observable mass, in keeping with emulation of the $\MSbar$ renormalisation scheme, the diagrams of \fref{fig:photonloops}(i) %
therefore make no contribution to $m_{\ell_i}^2$. 

It is complementary to note that when multiple photon loops exist, the summation over different pseudovacuum expansions of a diagram is equivalent to treating the background photon loop as being of a unique species distinguishable from any foreground photon loops, then summing over the contributions of the resulting effectively topologically distinct diagrams. Since the background loop thus behaves as a distinct species, the EM loop corrections to the background photon interaction are identical to the EM corrections to the background gluon interaction and do not attract additional symmetry factors on account of the foreground and background interactions both being photon interactions. Thus there are no relative terms of $\ILO{\alpha}$ on $5m_{\ell_i}^2/(m_c^*)^2$.

Next, consider the scalar boson loops of \fref{fig:photonloops}(ii). In these diagrams the intermediate lepton has one preon reversed, inverting its electromagnetic charge, and the first two diagrams therefore yield factors of $\alpha/(6\pi)$ rather than $\alpha/(2\pi)$. In conjunction with negation of the Standard Model corrections, they therefore yield a net correction to the background scalar term with weight
\begin{equation}
-\frac{2\alpha q_\ell^2}{3\pi e^2}.\label{eq:scalbosfacpt1}
\end{equation}
The third diagram is consistent with the Standard Model equivalent and so once again does not contribute to $m_{\ell_i}$.

Finally, there are two more diagrams from the scalar boson sector to consider. Moving a single scalar boson vertex onto the photon loop is prohibited as the diagram as a whole does not then leave the preon triplet unchanged (one preon gets replaced by an antipreon), but moving both vertices onto the loop is admissible. %
This results in the %
diagrams shown in \fref{fig:photonloops}(iii).
However, these loops contribute to the mass of the loop photon (which vanishes by gauge), and thus will necessarily cancel with other terms in which the loop photon couples to the background fields. They may therefore be disregarded.
The net expression for lepton mass thus far derived is therefore
\begin{align}
m_{\ell_i}^2 =\,& {\frac{f^2}{2}\left[k^{(\ell)}_1\right]^4}{\omega_0}^2{N_0}^8S_{18,147}\label{eq:lep1loopA}\\
&\begin{aligned}
\times\Bigg\{&\frac{q_\ell^2}{e^2}+\frac{5m_{\ell_i}^2}{(m_c^*)^2}+\frac{40m_{\ell_i}^2}{3m_\bmh^2\left[k^{(e)}_{1}N_0\right]^4}\left(1-\frac{2\alpha q_\ell^2}{3\pi e^2}\right)%
\left[1+\ILO{{N_0}^{-1}}+\OO{\frac{m_{\ell_i}^2}{m_\bmh^2}}+\ILO{\alpha^2}\right]
\Bigg\}\nn
\end{aligned}\\
&\times\left(1+\ldots\right)\nn
\end{align}
where ``$\ldots$'' represents massive loop corrections of order $\alpha$ which are derived starting in \sref{sec:f_gluonloopcorrs}. 

For a further amendment to this expression, also note that the extra photon making up the loop may potentially comprise some of the same preons as are involved in interactions elsewhere within the local correlation region, with effect on the FSF symmetry factors associated with the loop photon vertices. The resulting corrections comprise both massless and massive terms.
All relevant boson masses are of comparable magnitude, $m_c^2\sim m_W^2\sim m_Z^2\sim m_\bmh^2$, and gluon terms will be seen to outweigh weak sector terms, therefore write the leading elements of each term correcting $m_{e_i}$ as $\ILO{\alpha{N_0}^{-1}}$ and $\ILOO{\alpha{N_0}^{-1}m_{e_i}^2/(m_c^*)^2}$ respectively. Neither of these corrections have counterparts in the Standard Model. It is convenient to insert them into the final bracket of the above expression,
\begin{equation}
(1+\ldots)\longrightarrow \left\{1+\OO{\frac{\alpha}{N_0}}+\OOO{\frac{\alpha m_{\ell_i}^2}{N_0(m_c^*)^2}}+\ldots\right\}\label{eq:errwithalphaN0}
\end{equation}

\subsubsection{\prm{\ILO{{N_0}^{-1}}} correction to 1-loop EM corrections\label{sec:ONtoalpha}}

It is relatively straightforward to calculate the $\ILO{{N_0}^{-1}}$ corrections to the 1-loop EM corrections. First, recognise that these corrections only apply when a loop vertex is within the same correlation region as another vertex. If not, then only the usual EM symmetry factor of $S_\alpha$ applies, which is incorporated within the coefficient $\alpha$ associated with the loop vertices.

However, consider the first diagram of \fref{fig:photonloops}(i), in which the lower vertex on the foreground photon is bracketed by two correlated couplings to the pseudovacuum. Assuming the window approximation~\Peref{I}{eq:window}, in the isotropy frame of the pseudovacuum these bracketing vertices must lie within the autocorrelation distance and time of one another. In the dominant (on-shell) contribution, propagation from one background interaction vertex to the other is linear and hence the lower vertex of the foreground photon also lies within the same correlation region. However, as discussed in \aref{apdx:symloops}, in the dominant contribution to the loop correction the upper vertex does not.

Next, recognise that corrections of this form apply whenever the loop boson vertex in the autocorrelation region contains preons matching those present at the pseudovacuum vertices, regardless of the boson species involved. Consequently to overall order $\ILO{\alpha/{N_0}}$ these $\ILO{{N_0}^{-1}}$ corrections arise as amendments to the EM loop corrections to background photon interactions, and to order $\ILO{\alpha m_{\ell_i}^2/[N_0(m_c^*)^2]}$ they arise as amendments to (ii)~EM loop corrections to background gluon interactions, (iii)~gluon loop corrections to background photon interactions, and (iv)~weak boson loop corrections to background photon interactions. Terms~(i) and~(ii) are evaluated here, and terms~(iii)-(iv) in \srefs{sec:N0gluonEM}{sec:N0WZEM} respectively.

\paragraph[\prm{\ILO{{N_0}^{-1}}} correction to EM loop correction to background photon coupling]{$\ILO{{N_0}^{-1}}$ correction to EM loop correction to background photon coupling:\label{sec:N0EMEM}}
To evaluate term~(i), %
the $\ILO{{N_0}^{-1}}$ corrections arising due to FSF interchange when a photon loop corrects the coupling to the background photon field,
first recognise that
within a given foreground vertex there are two inbound preon lines and two outbound preon lines. 
It suffices to consider each separately.

Recognising that all preon lines undergo interactions with the background fields which may cause them to change colour, and that colour is only conserved separately within the foreground and background fields on average over scales large compared with $\mc{L}_0$, the colour of each inbound or outbound line both in this vertex and in the background field interaction vertices may be considered independently random. Thus there is a chance of $\frac{1}{3}$ that the colour of a given line at the foreground vertex will match the equivalent line at a given background field vertex.

In contrast, $A$-charge is guaranteed to match as all photons arise from the same fermion, which contains only one type of preon carrying a well-defined $A$-charge. However, these fermions emit bosons associated with representations containing both $\delta_{\dot cc}\bar\psi^{1\dot c}\bsm\psi^{1c}-\delta_{\dot c'c'}\bar\psi^{2\dot c'}\bsm\psi^{2c'}$ and $\delta_{\dot cc}\bar\psi^{1\dot c}\bsm\psi^{1c}+\delta_{\dot c'c'}\bar\psi^{2\dot c'}\bsm\psi^{2c'}$, with the former corresponding to photons and the latter to~$Z$ and $N$~bosons. By conservation of charge at vertices, and the requirement that the fermion line carries a well-defined $A$-charge, these two groups are emitted with equal weight. Consequently, when there is a match of $A$-charges at two vertices, there is only a 50\% chance that the line being matched in the $e_{ij}$ basis corresponds to a photon. Thus the $A$~sector provides a further factor of $\frac{1}{2}$.

There are four lines at the foreground (loop) vertex which is in the correlated region, and there are two background vertices from the leading-order diagram with which to seek symmetry matches, for factors of~4 and~2 respectively. However, the requirement that preons at a vertex exhibit appropriate correlations (i.e.~preons in photons must be pairwise correlated, those in fermions must form correlated triplets, etc.) implies that:
\begin{itemize}
\item A preon in a photon can only be matched with a preon in a photon, and a preon in a fermion can only be matched with a preon in a fermion.
\item Similarly, the orientations of preons at a vertex (inbound/outbound) must also match.
\item When the FSF associated with one preon in a photon is exchanged with that of a preon in a different photon, \emph{both} FSFs associated with those photons must (i)~carry matching charges, and (ii)~be exchanged. This halves the number of independent matches, but doubles the increment in FSF symmetry factor associated with a match. This effect may therefore be ignored.
\item A similar consideration applies for fermions, with all three FSFs being exchanged. However, for the FSFs of non-vertex fermions this is trivial and has no bearing on the calculation.
\end{itemize}
A given preon in the foreground (loop) vertex may therefore be characterised as (for example) ``photon, inbound to vertex'' and only yields an $\ILO{{N_0}^{-1}}$ increment of the $\ILO{\alpha}$ symmetry factor if it matches a ``photon, inbound to vertex'' preon on a background vertex (from the leading-order diagram).

Putting this all together, there are four preon lines in the foreground (loop) vertex, each of which may match one preon line in each of the two background (leading-order diagram) vertices, with a chance of $\frac{1}{6}$. When they match, this increments an FSF symmetry factor from $N_0+k$ to $N_0+(k+1)$, and this is equivalent to mutiplying by a correction factor of $[1+{N_0}^{-1}+\ILO{{N_0}^{-2}}]$.
Thus the overall net multiplicative factor arising from all lines in all vertices is
\begin{equation}
\left[1+\frac{4}{3N_0}+\OO{{N_0}^{-2}}\right].
\end{equation}

A similar analysis applies to the second diagram of \fref{fig:photonloops}(i) but not to the third,
yielding an overall loop correction factor to the tree level vertex, accurate to $\ILO{\alpha{N_0}^{-1}}$, %
of
\begin{equation}
\begin{split}
&\biggl[1+\frac{\alpha}{2\pi}\left(1+\frac{4}{3N_0}\right)+\frac{\alpha}{2\pi}\left(1+\frac{4}{3N_0}\right)+\frac{\alpha}{2\pi}%
+\OO{\frac{\alpha}{{N_0}^2}}+\OO{\alpha^2}\biggr]\\
=&\biggl[1+\frac{3\alpha}{2\pi}\left(1+\frac{8}{9N_0}\right)+\OO{\frac{\alpha}{{N_0}^{2}}}+\OO{\alpha^2}\biggr].
\end{split}
\end{equation}
This is compared with the Standard Model factor of $3\alpha/(2\pi)+\OO{\alpha^2}$ to yield a correction factor accurate to $\ILO{{N_0}^{-1}}$ and $\ILO{\alpha}$ of $(1+c_{\alpha/N_0})$ where
\begin{equation}
\begin{split}
c_{\alpha/N_0}:=\,&\frac{1+\frac{3\alpha}{2\pi}\left(1+\frac{8}{9{N_0}}\right)}{1+\frac{3\alpha}{2\pi}}-1=\frac{8\alpha}{3N_0(3\alpha+2\pi)},
\end{split}\label{eq:calphaN0v1}
\end{equation}
and a net expression for electron mass
\begin{align}
m_{\ell_i}^2 =\,& {\frac{f^2}{2}\left[k^{(\ell)}_1\right]^4}{\omega_0}^2{N_0}^8S_{18,147}\label{eq:lep1loopA2}\\
&\begin{aligned}
\times\Bigg\{&\frac{q_\ell^2(1+c_{\alpha/N_0})}{e^2}+\frac{5m_{\ell_i}^2}{(m_c^*)^2}\\
&+\frac{40m_{\ell_i}^2}{3m_\bmh^2\left[k^{(e)}_{1}N_0\right]^4}\left(1-\frac{2\alpha q_\ell^2}{3\pi e^2}\right)%
\left[1+\ILO{{N_0}^{-1}}+\OO{\frac{m_{\ell_i}^2}{m_\bmh^2}}+\ILO{\alpha^2}\right]
\Bigg\}\nn
\end{aligned}\\
&\times\biggl\{1\!+\!\OO{\frac{\alpha}{{N_0}^2}}\!+\!\OO{\frac{\alpha^2}{{N_0}}}\!+\!\OOO{\frac{\alpha m_{\ell_i}^2}{N_0(m_c^*)^2}}\!+\!\ldots\biggr\}\nn
\end{align}
where the correction of $\ILO{\alpha/N_0}$ in \Eref{eq:errwithalphaN0} has been evaluated to yield $c_{\alpha/N_0}$, %
and the next-highest unevaluated contributions are $\ILO{\alpha/{N_0}^2}+\ILO{\alpha^2/N_0}$. [As with corrections of $\ILO{\alpha}$, corrections of $\ILO{\alpha^2}$ are common to both the $\Cw{18}$ model and the Standard Model and thus %
do not need to be incorporated into the expression for the mass vertex.]

\paragraph[\prm{\ILO{{N_0}^{-1}}} correction to EM loop correction to background gluon coupling]{$\ILO{{N_0}^{-1}}$ correction to EM loop correction to background gluon coupling:\label{sec:N0EMgluon}}

The next corrections to be considered are the $\ILO{{N_0}^{-1}}$ corrections arising due to FSF interchange when a photon loop corrects the coupling to the background gluon field.
The principle is the same as for the photon loop, and calculation of the correction broadly follows \sref{sec:N0EMEM}. Again, one vertex from the loop correction lies between the two background field couplings, and on-shell, will be within the same correlation region. However:
\begin{itemize}
\item The preons in the background fields may each carry any of the three colour charges and any of the three $A$-charges. The chance of them matching a given preon in the loop photon is thus $\frac{1}{9}$, in contrast with $\frac{1}{6}$ in \sref{sec:N0EMEM} (where the $A$-charge of the background preons can only be~1 or~2).
\item In the background gluons, both preons carry the same $A$-charge but their $C$-charges may vary independently. However, the sum over two (non-identical) preons, each of which may potentially match the background preon, yields a factor of two. In contrast, in \sref{sec:N0EMEM} the preons in the loop photon both carry the same $C$-charge as well as $A$-charge; either both match or both do not. However, matching two preons again results in a factor of two. The independence of the $C$-charges in the background gluons therefore does not further change the correction factor.
\end{itemize}
The resulting correction is thus
\begin{equation}
\frac{5m_{\ell_i}^2}{(m_c^*)^2}\longrightarrow\frac{5m_{\ell_i}^2}{(m_c^*)^2}\left[1+\frac{2c_{\alpha/N_0}}{3}+\OO{{N_0}^{-2}}\right]
\end{equation}
which may be conveniently realised by redefining $c_{\alpha/N_0}$ in \Eref{eq:lep1loopA2} as
\begin{equation}
c_{\alpha/N_0}:=\frac{8\alpha}{3N_0(3\alpha+2\pi)}\left[1+\frac{10m_{\ell_i}^2}{3(m_c^*)^2}\right].\label{eq:calphaN0v2}
\end{equation}
The error terms in \Eref{eq:lep1loopA2} remain unchanged, as this correction does not exhaust the terms of $\ILOOO{{\alpha m_{\ell_i}^2}/[{N_0(m_c^*)^2}]}$. For further terms of this order, see \srefs{sec:N0gluonEM}{sec:N0WZEM}.

\subsubsection{1-loop gluon corrections\label{sec:f_gluonloopcorrs}}

The next loop corrections to consider are those arising from gluon loops, analogous to the diagrams of \fref{fig:photonloops}. These interactions occur below the fermion scale, at the level of preons, and thus participation of the ninth gluon (equivalently, $N_\mu$) is not prohibited by gauge. It then is convenient to work in the $e^C_{ij}$ basis, in which the interactions of all nine gluons are equivalent on the preon scale.

In the $e^C_{ij}$ basis, gluons are in general confined so cannot escape to the far field but must remain within the local autocorrelation region (within which the pseudovacuum %
is permitted to appear transiently colour-inhomogeneous). The exception to this is the superposition corresponding to the $N_\mu$ boson, but where gluons are emitted from fermions or vector bosons, this superposition is also prohibited from propagating into the far field (by gauge). Thus in the present context all nine species may be considered confined regardless of choice of basis.

In the context of fermion mass interactions, the loop gluons are foreground particles existing over scales of $\ILO{\mc{L}_0}$ which corresponds to the interaction scale for boson mass, and are therefore massive. %
Over these scales, gluons which form local closed loops are effectively shielded by the inhomogeneities in the pseudovacuum and exhibit at most only the bare gluon mass of \sref{sec:gluonmasses} (including the gluon deficit correction).

In the present chapter only loop corrections to the background photon and background gluon interactions need be considered: Corrections to the background scalar boson interaction are of $\ILOO{\alpha m_{\ell_i}^2/(m_c^*)^2}$ and thus are smaller than the terms of $\ILO{m_{\ell_i}^2/m_\bmh^2}$ in \Eref{eq:lep1loopA}.
First the requisite corrections are calculated while neglecting preon degeneracy, and then the $\ILO{{N_0}^{-1}}$ corrections to the corrections are computed similarly to those previously seen in \srefr{sec:N0EMEM}{sec:N0EMgluon}.

\begin{figure}
\includegraphics[width=\linewidth]{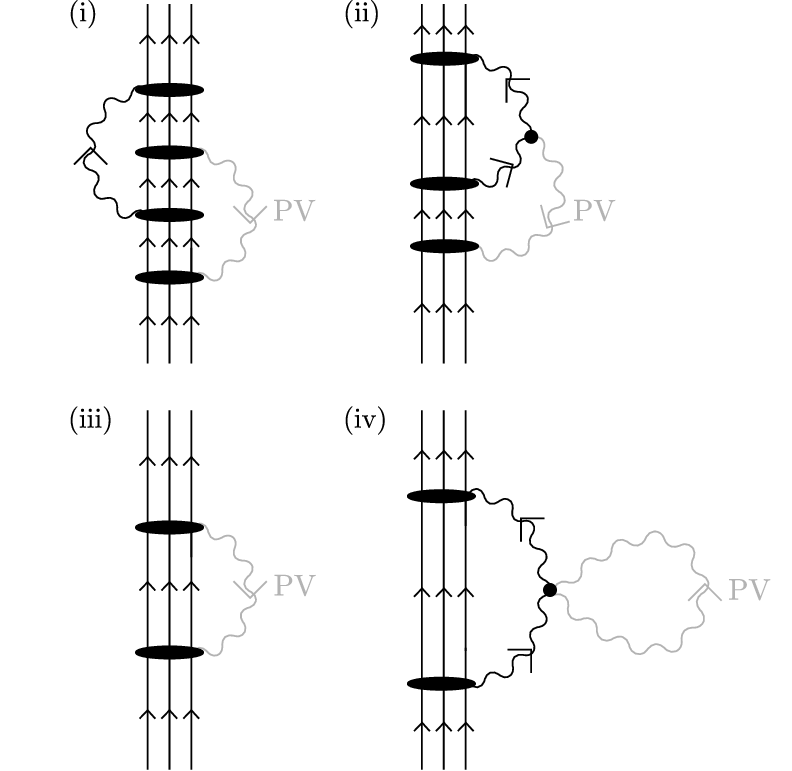}
\caption{(i)-(ii)~Gluon loop equivalents to the first diagram of \pfref{fig:photonloops}(i). When the fermion couples to the background photon field, only diagram~(i) may be constructed as the gluon does not carry an electromagnetic charge. When the fermion couples to the background gluon field, both diagram~(i) and diagram~(ii) may be constructed. Conservation of colour charge indicates that on summing the coupling of the background gluon to the fermion and the background gluon to the loop gluon, this is equal to the coupling of the background gluon to the fermion in the original leading-order diagram~(iii). Similarly, diagram~(iv)
is the counterpart to the third diagram of \pfref{fig:photonloops}(i). This diagram is especially interesting as it contributes both to the fermion mass and to the mass of the loop gluon. Note that the two background gluons must couple with the loop gluon at a single vertex, as discussed in \pt{\Psref{V}{sec:vecbosonmasses}}.
\label{fig:gluonloops}}
\end{figure}

\paragraph[Main correction]{Main correction:\label{sec:1-loopgluonmain}} Consider the gluon loop counterparts to \fref{fig:photonloops}(i). Where the fermion interacts with the background photon field, these corrections take on the form of \fref{fig:gluonloops}(i).
For interactions between the fermion and the background gluon fields they may take on the form of either \fref{fig:gluonloops}(i) or~(ii), and conservation of colour charge indicates that when these couplings are summed, this is equal to the coupling of the background gluon to the fermion in the original leading-order diagram [\fref{fig:gluonloops}(iii)]. Further, when the loop correction is evaluated, this collapses to a numerical multiplier on the original background interaction vertex and the customary application of spinor identities \Perefr{IV}{eq:unreducedeLmassint}{eq:reducedeLmassint} yields a non-vanishing contribution to $m_{\ell_i}^2$.
It then suffices to consider an example where the coupling of the background gluon and loop gluon vanishes, making it equivalent to the background photon case, to evaluate the correction to the interaction vertex, and to extrapolate this across all gluon colour combinations by $\GLTR$ symmetry.

\Fref{fig:gluonloops}(iv) shows a further diagram which may be constructed using gluon loop corrections. This is a counterpart to the gluon version of the third diagram of \fref{fig:photonloops}(i), again with vertices moved onto the loop boson, and also contributes to the mass of the loop gluon.

Note that in contrast to the photon terms, where the third diagram of \fref{fig:photonloops}(i) was accounted for in the Standard Model, for preon/gluon interactions all loop corrections must be evaluated as there are no corresponding Standard Model terms.

To evaluate these corrections begin with \fref{fig:gluonloops}(i), which is the gluon loop counterpart to \fref{fig:photonloops}(i). Start with the fermion coupling to the background photon field, and a specific choice of off-diagonal gluon. Compare with the equivalent EM loop figure and note the following changes:
\begin{itemize}
\item The vertex factors increase from $f^2/2$ to $f^2$, for a relative factor of~2.
\item The boson is off-diagonal, for a relative structural factor of $\frac{5}{3}$.
\item The photon source may be any of three charged preons, but the gluon may only be emitted by a preon of appropriate colour. However, any of the three preons may be the preon of that colour, for a net factor of one.
\item Emission of the off-diagonal gluon changes the colour of the emitting preon. There are then two preons of that colour which are capable of absorbing the loop gluon, and both of the resulting diagrams count as loop corrections due to the implicit exchange (not shown) of further gluons as a binding interaction sharing momentum between all members of the preon triplet. The choice of absorbing preons gives a factor of~2.
\item The loop gluon is massive, and is in the presence of a foreground fermion so experiences a gluon field mass deficit giving a loop factor of $m_{\ell_i}^2/(m_c^*)^2$.
\item These factors multiply the equivalent photon loop factor, which is $\alpha/(2\pi)$.
\end{itemize}
The per-gluon correction weight from this diagram is therefore
\begin{equation}
\frac{10\alpha}{3\pi}\frac{m_{\ell_i}^2}{(m_c^*)^2}.
\end{equation}
This calculation is repeated for the gluon counterparts to the other two figures of \fref{fig:photonloops}(i), and each of these figures may involve any of nine gluons, for a total weight of
\begin{equation}
\frac{90\alpha}{\pi}\frac{m_{\ell_i}^2}{(m_c^*)^2}.
\end{equation}
Correction of the background gluon interactions proceeds equivalently, and as noted above the correction to the scalar boson may be ignored at current precision, for a net lepton mass so far of
\begin{align}
m_{\ell_i}^2 =\,& {\frac{f^2}{2}\left[k^{(\ell)}_1\right]^4}{\omega_0}^2{N_0}^8S_{18,147}\label{eq:lep1loopA3}\\
&\begin{aligned}
\times\Bigg\{&\frac{q_\ell^2}{e^2}\left[1+c_{\alpha/N_0}+\frac{90\alpha m_{\ell_i}^2}{\pi(m_c^*)^2}\right]%
+\frac{5m_{\ell_i}^2}{(m_c^*)^2}\left[1+\frac{90\alpha m_{\ell_i}^2}{\pi(m_c^*)^2}\right]\\
&+\frac{40m_{\ell_i}^2}{3m_\bmh^2\left[k^{(e)}_{1}N_0\right]^4}\left(1-\frac{2\alpha q_\ell^2}{3\pi e^2}\right)%
\left[1+\ILO{{N_0}^{-1}}+\OO{\frac{m_{\ell_i}^2}{m_\bmh^2}}+\ILO{\alpha^2}\right]
\Bigg\}\nn
\end{aligned}\\
&\times\biggl\{1\!+\!\OO{\frac{\alpha}{{N_0}^2}}\!+\!\OO{\frac{\alpha^2}{{N_0}}}\!+\!\OOO{\frac{\alpha m_{\ell_i}^2}{N_0(m_c^*)^2}}\!+\!\ldots\biggr\}\nn
\end{align}
where the next major terms to be determined are the one-loop weak boson corrections.

\paragraph[\prm{\ILO{{N_0}^{-1}}} correction to gluon loop correction to background photon coupling]{$\ILO{{N_0}^{-1}}$ correction to gluon loop correction to background photon coupling:\label{sec:N0gluonEM}}

The next corrections to be considered are the $\ILO{{N_0}^{-1}}$ corrections arising due to FSF interchange when a {gluon} loop corrects the coupling to the background photon field.
The principle is the same as for the photon loop, and calculation of the correction broadly follows \srefr{sec:N0EMEM}{sec:N0EMgluon}. Again, one vertex from the loop correction lies between the two background field couplings, and on-shell, will be within the same correlation region. Again, matching is now between preons from a gluon and a photon, reducing the correction by a factor of $\frac{2}{3}$ relative to \sref{sec:N0EMEM}.
The resulting correction is thus
\begin{equation}
\frac{q_\ell^2}{e^2}\frac{90\alpha m_{\ell_i}^2}{\pi (m_c^*)^2}\longrightarrow\frac{q_\ell^2}{e^2}\frac{90\alpha m_{\ell}^2}{\pi (m_c^*)^2}\left[1+\frac{2c_{\alpha/N_0}}{3}+\OO{{N_0}^{-2}}\right]
\end{equation}
which may be conveniently realised by redefining $c_{\alpha/N_0}$ in \Eref{eq:lep1loopA3} as
\begin{equation}
\begin{split}
c_{\alpha/N_0}:=\,&\frac{8\alpha}{3N_0(3\alpha+2\pi)}\left[1+\frac{10m_{\ell_i}^2}{3(m_c^*)^2}+\frac{60\alpha m_{\ell_i}^2}{\pi(m_c^*)^2}\right]\\
=\,&\frac{8\alpha}{3N_0(3\alpha+2\pi)}\left[1+\frac{(10\pi+180\alpha)m_{\ell_i}^2}{3\pi(m_c^*)^2}\right].\label{eq:calphaN0v3}
\end{split}
\end{equation}
The error terms in \Eref{eq:lep1loopA3} remain unchanged, as this correction does not exhaust the terms of $\ILOOO{{\alpha m_{\ell_i}^2}/[{N_0(m_c^*)^2}]}$. For further terms of this order, see \sref{sec:N0WZEM}.

\subsubsection{1-loop weak force corrections\label{sec:f_weakloopcorrs}}

In the interest of brevity, this Section and those which follow specialise to the charged leptons only, denoted $e_i$ in lieu of $\ell_i$.

\begin{figure}
\includegraphics[width=\linewidth]{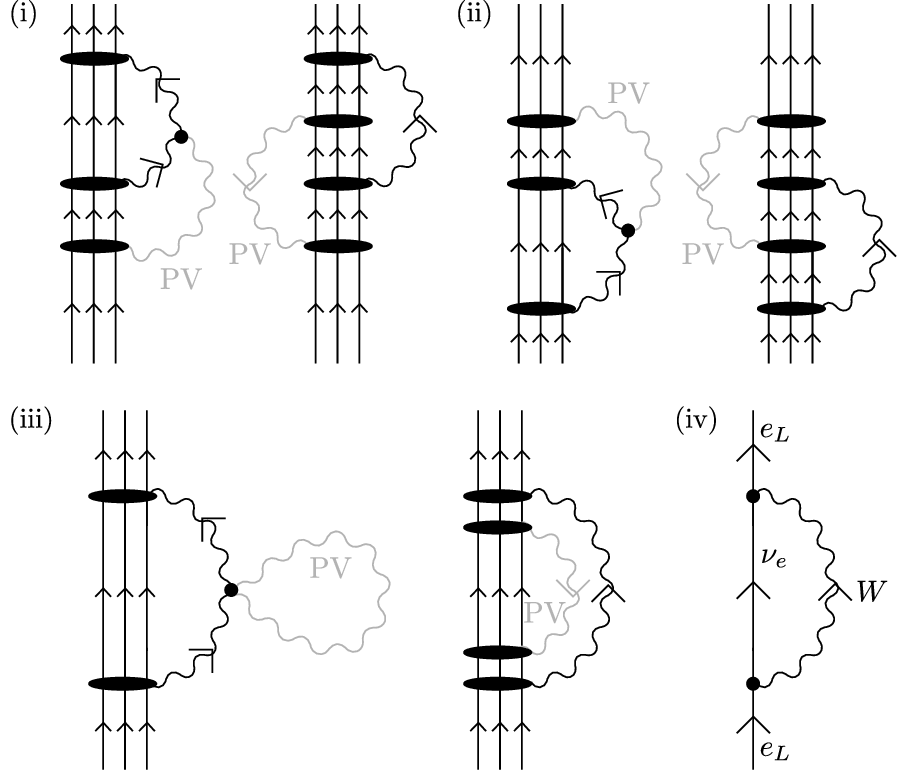}
\caption{(i)-(iii)~\pt{$\tW$}~boson loop corrections to the background vector boson interactions. 
Equivalent background photon and background gluon terms are grouped together; photon terms have no orientation arrows on the background boson lines. Note that background gluons do not couple to the \pt{$\tW$}~boson as the latter contains preons in a colour/anti-colour pair whose net gluon coupling therefore vanishes.
(iv)~Loop involving a \pt{$W$} boson and a neutrino, appearing in the proper self-energy corrections of %
the Standard Model%
.\label{fig:Wbosonloops_QLphoton}}
\end{figure}

\paragraph[Background photon interaction]{Background photon interaction:\label{sec:weakcorrtophoton}}
When the fermion interacts with the background photon field, $\tW$ boson loops (\sref{sec:tW}) may correct this interaction as shown in \fref{fig:Wbosonloops_QLphoton}.
As with the gluon loops in \sref{sec:f_gluonloopcorrs}, the boson loops may be collapsed to numerical factors on the vertices of the corresponding leading-order diagram, and the background field terms contracted using spinor/sigma matrix identities.

Some caution is required with sign---first consider diagram~(iii). Evaluating this diagram as a tensor network in the manner of \rcite{penrose1971,pfeifer2010,pfeifer2011}, it is readily seen to be an absolute square and therefore yields a contribution to $m_{e_i}^2$ which is additive to the leading order term. Diagrams~(i) and~(ii) contain as a subdiagram a correction to the EM emission process which is usually associated with the opposite sign to direct emission by the fermion; however, compared with diagram~(iii) they have also acquired an additional fermion propagator segment which offsets this, so they are also additive. Relative to the leading-order diagram, \freft{fig:Wbosonloops_QLphoton}(i)-(ii) acquire factors
\begin{itemize}
\item $(10/3)(m_{e_i}^2/m_\tW^2)$, being the usual (absolute) factor for a $W$-type loop correction to an EM vertex,
\item $1/2$ due to loss of exchange symmetry for the two background photon/composite fermion interactions. (Even with a composite fermion, an effective vertex is defined on summing over all possible preon/photon pairings and when identical, these effective vertices may still be exchanged for a symmetry factor of~2.)
\end{itemize}
Diagram~(iii) retains the exchange symmetry, this time on the two copies of the photon operator at the vertex, so attracts the factor of $(10/3)(m_{e_i}^2/m_\tW^2)$ only. These three diagrams sum to yield a net factor
\begin{equation}
\frac{\alpha}{2\pi}\frac{20m_{e_i}^2}{3m_\tW^2}.\label{eq:netfactor}
\end{equation}

Finally, consider the Standard Model counterpart: The charged lepton Proper Self-Energy (PSE) terms in the Standard Model give rise to the $W$/neutrino loop shown in \fref{fig:Wbosonloops_QLphoton}(iv). This gives rise to a term analogous to \fref{fig:Wbosonloops_QLphoton}(iii), but has no counterpart to the symmetry factor associated with the pair of identical photon operators on the pseudovacuum interaction vertex. Its associated factor is therefore $(5/3)(m_{e_i}^2/m_W^2)$, or in the $\Cw{18}$ model, $(5/3)(m_{e_i}^2/m_\tW^2)$. 
The PSE terms %
appear in the denominator of $m_{e_i}^2$, corresponding to an overall multiplicative factor %
of
\begin{equation}
\left(1+\ldots+\frac{\alpha}{2\pi}\frac{5 m_{e_i}^2}{3m_\tW^2}+\ldots\right)^{-1}.
\end{equation}
Recalling that mass-independent PSE terms cancel exactly between numerator and denominator (see also \sref{sec:1loopfgEM}) and that there are no $N_0$-dependent PSE terms, this particular term in the denominator may be approximately subsumed into the numerator by subtracting $[5\alpha/(6\pi)](m_{e_i}^2/m_\tW^2)$ from \Eref{eq:netfactor},
with an error of at most $\ILOO{{\alpha^2m_{e_i}^4}/{(m_c^*)^4}}$ on $m_{e_i}^2$.
The net weight of the $\tW$~boson loop corrections is thus
\begin{equation}
\frac{\alpha}{2\pi}\left(\frac{20m_{e_i}^2}{3m_\tW^2}-\frac{5 m_{e_i}^2}{3m_\tW^2}\right)=\frac{5\alpha m_{e_i}^2}{2\pi m_\tW^2}.
\end{equation}

Next considering the $Z$ boson loops, these take form directly analogous to \fref{fig:photonloops}, and by arguments similar to the above are also seen to all be additive. Noting that the $Z$ boson couples with opposite sign to $e_L$ and $e_R$, and that mass vertices in the Standard Model reverse electron spin whereas the pseudovacuum couplings do not, the $Z$ boson equivalents to the first two diagrams of \fref{fig:photonloops}(i) are therefore of opposite sign to their Standard Model counterparts and hence attract a factor of two, to first offset the Standard Model-derived term in the PSE, and then implement the actual correction. The third diagram of \fref{fig:photonloops}(i), on the other hand, has the same sign as its Standard Model counterpart and thus yields no net contribution to $m_{e_i}^2$. The net outcome from $Z$ boson loops is a term with weight
\begin{equation}
-\frac{4f_Z\alpha m_{e_i}^2}{2\pi m_\tW^2}
\end{equation}
where $f_Z$ is as defined in \Eref{eq:fZ2}. %
Overall, the weight of the weak boson corrections to the background photon interaction is seen to be
\begin{equation}
\frac{(5-4f_Z)\alpha m_{e_i}^2}{2\pi m_\tW^2}.
\end{equation}

\paragraph[Background gluon interactions]{Background gluon interactions:\label{sec:N0gluon}}
This term is readily obtained as the calculation proceeds similarly to \sref{sec:weakcorrtophoton}, and it is found to be right at the threshold of relevance at the precision employed the current chapter. 

As usual, start with the fermion interaction with a single off-diagonal background gluon channel and then use $\GLTR$ symmetry to multiply by nine for the contributions of the other eight gluons. For the gluon channels, the background gluons continue to couple to the fermion and not the looping $\tW$ boson, so when evaluating the equivalents of \freft{fig:Wbosonloops_QLphoton}(i)-(ii) there is no factor of $\frac{1}{2}$ for lost symmetry. The calculation is otherwise equivalent, for net $\tW$ and $Z$ corrections of
\begin{equation}
\frac{(25-12f_Z)\alpha m_{e_i}^2}{6\pi m_\tW^2}.\label{eq:gluongluon}
\end{equation}
Note that deductions are once again made for the Standard Model counterpart terms, as corrections to mass vertices in the Standard Model correspond to corrections to both %
the photon and gluon terms in the present model.

\paragraph[Background scalar boson interaction]{Background scalar boson interaction:}
These terms fall below the threshold of relevance, being smaller by a factor of $\ILO{\alpha}$ than the unevaluated correction of $\ILO{m_{e_i}^2/m_\bmh^2}$.

The cumulative expression for charged lepton mass incorporating weak boson corrections is therefore
\begin{align}
m_{e_i}^2 =\,& {\frac{f^2}{2}\left[k^{(e)}_i\right]^4}{\omega_0}^2{N_0}^8S_{18,147}\label{eq:elecmasses_pre}\\
&\begin{aligned}
\times\Bigg\{&\left[1+c_{\alpha/N_0}+\frac{90\alpha m_{e_i}^2}{\pi(m_c^*)^2}+\frac{(5-4f_Z)\alpha m_{e_i}^2}{2\pi m_\tW^2}\right]%
+\frac{5m_{e_i}^2}{(m_c^*)^2}\left[1+\frac{90\alpha m_{e_i}^2}{\pi(m_c^*)^2}+\frac{(25-12f_Z)\alpha m_{e_i}^2}{6\pi m_\tW^2}\right]\\
&+\frac{40m_{e_i}^2}{3m_\bmh^2\left[k^{(e)}_{1}N_0\right]^4}\left(1-\frac{2\alpha}{3\pi}\right)%
\left[1+\ILO{{N_0}^{-1}}+\OO{\frac{m_{e_i}^2}{m_\bmh^2}}+\ILO{\alpha^2}\right]
\Bigg\}\nn
\end{aligned}\\
&\times\biggl\{1\!+\!\OO{\frac{\alpha}{{N_0}^2}}\!+\!\OO{\frac{\alpha^2}{{N_0}}}\!+\!\OOO{\frac{\alpha m_{e_i}^2}{N_0(m_c^*)^2}}\!+\!\OOO{\frac{\alpha^2m_{e_i}^4}{(m_c^*)^4}}\!+\!\ldots\biggr\}\nn
\end{align}

\paragraph[\prm{\ILO{{N_0}^{-1}}} correction to weak loop corrections to background photon coupling]{$\ILO{{N_0}^{-1}}$ correction to weak loop corrections to background photon coupling:\label{sec:N0WZEM}}

The next corrections to be considered are the $\ILO{{N_0}^{-1}}$ corrections arising due to FSF interchange when a $\tW$ or $Z$~boson loop corrects the coupling to the background photon field.
The principle is similar to that of \sref{sec:N0gluonEM} where a gluon loop corrected the background photon interaction, with the only numerical difference being the chance that preons at the participating $\tW$ or $Z$ vertex match the preons in the background photon field.

First consider the $\tW$~loop and compare with the original weights in \sref{sec:N0EMEM}. If the first preon considered is the $\psi^2$ preon, this has a 50\% chance of matching the $A$-charge of the relevant preon in the background photon, which is the same as in \sref{sec:N0EMEM}. However, the second preon is of type $\psi^3$ which does not appear in the background photon field at all. For a $\tW$~loop the correction is therefore reduced by a factor of $\frac{1}{2}$. Similarly, for a $Z$~loop each preon has a 50\% chance of being of type $\psi^3$ at the $Z$~vertex and a 50\% chance of being in $\{\psi^1,\psi^2\}$. Thus the correction to the $Z$ boson loop is also reduced by a factor of two,
\begin{equation}
\frac{(5-4f_Z)\alpha m_{e_i}^2}{2\pi m_\tW^2}\longrightarrow\frac{(5-4f_Z)\alpha m_{e_i}^2}{2\pi m_\tW^2}\!\left[1\!+\!\frac{c_{\alpha/N_0}}{2}\!+\!\OO{{N_0}^{-2}}\right].
\end{equation}
This may be conveniently realised by redefining $c_{\alpha/N_0}$ as
\begin{align}
c_{\alpha/N_0}:=\,&\frac{8\alpha}{3N_0(3\alpha+2\pi)}\label{eq:calphaN0v4}
\left[1+\frac{(10\pi+180\alpha)m_{e_i}^2}{3\pi(m_c^*)^2}%
+\frac{(5-4f_Z)\alpha m_{e_i}^2}{4\pi m_\tW^2}\right].%
\end{align}
This exhausts the corrections of $\ILOOO{{\alpha m_{e_i}^2}/[{N_0(m_c^*)^2}]}$, with \Eref{eq:elecmasses_pre} becoming
\begin{align}
m_{e_i}^2 =\,& {\frac{f^2}{2}\left[k^{(e)}_i\right]^4}{\omega_0}^2{N_0}^8S_{18,147}\label{eq:elecmasses}\\
&\begin{aligned}
\times\Bigg\{&\left[1+c_{\alpha/N_0}+\frac{90\alpha m_{e_i}^2}{\pi(m_c^*)^2}+\frac{(5-4f_Z)\alpha m_{e_i}^2}{2\pi m_\tW^2}\right]%
+\frac{5m_{e_i}^2}{(m_c^*)^2}\left[1+\frac{90\alpha m_{e_i}^2}{\pi(m_c^*)^2}+\frac{(25-12f_Z)\alpha m_{e_i}^2}{6\pi m_\tW^2}\right]\\
&+\frac{40m_{e_i}^2}{3m_\bmh^2\left[k^{(e)}_{1}N_0\right]^4}\left(1-\frac{2\alpha}{3\pi}\right)%
\left[1+\ILO{{N_0}^{-1}}+\OO{\frac{m_{e_i}^2}{m_\bmh^2}}+\ILO{\alpha^2}\right]
\Bigg\}\nn
\end{aligned}\\
&\times\biggl\{1\!+\!\OO{\frac{\alpha}{{N_0}^2}}\!+\!\OO{\frac{\alpha^2}{{N_0}}}\!+\!\OOO{\frac{\alpha m_{e_i}^4}{N_0(m_c^*)^4}}\!+\!\OOO{\frac{\alpha^2m_{e_i}^4}{(m_c^*)^4}}\!+\!\ldots\biggr\}.\nn
\end{align}
Note that terms of $\ILOOO{\alpha^2 m_{e_i}^2/[N_0(m_c^*)^2]}$ and $\ILOOO{\alpha m_{e_i}^2/[{N_0}^2(m_c^*)^2]}$ also exist but are small compared with those of $\ILO{\alpha^2/N_0}$ and $\ILO{\alpha/{N_0}^2}$ already appearing in \Eref{eq:elecmasses}. The term $\ILOOO{\alpha m_{e_i}^4/[N_0(m_c^*)^4]}$ arises from $\ILO{{N_0}^{-1}}$ corrections to the gluon loop correction to the gluon mass interaction.

\subsubsection{1-loop scalar corrections\label{sec:1loopscalar}}
The largest of these terms are the scalar loop corrections to the background photon interaction. These are of comparable magnitude to the photon loop corrections to the background scalar interaction. Leaving these terms unevaluated replaces the factor of $\left[1-{2\alpha}/(3\pi)\right]$ on the scalar interaction term with an unspecified correction of $\left[1+\ILO{\alpha}\right]$.

\subsubsection{2-loop EM corrections}

By the same arguments as \sref{sec:1loopfgEM}, the massless $\ILO{\alpha^2}$ corrections are identical for the background photon and gluon interactions and coincide with their Standard Model counterparts. The massless 2-loop EM corrections therefore yield at most a correction of $\ILO{\alpha^2}$ to the scalar term, which is smaller than the $\ILO{\alpha}$ term from \sref{sec:1loopscalar}.

The next term which may differentially affect the photon and gluon terms, and which exhibits dependency on particle generation, is a two-loop massive contribution. All relevant masses are of comparable magnitude, $m_c^2\sim m_\tW^2\sim m_Z^2\sim m_\bmh^2$, and gluon terms have thus far outweighed weak sector terms, therefore write this correction in terms of the gluon mass. It is also useful to pull all higher-order corrections out into the final term, %
and
the cumulative expression for charged lepton mass incorporating all loop corrections evaluated in this paper is thus %
\begin{align}
m_{e_i}^2 =\,& {\frac{f^2}{2}\left[k^{(e)}_i\right]^4}{\omega_0}^2{N_0}^8S_{18,147}\label{eq:elecmasses2}\\
&\begin{aligned}
\times\Bigg\{&\left[1+c_{\alpha/N_0}+\frac{90\alpha m_{e_i}^2}{\pi(m_c^*)^2}+\frac{(5-4f_Z)\alpha m_{e_i}^2}{2\pi m_\tW^2}\right]\\
&+\frac{5m_{e_i}^2}{(m_c^*)^2}\left[1+\frac{90\alpha m_{e_i}^2}{\pi(m_c^*)^2}+\frac{(25-12f_Z)\alpha m_{e_i}^2}{6\pi m_\tW^2}\right]%
+\frac{40m_{e_i}^2}{3m_\bmh^2\left[k^{(e)}_{1}N_0\right]^4}
\Bigg\}%
\nn
\\
\times\bm{\Bigg(}&1+\OO{\frac{\alpha}{{N_0}^2}}+\OO{\frac{\alpha^2}{{N_0}}}+\OOO{\frac{\alpha m_{e_i}^4}{N_0(m_c^*)^4}}%
+\OOO{\frac{\alpha^2 m_{e_i}^2}{(m_c^*)^2}}\\&+\OOOOx{\frac{\alpha m_{e_i}^2}{m_\bmh^2\left[k^{(e)}_1{N_0}\right]^4}}%
+\OOOOx{\frac{m_{e_i}^2}{m_\bmh^2\left[k^{(e)}_1\right]^4{N_0}^5}}+\OOOOx{\frac{m_{e_i}^4}{m_\bmh^4\left[k^{(e)}_1{N_0}\right]^4}}
\bm{\Bigg)}.\nn
\end{aligned}
\end{align}

\subsection{Corrections to the lepton mass angle\label{sec:thetacorr}}

\subsubsection{Origin of corrections}

There remains one more substantial correction to evaluate. The leading-order contributions to the lepton mass vertices and the preon colour mixing matrix were evaluated in \sref{sec:leptonleadingorder}. Loop corrections to the mass vertices were evaluated in \sref{sec:leptonloops}. It remains now to determine how these loop corrections impact the preon colour mixing matrix $\K$.

When this matrix was constructed in \sref{sec:leptonKmatrix} there was an implicit assumption that all eigenvectors experience equal couplings to the background fields. The eigenvalues of matrix $\K$ then imparted different masses to the three eigenvectors, corresponding to the three members of a given fermion family. However, from \Eref{eq:elecmasses2} it is apparent that the higher-order corrections to the pseudovacuum coupling themselves depend on particle mass, leading to differential augmentation of the three eigenvectors (mass channels). As these eigenvectors correspond to different vectors in colour space, imparting different relative phases to the gluon couplings between the preon triplets, a change in the relative pseudovacuum couplings for the different eigenvectors will impact the colour mixing process. Conveniently, it proves possible to represent this modification by a correction to the value of $\theta_\ell$.

Recognising that the fermion masses vary between different particle families, it is necessary to specify the family for which the corrected value of $\theta_\ell$ is being evaluated. This is done by replacing $\ell$ with a member of the particle family in question, e.g.~$e$, $\mu$, or $\tau$ for the electron family.

\subsubsection{Preamble\label{sec:preamble}}

To obtain a corrected expression for the electron mass angle at some energy scale $\mc{E}$, first recall that the eigenvectors $\bm{\{}v%
_i|i\in\{1,2,3\}\bm{\}}$ of $\K$~\erefr{eq:v1}{eq:v3} are independent of the value of $\theta_\ell$. For the electron family, matrix $\K$ therefore always admits the decomposition
\begin{equation}
\K = \sum_{i=1}^3 k^{(\ell)}_i v_i%
v_i^\dagger\label{eq:Kdecomp}%
\end{equation}
where
\begin{align}
\label{eq:v1v1}v_1v_1^\dagger&=\frac{1}{3}\bgrid 1&1&1\\1&1&1\\1&1&1\egrid\\
\label{eq:v2v2}v_2v_2^\dagger&=\frac{1}{3}\bgrid 
1&e^{\frac{2\pi\rmi}{3}}&e^{-\frac{2\pi\rmi}{3}}\\
e^{-\frac{2\pi\rmi}{3}}&1&e^{\frac{2\pi\rmi}{3}}\\
e^{\frac{2\pi\rmi}{3}}&e^{-\frac{2\pi\rmi}{3}}&1\egrid\\
\label{eq:v3v3}v_3v_3^\dagger&=\frac{1}{3}\bgrid
1&e^{\frac{-2\pi\rmi}{3}}&e^{\frac{2\pi\rmi}{3}}\\
e^{\frac{2\pi\rmi}{3}}&1&e^{-\frac{2\pi\rmi}{3}}\\
e^{-\frac{2\pi\rmi}{3}}&e^{\frac{2\pi\rmi}{3}}&1\egrid.
\end{align}
Then note that the imaginary components of the off-diagonal matrix elements arise from components~2 and~3 only \erefr{eq:v2v2}{eq:v3v3}, with an increase in component~2 (corresponding, for charged leptons, to the muon) making a positive-signed contribution to the imaginary component in $[\K]_{12}$, $[\K]_{23}$, and $[\K]_{31}$. Examining the vectors on the complex plane associated with $\exp(\theta_e)=\exp(-3\pi\rmi/4)$ and a small correction proportional to $\exp(2\pi\rmi/3)$, it is seen that the action of this correction
is to
increase the magnitude of
$\theta_\ell$ (making it more negative). Similarly, an isolated increase in component~3 (the tau) decreases the magnitude of $\theta_\ell$. Finally, an isolated increase in component~1 (the electron) affects only the real component of an entry such as $[\K]_{12}$ and thus when $[\K]_{12}$ is complex, once again affects the value of $\theta_\ell$.
When $\theta_\ell$ is no longer fixed to be $-3\pi/4$, it is labelled by the relevant particle family, e.g.~$\theta_e$, and it becomes necessary to specify $\K$ as $\K(\theta_\ell)$ for some family $\ell$.

Next, recognise that it is convenient to write the mass interaction as a leading-order term derived from the background photon field (and associated with colour mixing matrix $\K$ as derived %
\sref{sec:leptonKmatrix}) plus corrections. For the charged lepton masses, and expanding $c_{\alpha/N_0}$ as per \Eref{eq:calphaN0v4}, this corresponds to rewriting \Eref{eq:elecmasses2} as
\begin{align}
m_{e_i}^2 
=\,&{\frac{f^2}{2}\left[k^{(e)}_i\right]^4}{\omega_0}^2{N_0}^8S_{18,147}%
\,\left[1+\Delta_{e}(m_{e_i})\right]\left[1+\mc{O}_{e}(m_{e_i})\right]\\
=\,&\left[m_{e_i}^{(0)}\right]^2\left[1+\Delta_{e}(m_{e_i})\right]\left[1+\mc{O}_{e}(m_{e_i})\right]\label{eq:mDelta}
\end{align}
\begin{align}
\begin{split}
\label{eq:Delta}\Delta_{e}(m_{e_i})%
=\,&%
\frac{8\alpha}{3N_0(3\alpha+2\pi)}%
\left[1+\frac{(10\pi+180\alpha)m_{e_i}^2}{3\pi(m_c^*)^2}%
+\frac{(5-4f_Z)\alpha m_{e_i}^2}{4\pi m_\tW^2}\right]%
+\frac{90\alpha m_{e_i}^2}{\pi(m_c^*)^2}\\
&%
+\frac{(5-4f_Z)\alpha m_{e_i}^2}{2\pi m_\tW^2}%
+\frac{5m_{e_i}^2}{(m_c^*)^2}\left[1%
+\frac{90\alpha m_{e_i}^2}{\pi(m_c^*)^2}+\frac{(25-12f_Z)\alpha m_{e_i}^2}{6\pi m_\tW^2}\right]\\&
+\frac{40m_{e_i}^2}{3m_\bmh^2\left[k^{(e)}_{1}N_0\right]^4}
\end{split}\\
\begin{split}
\mc{O}_e(m_{e_i})
=\,&\OO{\frac{\alpha}{{N_0}^2}}+\OO{\frac{\alpha^2}{{N_0}}}+\OOO{\frac{\alpha m_{e_i}^4}{N_0(m_c^*)^4}}+\OOO{\frac{\alpha^2 m_{e_i}^2}{(m_c^*)^2}}\label{eq:Oe}\\
&+\OOOOx{\frac{\alpha m_{e_i}^2}{m_\bmh^2\left[k^{(e)}_1{N_0}\right]^4}}+\OOOOx{\frac{m_{e_i}^2}{m_\bmh^2\left[k^{(e)}_1\right]^4{N_0}^5}}+\OOOOx{\frac{m_{e_i}^4}{m_\bmh^4\left[k^{(e)}_1{N_0}\right]^4}}
\end{split}
\end{align}
where $\{e_1,e_2,e_3\}=\{e,\mu,\tau\}$ and the subscript $e$ on $\Delta_e$ indicates the family $\{e,\mu,\tau\}$ rather than the electron in particular.
As can be readily seen from \Eref{eq:Delta}, the amplitudes of these corrections to particle mass are dependent upon the squared lepton masses $m_{e_i}^2$, with the largest corrections being those associated with the tau. The charged lepton mixing matrix $\Ke$ is implicitly dependent on the same corrections, as changes to the particle mass ratios affect colour mixing, and hence the co-ordinate transformation required to restore colour neutrality. 
Allowing the corrections to augment mass and also to adjust $\Ke$ is not double-counting, %
as the mass correction \emph{induces} the corresponding adjustment in the co-ordinate transformation implemented by $\Ke$. The co-ordinate transformation does, however, have an effect on the magnitude of the mass correction term, and this effect must be compensated for---see \sref{sec:rescaling}.

Note that $\Delta_e$ and $\mc{O}_e$ are also dependent on $k^{(e)}_i$ which in turn is dependent on $\Ke$ which is dependent on $\theta_e$ which is dependent on $\Delta_e$ which is dependent on $m_{e_i}$. No attempt is made to express full dependencies of parameters save through the equations defining them, and going forward, notation such as $\Delta_e(m_{e_i})$ or $\Delta_e(\mc{E})$ is merely a heuristic to remind the reader of a parameter's dependency on some stated energy scale. Similarly, for $\Ke$, dependency on $\theta_e$ is shown to clearly distinguish the mass-dependent $\Ke$ from the mass-independent $\K$.

\subsubsection{First-order correction to $\K^4$ from the tau channel\label{sec:1storderK}}

To understand how these corrections to lepton masses affect the matrices $\K$, consider the leading order term $\bigl[m_{e_i}^{(0)}\bigr]^2$ which arises from the set of nine diagrams described in the caption of \fref{fig:ActionOfK}, and may be thought of as the action of an operator $\left[\hat{m}^{(0)}\right]^2$ on the preon triplet comprising a sum over nine terms. The action of this operator on the colour space of the preon triplet breaks colour neutrality, but this is then corrected by application term-by-term of four matrices (or operators) $\K$, heuristically $\K^4$ (with implicit action of each operator on the appropriate Hilbert space as per the appropriate diagram of \fref{fig:ActionOfK}), such that $\K^4\left[\hat{m}^{(0)}\right]^2$ as a whole leaves the colour of the preon triplet unchanged up to a sum over cycles $r\rightarrow g\rightarrow b\rightarrow r$.

As seen in \Eref{eq:mDelta}, higher-order corrections enhance the action of $\left[\hat{m}^{(0)}\right]^2$ on eigenstate $e_i$ by a factor of $\left[1+\Delta_{e}(m_{e_i})\right]$. If colour neutrality is to be conserved, then for colour-changing diagrams [including those which would mix colours under some global $\SU{3}_C$ transformation] %
there must be an equivalent enhancement of the matrices $\K$, 
\begin{equation}
\K\rightarrow\Ke
\end{equation}
[where dependence on $\theta_e(\bm{\{}m_{e_i}|i\in\{1,2,3\}\bm{\}})$ is assumed but will subsequently be shown] such that the action of $\Kee{4}$ on lepton $e_i$ is equivalent to the action of 
\begin{equation}
\K^4\left[1+\Delta_{e}(m_{e_i})\right].\label{eq:K4Delta}
\end{equation}

\begin{figure}
\includegraphics[width=\linewidth]{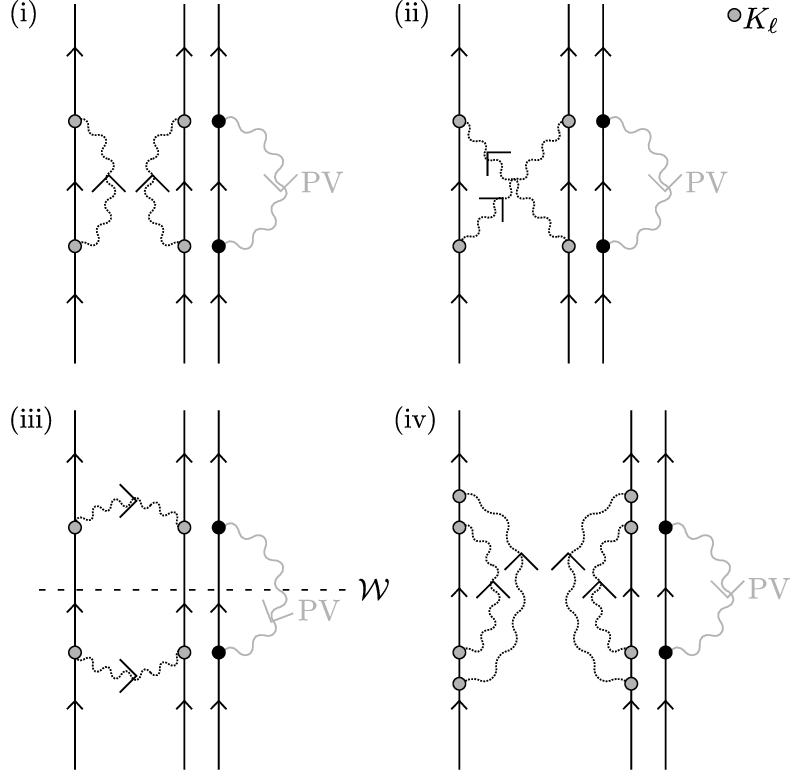}
\caption{When the matrices \pt{$\K$} are replaced by \pt{$\Ke$}, which depends on energy scale, this induces a pair of gauge bosons which may in principle connect the four matrices \pt{$\Ke$} in any of three ways [shown for \pfref{fig:ActionOfK}(i) in diagrams~(i)-(iii)]. In diagram~(i), the bosons each connect to two matrices \pt{$\Ke$} on the same preon and thus enact a co-ordinate transformation followed by its inverse. In diagram~(ii), where the bosons are crossed, there may be a permutation of the preon colour labels but overall colour neutrality is nevertheless conserved. Diagram~(iii) contrasts with the other two in that the boson trajectories are orthogonal to the preon trajectories. For a worldsheet such as that shown (labelled \pt{$\mc{W}$}), which is equidistant between the boson vertices and isochronous in the rest frame of the particle, the total number of oriented crossings for each of the bosons in diagram~(iii) is zero regardless of the trajectory followed. This implies that these bosons do not describe a co-ordinate transformation extant at this worldsheet, and thus this diagram does not contribute to \pt{$\Ke$}. Diagram~(iv) is an example second-order diagram having four bosons arising from the co-ordinate transformation and spanning from the lower to the upper collection of matrices \pt{$\Ke$}. This diagram represents the consecutive application of two (possibly different) transformations in \pt{$\SU{3}_C$} to each preon. 
\label{fig:ActionOfKcorr}}
\end{figure}

As the largest such correction arises from the most massive particle, begin by considering $e_3$, the tau.
First note that introduction of a mass dependency for the matrices $\Ke$ implies that the bosons associated with the colour-neutrality-preserving co-ordinate transformation are no longer parameterless, and may therefore carry momentum in the higher-order diagrams. They must therefore be represented explicitly as per \fref{fig:ActionOfKcorr}.
By construction these bosons only interact at the boundary of the transformed co-ordinate patch, and are consequently both foreground and massless. Let the diagrams of \fref{fig:ActionOfKcorr}(i)-(iii), having two co-ordinate transformation bosons, be termed ``first order'', and for now ignore diagrams having four co-ordinate transformation bosons (``second order'') or more.

Regarding the two co-ordinate transformation bosons of \fref{fig:ActionOfKcorr}(i)-(ii), their vertex factors are incorporated into the leading order matrix $\K$. Integration over their degrees of freedom then yields a factor per diagram of $(4\pi)^{-2}$ [not $(2\pi)^{-2}$, as they are not self-dual and thus their loop is less symmetric than that for the photon]. Note that \fref{fig:ActionOfKcorr}(ii) counts as a loop diagram as there are implicit massless gluon-mediated couplings between the preonic constituents of the lepton %
which %
complete closure of the loop.

Recognising that the leading term calculation does not perform any net colour mixing beyond the unmodified, constant factors of $\K^4$, the resulting factor of $2\cdot(4\pi)^{-2}$ does not appear in the leading term calculation. Its effect is seen only on the diagrams in $\Delta_\taue(m_\tau)$, where it %
represents variations in colour mixing relative to $\K^4$, and multiplies any correction to $\K^4$ by a factor of $2\cdot(4\pi)^{-2}$. %

Next, recognise that the components of each matrix $\Ke$ which perform rotations on $\SU{3}_C$ are composed from a weighted superposition of some set of representation matrices $\lambda'_i$ [possibly, but not necessarily the rescaled Gell-Mann matrices $\lambda_i$ in \PErefr{II}{eq:Cbasis1}{eq:Cbasis2}], of which there are eight. (The ninth basis element is associated with the $N$~boson and the trivial representation, but does not change colour and is %
independent of $\theta_\ell$ so is not yet of interest here.) An appropriate choice of basis for the coming calculation is one which permits every off-diagonal element in $\Ke$ to be associated with a single basis element \{for example, in $\K$, entry $[\K]_{12}$ is associated with $(\lambda_1+\lambda_2)/\sqrt{2}$\}. A change to $\theta_e$ may correspond to a change in choice of representation matrices [e.g.~to $\lambda_1\cos\beta+\lambda_2\sin\beta$ for some angle $\beta$] but this ability to construct a single basis element corresponding to a particular entry in $\Ke$ persists for any reasonably small perturbation around $\theta_e=-3\pi/4$.

Now consider a preon in \fref{fig:ActionOfK}(i) or~(ii) which has two matrices $\Ke$ acting on it, and recognise that if the portion of $\Ke$ which performs colour transformations is decomposed into these eight components, then if a given component $\lambda'_i$ acts on the preon in the lower position, then its conjugate must act in the upper position for the foreground preon colour to be left invariant overall.\footnote{Note that if one diagram family as per \pfref{fig:ActionOfK} can be made to have no net effect on colour, then all paired background bosons have no net effect on colour. Further, as unpaired bosons (as well as being exceedingly rare) have on average no net effect on colour, these may also be grouped into clusters having zero net colour effect overall, and an analogous co-ordinate transformation performed such that their effect on colour vanishes instant-by-instant as well as on average, with the associated bosons vanishing over distances large compared with \pt{$\mc{L}_0$}. Thus no error is made in requiring that colour is left unchanged on a per-diagram basis. Also note that once the mechanism for corrections to \pt{$\theta_e$} is fully elucidated, the co-ordinate transformations arising from the unpaired bosons yield no systematic effect on \pt{$\theta_e$} and thus their effect may be ignored over macroscopic scales.} The action of the \emph{pair} of matrices $\Ke$ on this boson consequently admits decomposition into eight channels, enumerated by the family of orthogonal representation matrices $\lambda'_i$ appearing at the lower matrix $\Ke$.

Next, consider the remaining two matrices $\Ke$ when these are not on the same preon. In this situation %
the representation matrices present %
will be conjugate up to a colour cycle $r\rightarrow g\rightarrow b\rightarrow r$ or its inverse. As the only colour-changing bosons in the pseudovacuum are gluons, and $\SU{3}_C$ symmetry is preserved on the gluon sector and in the definition of the charged lepton, evaluation of any diagram where the entries from $\Ke$ are offset by a colour cycle is necessarily equivalent to evaluation of a diagram where they are not offset. It is therefore acceptable to evaluate entries in $\Ke$ under the assumption that all matrices $\Ke$ appear in conjugate pairs.
In addition, in all diagrams all representation matrices are multiplied by their conjugate (whether independently or in conjunction with the representation matrices associated with the bosonic vertices) and thus the terms arising from %
all 64 channels contributing to the actions of the matrices $\Ke$ on the preon triplet %
are additive on the same (net trivial) charge sector, and therefore summed. %
By $\SU{3}_C$ symmetry, 
each channel will contribute equally to the overall correction to the leading order diagram. %

Now consider the single channel associated with a specific off-diagonal entry in at least one lower matrix $\Ke$. For definiteness, let this be $\Keij{12}$. As there are, overall, 64 contributing channels, each channel must produce a net correction of $\Delta_\taue(m_\tau)/64$. However, the synthetic bosons multiply all correction diagrams by a factor of $2\cdot(4\pi)^{-2}$ and thus $\Keij{12}$ and $\Keij{21}$ must satisfy
\begin{align}
\begin{split}
\Keij{12}^2\Keij{21}^2& = \left[\K\right]_{12}^2\left[\K\right]_{21}^2\left[1+\frac{(4\pi)^2}{2\cdot 64} \Delta_\taue(m_\tau)\right]\\
& = \left[\K\right]_{12}^2\left[\K\right]_{21}^2\left[1+\frac{\pi^2}{8}\Delta_\taue(m_\tau)\right]
\end{split}
\end{align}
\begin{align}
\Rightarrow\Keij{12}\Keij{21} &= \left[\K\right]_{12}\left[\K\right]_{21}\sqrt{1+\frac{\pi^2}{8}\Delta_\taue(m_\tau)}.
\end{align}
Noting that $\left[\K\right]_{12}=\left[\K\right]_{21}^\dagger$, it is convenient to write
\begin{align}
\delta_e(n) &= \sqrt{1+\frac{\pi^2n}{8}}-1\\
[1+\delta_e(n)] &= [1+\rmi\sqrt{\delta_e(n)}][1-\rmi\sqrt{\delta_e(n)}]
\end{align}
and assign the corresponding off-diagonal entries of $\Ke$ to be
\begin{align}
\begin{split}
\Keij{12} &= \left[\K\right]_{12}[1\pm\rmi\sqrt{\delta_e[\Delta_\taue(m_\tau)]}]\\
\Keij{21} &= \left[\K\right]_{21}[1\mp\rmi\sqrt{\delta_e[\Delta_\taue(m_\tau)]}],
\end{split}\label{eq:sqrtdeltacorrs}
\end{align}
preserving the hermeticity of $\Ke$. 

By the factor of $\pm\rmi$ on the correction term, this correction is orthogonal to the leading-order value of $[\K]_{12}$. As this orthogonality is independent of the value of $\theta_e$, %
this implies that for a non-infinitesimal correction, $[\K]_{12}\sqrt{\delta_e[\Delta_\taue(m_\tau)]}$ is the length of an arc. 
The corresponding correction to $\theta_\ell$ is %
\begin{equation}
\theta_\ell\rightarrow\theta_\ell\pm\sqrt{%
\delta_e[\Delta_\taue(m_\tau)]}.
\end{equation}
In the vicinity of $\theta_\ell=-3\pi/4$, this may be rewritten
\begin{equation}
\theta_\ell\rightarrow\theta_\ell\left\{1\mp\frac{4\sqrt{%
\delta_e[\Delta_\taue(m_\tau)]}}{3\pi}%
\right\}.
\end{equation}
As per the discussion under \Erefr{eq:v1v1}{eq:v3v3}, the tau correction is known to decrease the magnitude of $\theta_\ell$, giving
\begin{align}
\Keij{12} &= \frac{e^{\rmi\theta_e}}{\sqrt{2}}=\left[\K\right]_{12}\left\{1+\rmi\sqrt{%
\delta_e[\Delta_\taue(m_\tau)]}+\ldots
\right\}
\label{eq:Keij12}
\end{align}
\begin{align}
\theta_e &= -\frac{3\pi}{4}\left\{1-\frac{4\sqrt{%
\delta_e[\Delta_\taue(m_\tau)]}}{3\pi}%
\right\}\label{eq:thetae}
\end{align}
where the uncorrected value of $\theta_\ell$ has been written explicitly as $-\frac{3\pi}{4}$.

\subsubsection{Second-order correction to $\K^4$ from the tau channel\label{sec:2ndorderK}}

The first-order correction to $\K^4$ is applied to all diagrams which modify the colour mixing process, i.e.~all diagrams contributing to $\Delta_\taue$. For these diagrams, this correction is equivalent by construction to enacting the transformation
\begin{equation}
\K^4\rightarrow\K^4\left[1+\Delta_\taue(m_\tau)\right],
\end{equation}
resulting in a relative increase in these diagrams' contribution to particle mass equivalent to 
\begin{equation}
\Delta_\taue(m_\tau)\rightarrow\Delta_\taue(m_\tau)[1+\Delta_\taue(m_\tau)].
\end{equation}
Any species-dependent increase in mass affects colour mixing in precisely the way described above for correction $\Delta_\taue(m_\tau)$, regardless of whether this increase is diagrammatic (as in the first-order component) or gauge-dependent and derived from a change to $\theta_e$ (as here).
This correction therefore attracts a further smaller correction to $\K$, enhancing %
the first-order correction calculated above.
While this series may be continued indefinitely, it is convenient to truncate at second order for a precision of $\ILOO{\Delta_\taue^2(m_\tau)}$, ensuring that the error in numerical calculations is dominated by the error in $\Delta_\taue(m_\tau)$ itself, and not in the evaluation of $\delta_\taue[\Delta_\taue(m_\tau)]$.

To implement the corresponding adjustment to $\Ke$, let the $\ILOO{\Delta_\taue^2(m_\tau)}$ term be compensated by the second-order diagrams of which \fref{fig:ActionOfKcorr}(iv) is a prototype. In these diagrams, a first-order correction [\fref{fig:ActionOfKcorr}(i)-(ii)] is supplemented by a further, smaller correction.

For \freft{fig:ActionOfKcorr}(i)-(ii) there is a symmetry factor of two corresponding to crossing or not crossing the bosons. For \fref{fig:ActionOfKcorr}(iv),
\begin{itemize}
\item the inner pair may be crossed (factor of two),
\item the outer pair may be crossed (factor of two),
\item the inner and outer bosons on the left may be exchanged (factor of two), and
\item the inner and outer bosons on the right may be exchanged (factor of two). %
\end{itemize}
The factor of $2\cdot (4\pi)^{-2}$ associated with the set of diagrams \freft{fig:ActionOfKcorr}(i)-(ii) becomes a factor of $16\cdot (4\pi)^{-4}$ for the set of diagrams derived from \fref{fig:ActionOfKcorr}(iv). The expression for $\Keij{12}^2\Keij{21}^2$ then becomes
\begin{align}
\begin{split}
&\Keij{12}^2\Keij{21}^2\\
& = \left[\K\right]_{12}^2\left[\K\right]_{21}^2\left[1+\frac{(4\pi)^2}{2\cdot 64} \Delta_\taue(m_\tau)+\frac{(4\pi)^4}{16\cdot 64^2}\Delta_\taue^2(m_\tau)\right]\\
& = \left[\K\right]_{12}^2\left[\K\right]_{21}^2\left\{1+\frac{\pi^2}{8}\Delta_\taue(m_\tau)\left[1+\frac{\pi^2}{32}\Delta_\taue(m_\tau)\right]\right\}
\end{split}
\end{align}
for a redefinition of $\delta_e(n)$ in \Eref{eq:thetae} to
\begin{align}
\delta_e(n) &= \sqrt{1+\frac{\pi^2n}{8}\left(1+\frac{\pi^2n}{32}\right)}-1.\label{eq:delta2nddeg}
\end{align}
The next order term extends $\delta_e(n)$ to
\begin{equation}
\delta_e(n) = \sqrt{1+\frac{\pi^2n}{8}\left[1+\frac{\pi^2n}{32}\left(1+\frac{\pi^2n}{72}\right)\right]}-1\label{eq:delta3rddeg}
\end{equation}
but its effects are verified to fall below the threshold of relevance for the present paper when evaluated in \sref{sec:errors}. 

\subsubsection{Reparameterisation of \prm{\theta_e}\label{sec:rescaling}}

The above transformations have maintained colour neutrality, but at the expense of introducing an additional multiplicative factor of $[1+\Delta_\taue(m_\tau)]$ on all diagrams which affect the colour mixing process [including all diagrams in $\Delta_\taue(m_\tau)$, by virtue of matrices $\K$], as per \Eref{eq:K4Delta}. In effect this is equivalent to replacing any instance of $\Delta_\taue(m_\tau)$ with $\Delta_\taue(m_\tau)\cdot[1+\Delta_\taue(m_\tau)]$. This replacement, in turn, induces further corrections to $\K$, which then induces further corrections to the effective value of $\Delta_\taue(m_\tau)$, and so forth. The largest correction to $\Delta_\taue(m_\tau)$ arising from this effect is of order $m_{\tau}^4/(m_c^*)^4$, which is large compared with $\mc{O}_e(m_{\tau})$. It is therefore necessary to evaluate how this series in $\Delta_\taue(m_\tau)$ affects $m_\tau^2$, but this will be done indirectly. %

First, note that convergence of the series in $\Delta_\taue(m_\tau)$ yields a unique value of $m^2_\tau$ with associated unique values of $\theta_e(\Delta_\taue)$ and $\Delta_\taue(m_\tau)$, and is identified by consistency of equations in $\Delta_\taue(m_\tau)$~\eref{eq:Delta}, $\theta_e(\Delta_\taue)$~\eref{eq:thetae}, and $k^{(e)}_3$~\eref{eq:kell} [which is now a function of $\theta_e(\Delta_\taue)$]:
\begin{align}
\label{eq:mtausq1}m_{\tau}^2%
=\,&{\frac{f^2}{2}\left[k^{(e)}_3(\theta_e) \right]^4}{\omega_0}^2{N_0}^8S_{18,147}%
\left[1+\Delta_{e}(m_{\tau})+\ldots\right]\left[1+\mc{O}_{e}(m_{\tau})\right]\\
\begin{split}
\Delta_{e}(m_\tau)%
=\,&
\frac{8\alpha}{3N_0(3\alpha+2\pi)}%
\,\left[1+\frac{(10\pi+180\alpha)m_{\tau}^2}{3\pi(m_c^*)^2}%
+\frac{(5-4f_Z)\alpha m_{\tau}^2}{4\pi m_\tW^2}\right]+\frac{90\alpha m_{\tau}^2}{\pi(m_c^*)^2}\\
&%
+\frac{(5-4f_Z)\alpha m_{\tau}^2}{2\pi m_\tW^2}%
+\frac{5m_{\tau}^2}{(m_c^*)^2}\left[1+\frac{90\alpha m_{\tau}^2}{\pi(m_c^*)^2}+\frac{(25-12f_Z)\alpha m_{\tau}^2}{6\pi m_\tW^2}\right]\\
&+\frac{40m_{\tau}^2}{3m_\bmh^2\left[k^{(e)}_{1}N_0\right]^4}
\end{split}\\
\theta_e%
=\,& -\frac{3\pi}{4}\left\{1-\frac{4\sqrt{%
\delta_e[\Delta_\taue(m_\tau)]}}{3\pi}%
\right\}\tagref{eq:thetae}\\
k^{(e)}_n%
=\,& 1+\sqrt{2}\cos{\left[\theta_e-\frac{2\pi(n-1)}{3}\right]}.\label{eq:kenew}
\end{align}

Given the value of $m_\tau$ and thus $\Delta_\taue(m_\tau)$ for which these equations are mutually consistent, now reparameterise by writing
\begin{equation}
\Delta_\taue(m_{\tau}) = \Delta'_\taue(m_{\tau})[1-\Delta'_\taue(m_{\tau})]\label{eq:newDelta}
\end{equation}
such that prior to iterating, the tau mass equation is
\begin{align}
\begin{split}
m_{\tau}^2 
=\,&{\frac{f^2}{2}\left[k^{(e)}_i(\theta_e) \right]^4}{\omega_0}^2{N_0}^8S_{18,147}\\
&\times\left\{1+\Delta'_\taue(m_{\tau})[1-\Delta'_\taue(m_{\tau})]\right\}\left[1+\mc{O}_{e}(m_{\tau})\right].
\end{split}
\end{align}
On implementing the associated correction to $\theta_e$ as before, all occurrences of $\Delta_\taue(m_\tau)$ are incremented to $\Delta_\taue(m_\tau)[1+\Delta_\taue(m_\tau)]$ and thus
the factor of 
\begin{equation}
\{1+\Delta'_\taue(m_{\tau})[1-\Delta'_\taue(m_{\tau})]\}\label{eq:Dtprimefac}
\end{equation}
in $m_\tau^2$ becomes
\begin{align}
&1+\Delta'_\taue(m_{\tau})[1-\Delta'_\taue(m_{\tau})]\{1+\Delta'_\taue(m_{\tau})[1-\Delta'_\taue(m_{\tau})]\}\nn\\
&%
=1+\Delta'_\taue(m_{\tau})+\ILOOO{[\Delta'_\taue(m_{\tau})]^3}.
\end{align}
Under this reparameterisation there is no $\ILOOO{[\Delta'_\taue(m_{\tau})]^2}$ term in $m_\tau^2$.

But what, then, does $\theta_e(\Delta_\taue)$ look like when written in terms of $\Delta'_\taue$?
Observe that $\theta_e(\Delta'_\taue)$ yields a correction to $\Delta_\taue$ in %
\Eref{eq:mtausq1} of 
\begin{equation}
\{1+\Delta'_\taue(m_{\tau})[1-\Delta'_\taue(m_{\tau})]\}\tag{\ref{eq:Dtprimefac}}
\end{equation}
in place of
$[1+\Delta_\taue(m_\tau)]$, consistent with \Eref{eq:newDelta}.
However, this correction to \Eref{eq:mtausq1} 
arises from the application of four $\Ke$ matrices as per \fref{fig:ActionOfK}, which may be thought of as two channels each containing two $\Ke$ matrices. In \fref{fig:ActionOfK} this may nominally be identified as one channel per participating preon, though in practice an arbitrary basis may be chosen across these channels and this anyway undergoes mixing during propagation due to the preon binding couplings (not shown). Indeed, because of this mixing, \fref{fig:ActionOfK} may likewise be considered to contain two such channels. In each case, the correction must be factorised across these two channels.

Within each of these channels are a further nine subchannels corresponding to the nine entries in $\Ke$ identified with the nine basis elements of $\GLTR_C$. In contrast, the calculation of $\theta_e$ is performed on a single channel of $\SU{3}_C$, and thus on a single channel of $\GLTR_C$. It therefore follows that the calculation of $\theta_e$ must be performed with respect to %
only one ninth of the correction to $\Delta'_\taue(m_\tau)$, so as to deliver one ninth of the channel's total correction in each subchannel. 

The net effect of both of these considerations is to define
\begin{align}
\theta_e(\Delta'_\taue) = -\frac{3\pi}{4}\bm{\Biggr(}1&-\frac{4\sqrt{\delta_e\{r[\Delta'_\taue(m_\tau)]\}}}{3\pi}
\bm{\Biggr)}
\label{eq:newtheta}\\
r(n) &= n\cdot \sqrt{1-\frac{n}{9}\;}. %
\label{eq:rscaler}
\end{align}
which, taken in conjunction with \Eref{eq:mtausq1} without requiring any higher-order terms,
yields a particle mass accurate to a precision of $\ILOOO{[\Delta_e(m_\tau)]^2}$.

Dropping the primes, it follows that if $\theta_e(\Delta_\taue)$ is redefined as per \Eref{eq:newtheta} and \emph{no} corrective terms are introduced on \Eref{eq:mtausq1}, the value of $m_\tau^2$ thus obtained will be consistent to $\ILOOO{[\Delta_e(m_\tau)]^2}$ with those obtained from Eqs.~(\ref{eq:thetae}) and \erefr{eq:mtausq1}{eq:kenew} on iteration as described earlier. 
Recognising that the leading order term of $c_{\alpha/N_0}$ is of $\ILO{\alpha/N_0}$, 
the next order of corrections scales as
\begin{equation}
\left\{\OO{\frac{\alpha}{N_0}}+\OOO{\frac{m_{e_i}^2}{(m_c^*)^2}}\right\}^3.
\end{equation}
All terms in this expansion are dominated by terms already present in $\mc{O}_e$~\eref{eq:Oe}
with the exception of $\OOO{m_{e_i}^6/(m_c^*)^6}$. Incorporating this additional term into $\mc{O}_e(m_{e_i})$ gives
\begin{align}
\begin{split}
\mc{O}_e(m_{e_i})
=\,&\OO{\frac{\alpha}{{N_0}^2}}+\OO{\frac{\alpha^2}{{N_0}}}+\OOO{\frac{\alpha m_{e_i}^4}{N_0(m_c^*)^4}}%
+\OOO{\frac{\alpha^2 m_{e_i}^2}{(m_c^*)^2}}+\OOOOx{\frac{\alpha m_{e_i}^2}{m_\bmh^2\left[k^{(e)}_1{N_0}\right]^4}}%
\\
&+\OOO{\frac{m_{e_i}^6}{(m_c^*)^6}}+\OOOOx{\frac{m_{e_i}^2}{m_\bmh^2\left[k^{(e)}_1\right]^4{N_0}^5}}%
+\OOOOx{\frac{m_{e_i}^4}{m_\bmh^4\left[k^{(e)}_1{N_0}\right]^4}}.\label{eq:Oe2}
\end{split}
\end{align}

\subsubsection{Corrections to $\Kee{4}$ from the muon channel}

In the above it has been assumed that $\Ke$ attracts corrections from the tau channel only, corresponding to projection of $\Ke$ onto its $v_3v_3^\dagger$ component~\eref{eq:v3v3}. However, the term $[1+\Delta_e(m_{e_i})]$ in \Eref{eq:mDelta} corrects all masses $m_{e_i}$ and thus also affects the $v_1v_1^\dagger$ and $v_2v_2^\dagger$ components. 

Also note that a propagating fermion undergoes multiple scatterings off the background fields. These scatterings may impart energy to or take energy from the fermion. They are generally ignored as their average contributions vanish over length scales large compared with $\mc{L}_0$, but at any instant they may impart energy to the fermion in the range $-\frac{1}{2}\mc{E}_\Omega$ to $+\frac{1}{2}\mc{E}_\Omega$. In conjunction with this, background field interactions may also transiently modify the relative phases and amplitudes of the fermion's three constituent preons.

In consequence, although in the low-energy (large probe) limit a fermion occupies a definite eigenstate of the net mass matrix, at any given interaction a fermion with energy small compared with $\frac{1}{2}\mc{E}_\Omega$ (corresponding, close to the isotropy frame, to rest mass small compared to $\frac{1}{2}\mc{E}_\Omega c^{-2}$) will, in general, be arbitrarily off-shell in a quasi-random superposition of eigenstates. Such a fermion therefore attracts corrections from all eigensectors of $\Ke$.

To evaluate the correction arising from the $e_2$ or muon channel, recognise from \Erefr{eq:v2v2}{eq:v3v3} that the effect of this correction is in opposition to the tau correction, and increases the magnitude of $\theta_e$. As its scale is seen from \Eref{eq:Delta} to be small compared with the tau correction, and from the construction of \Eref{eq:sqrtdeltacorrs} it is seen to act in direct opposition to the tau correction, its effects are conveniently represented %
at energy scale $\mc{E}$ by subtracting the muon correction from the tau correction at the level of the interaction diagrams. 
This yields
\begin{align}
\theta_e(\mc{E}) = -\frac{3\pi}{4}\bm{\left(}1-\frac{4\sqrt{%
\delta_e\{r[\Delta_\taue(m_\tau,\mc{E})-\Delta_\taue(m_\mu,\mc{E})]\}}}{3\pi}
\bm{\right)}
\label{eq:thetacorrmutau}
\end{align}
where $\theta_e(\mc{E})$ is the effective value of $\theta_e$ experienced by a particle whose energy is $\mc{E}$ (i.e.~if at rest, a particle with rest mass $\mc{E}c^{-2}$), and henceforth all instances of $\theta_e$ are acknowledged to depend, either explicitly or implicitly, on a particle's energy scale $\mc{E}$.

The correction from the $e_1$ or electron channel is smaller, but not negligible at the level of precision employed in this paper. However, introducing this correction results in an electron rest mass which runs with energy scale, including when the electron is in motion with respect to the isotropy frame. It is desirable that the electron rest mass behave as a fixed reference point which can be calibrated against experimental observation independent of electron velocity in the isotropy frame, allowing it to be consistent with observation and to be taken as one of the input parameters of the model. Therefore, before addressing the electron channel correction to $\Ke$, it is first necessary to revisit the local scaling symmetry discarded in \Psref{III}{sec:gaugechoicewhichavailable}.

\subsubsection{Local scaling symmetry\label{sec:gss}}

As noted above, recognise that regardless of whether the mass is evaluated through the series in $\Delta_\taue(m_e)$, or whether $\theta_e$ is redefined to absorb the $[\Delta_\taue(m_e)]^2$ term in $m_\tau^2$, the value of $\theta_e$ necessarily runs with energy scale, and thus so too do the values of the colour mixing matrix eigenvalues $k^{(e)}_i$, and hence all lepton masses. This is undesirable if the model is to emulate the observable universe, in which electron rest mass is frame-independent. %
This issue with the model may be addressed by revisiting the discussion of local scaling symmetry. %

In \Psref{III}{sec:gaugechoicewhichavailable} it was stated that global scaling symmetry was broken by the introduction of a pseudovacuum with definite energy scale $\mc{E}_0$. However, an alternative perspective now proves more fruitful: On the introduction of any nonvanishing vacuum excitations, let the scaling degree of freedom associated with symmetry subgroup 
\begin{equation}
\bm{1}_A\otimes\bm{1}_C\otimes\mbb{R}^+ \label{eq:scalargaugegroup}
\end{equation}
be gauged such that on average, over regions of some length scale $\ILO{\mc{L}}$, the electron mass is constant. Since the electron mass is a function of the background photon energy scale, this largely corresponds to requiring that the
average energy scale of the pseudovacuum %
($\mc{E}_0$) %
is constant, at least in the presence of first-generation particles close to rest in the isotropy frame of the pseudovacuum. The assumption of maximum entropy of the pseudovacuum then ensures that $\mc{L}=\mc{E}^{-1}$, and that the net spacetime dilation factor is homogeneous and undetectable at probe scales large compared with $\mc{L}$. As in \sref{sec:pushlimits}, while this property is readily seen to hold for $\mc{L}=\mc{L}_0$, the density of FSF inflection points permits extension to smaller $\mc{L}$, $\mc{L}\sim\mc{L}_\Omega$.

Relative to this fixed reference value of $m_e^2$, computation of the lepton mass ratios then corresponds to computation of the ratios of the eigenvalues of the pseudovacuum mass coupling. 
Now that there are multiple energy scales involved, it is advisable to be more careful with notation.
Noting that a lepton may be off-shell (or in motion) and thus its energy scale $\mc{E}$ may not coincide with $m_{e_i}c^2$, therefore write 
\begin{align}
\label{eq:meisq1}m_{e_i}^2(\mc{E})%
=\,&{\frac{f^2}{2}\left[k^{(e)}_i(\mc{E}) \right]^4}{\omega_0}^2{N_0}^8S_{18,147}%
\,\left[1+\Delta_{e}(m_{e_i},\mc{E})\right]\left[1+\mc{O}_{e}(m_{e_i})\right]\\
\begin{split}
\Delta_{e}(m_{e_i},\mc{E})%
=\,&%
\frac{8\alpha}{3N_0(3\alpha+2\pi)}\label{eq:DeltaE}%
\left\{1+\frac{(10\pi+180\alpha)m_{e_i}^2}{3\pi\left[m_c^*(\mc{E})\right]^2}%
+\frac{(5-4f_Z)\alpha m_{e_i}^2}{4\pi m_\tW^2}\right\}+\frac{90\alpha m_{e_i}^2}{\pi\left[m_c^*(\mc{E})\right]^2}\\
&+\frac{(5-4f_Z)\alpha m_{e_i}^2}{2\pi m_\tW^2}%
+\frac{5m_{e_i}^2}{[m_c^*(\mc{E})]^2}\left\{1%
+\frac{90\alpha m_{e_i}^2}{\pi\left[m_c^*(\mc{E})\right]^2}+\frac{(25-12f_Z)\alpha m_{e_i}^2}{6\pi m_\tW^2}\right\}\\
&+\frac{40m_{e_i}^2}{3m_{\bmh}^2\left[k^{(e)}_1(\mc{E}_e)\,{N_0}\right]^4}%
\end{split}\\
[m_c^*(\mc{E})]^2 =\,& m_c^2\left(1-\frac{27}{10}\frac{\mc{E}^2}{m_c^2c^4}\right)%
\label{eq:mc*b}\\
\theta_e(\mc{E}) = -\frac{3\pi}{4}&\bm{\Biggl(}1-\frac{4\sqrt{%
\delta_e\{r[\Delta_\taue(m_\tau,\mc{E})-\Delta_\taue(m_\mu,\mc{E})]\}}}{3\pi}\bm{\Biggr)}
\label{eq:thetaeEpre}\\
k^{(e)}_n(\mc{E})
=\,& 1+\sqrt{2}\cos{\left[\theta_e(\mc{E})-\frac{2\pi(n-1)}{3}\right]}.\label{eq:keE}
\end{align}
where previous occurrences of these parameters in \srefr{sec:1storderK}{sec:rescaling} are noted to implicitly be dependent on energy scale, e.g.~
\begin{equation}
\Delta_e(m_\tau)\longrightarrow\Delta_e(m_{\tau},m_{\tau} c^2) 
\end{equation}
for an on-shell tau close to rest in the isotropy frame. (Note also that when $k^{(e)}_n$ takes a parameter with values of energy, this is the energy of the relevant foreground particle. There is no need for a factor of two; this factor applies only when converting relating the energy scales of the foreground particles and the pseudovacuum as discussed in e.g.~\sref{sec:meaningofEOmega}.) %

Calculation of higher-generation masses then corresponds to identification of energy scales $\mc{E}_{e_i}$ at which \Erefr{eq:meisq1}{eq:keE} are simultaneously consistent for rest mass $m_{e_i}c^2=\mc{E}_{e_i}$, taking into account the correction to $\theta_e(\mc{E})$ discussed in \sref{sec:rescaling} and the running of all parameters with energy. Prior to gauging of the scaling symmetry~\eref{eq:scalargaugegroup}, the electron rest mass satisfies
\begin{equation}
\begin{split}
\label{eq:mesq1}m_{e}^2=\,&{\frac{f^2}{2}\left[k^{(e)}_1(\mc{E}) \right]^4}{\omega_0}^2{N_0}^8S_{18,147}\\
&\times\left[1+\Delta_{e}(m_e,\mc{E})+\ldots\right]\left[1+\mc{O}_{e}(m_{e})\right]
\end{split}
\end{equation}
and runs with energy, but fixing a gauge to hold $m_e^2$ constant then corresponds to a rescaling of $\omega_0$ such that the product $\big[k^{(e)}_1\big]^4{\omega_0}^2$ remains fixed. Denoting the scaling field $\kappa(\mc{E})$, the electron mass becomes
\begin{equation}
\begin{split}
\label{eq:mesq2}m_{e}^2=\,&{\frac{f^2}{2}\left[k^{(e)}_1(\mc{E}) \right]^4}\left[\kappa(\mc{E})\,\omega_0\right]^2{N_0}^8S_{18,147}\\
&\times\left[1+\Delta_{e}(m_e,\mc{E})+\ldots\right]\left[1+\mc{O}_{e}(m_{e})\right].
\end{split}
\end{equation}
which is constant by definition of $\kappa(\mc{E})$. The general lepton mass at energy $\mc{E}$ then becomes
\begin{align}
\label{eq:mesq3}m_{e_i}^2(\mc{E})=\,&{\frac{f^2}{2}\left[k^{(e)}_i(\mc{E}) \right]^4}\left[\kappa(\mc{E})\,\omega_0\right]^2{N_0}^8S_{18,147}\\
&\times\left[1+\Delta_{e}(m_{e_i},\mc{E})+\ldots\right]\left[1+\mc{O}_{e}(m_{e_i})\right].\nn
\end{align}
Changes in energy scale modulate the ratio 
\begin{equation}
\frac{k^{(e)}_i(\mc{E})}{k^{(e)}_1(\mc{E})}\label{eq:kratio}
\end{equation}
and thus modulate $m_{e_i}^2/m_e^2$, while choice of gauge keeps $m_e^2$ unchanged. Valid ratios~\eref{eq:kratio} directly yield consistency of \Erefr{eq:DeltaE}{eq:keE} and~\eref{eq:mesq3}. %
For definiteness, let $\kappa(\mc{E}_e)=1$.

The search for consistent ratios is much simplified by writing
\begin{align}
\frac{m_{e_i}^2}{m_e^2}&=\frac
{\left[k^{(e)}_i(\mc{E}_{e_i})\right]^4\left[1+\Delta_e(m_{e_i},\mc{E}_{e_i})\right]}
{\left[k^{(e)}_1(\mc{E}_{e_i})\right]^4\left[1+\Delta_e(m_e,\mc{E}_{e_i})\right]}\left[1+\mc{O}_e(m_{e_i})\right], %
\label{eq:eieratio}
\end{align}
regarding which, note the following:
\begin{itemize}
\item Identification of a mass eigenstate corresponds to identification of an eigenvalue ratio \emph{at a particular energy scale} which yields consistency of \Erefr{eq:DeltaE}{eq:keE} and~\eref{eq:mesq3}. Thus the energy scale is $\mc{E}_{e_i}$ in both the numerator and the denominator.
\item The electron mass is a constant of the model, with $\kappa(\mc{E}_{e_i})$ offsetting the energy dependency of $k^{(e)}_1(\mc{E}_{e_i})$ and $\Delta_e(m_e,\mc{E}_{e_i})$ in \Eref{eq:mesq3}.
\item Conveniently, on evaluating numerator and denominator at a common energy scale, gauge parameter $\kappa(\mc{E}_{e_i})$ is identical in numerator and denominator and thus does not appear in \Eref{eq:eieratio}.
\end{itemize}

\subsubsection{Corrections to $\Kee{4}$ from the electron channel}

Having addressed frame and energy scale invariance of $m_e^2$, the correction to $\Ke$ from the $e_1$ or electron channel may now be determined. This channel is seen from \Eref{eq:v1v1} to contribute purely to the real part of $\Keij{12}$. For the imaginary part, all positive contributions to the imaginary portion of $\Keij{12}$ arise from the muon channel and all negative contrinutions from the tau channel, enabling the relatively simple form of \Eref{eq:thetacorrmutau}. In contrast, while the electron channel contributes to the real part of $\Keij{12}$ the bulk of the real part arises instead from the muon and tau channels. (Indeed, for $\theta_\ell:=-3\pi/4$ the entirety of the real part arises from the muon and tau channels, though the corrections from the tau and mu sectors already discussed ensure that this particular situation is avoided.)

To evaluate the electron-mass-induced corrections to the electron mass, consider first an electron mass not incorporating these effects,
\begin{equation}
m_e^2=\bigl[m_e^{(0)}\bigr]^2[1+\Delta_e(0,\mc{E})]
\end{equation}
for some $m_e^{(0)}$. Turning on these effects causes
the electron mass to undergo a rescaling 
\begin{equation}
m_e^2\rightarrow m_e^2\left[\frac{1+\Delta_e(m_e,\mc{E})}{1+\Delta_e(0,\mc{E})}\right]
\end{equation}
which may be rewritten as
\begin{align}
m_e^2&=m_e^2\left[1+\underline{\Delta}_e(m_e,\mc{E})\right]\\
\underline{\Delta}_e(m_e,\mc{E})&:=\left[\frac{1+\Delta_e(m_e,\mc{E})}{1+\Delta_e(0,\mc{E})}\right]-1. 
\end{align}
This rescaling of electron mass results in turn in some %
rescaling $(1+\varepsilon)$ of $k_1^{(e)}(\mc{E})$. 
As the real and imaginary components of $\exp(\rmi\theta_\ell)$ are both negative, the increase in the real component of $[v_1v_1^\dagger]_{12}$ associated with this correction will decrease the magnitude of $\theta_\ell$. 
However, by choice of gauge in \sref{sec:gss} the electron mass is one of the fixed input parameters to the model, and consequently %
this rescaling (and its associated effect on $\theta_\ell$) must be offset by a decrease in magnitude of the associated energy scale $\mc{E}_0$ \{and thus of $\bigl[m_e^{(0)}\bigr]^2$\} corresponding to multiplication by $(1+\varepsilon)^{-1}$. 
If this value is pulled out as an independent factor, in the electron mass diagram it cancels with the $(1+\varepsilon)$ arising from the electron mass rescaling, leaving $m_e$ and $k_1^{(e)}(\mc{E})$ unchanged (and eliminating any need to evaluate colour effects from this sector). 
However, for the muon and the tau it may be seen as a rescaling of $k_2^{(e)}(\mc{E})$ or $k_3^{(e)}(\mc{E})$ respectively %
by a factor of $(1+\varepsilon)^{-1}$. 
To elucidate the effect of this scaling factor on $\theta_\ell$, restore colour neutrality by likewise holding the muon and tau masses fixed, corresponding to enacting a transformation
\begin{equation}
k_i^{(e)}(\mc{E})\rightarrow k_i^{(e)}(\mc{E})~(1+\varepsilon)\quad | \quad i\in\{2,3\}.
\end{equation}
The leading colour effect arising from this transformation is equivalent to a rescaling of $k_3^{(e)}(\mc{E})$ (which generates the leading correction to $\theta_\ell$) by $(1+\varepsilon)$ and is therefore associated with an \emph{increase} in the magnitude of $\theta_\ell$. The rescaling of the muon term is by the same factor.
This multiplier is independent of the existing muon and tau corrections, having its origins on the real rather than the imaginary portion of $\Keij{12}$, and therefore must be evaluated as a separate correction to $\theta_\ell:=-3\pi/4$ rather than being conflated into a single term (as was possible for the muon and tau channels by their linearity on the imaginary component). 
In effect this correction acts on both the muon and the tau channel to yield
\begin{align}
\nn\theta_e(\mc{E}) = -\frac{3\pi}{4}&\bm{\Biggl(}1-\frac{4\sqrt{%
\delta_e\{r[\Delta_\taue(m_\tau,\mc{E})-\Delta_\taue(m_\mu,\mc{E})]\}}}{3\pi}\bm{\Biggr)}\\
\times&\bm{\left(}1+\frac{4\sqrt{%
\delta_e\left\{r\left[\underline{\Delta}_\taue(m_e,\mc{E})\right]\right\}}}{3\pi}\bm{\right)}
.
\label{eq:thetacorrmutaue}
\end{align}
This completes corrections to $\theta_e(\mc{E})$ at the level of precision employed in the present paper, with the form of $\theta_e(\mc{E})$ chosen to avoid the need for corrections to \Eref{eq:meisq1} at $\ILOOO{[\Delta_\taue(m_{e_i},\mc{E})]^2}$.

\subsubsection{Species and energy dependence of \prm{\K}-matrix eigenvalues in boson mass interactions\label{sec:bosonkmatrixdependencies}}

\paragraph[Species dependence]{Species dependence:\label{sec:bosonkmatrixspeciesdependency}}
As noted in \sref{sec:leptonKmatrix}, corrections to $\theta_\ell$ are dependent on the mass of the fermion, and the leading order contributions to fermion mass are charge-dependent so vanish for the neutrino. Consequently, even if massive, the neutrino would be anticipated to receive a much smaller correction to the fermion mass angle than does the electron, with consequence that $k^{(\nu)}_1\ll k^{(e)}_1$. When evaluating the generation-1 $W$~boson mass in \sref{sec:Wmass}, the $\K$-matrix interactions may be placed either slightly to the electron side or slightly to the neutrino side of the $W\nu_e\bar{e}_L$ vertex, corresponding to the location of the implicit background photon and gluon couplings.
A sum over these possibilities is overwhelmingly dominated by contributions where $\K$ acts on the electron, for eigenvalue $k^{(e)}_1$.

For higher generations, recognise that the $\K$-matrices are co-ordinate transformations on a region of manifold $M$. They are a freedom of the model, but are fixed up to physically irrelevant perturbations by the requirement that the $A$-sector bosons act on fermions as a representation of $\SU{3}_A$.
The $A$-sector components of these transformations may equivalently be represented as vertices of boson field couplings whose associated numerical factor is set to~1. In the boson mass interaction, these induced bosons form additional loop corrections.

Loop corrections are always a sum of terms proportional to the ratios of the masses of two of the participating species, bounded from above by~1. %
In \sref{sec:neutrinos} %
it is shown that all neutrinos in the $\Cw{18}$ analogue model are massless (and \cref{ch:LHCb} %
describes, in passing, how the $W/\nu$ coupling in the $\Cw{18}$ model is modified to distribute neutrinos across generations and eliminate the need for a seesaw mechanism). Consequently there are no contributions to the $\K$-matrix eigenvalues from the neutrino branch of the $W\nu_e\bar{e}_L$ interactions in \fref{fig:evalWNsyms} regardless of boson generation.

Subject to this inference of non-contribution from the neutrino sector, these masses are calculated in \sref{sec:heavyWZH}, %
with the lightest higher-generation boson being the second-generation~$W$ at $16.61320(46)~\TeV$. %
When seeking evidence of this boson, it must be remembered that there is a pseudovacuum energy cutoff of $\frac{1}{2}\mc{E}_\Omega$ in the $\Cw{18}$ model (not $\mc{E}_\Omega$---for a discussion of the factor of~$\frac{1}{2}$ see \cref{ch:CDF2}%
), and this cutoff is seen from \aref{apdx:accessory} to be approximately $3.1~\TeV$. %
This limits the amount of energy which may be borrowed when emulating QFT effects, and thus the threshold for observation of effects attributable to second-generation $W$~bosons in the $\Cw{18}$ model is anticipated to be no less than $13.5~\TeV$. %

\paragraph[Energy dependence]{Energy dependence:\label{sec:bosonkmatrixenergydependency}}
In evaluating the leading-order boson mass interaction of \fref{fig:Wmassbosoncorrsnew}(i), the pseudovacuum is partitioned into a background lepton associated with some energy scale $\mc{E}_\ell$ %
and the remaining pseudovacuum, of energy $\mc{E}_0-\mc{E}_\ell$. Recognising that this partitioning is %
dependent on the random fluctuations of energy distribution within the pseudovacuum, all expressions are averaged over all energies from $-\infty$ to $+\infty$ with appropriate weighting. Under the window approximation this reduces to $[-\frac{1}{2}\mc{E}_0,\frac{1}{2}\mc{E}_0]$, though on relaxation of this approximation as per \sref{sec:pushlimits} it is recognised that the true range of fluctuations is $[-\frac{1}{2}\mc{E}_\Omega,\frac{1}{2}\mc{E}_\Omega]$. The value of $\mc{E}_0$ is evaluated in \aref{apdx:accessory} as approximately $3.6~\GeV$, while $\mc{E}_\Omega$ evaluates to approximately $6.2~\TeV$. %
Certainly for generation one and likely for generation two it is reasonable to ignore the finite nature of this cutoff, so that the usual outcome applies where evaluation of interactions is dominated by the on-shell term. In \Ereft{eq:Wwithk2}, \eref{eq:Zwithk2}, \eref{eq:gluonbaremass}, and~\eref{eq:Hwithk2}%
, $k^{(e)}_i$ is therefore evaluated at $\mc{E}_{e_i}$ for $i\in\{1,2\}$. For generation three, if comparing the tau rest mass with $\frac{1}{2}\mc{E}_0$ it appears possible that cancellation of contributions from $|\mc{E}|<\mc{E}_\tau$ and $|\mc{E}|>\mc{E}_\tau$ could be sufficiently incomplete as to yield a significant correction to the third-generation boson masses. This is, however, unlikely when the pseudovacuum is capable of generating energy fluctuations on scales up to and including $\mc{E}_\Omega$, and
the possibility is not evaluated further here.

\section{Predictions of the $\Cw{18}$ analogue model\label{sec:relationships}}

\subsection{Mass relationships\label{sec:massrelationships}}

As seen in the above exploration of the $\Cw{18}$ analogue model, the increased structure of the $\Cw{18}$ free field model as compared with the Standard Model provides for complex interrelationships between the particle masses. When these relationships are collected together, it transpires that it suffices to take three input parameters from
\begin{equation}
\alpha,~m_e,~m_\mu,~m_\tau,~m_W,~m_Z,~m_\bmh
\end{equation}
with the remaining four parameters (and any other symbols such as $m_\tW$, $m_{W^{\dot cc}}$, etc.) then being derived quantities. It is, therefore, an obvious test of the $\Cw{18}$ model to evaluate these derived quantities. As is seen in \sref{sec:results}, the calculated quantities are in %
excellent agreement with observation. %

\newgeometry{left=2cm,right=2cm,top=2cm}
To obtain these predictions, combine the mass equations for the various particle species developed above, and explicitly expand $S_{6,13}$ and $S_{18,147}$ to yield the relationships
{\small
\begin{align}
\frac{m_{e_i}^2}{m_e^2}&\left.=\frac
{\left[k^{(e)}_i(\mc{E}_{e_i})\right]^4\left[1+\Delta_e(m_{e_i},\mc{E}_{e_i})\right]}
{\left[k^{(e)}_1(\mc{E}_{e_i})\right]^4\left[1+\Delta_e(m_e,\mc{E}_{e_i})\right]}\left[1+\mc{O}_e(m_{e_i},\mc{E}_{e_i})\right]\quad\right|\quad e_i\in\{e,\mu,\tau\}\tagref{eq:eieratio}
\\
\frac{m_\tW^2}{m_e^2}&=18{N_0}^{4}\left(1+\frac{2}{N_0}\right)^{-4}\left(1+\frac{1}{N_0}\right)^{-4}\frac
{\Biggl[1+\left(64+\frac{3}{2\pi}-f_Z\right)\frac{\alpha}{2\pi}\Biggr]\left\{1+\frac{51}{18\left[k^{(e)}_1(\mc{E}_e)\,{N_0}\right]^4}\right\}%
}
{\left[1+\Delta_{e}(m_e,\mc{E}_e)\right]%
}
\left[1+\mc{O}_b+\mc{O}_e(m_e,\mc{E}_e)\right]\label{eq:Weratio}
\\
\frac{m_\tW^2}{m_Z^2}&=\frac
{3\left[1+\left(64+\frac{3}{2\pi}-f_Z\right)\frac{\alpha}{2\pi}\right]\left\{1+\frac{51}{18\left[k^{(e)}_1(\mc{E}_e)\,{N_0}\right]^4}\right\}}
{4\left[1+\left(\frac{401}{12}+\frac{3}{2\pi}\right)\frac{\alpha}{2\pi}\right]\left\{1+\frac{55}{18\left[k^{(e)}_{1}\,(\mc{E}_e)\,{N_0}\right]^4}\right\}}\left(1+\mc{O}_b\right)
\label{eq:WZratio}
\\
\frac{m_\tW^2}{m_{\bmh}^2}&=\frac
{9\left[1+\left(64+\frac{3}{2\pi}-f_Z\right)\frac{\alpha}{2\pi}\right]\left\{1+\frac{51}{18\left[k^{(e)}_1(\mc{E}_e)\,{N_0}\right]^4}\right\}}
{20\left\{\left(1-\frac{2}{3N_0}+\frac{1}{3{N_0}^2}\right)\left[1+\frac{30\alpha}{9\pi}\left(1+\frac{1}{3N_0}\right)\right]+\frac{1}{2\pi}\left[1+\frac{30\alpha}{\pi}\left(1-\frac{1}{3N_0}\right)\right]\right\}\left\{1+\frac{39}{18\left[k^{(e)}_1(\mc{E}_e)\,{N_0}\right]^4}\right\}}
\left(1+\mc{O}_b\right)
\label{eq:WHratio}
\end{align}
\begin{align}
\frac{m_\tW^2}{m_c^2}&=\frac
{1+\frac{51}{18\left[k^{(e)}_1(\mc{E}_e)\,{N_0}\right]^4}}
{1+\frac{131}{18\left[k^{(e)}_1(\mc{E}_e)\,{N_0}\right]^4}}
\left(1+\mc{O}_b\right) %
\qquad\qquad
\frac{m_W^2}{m_\tW^2}=\frac
{1+\frac{19}{18\left[k^{(e)}_1(\mc{E}_e)\,{N_0}\right]^4}}
{1+\frac{51}{18\left[k^{(e)}_1(\mc{E}_e)\,{N_0}\right]^4}}
\left(1+\mc{O}_b\right) %
\\
\frac{m_{W^{\dot cc}}^2}{m_\tW^2}&=\frac
{1+\frac{281}{36\left[k^{(e)}_1(\mc{E}_e)\,{N_0}\right]^4}}
{1+\frac{51}{18\left[k^{(e)}_1(\mc{E}_e)\,{N_0}\right]^4}}
\left(1+\mc{O}_b\right) %
\qquad\qquad
\frac{m_{Z^{\dot cc}}^2}{m_Z^2}=\frac
{1+\frac{491}{36\left[k^{(e)}_1(\mc{E}_e)\,{N_0}\right]^4}}
{1+\frac{144}{18\left[k^{(e)}_1(\mc{E}_e)\,{N_0}\right]^4}}
\left(1+\mc{O}_b\right) %
\\
m_{W^\tc}^2&=m_W^2 + \frac{9}{8}\left(m_{W^{\dot cc}}^2-m_W^2\right)\quad(\ref{eq:Wcmass})%
\quad\qquad
m_{Z^\tc}^2=m_Z^2 + \frac{9}{8}\left(m_{Z^{\dot cc}}^2-m_Z^2\right)\tagref{eq:Zcmass}
\end{align}
\begin{align}
\begin{split}
\Delta_{e}(m_{e_i},\mc{E})=\,&\nn
\frac{8\alpha}{3N_0(3\alpha+2\pi)}\left\{1+\frac{(10\pi+180\alpha)m_{e_i}^2}{3\pi\left[m_c^*(\mc{E})\right]^2}+\frac{(5-4f_Z)\alpha m_{e_i}^2}{4\pi m_\tW^2}\right\}%
+\frac{90\alpha m_{e_i}^2}{\pi\left[m_c^*(\mc{E})\right]^2}+\frac{(5-4f_Z)\alpha m_{e_i}^2}{2\pi m_\tW^2}
\\
&+\frac{5m_{e_i}^2}{[m_c^*(\mc{E})]^2}\left\{1+\frac{90\alpha m_{e_i}^2}{\pi\left[m_c^*(\mc{E})\right]^2}+\frac{(25-12f_Z)\alpha m_{e_i}^2}{6\pi m_\tW^2}\right\}+\frac{40m_{e_i}^2}{3m_{\bmh}^2\left[k^{(e)}_1(\mc{E}_e)\,{N_0}\right]^4}
+\mc{O}_e(m_{e_i},\mc{E})
\end{split}\tagref{eq:DeltaE}
\\
\theta_e(\mc{E}) =\,& -\frac{3\pi}{4}\bm{\Biggl(}1-\frac{4\sqrt{%
\delta_e\{r[\Delta_\taue(m_\tau,\mc{E})-\Delta_\taue(m_\mu,\mc{E})]\}}}{3\pi}\bm{\Biggr)}
\bm{\left(}1+\frac{4}{3\pi}\sqrt{\delta_e\left\{r\left[\frac{1+\Delta_\taue(m_e,\mc{E})}{1+\Delta_\taue(0,\mc{E})}-1\right]\right\}}\bm{\right)}
\label{eq:thetacorrmutaue2}
\\&
f_Z=\frac{1}{3}\left(4-24\frac{m_\tW^2}{m_Z^2}+16\frac{m_\tW^4}{m_Z^4}\right)
\qquad(\ref{eq:fZ2})\qquad\qquad
\left[m_c^*(\mc{E})\right]^2=m_c^2\left(1-\frac{27}{10}\frac{\mc{E}^2}{m_c^2c^4}\right)\tagref{eq:mc*b}
\\&
\mc{E}_\ell = m_\ell c^2
\qquad\qquad\qquad\qquad\qquad\qquad\qquad
k^{(\ell)}_n(\mc{E}) = 1+\sqrt{2}\cos{\left[\theta_\ell(\mc{E})-\frac{2\pi(n-1)}{3}\right]}\label{eq:kell2}
\\&
r(n) = n\cdot \sqrt{1-\frac{n}{9}\;}%
\qquad(\ref{eq:rscaler})\qquad\qquad\qquad
\delta_e(n) = \sqrt{1+\frac{\pi^2n}{8}\left[1+\frac{\pi^2n}{32}\right]+\ILO{n^3}}-1\label{eq:deltawitherr}
\\&
\mc{O}_b=\OOOO{\frac{\alpha}{\pi}\left[k^{(e)}_1(\mc{E}_e)\,{N_0}\right]^{-4}}+\OO{\frac{\alpha^2}{\pi^2}}
\\
\begin{split}
\mc{O}_e(m_{e_i},\mc{E})
=\,&\OO{\frac{\alpha}{\pi{N_0}^2}}+\OO{\frac{\alpha^2}{\pi^2{N_0}}}+\OOO{\frac{\alpha m_{e_i}^4}{\pi N_0[m_c^*(\mc{E})]^4}}+\OOOO{\frac{\alpha^2 m_{e_i}^2}{\pi^2\left[m_c^*(\mc{E})\right]^2}}
+\OOOO{\frac{\alpha m_{e_i}^2}{\pi m_\bmh^2\left[k^{(e)}_1(\mc{E}_e)\,{N_0}\right]^4}}\\&
+\OOOO{\frac{m_{e_i}^6}{\left[m_c^*(\mc{E})\right]^6}}
+\OOOO{\frac{m_{e_i}^2}{m_\bmh^2\left[k^{(e)}_1(\mc{E}_e)\right]^4{N_0}^5}}
+\OOOO{\frac{m_{e_i}^4}{m_\bmh^4\left[k^{(e)}_1(\mc{E}_e)\,{N_0}\right]^4}}.
\end{split}\label{eq:Oe3}
\end{align}
}%
\restoregeometry
Note that $r(n)$ is taken to be exact by definition. The finite length of the series in $n$ is associated with an error in the lepton masses, but
this is incorporated into $\mc{O}_e(m_{e_i},\mc{E}_{e_i})$.
Also note that as in \rcite{aoyama2019}, factors of $\pi^{-1}$ %
are displayed explicitly in the terms of $\mc{O}_b$ and $\mc{O}_e(m_{e_i},\mc{E}_{e_i})$.

\subsection{Results\label{sec:results}}

Taking $m_e$, $m_\mu$, and $\alpha$ as input parameters \cite{workman2022,tiesinga2018}, 
\begin{align}
m_e&=0.51099895000(15)~\MeV/c^2\\
m_\mu&=105.6583755(23)~\MeV/c^2\\
\alpha&=7.2973525693(11)\times 10^{-3},
\end{align}
the relationships of \sref{sec:massrelationships} may be solved numerically %
to yield %
the results given in \tref{tab:VI:results}.
The calculated values obtained for fundamental constants $m_W$, $m_Z$, $m_\bmh$, and $m_\tau$ are all within $0.2\,\sigma_\mrm{exp}$ %
of the experimental results.
For a discussion of the numerical methods used to solve the equations of \sref{sec:massrelationships}, see \aref{apdx:solve}.

For purpose of comparing the scale of terms in $\mc{O}_e(m_{e_i},\mc{E})$, it is noted that the value of $k^{(e)}_1(\mc{E}_e)$ evaluates as
\begin{equation}
k^{(e)}_1(\mc{E}_e)=0.04035007804(41).\label{eq:ke1} %
\end{equation}

For purpose of considering the breakdown scale of the $\Cw{18}$ analogue model, in a gauge in which $\kappa(\mc{E}_e)=1$ it follows from \Ereft{eq:V:EOmega}, \eref{eq:meisq1}, and~\eref{eq:calcf} that
\begin{align}
\frac{1}{2}\mc{E}_\Omega^2\approx&\left.\frac{\N^2{N_0}^2m_e^2}{2\alpha\Bigl[k_1^{(e)}(\mc{E}_e)\Bigr]^4}\quad\right|\quad\N=9.\label{eq:EOmegavalue}
\end{align}
The value of $\mc{E}_\Omega$ is evaluated more precisely in 
\aref{apdx:accessory}, and
gives a UV cutoff of $\pm 3.1~\TeV$. %

For completeness it is also interesting to note the value of $f$, which is also evaluated in \aref{apdx:accessory} and is found to satisfy
\begin{align}
f^2&=\frac{2\alpha}{{N_0}^6S_\alpha(1+a_e)^2}\label{eq:calcf}\\
\Rightarrow~f&=1.670144(95)\times 10^{-8}.\label{eq:valuef} %
\end{align}

\begin{table}
\caption{Calculated values of particle masses in the \pt{$\Cw{18}$} analogue model. Quantity \pt{$m_c$} is the bare gluon mass, which is not observable at the confinement scale but enters some calculations in this chapter as a collective property of momentum transfer in the colour sector. Quantities \prm{m_{W^\tc}} and \prm{m_{Z^\tc}} are the masses of coloured counterparts to the \prm{W} and \prm{Z} boson respectively, while \prm{m_N} is the mass of a dark matter candidate which interacts primarily with the scalar boson and has a gravitational mass \prm{1381.486(37)~m_e}. The notation \prm{\sigma_\mrm{exp}} corresponds to the uncertainty in the observed values of the particle masses.
\label{tab:VI:results}}
~\\
\begin{center}
\begin{tabular}{c|c|c|c} %
Parameter & Calculated value & Observed value & Discrepancy\\
 & ($\GeV/c^2$) & ($\GeV/c^2)$ \cite{workman2022,aad2023} & \\
\hline\hline
$m_\tau$ & $1.776867413(43)$ & $1.77686(12)$ & $\p{<}0.1\,\sigma_\mrm{exp}$\\
$m_W$ & $80.3587(22)$ & $80.360(16)$ & $\p{<}0.1\,\sigma_\mrm{exp}$\\
$m_Z$ & $91.1877(35)$ & $91.1876(21)$ & $<0.1\,\sigma_\mrm{exp}$\\
$m_{\bmh}$ & $125.1261(48)$ & $125.11(11)$ & $0.1\,\sigma_\mrm{exp}$\\
$m_{W^\tc}$ & $80.4434(22)$ & See \cref{ch:CDF2} & See \cref{ch:CDF2}\\
$m_{Z^\tc}$ & $91.1928(35)$ & See \cref{ch:CDF2} & See \cref{ch:CDF2}\\
$m_c$ & 0 or $80.42815(42)^*$ & 0 (theory) & 0\\
$m_N$ & 0 or $80.3810(22)^\dagger$ & -- & --
\end{tabular}
\end{center}
\small{$^*$~{Not directly observable}\\$^\dagger$~Dependent on scale of observation}
\end{table}

\subsection{Sources of numerical error\label{sec:errors}}

The relationships between the fundamental constants presented in \sref{sec:massrelationships} %
incorporate numerous truncations. These are listed in \tref{tab:VI:errorlist}, with their impacts on the numerical results being presented in \tref{tab:VI:erroreffects}.
\begin{table}
\caption{\label{tab:VI:errorlist}List of sources of error in the results of \ptref{tab:VI:results}. Note that coefficients of \prm{\mc{O}_b} are assumed to vary independently in each of the boson masses, and hence in each of the listed mass ratios.}
~\\
\begin{center}
\begin{tabular}{c|l}
Label~ & \multicolumn{1}{c}{Description}\\\hline\hline
\E{1}&~ Term~1 of $\mc{O}_b$ applied to $m_c^2/m_W^2$\\
\E{2}&~ Term~2 of $\mc{O}_b$ applied to $m_c^2/m_W^2$\\
\E{3}&~ Term~1 of $\mc{O}_b$ applied to $m_W^2/m_Z^2$\\
\E{4}&~ Term~2 of $\mc{O}_b$ applied to $m_W^2/m_Z^2$\\
\E{5}&~ Term~1 of $\mc{O}_b$ applied to $m_W^2/m_\bmh^2$\\
\E{6}&~ Term~2 of $\mc{O}_b$ applied to $m_W^2/m_\bmh^2$\\
\E{7}&~ Term~1 of $\mc{O}_b$ applied to $m_W^2/m_e^2$\\
\E{8}&~ Term~2 of $\mc{O}_b$ applied to $m_W^2/m_e^2$\\
\E{9}&~ Term~1 of $\mc{O}_e$ applied to $\Delta_e(m_{e_i},\mc{E})$\\
\E{10}&~ Term~2 of $\mc{O}_e$ applied to $\Delta_e(m_{e_i},\mc{E})$\\
\E{11}&~ Term~3 of $\mc{O}_e$ applied to $\Delta_e(m_{e_i},\mc{E})$\\ %
\E{12}&~ Term~4 of $\mc{O}_e$ applied to $\Delta_e(m_{e_i},\mc{E})$\\
\E{13}&~ Term~5 of $\mc{O}_e$ applied to $\Delta_e(m_{e_i},\mc{E})$\\
\E{14}&~ Term~6 of $\mc{O}_e$ applied to $\Delta_e(m_{e_i},\mc{E})$\\
\E{15}&~ Term~7 of $\mc{O}_e$ applied to $\Delta_e(m_{e_i},\mc{E})$\\
\E{16}&~ Term~8 of $\mc{O}_e$ applied to $\Delta_e(m_{e_i},\mc{E})$\\ %
\E{17}&~ The next term in $\delta_e(n)$\\
\E{18}&~ Uncertainty in $\alpha$\\
\E{19}&~ Uncertainty in $m_e$\\
\E{20}&~ Uncertainty in $m_\mu$
\end{tabular}
\end{center}
\end{table}
\newgeometry{left=2cm,right=2cm,top=2cm}
\begin{table*}[p]
\caption{\label{tab:VI:erroreffects}Contributions of different sources of error to the main results of \ptref{tab:VI:results}, expressed in units of energy, and relative to the experimental error. Labels are enumerated in \ptref{tab:VI:errorlist}. For purposes of summation, sources of error are assumed independent.}
{\small
\begin{center}
\begin{tabular}{c|c|c|c|c|c|c|c|c|c}
Label&Coefficient &\multicolumn{2}{c|}{Uncertainty in}		&\multicolumn{2}{c|}{Uncertainty in}&
	   \multicolumn{2}{c|}{Uncertainty in}	&\multicolumn{2}{c}{Uncertainty in}
	   \\
	 &&\multicolumn{2}{c|}{$\tau$ mass}			&\multicolumn{2}{c|}{$Z$ mass}&
	   \multicolumn{2}{c|}{$W$ mass}	&\multicolumn{2}{c}{$\bmh$ mass}
	   \\\cline{3-10}
	 &&$\eV/c^2$		&$10^{-4}\,\sigma_\mrm{exp}$				&$\MeV/c^2$		&$\sigma_\mrm{exp}$			&
	   $\MeV/c^2$		&$\sigma_\mrm{exp}$				&$\MeV/c^2$		&$10^{-2}\,\sigma_\mrm{exp}$			
	   \\
\hline\hline %
\E{1} & $\pm10$  & 0.1   & 0.01  & 0.29 & 0.14 & 0.26 & 0.02 & 0.40 & 0.37 \\
\E{2} & $\pm10$  & 0.8   & 0.07  & 2.46 & 1.17 & 2.16 & 0.14 & 3.37 & 3.07 \\
\E{3} & $\pm10$  & 0.0   & 0.00  & 0.29 & 0.14 & 0.00 & 0.00 & 0.00 & 0.00 \\
\E{4} & $\pm10$  & 0.1   & 0.00  & 2.46 & 1.17 & 0.00 & 0.00 & 0.00 & 0.00 \\
\E{5} & $\pm10$  & 0.0   & 0.00  & 0.00 & 0.00 & 0.00 & 0.00 & 0.40 & 0.37 \\
\E{6} & $\pm10$  & 0.1   & 0.01  & 0.00 & 0.00 & 0.00 & 0.00 & 3.37 & 3.07 \\
\E{7} & $\pm10$  & 0.0   & 0.00  & 0.00 & 0.00 & 0.00 & 0.00 & 0.00 & 0.00 \\
\E{8} & $\pm10$  & 0.1   & 0.01  & 0.00 & 0.00 & 0.00 & 0.00 & 0.01 & 0.01 \\
\E{9} & $\pm10$  & 1.4   & 0.12  & 0.00 & 0.00 & 0.00 & 0.00 & 0.00 & 0.00 \\
\E{10} & $\pm10$ & 0.6   & 0.05  & 0.00 & 0.00 & 0.00 & 0.00 & 0.00 & 0.00 \\
\E{11} & $\pm10$ & 0.0   & 0.00  & 0.00 & 0.00 & 0.00 & 0.00 & 0.00 & 0.00 \\
\E{12} & $\pm10$ & 24.3  & 2.02  & 0.47 & 0.22 & 0.42 & 0.03 & 0.65 & 0.59 \\
\E{13} & $\pm10$ & 2.4   & 0.20  & 0.02 & 0.01 & 0.02 & 0.00 & 0.03 & 0.03 \\
\E{14} & $\pm10$ & 1.1   & 0.09  & 0.02 & 0.01 & 0.02 & 0.00 & 0.03 & 0.03 \\
\E{15} & $\pm10$ & 5.4   & 0.45  & 0.05 & 0.03 & 0.05 & 0.00 & 0.07 & 0.07 \\
\E{16} & $\pm10$ & 0.2   & 0.02  & 0.00 & 0.00 & 0.00 & 0.00 & 0.00 & 0.00 \\
\E{17} & 1       & 0.0   & 0.00  & 0.00 & 0.00 & 0.00 & 0.00 & 0.00 & 0.00 \\
\E{18} & $\pm1$  & 0.0   & 0.00  & 0.00 & 0.00 & 0.00 & 0.00 & 0.00 & 0.00 \\
\E{19} & $\pm1$  & 0.0   & 0.00  & 0.00 & 0.00 & 0.00 & 0.00 & 0.00 & 0.00 \\
\E{20} & $\pm1$  & 35.3  & 2.94  & 0.00 & 0.00 & 0.00 & 0.00 & 0.00 & 0.00 \\\hline %
Total &          & 43.3  & 3.61  & 3.53 & 1.68 & 2.22 & 0.14 & 4.85 & 4.41 %
\end{tabular}
\end{center}}
\end{table*}
\restoregeometry
In determining the error bars on the results of \tref{tab:VI:results}, all sources of error have been assumed independent. Regarding the coefficients column of \tref{tab:VI:erroreffects}, it would be easy to assume that unevaulated higher-order terms such as those in $\mc{O}_e$ and $\mc{O}_b$ attract coefficients of $\ILO{1}$ or less. However, in CASMIR situations frequently arise where $\ILO{10}$ degenerate channels may reinforce one another (e.g.~equivalent loop correction diagrams due to eight species of coloured $W$ bosons and one colourless $W$ boson). The confidence intervals of \tref{tab:VI:erroreffects} are therefore evaluated for coefficients in the range $\pm10$. While this is a reasonable precaution in the absence of any exploration of these higher-order terms, it may potentially overestimate the uncertainty associated with the CASMIR values of the calculated parameters, artificially reducing tension. As a check on the values calculated, an upper bound on tension between experiment and the CASMIR expressions to present order may be obtained by taking the CASMIR uncertainty to zero. 
On the rare occasion within this monograph that a discernable impact occurs, this is %
no larger than $0.3\,\sigma$, %
and is remarked on in the associated text and/or table caption. 
These changes have minimal consequence for interpretation of the results presented.

\section{Conclusion}

The $\Cw{18}$~analogue model has many elements in common both with the Standard Model of particle physics and with the observable universe. As it has only three tunable parameters, which must be set by reference to physical constants, there exist many opportunities for testing the $\Cw{18}$~analogue model against observation.

In the present chapter, calculation of higher-order terms in the mass relationships of the $\Cw{18}$~model permitted the values of four fundamental constants, the masses of the $W$~boson, $Z$~boson, Higgs boson, and tau particle, to be predicted (retrodicted) with precision comparable to current experimental observation. The results were found to be in agreement with observation within the limits of experimental precision (tensions all $<0.2\,\sigma$). %
Although the conceptual underpinnings of the $\Cw{18}$~analogue model are radically different to those of the Standard Model, and only emulating a quantum field theory on Minkowski (or, in \cref{ch:gravity}, pseudo-Riemannian) space--time in an appropriate regime, this chapter nevertheless demonstrates that the $\Cw{18}$ model may yield numerically appropriate results under appropriate circumstances.

Emergent properties of the $\Cw{18}$~model not shared by the Standard Model include
\begin{itemize}
\item a ``neutral gluon'' $N_\mu$, which couples primarily to the scalar boson, and is a dark matter candidate with a mass of $80.3810(22)~\GeV/c^2$, %
\item coloured counterparts to the $W$ and $Z$ bosons with masses $80.4434(22)~\GeV/c^2$ and $91.1928(35)~\GeV/c^2$ respectively, and %
\item second and third generation massive weak bosons, with the lightest being the second-generation $W$-analogue at $16.61320(46)~\TeV$. %
\end{itemize}

The model also exhibits two further particles, weakly-interacting right-handed weak bosons denoted $G_\mu$ and $G^\dagger_\mu$. However, detection of these particles should not be anticipated as they are eliminated in \cref{ch:gravity}, %
in which the $\Cw{18}$~analogue model is extended to curved space--times and is shown to reproduce gravitational metrics consistent with General Relativity (again, in appropriate regimes).

\appendix

\section{Mass-related loop coefficients\label{apdx:massloops}}

In \sref{sec:Wmassbosonloops} it was observed that the one-photon-loop correction to lepton magnetic moment is associated with a factor of
\begin{equation}
2\alpha\cdot \bmf{\frac{m_\ell^2}{m_A^2}}\cdot \frac{1}{4\pi}\tagref{eq:Aloopcorr}
\end{equation}
while the one-$W$-loop correction is associated with \cite{peskin1995}
\begin{equation}
-\frac{10\alpha}{3}\left[1+\OO{\alpha}\right]\cdot\bmf{\frac{m_\ell^2}{m_W^2}}\cdot\frac{1}{4\pi}\tagref{eq:Wloopcorr}
\end{equation}
where the mass dependency $\bmfcdot$ evaluates as
\begin{equation}
\bmf{\frac{m_\ell^2}{m_b^2}}\longrightarrow\left\{\begin{aligned}1~~~~\quad&\textrm{if }m_b^2=0\\
\frac{m_\ell^2}{4\pi m_b^2}\quad&\textrm{if }m_b^2\gg m_\ell^2.\end{aligned}\right.
\end{equation}
All bosons in these interactions are foreground fields.

In \sref{sec:AcQLcouple} a similar situation is encountered save that the loop bosons are background fields. That is, their values are evaluated as expansions around the mean field value, with terms beyond the first making negligible numerical contribution provided the foreground fields are small compared with $\frac{1}{2}\mc{E}_\Omega$~(see \Psref{I}{sec:pushlimits}). Although not making significant contribution to the overall value of the diagram, these higher-order contributions are conceptually important, representing 4-momentum transferred to and then recovered from the background fields, analogous %
to the foreground loops associated with the corrections to the lepton magnetic moment.

There is, however, a critical difference in the value of $\bmfcdot$ for massive foreground and background fields. 
For foreground fields, emission and reabsorption of  a single particle forms a loop, and integration over this loop yields a factor of $(4\pi)^{-1}$. For background fields, the loop is lost and it is replaced by two randomly oriented interactions with photons from the pseudovacuum. It may be anticipated that some other factor will then replace $(4\pi)^{-1}$. The relative weights of these two distinct scenarios may be evaluated by considering the semiclassical emission and absorption process on $\Cw{18}$ from a geometric perspective.

First examining the massive foreground scenario, if the lepton is stationary prior to emitting the loop particle, and the emitted particle has nonzero spatial momentum, there exists a preferred direction from which to recover the loop particle. %
Barring external interference, preferred absorption is from the same direction that emission was towards. This yields a function in solid angle which approaches a delta function in the classical limit.
Similarly, if the particle emitted is stationary in the rest frame of the lepton (this yielding a trajectory in some sense ``closest to classical'', about which other trajectories form symmetric perturbations), then change frames such that both are in motion. The preferred direction of emission is along the future trajectory of the lepton, and the preferred direction for absorption is from the past trajectory of the lepton, again yielding a delta function in solid angle.

If, however, the two vertices represent interactions with the background field, then provided the foreground momentum superimposed upon this is sufficiently small, the returning boson at the inbound vertex is equally likely to come from any spatial orientation. The delta function is lost, and integration over solid angle yields a relative factor of $4\pi$.

Finally, if the boson is massless, then it is either a photon or a $G^\bdag$ boson. The $G^\bdag$ bosons are neglected in the current chapter and eliminated in \cref{ch:gravity}, therefore assume the boson is a photon. All charged particles in the $\Cw{18}$ model (except for the $G^\bdag$ boson) are massive, and so regardless of whether the photon is foreground or background the same argument applies: The emitting particle is massive and the photon is massless, so there is no comoving choice of spatial momentum. The closest-to-classical trajectory is therefore that in which the loop contracts down to a point. There is no delta function over solid angle.

The corresponding values of $\bmfcdot$ are thus %
\begin{equation}
\bmf{\frac{m_\ell^2}{m_b^2}}\longrightarrow\left\{\begin{aligned}1~~~~\quad&\textrm{if }m_b^2=0\\
\frac{m_\ell^2}{4\pi m_b^2}\quad&\textrm{if }m_b^2\gg m_\ell^2\textrm{ and $b$ is $\fgfield{b}$}\\
\frac{m_\ell^2}{m_b^2}\quad&\textrm{if }m_b^2\gg m_\ell^2\textrm{ and $b$ is $\bgfield{b}$}.\end{aligned}\right.
\end{equation}

It is also worth elaborating on what it means for photon loops to be 
contracted ``down to a point''. In the limit of any loop being sufficiently large, emission and absorption approach being equally likely to be in phase or antiphase, and the long-range contribution of the loop to the vertex correction vanishes. Over distances which are sub-wavelength, however, correlated emission and absorption dominate, yielding a systematic contribution to the loop correction. Recognising that interactions with the background field emulate quantum uncertainty, for a photon with energy below $\frac{1}{2}\mc{E}_\Omega$ an interaction is functionally pointlike if it takes place over a distance of less than half a wavelength. Within the subset of all such contributions, other process considerations (such as the energy scale of the particles involved) will determine which dominate. If the dominant loop energy is small compared with $\frac{1}{2}\mc{E}_0$, then a loop contracted down ``to a point'' still spans a region large compared with the autocorrelation length of the pseudovacuum, $\mc{L}_0$. This is seen to have relevance for the calculation of some symmetry factors for loop corrections in \cref{ch:gravity} and \aref{apdx:symloops}.

\section{Symmetry factors and photon loops\label{apdx:symloops}}

\begin{figure}
\includegraphics[width=\linewidth]{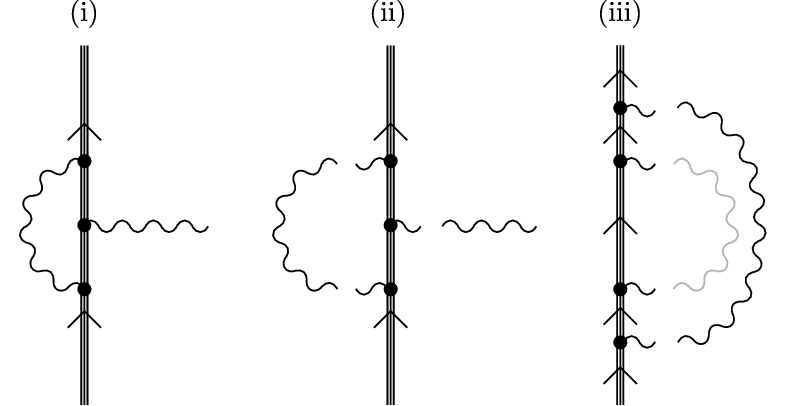}
\caption{Evaluation of symmetry factors of \prm{\ILO{\alpha}}~loop corrections. Background (pseudovacuum) photons are shown in grey. (i)~The one-loop electromagnetic correction to the EM~vertex. (ii)~This diagram may be decomposed as a series of interactions between a propagating electron and a set of photons in its vicinity. The arc of the loop may be understood as a correlated photon pair. (iii)~A similar decomposition of the third diagram of \pfref{fig:photonloops}(iii).\label{fig:loopsyms}}
\end{figure}

When evaluating $\ILO{\alpha}$ loop diagrams such as those shown in \fref{fig:photonloops}, it can be convenient to have a quick method of evaluating symmetry factors relative to the one-loop electromagnetic correction to the EM~vertex of \fref{fig:loopsyms}(i).
To this end, consider diagram~(i) as constituting an electron which, as it propagates, interacts with two correlated photons and one uncorrelated photon. The components of this interaction are shown in diagram~(ii). The first interaction is with a correlated photon, with two to choose from. Next there is the interaction being corrected, which is with the uncorrelated photon, and the final interaction is with the photon correlated with the first photon, choice of one, for a relative factor of two.

This may be contrasted with diagram~(ii) where the loop photon has been replaced by a $W$~boson. The interactions of the loop correction are now with $W$ and $W^\dagger$, so each attracts a factor of one, for an overall relative symmetry factor of one.

Next, consider diagram~(iii). In this diagram, which corresponds to the third diagram of \fref{fig:photonloops}(i), the electron interacts with two pairs of correlated photons. However, these pairs are distinguishable by the characteristics of their correlations---for the background pair, these correlations vanish over length scales large compared with $\mc{L}_0$, whereas for the foreground pair they do not. The diagram being constructed mandates whether each interaction is with a foreground or background boson, and thus the first interaction is with one of the two correlated foreground photons for a symmetry factor of two. This is then followed by the pair of interactions being corrected, and finally by the interaction with the second boson of the correlated foreground pair. %
The symmetry factor associated with the loop is thus two, and like \fref{fig:SMphotonloops}, the diagrams of \fref{fig:photonloops}(i) attract loop correction factors of $\alpha/(2\pi)$ apiece.

Although not encountered in this chapter, an example will be seen in \cref{ch:gravity} in which the loop correction boson of \fref{fig:loopsyms}(iii) is also constrained to exhibit rapidly vanishing contributions over length scales large compared with $\mc{L}_0$. In this context the two correlated pairs of \fref{fig:loopsyms}(iii) become indistinguishable and the first interaction attracts a symmetry factor of four as it may be with either constituent of either pair. The next two interactions comprise the process receiving the loop correction, and the final interaction attracts a symmetry factor of one, for a total symmetry factor of four. Such a loop thus attracts a factor of $4\cdot \alpha/(4\pi) = \alpha/\pi$ rather than the usual $2\cdot\alpha/(4\pi)=\alpha/(2\pi)$. As discussed in \aref{apdx:massloops}, however,
most loop corrections do not exhibit such a constraint, and are dominated by terms which generate the symmetry factor of $2\cdot\alpha/(4\pi)=\alpha/(2\pi)$ seen in the Standard Model.

It is also worth making an observation regarding the sign of loop corrections, and the number of fermion lines present. A loop correction to a single photon emission process, such as \fref{fig:loopsyms}(i), is positive when the loop incorporates an even number of fermion lines. In contrast, the one-photon loop in the PSE expansion of the electron propagator yields a positive correction when the number of fermion lines is odd. On closing the background photon loop to obtain the parent diagram (prior to mean field expansion), the relationship between diagram~(iii) and the PSE is observed. Although diagram~(iii) is not part of the fermion PSE of the Standard Model, as there are no pseudovacuum photons in the Standard Model, this diagram is nevertheless seen to be positive with an odd number of fermion lines.

\section{Solving the particle mass relationships\label{apdx:solve}}

To solve the particle mass relationships of \sref{sec:massrelationships}, proceed as follows:

Take $\alpha$, $m_e$, and $m_\mu$ as input parameters. 
To set up the initial value of $m_\tau$, take the uncorrected angle $\theta_\ell=-3\pi/4$, construct the uncorrected $\K$~matrix as per \Eref{eq:VI:Kmatrix}, and diagonalise to obtain initial values for $k^{(e)}_i$, then write the initial value of $m_\tau$ as
\begin{equation}
[m_\tau]_0 = \frac{\left[k^{(e)}_3\right]^2}{\left[k^{(e)}_2\right]^2}\,m_\mu.
\end{equation}
Let the initial value of $m_\tW$ be arbitrary and large, e.g.~$10^4~\GeV/c^2$.
Let initial values of $m_c$, $m_\bmh$, and $m_Z$ be computed from $m_\tW$ using tree-level calculations. Determine an initial value for $N_0$ using \PEref{V}{eq:N0value}. Determine also an initial value for $\ke{1}{e}$; when assigning this initial value it is convenient to neglect terms in $\theta_e$ involving $\bigl[\ke{1}{e}N_0\bigr]^{-4}$.

Expand \Eref{eq:eieratio} for $i=2$ (muon channel, energy scale $\mc{E}_\mu$) and substitute $m_\tW^2$ for $(m_c^*)^2$ using the expressions for $m_\tW^2/m_c^2$~\eref{eq:WZratio} and $[m_c^*(\mc{E})]^2$~\eref{eq:mc*b}. %
Rewrite as
{\small
\begin{align}
f^\ell_{\tW c}=\,&\frac{1+\frac{51}{18\left[k^{(e)}_1(\mc{E}_e)N_0\right]^4}}
{\left\{1+\frac{131}{18\left[k^{(e)}_1(\mc{E}_e)N_0\right]^4}\right\}\left(1-\frac{27m_\ell^2}{10m_c^2}\right)}\qquad\quad\tilde{f}^\ell_{\tW c}=f^\ell_{\tW c}\left\{1+\frac{\E{1}\,\alpha}{\pi\left[k^{(e)}_1(\mc{E}_e)N_0\right]^4}+\frac{\E{2}\,\alpha^2}{\pi^2}\right\}
\end{align}
\begin{align}
\nn0=\left[m_\tW^4\right]_{n+1}\bm{\Biggl(}&\left\{\left[\ke{1}{\mu}\right]^4m_\mu^2-\left[\ke{2}{\mu}\right]^4m_e^2\right\}\left[1+\frac{8\alpha}{3N_0(3\alpha+2\pi)}+\frac{\E{9}\,\alpha}{\pi{N_0}^2}+\frac{\E{10}\,\alpha^2}{\pi^2N_0}\right] \\
\nn&+\left\{\left[\ke{1}{\mu}\right]^4-\left[\ke{2}{\mu}\right]^4\right\}\Biggl\{\frac{40m_e^2m_\mu^2}{3m_\bmh^2\bigl[\ke{1}{e}N_0\bigr]^4}+\left(\frac{\E{13}\,\alpha}{\pi}+\frac{\E{15}}{N_0}\right)\frac{m_e^2m_\mu^2}{m_\bmh^2\bigl[\ke{1}{e}N_0\bigr]^4}\Biggr\}\\
\nn&+\left\{\left[\ke{1}{\mu}\right]^4m_e^2-\left[\ke{2}{\mu}\right]^4m_\mu^2\right\}\frac{\E{16}\,m_e^2m_\mu^2}{m_\bmh^4\bigl[\ke{1}{e}N_0\bigr]^4}
\bm{\Biggr)}\\
\nn+\left[m_\tW^2\right]_{n+1}\bm{\Biggl(}&\left\{\left[\ke{1}{\mu}\right]^4-\left[\ke{2}{\mu}\right]^4\right\}m_e^2m_\mu^2\\
\nn&\times\Biggl\{\frac{8\alpha}{3N_0(3\alpha+2\pi)}\left[\frac{(10\pi+180\alpha)\tilde{f}^\mu_{\tW c}}{3\pi}\!+\!\frac{(5-4f_Z)\alpha}{4\pi}\right]\!+\! 
\frac{90\alpha\tilde{f}^\mu_{\tW c}}{\pi} \!+\! \frac{(5-4f_Z)\alpha}{2\pi}\!+5\tilde{f}^\mu_{\tW c} \!+\! \frac{\E{12}\,\alpha^2{f}^\mu_{\tW c}}{\pi^2}
\Biggr\}\bm{\Biggr)}
\\
\nn+\p{\left[m_\tW^2\right]_{n+1}}\bm{\Biggl(}&
\left\{\left[\ke{1}{\mu}\right]^4m_e^2-\left[\ke{2}{\mu}\right]^4m_\mu^2\right\}m_e^2m_\mu^2
\left\{5\tilde{f}^\mu_{\tW c}\left[\frac{90\alpha\tilde{f}^\mu_{\tW c}}{\pi}+\frac{(25-12f_Z)\alpha}{6\pi}\right]+\frac{\E{11}\,\alpha(\tilde{f}^\mu_{\tW c})^2}{\pi N_0}\right\}
\\
&+\left\{\left[\ke{1}{\mu}\right]^4m_e^4-\left[\ke{2}{\mu}\right]^4m_\mu^4\right\}\frac{\E{14}\,m_e^2m_\mu^2(\tilde{f}^\mu_{\tW c})^3}{\left[m_\tW^2\right]_{n}} \bm{\Biggr)}
\label{eq:nextW}
\end{align}}
$\!\!\!$In correction $\E{14}$ it is convenient to use the value of the $\tW$~mass from the previous iteration such that the equation as a whole remains quadratic in $[m_\tW^2]_{n+1}$. %
For numerical stability, terms involving $m_\bmh^2$ are only evaluated provided the working value of $m_\bmh^2$ satisfies $m_\bmh^2>m_\mu^2$.
Values on convergence are unaffected by this precaution. 

To update $m_\tW^2$, %
iterate across trial values of $\theta_e$ in the vicinity of $-3\pi/4$. Note that as per \sref{sec:preamble}, the leading correction to $\theta_e$ (from the tau channel) makes it less negative so only values $\theta_e>-3\pi/4$ need be considered. For each value of $\theta_e$, compute the associated eigenvalues of the $K$-matrix at energy scale $\mc{E}_\mu$ using \Eref{eq:kell2}, then solve \Eref{eq:nextW} for an updated value of $m_\tW^2$. Use this to compute associated values of $m_c^2$, $m_\bmh^2$, and $m_Z^2$ using the mass ratios of \Erefr{eq:WZratio}{eq:WHratio}, and $N_0$ using \Eref{eq:Weratio} rearranged to solve for ${N_0}^4$,
\begin{align}
\big[{N_0}^4\big]_{n+1}=\,&\frac{m_\tW^2\left(1+\frac{2}{[N_0]_n}\right)^4\left(1+\frac{1}{[N_0]_n}\right)^4%
\left[1+\Delta_e(m_e,\mc{E}_e)\right]}
{18m_e^2\left[1+\left(64+\frac{3}{2\pi}-f_Z\right)\frac{\alpha}{2\pi}\right]%
\bm{\left(}1+\frac{51}{18\left\{\ke{1}{e}[N_0]_n\right\}^4}\bm{\right)}}
\left(1+\mc{O}_N\right)\label{eq:N04}
\\
\begin{split}
\mc{O}_N:=\,&%
\frac{\E{7}\,\alpha}{\pi\left\{\ke{1}{e}[N_0]_n\right\}^4}
+\frac{\E{8}\,\alpha^2}{\pi^2} +\frac{\E{9}\,\alpha}{\pi {[N_0]_n}^2}+\frac{\E{10}\,\alpha^2}{\pi^2 {[N_0]_n}}+\left\{\frac{\E{11}\,\alpha m_e^2}{\pi [N_0]_n\left[m_c^*(\mc{E}_e)\right]^2}+\frac{\E{12}\,\alpha^2}{\pi^2}\right\}\frac{m_e^2}{[m_c^*(\mc{E}_e)]^2}\\
&\qquad+\frac{\E{13}\,\alpha m_e^2}{\pi m_\bmh^2\left\{\ke{1}{e}[N_0]_n\right\}^4}
+\frac{\E{14}\,m_e^6}{[m_c^*(\mc{E}_e)]^6}+\frac{\E{15}\,m_e^2}{m_\bmh^2\left[\ke{1}{e}\right]^4\left[{N_0}^5\right]_n}
+\frac{\E{16}\,m_e^4}{m_\bmh^4\left\{\ke{1}{e}[N_0]_n\right\}^4}%
\end{split}\label{eq:ON}
\end{align}
where $\Delta_e(m_e,\mc{E}_e)$ is evaluated using $[N_0]_n$, and 
the error terms in \Eref{eq:DeltaE} are ignored when computing this value
as these terms are all accounted for explicitly in \Erefs{eq:N04}{eq:ON}.
Using these updated values, then attempt to recover $\theta_e$ by evaluating \Eref{eq:thetacorrmutaue2}. 
If the trial value is denoted $\theta_e^T$ and the recovered value is denoted $\theta_e^R$, then each trial value of $\theta_e$ is assigned a score based on how well $\theta_e^R$ reproduces $\theta_e^T$. Repeat and refine the trial values of $\theta_e$ until a best fit is obtained. Update the values of $m_c^2$, $m_\bmh^2$, $m_Z^2$, and $N_0$ accordingly.

Since the value of $m_\tW^2$ depends on $m_c^2$, $m_\bmh^2$, and $N_0$, and the values of $m_c^2$, $m_\bmh^2$, and $N_0$ depend on $m_\tW$, it is advisable to iterate between updating $m_\tW^2$ and updating $m_c^2$, $m_\bmh^2$, $m_Z^2$, and $N_0$ until these parameters are satisfactorily converged. In practice, performing one further update to $m_\tW^2$ and one further update to $m_c^2$, $m_\bmh^2$, $m_Z^2$, and $N_0$ suffices.

Next, expand \Eref{eq:eieratio} for $i=3$ (tau channel, energy scale $\mc{E}_\tau$) and rewrite as
{\small
\begin{align}
\nn
0=& \left[m_\tau^4\right]_{n+1}\Biggl\{\frac{450\alpha}{\pi[m_c^*(\mc{E}_\tau)]^4}
+\frac{5(25-12f_Z)\alpha}{6\pi[m_c^*(\mc{E}_\tau)]^2m_\tW^2}
+\frac{\E{11}\,\alpha}{\pi N_0\left[m_c^*(\mc{E}_\tau)\right]^4}
+\frac{\E{14}\,\left[m_\tau^2\right]_n}{[m_c^*(\mc{E}_\tau)]^6}
+\frac{\E{16}}{m_\bmh^4\bigl[\ke{1}{e}N_0\bigr]^4}\Biggr\}\\
\nn+&\left[m_\tau^2\right]_{n+1}\bm{\Biggl(}\frac{90\alpha}{\pi[m_c^*(\mc{E}_\tau)]^2}+\frac{(5-4f_Z)\alpha}{2\pi m_\tW^2}
+\frac{5}{[m_c^*(\mc{E}_\tau)]^2}+\frac{40}{3m_\bmh^2\bigl[\ke{1}{e}N_0\bigr]^4}-\frac{\bigl[\ke{1}{\tau}\bigr]^4\left[1+\Delta_e(m_e,\mc{E}_\tau)\right]}{\bigl[\ke{3}{\tau}\bigr]^4m_e^2}\\
\nn&~~~~~~~~~~~~~~~~+\frac{8\alpha}{3N_0(3\alpha+2\pi)}\left\{\frac{10\pi+180\alpha}{3\pi[m_c^*(\mc{E}_\tau)]^2}+\frac{(5-4f_Z)\alpha}{4\pi m_\tW^2}\right\}
-\Biggl(\frac{\E{9}\,\alpha}{\pi{N_0}^2m_e^2}+\frac{\E{10}\,\alpha^2}{\pi^2N_0m_e^2}\Biggr)\Biggl[\frac{\ke{1}{\tau}}{\ke{3}{\tau}}\Biggr]^4
\\
\nn&~~~~~~~~~~~~~~~~-\Biggl\{\frac{\E{11}\,\alpha}{\pi N_0\left[m_c^*(\mc{E}_\tau)\right]^4}+\frac{\E{14}\,m_e^2}{[m_c^*(\mc{E}_\tau)]^6}+\frac{\E{16}}{m_\bmh^4\bigl[\ke{1}{e}N_0\bigr]^4}\Biggr\}m_e^2\Biggl[\frac{\ke{1}{\tau}}{\ke{3}{\tau}}\Biggr]^4\\
\nn&~~~~~~~~~~~~~~~~+\Biggl\{\frac{\E{12}\,\alpha^2}{\pi^2[m_c^*(\mc{E}_\tau)]^2}+\frac{\E{13}\,\alpha}{\pi m_\bmh^2\bigl[\ke{1}{e}N_0\bigr]^4}+\frac{\E{15}}{m_\bmh^2\bigl[\ke{1}{e}\bigr]^4{N_0}^5}\Biggr\}\Biggl\{1-\Biggl[\frac{\ke{1}{\tau}}{\ke{3}{\tau}}\Biggr]^4\Biggr\}\bm{\Biggr)}
\\
+&1+\frac{8\alpha}{3N_0(3\alpha+2\pi)}+\frac{\E{9}\,\alpha}{\pi{N_0}^2}+\frac{\E{10}\,\alpha^2}{\pi^2N_0}.
\end{align}}
$\!\!\!$The value of $\mc{E}_\tau$ is evaluated using $\left[m_\tau^2\right]_n$, and correction~$\E{14}$ is likewise partially evaluated using $\left[m_\tau^2\right]_n$, such that this expression 
is quadratic in $\left[m_\tau^2\right]_{n+1}$.
Again, iterate over trial values of $\theta_e$, denoted $\theta_e^T$, this time solving for $\left[m_\tau^2\right]_{n+1}$ and again computing the recovered value $\theta_e^R$ using \Eref{eq:thetacorrmutaue2}. Again refine and reiterate until the optimal value of $\theta_e$ is identified, this time with all other parameters held constant except $m_\tau^2$.

Repeatedly alternate between updating $m_\tW^2$ (and other boson masses and $N_0$) and $m_\tau^2$ until convergence.

As a couple of practical notes:
\begin{itemize}
\item Close to $\theta_e=-3\pi/4$, $k^{(e)}_1(\mc{E}_e)$ as obtained from \Eref{eq:kell2} becomes very small and it may be preferable to write 
\begin{equation}
\left[k^{(e)}_1(\mc{E}_e)\right]^2 \approx \frac{m_e^2}{m_\mu^2}\left[k^{(e)}_2(\mc{E}_e)\right]^2,
\end{equation}
this approximation being accurate to $\OOO{\alpha m_{e_i}^2/(m_c^*)^2}$. This approximation should not be employed close to convergence.
\item On the first pass for updating $m_\tW^2$, it may be convenient to suppress terms in $\bigl[k^{(e)}_1(\mc{E}_e)\,N_0\bigr]^4$. If omitted, these terms can be incorporated for a small correction from the second pass onward, once initial updates to $m_\tW^2$ and $N_0$ have been obtained.
\item Negative solutions for $m_\tW^2$ or $m_\tau^2$ may be discarded.
\item The algorithm presented here is vulnerable to becoming trapped in pseudominima. These can be recognised by inspecting the value of $\|\theta_e^T-\theta_e^R\|$, which converges to far smaller values at a true minimum (always under $10^{-10}$ and generally under $10^{-13}$ %
in the reference implementation). Pseudominima may be avoided by changing the sampling density over $\theta_e$.
\end{itemize}

Octave/C++ code implementing the above algorithm may be found accompanying this volume~\cite{code2022}. It has been tested under \mbox{macOS~11.7.4} in Octave~6.2.0 with symbolic package~2.9.0, Clang~11.0.0, and Python~3.9.4. %

\section{Evaluation of accessory results\label{apdx:accessory}}

Although not yet determined to be directly observable, it is also useful to evaluate the values of the model parameters $N_0$, $f$, $\omega_0$, $\mc{E}_0$, and $\mc{E}_\Omega$.
Allowing the coefficients $\E{1}$-$\E{20}$ to range as per \tref{tab:VI:erroreffects} and assuming that all error terms are independent, solving \Eref{eq:N04} yields
\begin{equation}
N_0=191.9470(37). \label{eq:valueN0} %
\end{equation}
Rearranging \PEref{III}{eq:f(alpha)} gives
\begin{equation}
f^2=\frac{2\alpha}{{N_0}^6S_\alpha(1+a_e)^2}
\end{equation}
where
\begin{equation}
S_\alpha:={N_0}^{-6}(N_0+\tfrac{5}{4})(N_0+1)^3(N_0+2)^2.\Ptagref{III}{eq:defSalpha}
\end{equation}
Now note the equivalence of $a_e$ in the Standard Model and in the $\Cw{18}$ model, valid at least up to the mass-independent terms of $\ILO{\alpha^2/\pi^2}$, and the existence of an uncertainty of $\ILO{\alpha^2/\pi^2}$ in the calculation of $N_0$~\eref{eq:N04}. The magnitude of this uncertainty justifies the use of either the observed or the Standard Model value of $a_e$ when evaluating~\Eref{eq:calcf}---if there is any error associated with doing so, it is below the threshold of relevance for the calculation. To minimise the number of measured parameters in the software, the reference implementation takes 
\begin{equation}
\begin{split}
a_e&=\frac{\alpha}{2\pi}+\left\{\frac{197}{144}+\frac{\pi^2[1-6\,\mrm{ln}(2)]}{12}+\frac{3\,\zeta(3)}{4}\right\}\frac{\alpha^2}{\pi^2}\\
&\,\p{\equiv}\,+\OO{\frac{\alpha^2m_e^2}{m_\mu^2}}\\
&\approx\frac{\alpha}{2\pi}-0.328478965579193\,\frac{\alpha^2}{\pi^2}+\OO{\frac{\alpha^2m_e^2}{m_\mu^2}}\label{eq:approxae}
\end{split}
\end{equation}
where the numeric coefficient on $\alpha^2/\pi^2$ is as per \rcitess{petermann1958,sommerfield1958,aoyama2019}, %
and the uncertainty term, designated $\E{21}$, is evaluated as
\begin{equation}
\left.\frac{\E{21}\,\alpha^2 m_e^2}{\pi^2m_\mu^2}\qquad \right| \qquad\E{21}=\pm 10\label{eq:E19}
\end{equation}
in keeping with the approach employed for other unevaluated higher-order terms in the present paper. 
For $a_e$ this is an extremely conservative approach, with the Standard Model value of $\E{21}$ being approximately $0.022220$ \cite{aoyama2019}.
Directly evaluating the effect of each error term on the computed value of $f$ yields
\begin{equation}
f=1.670144(95)\times 10^{-8}\tagref{eq:valuef} %
\end{equation}
where independence of sources of uncertainty $\E{1}$-$\E{21}$ has been assumed.

The value of $\omega_0$ is given by rearranging \Eref{eq:meisq1} and substituting \Eref{eq:calcf},
\begin{align}
\omega_0=&\left\{\frac{m_e^2S_\alpha(1+a_e)^2}{\alpha\Bigl[k_1^{(e)}(\mc{E}_e)\Bigr]^4\!{N_0}^2S_{18,147}%
\left[1+\Delta_e(m_e,\mc{E}_{e})\right]}\right\}^\frac{1}{2}\nn\\
&\times\left[1+\tfrac{1}{4}\mc{O}_N+\tfrac{1}{2}\mc{O}_e(m_e,\mc{E}_e)\right]\label{eq:calcomega0}
\end{align}
\begin{align}
S_{18,147}&:={N_0}^{-12}(N_0+2)^6(N_0+1)^6\label{eq:redefS18147}
\end{align}
where the definition of $S_{18,147}$ agrees with \PEref{IV}{eq:defS18147},
while
\begin{align}
\mc{E}_0&:=N_0\omega_0\,(1-\tfrac{1}{4}\mc{O}_N)\\
\mc{E}_\Omega&:=\N{N_0}(N_0-\tfrac{1}{2})\,\omega_0\quad|\quad\N=9\Ptagref{V}{eq:V:EOmega}
\end{align}
where the coefficients on $\mc{O}_e$ and $\mc{O}_N$ reflect rescalings of the range of the error coefficients $\E{1}$-$\E{20}$ relative to their original appearances in \Erefr{eq:eieratio}{eq:Oe3}. Once again allow the coefficients $\E{1}$-$\E{20}$ to range as per \tref{tab:VI:erroreffects} prior to this rescaling, and $\E{21}$ as noted above. The errors associated with each of $\E{1}$-$\E{21}$ are again evaluated directly, then combined under the assumption of independence.

Note that the terms in $\mc{O}_N$ disappear on $\mc{E}_0$. However, $\mc{O}_e$ contains corrections of $\ILOO{\alpha/(\pi N_0)}$ which are large enough to permit the use of \Eref{eq:approxae} %
for $a_e$ in \Eref{eq:calcomega0} without discernible error.

On evaluation these equations %
yield
\begin{align}
\omega_0&=18.68478(35)~\MeV\\
\mc{E}_0&=3.5864883(17)~\GeV\\
\mc{E}_\Omega&=6.17960(12)~\TeV.  %
\end{align}

\section{Further notes on the colour sector}

\subsection{Length scales and gluon masses\label{apdx:gluonmass}}

All bosons in the $\Cw{18}$ Classical Analogue to the Standard Model exhibit a scale-dependent mass interaction. However, for $A$-sector bosons this phenomenon is obscured by the fact that $A$-sector phenomena are customarily observed over length scales large compared with the largest length scale of the model, $\mc{L}_0$. In contrast, gluon-related phenomena occur over a range of length scales. The critical scales are as follows:
\begin{itemize}
\item[$\mc{L}<\mc{L}_\preon$:] Gluon exchange---more correctly at this scale, coloured preon pair exchange---binds preons into triplets or preon/antipreon doublets. On this length scale, which is much smaller than the length scale of the mass interaction, the gluon is massless.
\item[$\mc{L}_\preon<\mc{L}<\mc{L}_0$:] On this length scale, the pseudovacuum is inhomogeneous. Under appropriate conditions, and particularly for $2\mc{L}_\Omega<\mc{L}<\mc{L}_0$, colour shielding effects may permit the gluon to propagate as a free massive particle (see \cref{ch:CDF2} and \sref{sec:Wmass5v} %
for more details). Loop corrections to particle masses involve composite quasiparticles and thus implicitly occur at length scales large compared with $\mc{L}_\preon$, but occur over length scales associated with the mass interaction so are bounded from above by $\mc{L}_0$. Loop corrections to particle masses therefore involve gluons with masses of order $m_c^2$. A small correction described in \sref{sec:gluonscalarmassdeficit} introduces an energy dependency denoted by the replacement $m_c^2\rightarrow [m_c^*(\mc{E})]^2$.
\item[$\mc{L}>\mc{L}_0$:] Due to confinement, free gluons are not observed at these length scales---with the exception of the stabilised gluon holes in the pseudovacuum described in \sref{sec:gluonscalarmassdeficit}, which again exhibit an effective mass $(m_c^*)^2$.
\end{itemize}

\subsection{Separability of~$A$ and~$C$ charge sectors\label{apdx:ACseparability}}

In \sref{sec:symC18}, the $\SU{9}_{\bm{80}}\oplus\GL{1}{R}_{\bm{1}}$ local symmetry of the $\mbb{C}^{\wedge18}$ model was factorised to yield $\{[\SU{3}_A]_{\bm{8}}\oplus[\GL{1}{R}_A]_{\bm{1}}\}\otimes\{[\SU{3}_C]_{\bm{8}}\oplus[\GL{1}{R}_C]_{\bm{1}}\}$, reducing the pre-gauging number of bosons from 81 to 18. In the context of a classical model this decomposition is wholly satisfactory, as any boson field in $\SU{9}_{\bm{80}}\oplus\GL{1}{R}_{\bm{1}}$ may be reconstructed as the product of a field from $\{[\SU{3}_A]_{\bm{8}}\oplus[\GL{1}{R}_A]_{\bm{1}}\}$ and a field from $\{[\SU{3}_C]_{\bm{8}}\oplus[\GL{1}{R}_C]_{\bm{1}}\}$. However, in the quantised effective description which emerges from the $\Cw{18}$ model, the local symmetry group $\SU{9}_{\bm{80}}\oplus\GL{1}{R}_{\bm{1}}$ permits a single boson to carry both an $A$-sector and a $C$-sector charge whereas symmetry groups $\{[\SU{3}_A]_{\bm{8}}\oplus[\GL{1}{R}_A]_{\bm{1}}\}\otimes\{[\SU{3}_C]_{\bm{8}}\oplus[\GL{1}{R}_C]_{\bm{1}}\}$ require the exchange of two quanta (one carrying an $A$-sector charge and one carrying a $C$-sector charge) to achieve the same effect.

This distinction may be overlooked in situations where one or both of the charge sectors interacts in a manner approaching the continuum. Thus, for example, during tree-level lepton mass interactions (\srefs{sec:fermionmasses}{sec:leptonleadingorder}), while the $A$-sector interactions are discrete, the $C$-sector interactions are represented cumulatively by the $\K$-matrices.
Similarly, during loop corrections to these interactions (\sref{sec:leptonloops}), all particles are assumed massive (with solution by consistency) and thus both $A$ and $C$ sector interactions occur continuously in the vicinity of the explicit loop corrections. Thus electroweak sector loop corrections may be considered without attaching $C$-sector charges (due to the rescaling of the fermion fields described in \aref{apdx:gaugeSU9} and the presence of $C$-sector $\K$-matrices in the implicit mass vertices of all propagators). Effective separability of the $A$-sector charges also necessarily implies effective separability of the $C$-sector charges. %

For boson mass interactions, in the dominant (pseudovacuum fermion) terms the $A$-sector interactions are discrete and the $C$-sector interactions are represented cumulatively by the $\K$-matrices. The pseudovacuum boson terms vary depending on the charges of the boson whose mass is being determined, and it is necessary to take both $A$- and $C$-sector corrections to the leading-order term into account.

\appendixend

\chapter{{Curved} space--times from \protect{$\mathbb{C}^{\wedge 18}$}\label{ch:gravity}}

\begin{abstract}
The $\Cw{18}$ analogue model is a classical model of free fields on a manifold with anticommuting co-ordinates which emulates the quantum field theory of the Standard Model. In some ways this emulation is remarkably accurate, predicting masses for the weak bosons and the tau which are in precise agreement with observation. However, the model also predicts a right-handed weak interaction which has not been observed. In this chapter the final ungauged freedom of the $\Cw{18}$ analogue model is used to eliminate the right-handed weak interaction, while simultaneously introducing space--time curvature, and a gravitational interaction which emulates general relativity in experimentally observed regimes. The model is predictive of the value of Newton's constant, yielding $G_N=6.67426(230)\times 10^{-11}~\mathrm{m}^3\mathrm{kg}^{-1}\mathrm{s}^{-2}$. Although the error bars on this calculated value are quite large, the central value is in agreement with the observed value of $G_N=6.67430(15)\times 10^{-11}~\mathrm{m}^3\mathrm{kg}^{-1}\mathrm{s}^{-2}$ to a precision of $0.3\,\sigma_\mathrm{exp}$.
\end{abstract} %

\section{Introduction}

Introduced in \crefr{ch:simplest}{ch:detail}, the $\Cw{18}$ analogue model is a classical model on a manifold with anticommuting co-ordinates which is capable of supporting a quasiparticle spectrum analogous to the Standard Model.
Manifold $\Cw{18}$ is taken to support an infinite number of scalar fields, and
on mapping a subspace $M\subset\Cw{18}$ onto $\RM$, 
\begin{equation}
\G(M)\cong\RM,
\end{equation}
the product of these fields behaves (in appropriate limits) as a pseudovacuum. Solitonic excitations about the pseudovacuum state then behave as preonic quasiparticles from which the species of the Standard Model are constructed. These are necessarily supplemented by a weakly-interacting massive vector boson denoted $N_\mu$, higher-generation counterparts to the weak sector bosons (starting with $W_2$ at $16.61320(46)~\TeV/c^2$), and an attenuated right-handed weak interaction mediated by a pair of bosons denoted $G_\mu$ and $G^\dagger_\mu$. %

An essential part of relating the $\Cw{18}$ analogue model to the Standard Model is the choice of a gauge on the %
local symmetry of the analogue model, described in \Psref{III}{sec:GL18Cgauge}. The local symmetries which are gauged are an emergent property of the model arising in the limit of low quasiparticle energy, and originate in the freedom to choose an arbitrary co-ordinate frame on manifold $\Cw{18}$. Of the gaugeable degrees of freedom described in \Psref{III}{sec:gaugechoicewhichavailable}, the vast majority are fixed in \crefs{ch:SM}{ch:detail} on physically motivated grounds. The sole exception is the gauging of the $\bm{1}_A\otimes\bm{1}_C\otimes\SLTC$ subgroup associated with the space--time connections on $\G(M)$. 
These connections are fixed (up to a change of co-ordinates on a fixed manifold) by requiring, arbitrarily, that the target of mapping $\G$ in \cref{ch:SM} be flat.

In the present chapter it is seen that relaxation of this requirement, such that the target of mapping $\G$ need only be locally Minkowski, permits a choice of gauge on $\bm{1}_A\otimes\bm{1}_C\otimes\SLTC$ which eliminates the $G^\bdag_\mu$ bosons (and also some other beyond-Standard-Model effects). Adopting this gauge uniquely fixes the geometry of $\G(M)$, and in the vicinity (but outside the Schwarzschild radius) of a nonrotating electrically neutral massive body this geometry is seen to correspond to the Schwarzschild metric. The Kerr metric for a rotating source is anticipated to follow in a similar regime from the usual classical arguments~\cite{kerr1963}. Further, the effective value of Newton's constant in this gravitational analogue is uniquely determined by the structure of the $\Cw{18}$ model and is found to be $G_N=6.67426(230)\times 10^{-11}~\mathrm{m}^3\mathrm{kg}^{-1}\mathrm{s}^{-2}$, in excellent agreement with observation. 

This is not the first attempt to calculate of the value of Newton's constant. Acknowledgement must be made of the %
efforts of \citeauthor{di-mario2003} in a series of articles made available online from~2003~\cite{di-mario2003}, whose equations reduce to the relationship
\begin{equation}
G_N=\frac{\alpha^2h^3\omega_1^2}{4\pi c^3m_e^4}\left[1+\OO{\alpha}\right].\label{eq:dimarioG}
\end{equation}
However, the parameter $\omega_1$ in \Eref{eq:dimarioG} is a unitful constant taking the value of exactly $1~\mrm{s}^{-1}$, and %
is never satisfactorily explained. 
This equation must therefore be considered 
coincidental.

The structure of this chapter is as follows: \Sref{sec:terminology} recaps and references the conventions of notation and terminology employed in this chapter. \Sref{sec:QCcorr} then introduces a  convenient relationship between Feynman diagrams and classical fields termed a \emph{quantum/classical correspondence}. 
\Srefs{sec:curvedasgauge}{sec:ambientGG} begin the mapping of beyond-Standard-Model effects to space--time curvature. The most ubiquitous interaction involving $G^\bdag$~bosons is identified as involving a spin-2 pair $G_\mu G^\dagger_\nu$, and elimination of this process (and other related beyond-Standard-Model processes) through modification of mapping~$\G$ is compared with choosing a gauge in the low-energy regime. Calculation of the resulting metric is performed in \sref{sec:curvature}, and is shown to be predictive of the value of Newton's constant. \Sref{sec:moreprocesses} extends this treatment to include all sources of $G^\bdag$~bosons, including single-boson (spin-1) interactions%
. Numerical results are presented in \sref{sec:valueGN}, with some discussion of implications in \sref{sec:qual}, and future tests of the $\Cw{18}$ model are discussed in \sref{sec:conclusions}.

\section{Conventions\label{sec:VII:conventions}}

\subsection{Notation and terminology\label{sec:terminology}}

This chapter follows the same conventions as \crefr{ch:simplest}{ch:detail}. 
\standalone{
As in previous chapters, units are chosen such that $c=1,~h=1$.
When equations and lemmas from \crefr{ch:simplest}{ch:detail} are referenced, they take the forms (\textbf{1}.1), (\textbf{2}.1), %
etc.

When referring to uncertainty in results, experimental uncertainties will be denoted $\sigma_\mrm{exp}$, and uncertainties in the theoretical calculation will be denoted $\sigma_\mrm{th}$.
}
As in \cref{ch:detail}, it is generally assumed that any particle under study in this chapter is at rest or near-rest with respect to the isotropy frame of the pseudovacuum. 

\standalone{
Regarding terminology around Feynman diagrams and symmetry factors:
\begin{itemize}
\item Where there exist multiple ways to connect up sources, vertices, and sinks to obtain equivalent diagrams up to interchange of non-distinguishable co-ordinates, the same term is obtained from the generator $\Z$ in multiple different ways and thus the diagram acquires a multiplicative factor. This is referred to in the present volume as a \emph{symmetry factor}.
\item Where integration over the parameters of a diagram (for example, over source/sink co-ordinates) 
yields the same diagram multiple times up to interchange of labels on these parameters, 
this represents a double- (or multiple-)counting of physical processes. It is then necessary to eliminate this multiple-counting by dividing by the appropriate symmetry factor. This is referred to in the present volume as \emph{diagrammatic redundancy} or \emph{double- (multiple-)counting.}
\end{itemize}
}

\subsection{Quantum/classical correspondence\label{sec:QCcorr}}

In the limit of large particle numbers it is convenient to introduce a quantum/classical correspondence which permits classical field profiles to be obtained from the Feynman diagrams describing the processes taking place within these fields. To my knowledge this correspondence has not been formalised to the extent employed here, %
and so it must be conjectured, with the results obtained being dependent upon it.

\begin{figure}
\includegraphics[width=\linewidth]{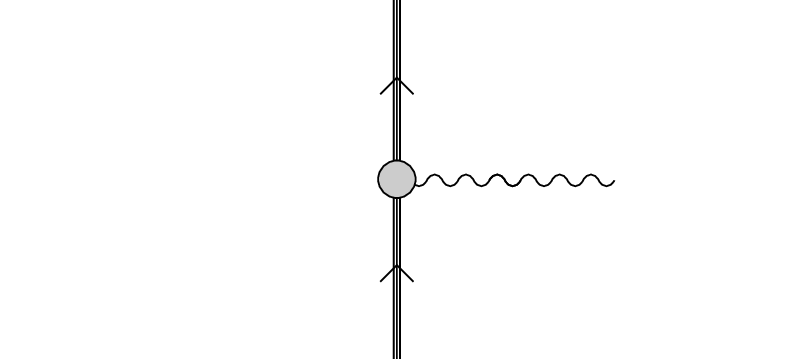}
\caption{Emission of a single photon by a single electron. The grey circle is the all-orders vertex incorporating loop corrections. The triple line represents the existence of a preonic substructure, consistent with Chapters~\ref{ch:simplest}-\ref{ch:detail}.\label{fig:EMvertex}}
\end{figure}

As a specific example, consider the emission of a photon by an electron as shown in \fref{fig:EMvertex}.
Although a classical-regime $A$~field is assumed to be made up of many, many virtual photons, these are all emitted in accordance with \fref{fig:EMvertex} and it is customary to associate features of this diagram with elements of the classical expression
\begin{equation}
A=\frac{Q}{|e|}\frac{\alpha}{r}\label{eq:Aclassicalunsym}
\end{equation}
(in which only the source is associated with a factor of $\alpha$) or with its symmetrised form
\begin{equation}
A=\frac{Q}{|e|}\frac{\sqrt{\alpha}}{r}\label{eq:Aclassical}
\end{equation}
in which a factor of $\sqrt{\alpha}$ appears at both source and sink.
To set up the quantum/classical correspondence used in the present chapter: 
\begin{itemize}
\item First, as is customary, 
associate the all-orders vertex in \fref{fig:EMvertex} with a factor of $\sqrt{\alpha}$, the classical electromagnetic field coupling. This is consistent with the quantum field theory treatment of \Psref{III}{sec:photonint}, and in particular with \PEref{III}{eq:f(alpha)}. When a photon is exchanged between two charged particles, this coefficient is squared to yield $\alpha$, and incorporates a factor of $1/(4\pi)$ which is associated with directionality of emission.
\item Second, normalise the fermion fields such that the fermion line and associated operators (e.g.~$\bar e_L\bsm e_L$) may be identified as a number operator. This is to be understood in the sense that (for example) given a collection of fermions at rest, $\|\bar e_L\bsm e_L\|$ counts the number of left-handed electrons per unit volume.
\item Third, it is necessary to relate a single boson to a field profile. This is again a choice not just in the quantum/classical correspondence, but also in quantum field theory, and must reproduce the geometric factor of $1/r$ in \Eref{eq:Aclassical}. Thus the photon line emerging from the interaction vertex for %
a spherically symmetric point source, and hence the associated photon operator at that vertex, must be associated with a factor of $r^{-1}$.
(It should be noted that this %
latter association applies only to radially propagating fields. In contrast,
for example, the mean energy per effective excitation in the background field is known to be $\omega_0=\mc{E}_0/N_0$ everywhere.)
\end{itemize}
Having imposed this %
relationship between Feynman diagrams and classical fields, which is almost trivial in the $\MSbar$ renormalisation scheme, it is now possible to use Feynman diagrams to calculate the profiles of classical fields in the presence of perturbing processes. 
The above three choices calibrate the electromagnetic 4-potential of a single electron against the emission diagram of \fref{fig:EMvertex}, and application of the same calibration to the diagrams for other processes then permits determination of their effects on fields in the classical limit. It is worth noting that the third step in this choice of correspondence introduces an arbitrary relationship between boson number $N_b$ (unitless) and field intensity (unitful), such that $N_b=r^{-1}|_{r=1}$ for $r$ in some favoured units, equivalently $N_b=kr^{-1}|_{r=k}$ for some unitful $k$. Independence of the physical quantities $N_b$ and $N_b A_r = r^{-1}$ from the value of~$k$ follows immediately, with the consequence that choice of~$k$ is arbitrary and makes no appearance in expressions evaluated in the classical limit.

The remainder of this chapter is essentially a calculation performed using the quantum/classical correspondence specified above, taking into account additional effects due to the presence of the pseudovacuum and the fundamental scalar field (FSF) symmetry factors.

\section{\prm{\Cw{18}} with curved space--time\label{sec:curved}}

\subsection{Curvature and choice of gauge\label{sec:curvedasgauge}}

As noted in the introduction, the choice of co-ordinate frame on the $\bm{1}_A\otimes\bm{1}_C\otimes\SL{2}{C}$ subgroup of $\GL{18}{C}$ is a freedom of the $\Cw{18}$ model. In choosing this frame the 4-volume form must be preserved under translations, as the scaling symmetry $\bm{1}_A\otimes\bm{1}_C\otimes\mbb{R}^+$ is fixed in \Psref{VI}{sec:gss}. Once a prescription has been described for choosing such a frame based on the field content of the model on $\Cw{18}$, the target manifold of mapping $\G$ must then be chosen to %
support the associated metric, with this target manifold being the manifold on which the $\Cw{18}$ model emulates particle dynamics. 

Recognising that all choices of gauge for local symmetries of the low-energy effective description of the $\Cw{18}$ model similarly correspond to choices of co-ordinate frame on $\Cw{18}$, it is natural to refer to this particular choice of co-ordinate frame in the same language, loosely describing the degrees of freedom associated with this co-ordinate frame as (analogous to) gauge degrees of freedom, and the
prescription for choosing the co-ordinate frame as being (analogous to) a choice of gauge.

\subsection{%
Beyond-Standard-Model processes in the photon pair field\label{sec:ambientGG}}

\subsubsection{Origin of the photon pair field\label{sec:origphotpair}}

The governing principle behind the choice of co-ordinate frame on $\SLTC$ is that it should eliminate as many beyond-Standard-Model effects as possible, including the $G^\bdag$ fields and thus the right-handed weak interaction. As a first step, it is necessary to identify the most prevalent $G^\bdag$-boson-mediated interactions appearing in the $\Cw{18}$ analogue model. 
These interactions take place within a photon pair field and have no counterpart in the Standard Model.

\begin{figure}
\includegraphics[width=\linewidth]{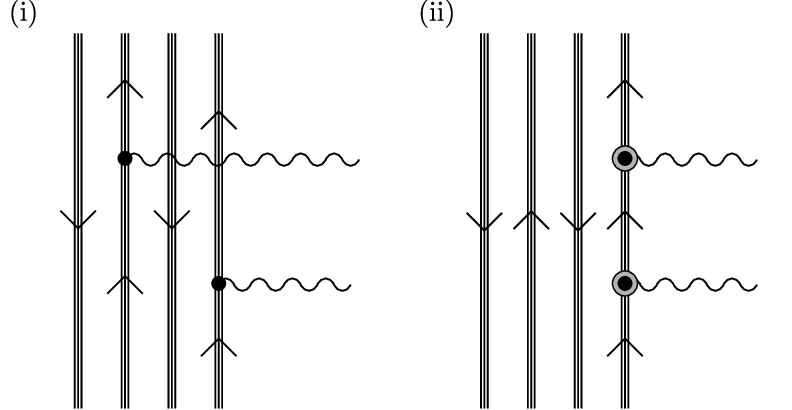}
\caption{Emission of a pair of photons, (i)~with standard electromagnetic vertices, and (ii)~with \prm{K}-augmented vertices. In diagram~(i), emission may be from any fermion or pair of fermions in the matter source. In diagram~(ii), for the background interactions which give rise to the \prm{K}~matrices to be on average nonvanishing, emission of the pair must be from the same fermion.\label{fig:photonpairemission}}
\end{figure}

As a prototype for fermionic matter, consider a collection of positrons and electrons with zero net spin and zero net charge (and see \sref{sec:principequiv} for discussion of more general matter sources).
Let this source emit a pair of photons as per \fref{fig:photonpairemission}(i), either with both being emitted from the same fermion, or with both being emitted from different fermions. 
There are precisely as many diagrams for emission of pairs from sources with charges of the same sign as there are for emissions of pairs from sources with different sign, and in the far field, being the regime in which all source elements may be treated as effectively collocated, these diagrams cancel to yield no net photon pair field $A_\mu A_\nu$.

However, in the $\Cw{18}$ model there is a further emission channel to consider. In addition to the simple photon pair emission channel described above, if both sources are correlated and both vertices are within $\mc{L}_0$ of one another (with these conditions in practice implying that both emission vertices are on the same fermion) there is also a process of emission augmented by background colour field interactions.

In the mass calculations of \crefs{ch:fermion}{ch:detail} it was convenient to separate the description of interactions with the background gluon fields into colourless boson couplings (implicitly exploiting symmetry of the $C$~sector under colour mixing) and independent colour transformations (represented by $K$~matrices), and then to write the effectively colourless gluon contributions to particle mass as small corrections to the photon contributions. This approach is once again useful here, although it is not the emission of the gluon pairs themselves which is important this time, since the gluons are (i)~massive and (ii)~confined. This time, it is the effects of these gluon interactions on preon {colour} (as represented by the $K$~matrices) which {are} of interest, as these effects may be nonvanishing over length scales of up to $\ILO{\mc{L}_0}$. %
Physically these diagrams represent processes whereby preons scatter off background gluons one or more times to change colour, then scatter again to change back again, contemporaneous with the emission of a photon pair. The scattering process itself is absorbed into the normalisation of the fermion propagator, but the effects on preon colour provide physically distinguishable alternative channels for photon emission. The pair of background field
scatterings must be correlated, and in matter of normal densities, this necessitates the involvement of the same fermion at each background interaction. 

In principle these $K$~matrices are associated with explicit gluon couplings as per \PEref{IV}{eq:defMCK}, but these explicit field couplings are subsumed into the fermion propagator as indicated in \fref{fig:interactingpreon}, and hence into the fermion mass term. However, as per \Psref{IV}{sec:Csector}, knowing that the application of $K$~matrices is in 1:1~correspondence with $A$-sector interactions permits these matrices to be equivalently associated with the non-subsumed photon interaction vertices. Proper accounting for colour effects is then obtained if up to one set of $K$~matrices is applied for every $A$-sector interaction (as discussed in \Psref{IV}{sec:Csector}), with the matrices being applied to the same fermion as participates in the $A$-sector interaction. (Applying the $K$~matrices elsewhere has no effect on the emission process.)
An example diagram for the resulting process is shown in \fref{fig:photonpairemission}(ii). As emission is constrained to always be of two photons from the same fermion, there is no cancellation of diagrams and a nonvanishing photon pair field is supported. In flat space--time, both the intensity of the foreground photon pair field $\fgfield{A_\mu A_\nu}$ and the number of pairs traversing unit area of a spherical shell scale as $r^{-2}$. This is a convenient feature of the photon pair field in the position basis, and provided all subsequent processes act equivalently on all photon pairs, it allows any relative (i.e.~multiplicative) attrition of the field to be understood equivalently either in terms of a relative reduction in the field amplitude, or a relative reduction in the number of participating pairs.

Further note that as in \srefs{sec:universalcorrs}{sec:Zuniversalcplg}, when foreground bosons interact with fermions via a pair of vertices which may involve $K$-matrices, it is also possible to rewrite the pair of fermion triplets as a triplet of boson pairs, which do not attract $K$-matrices, and then expand two of these pairs about the mean field value and integrate out to obtain a boson/boson correction termed the indirect (universality) coupling. This effect further augments the beyond-Standard-Model photon pair field and is discussed in further detail in \sref{sec:pairdecay}.

\subsubsection{Interactions of the photon pair field\label{sec:pairfieldint}}

\begin{figure}
\includegraphics[width=\linewidth]{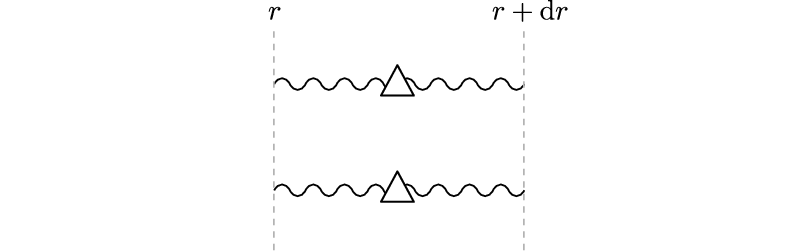}
\caption{Free propagation of a pair of photons across interval \prm{[r,r+\rmd r)}, with the propagator term of the Lagrangian represented as an interaction vertex~\prm{\triangle}.\label{fig:pairfield}}
\end{figure}

Once a pair of foreground photons has been emitted by a matter source, that pair may propagate radially unchanged, or may undergo interactions. To understand these processes, it is helpful to study the behaviour of a photon pair on propagation across an infinitesimal interval $[r,r+\rmd r)$. %
By the requirement that the photon be capable of interacting, but that an infinitesimal interval be unsubdividable, each photon undergoes one interaction within such an interval. Free propagation is represented by treating the propagator term of the Lagrangian, $\frac{1}{2}A_\mu\triangle^{\mu\nu}A_\nu$, as a vertex. This is represented diagrammatically in \fref{fig:pairfield}.

\begin{figure}
\includegraphics[width=\linewidth]{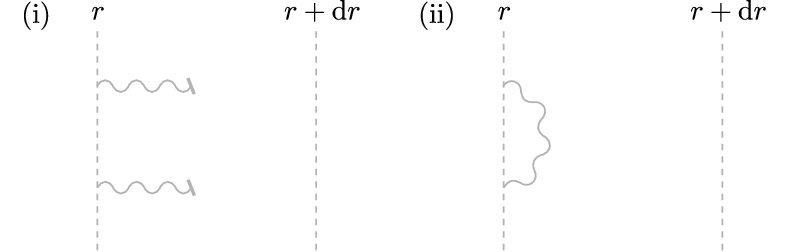}
\caption{(i)~A pair of photons arising from the pseudovacuum in interval \prm{[r,r+\rmd r)} propagates radially inward and interacts with a fermion in the matter source of the foreground photon pair field. The symbol \prm{\backslash} denotes origin within the local pseudovacuum. (ii)~The parent figure for diagram~(i) reveals this to be the mean field expansion of a term in the Proper Self Energy~(PSE) of the source fermion. Pseudovacuum photons are shown in grey.\label{fig:BGpairs}}
\end{figure}

In addition to the emission of foreground photon pair fields, a charged fermion
necessarily also interacts with background photon pairs to yield the mass interactions discussed in \crefs{ch:fermion}{ch:detail}. \Fref{fig:BGpairs}(i) shows a pair of photons arising from the background field at a radius between $r$ and $r+\rmd r$ and propagating inward to interact with a fermion in the source. 
In general such diagrams make vanishingly small contributions to particle mass for $r\gg\mc{L}_0$, but it must nevertheless be asked whether they contribute to the photon pair field at $r'<r$. This may be answered in the negative by considering emulation of foreground fields under the $\MSbar$ normalisation scheme, and recognising that \fref{fig:BGpairs}(i) is the zeroth order term in the mean field expansion of \fref{fig:BGpairs}(ii), which belongs to the Proper Self Energy (PSE) series of the source fermion. All such diagrams are accounted for in the renormalised foreground fermion propagator, and so may be discounted, as per \Psref{IV}{sec:generalconsider}.

\begin{figure*}
\includegraphics[width=\linewidth]{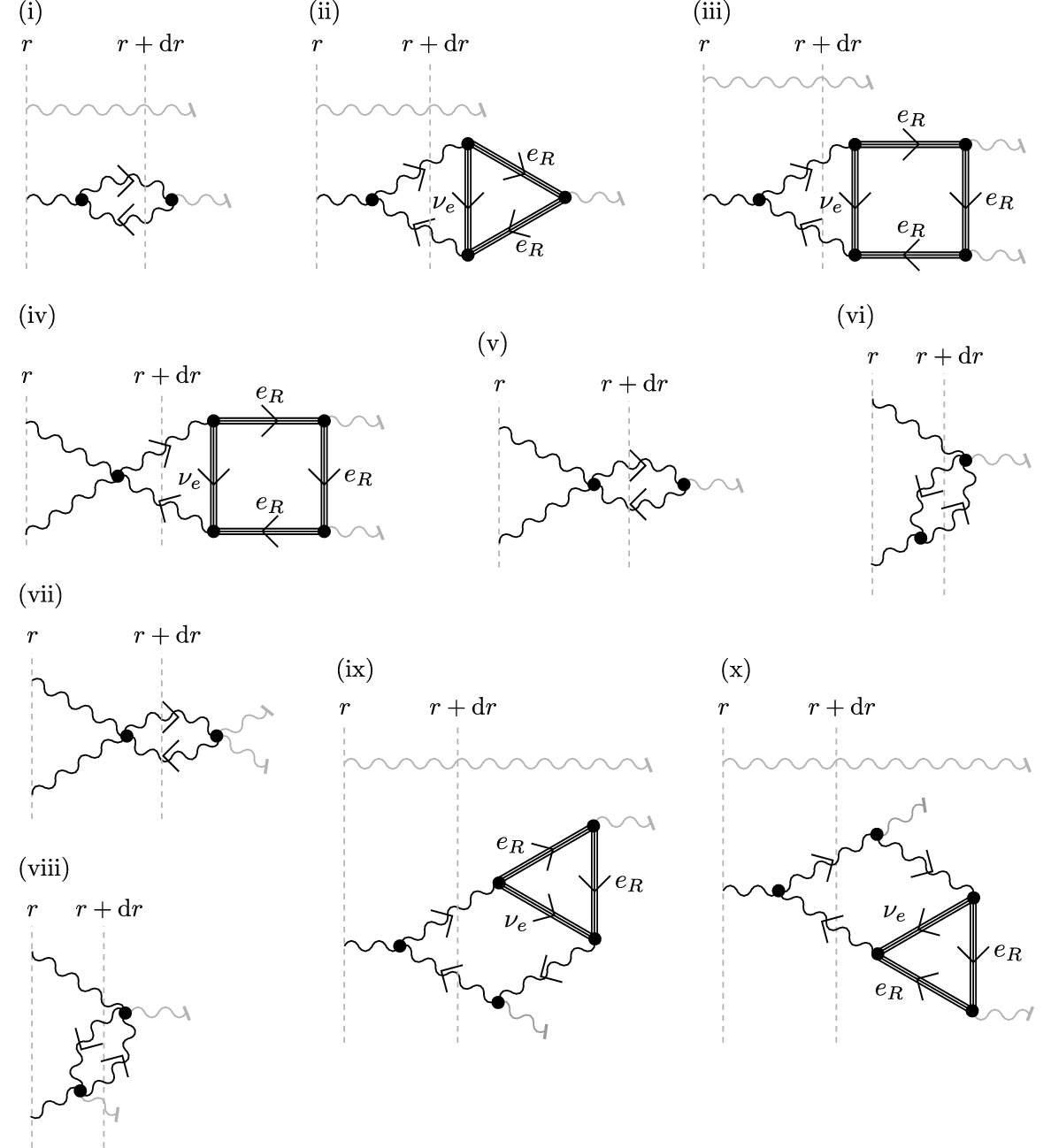} %
\caption{Candidate diagrams for processes occurring within the photon pair field. Boson lines without an arrow represent photons, and boson lines with an arrow represent \prm{G}~bosons. Disruption of the pair occurs within interval \prm{[r,r+\rmd r)}. Species may be massive with generations, or if existing over time and distance short compared with \prm{{\mc{L}_0}} 
they {may} %
lack interactions with the pseudovacum and be massless and without generation. Note that by \protect{\Psref{I}{sec:normWrtBgFields}}, all interactions involving only background fields may be ignored. Consequently, background photons attract no vertex in interval \prm{[r,r+\rmd r)}. When the propagation term is treated as a vertex, it is necessarily included in this treatment.
\label{fig:pairinteractions}}
\end{figure*}

It is next necessary to understand what further processes do take place within the propagating foreground photon pair field. A great many of these processes may be discounted for a wide variety of reasons, including %
the following:
\begin{itemize}
\item As in \fref{fig:BGpairs}, any diagram which is itself part of the PSE series of the source fermion, or whose parent diagram is part of the PSE series of the source fermion, may be discounted as (regardless of $r$) it is accounted for in the propagator of the source fermion. Further examples include \freft{fig:pairinteractions}(i)-(ii), (iv), and~(vii)-(viii).
\item Any diagram for which the only possible parent diagrams are tadpole diagrams must also vanish. (No examples shown.) %
\item Any diagram in which a $G^\bdag$~boson couples to a background photon must have vanishing net effect (whether singly or pairwise) when averaged over length scales greater than $\mc{L}_0$. Otherwise background photons could generate $G^\bdag$ boson pairs, and these in turn would necessarily have background character, in violation of gauge choice~\Peref{III}{eq:bga45gauge}. Examples include \freft{fig:pairinteractions}(i) and (v)-(x).
\item Diagrams involving an odd number of fermion lines may be eliminated using a generalisation of Furry's theorem. If a diagram or set of diagrams remain invariant when left- and right-handed representations are interchanged, corresponding to reversal of all arrows on the diagram and replacing
\begin{equation}
\bar\psi\bsm\psi\rightarrow \psi\sigma^\mu\bar\psi = -\bar\psi\bsm\psi,
\end{equation}
it follows that that set of diagrams must vanish. Examples include \freft{fig:pairinteractions}(ii) and~(ix)-(x), with diagrams~(ix)-(x) forming a set which map into one another on arrow reversal, and therefore collectively cancel.
\item Any diagram which gives mass to the photon may be discounted by gauge choice~\eref{eq:ma3gaugenew}. %
Examples include \freft{fig:pairinteractions}(vii)-(viii).
\item Any diagram in which pseudovacuum couplings explicitly give rise to mass for any of the constituent particles may also be ignored. Examples include \freft{fig:pairinteractions}(iii)-(iv), in which pseudovacuum photon coupling gives mass to the electron. [As with many of the examples in \fref{fig:pairinteractions} any give diagram may have multiple reasons to be ignored---%
for example, the parent of diagram~(iv) belongs to the PSE expansion of the source fermion propagator.]

Regarding why diagrams containing mass interactions may be ignored, recognise that foreground species other than photons and $G^\bdag$ bosons
may either acquire mass as a result of interaction with the background fields, placing them in any generation, or they may be ephemeral---generationless and massless---if they only exist over scales of $\mc{L}_0$ 
or less in the isotropy frame (and must be ephemeral if only existing over scales ${2\mc{L}_\Omega}$ or less in the isotropy frame%
.) %
Both situations are in principle admissible in photon pair decay processes, and thus both massive and massless propagators must be taken into account when evaluating decay processes, unless otherwise constrained. Diagrams in which the foreground propagators are massive implicitly incorporate all diagrams in which pairs of background photon interactions behave as mass vertices as per \crefr{ch:fermion}{ch:detail}, and thus these pseudovacuum interactions do not need to also be considered separately.
\end{itemize}

\begin{figure}
\begin{center}
\includegraphics[width=\linewidth]{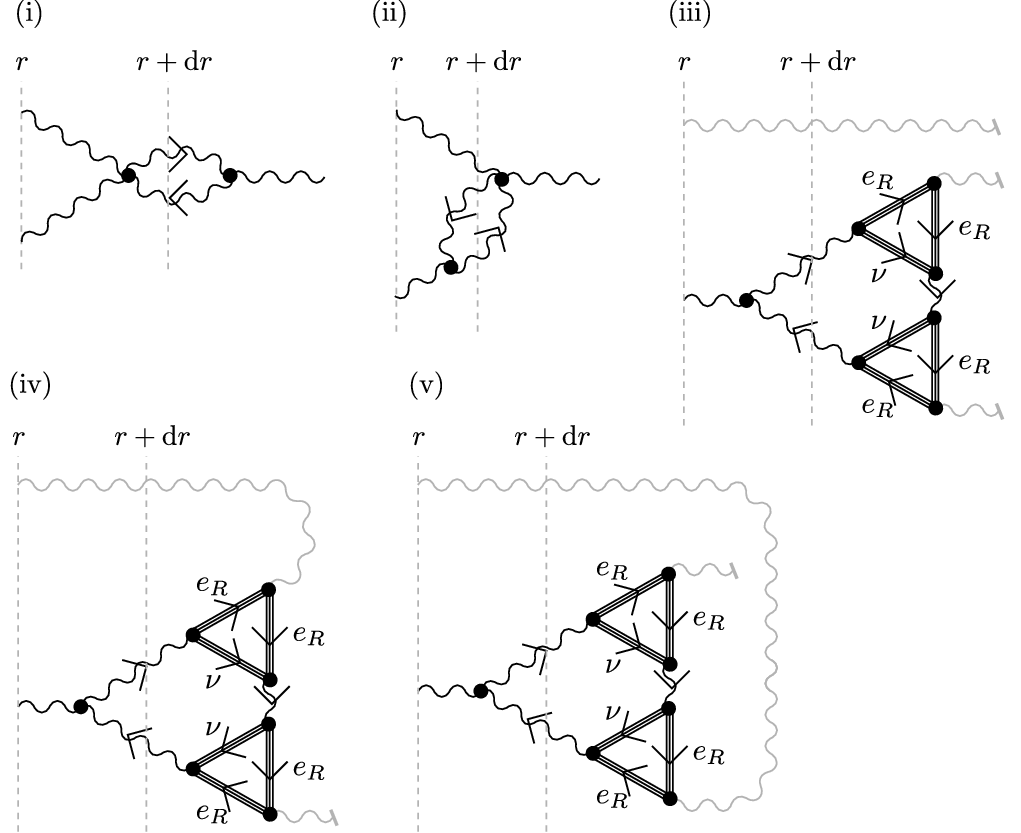} %
\end{center}
\caption{Example diagrams not eliminated by the considerations of \psref{sec:pairfieldint}. Bosons with no arrows are photons, and bosons with arrows are \prm{G^\bdag} bosons. Background bosons are shown in grey. Diagrams~(i)-(ii) are \prm{\ILO{f^3}} in vertex coefficients, but scale as \prm{r^{-3}} so in the far field they are supplanted by diagram~(iii), which is of \prm{\ILO{f^7}} but is seen in \psref{sec:pairdecay} to scale as \prm{r^{-1}}. Diagrams~(iv)-(v) are parent diagrams for diagram~(iii), meaning that they yield it as the zeroth-order term of mean-field expansion about the pseudovacuum state for the indicated photons. 
\label{fig:persistentfigures}}
\end{figure}

Now consider only diagrams not eliminated by the above considerations. In the large-$r$ limit the most important diagrams in evaluation of the foreground pair field are those with the least-negative exponent in the radial co-ordinate. As examples of candidate processes, consider the $G^\bdag$-mediated diagrams shown in \fref{fig:persistentfigures}. 
The presence of a radially outbound photon indicates that these diagrams do not belong to the PSE of the source fermion propagator, and although neutrality of the matter source implies that the \emph{average} value of the outbound foreground photon field must vanish, for any individual photon pair it may be non-zero. Furthermore, in contrast with \freft{fig:pairinteractions}(i) and~(v)-(x), there is no gauge-induced requirement that this process have vanishing impact on the photon pair field as a whole.
Indeed, if the source is examined fermion by fermion, these processes may be understood instead as making a fairly high-order contribution to the gyromagnetic anomaly series for the electron.

Although shown as $G^\bdag$ bosons in \freft{fig:pairinteractions}-\ref{fig:persistentfigures}, the circulating off-diagonal boson in these processes may in principle be either %
$W$ or~$G$. %
The processes associated with both choices are potentially non-vanishing, and are all beyond-Standard-Model processes due to the presence of $K$-matrices in the emission vertices as per \fref{fig:photonpairemission}(ii).

Specialising first to the $G^\bdag$ bosons,
the requirement that these bosons do not couple to the background photon field implies that all photons in \freft{fig:persistentfigures}(i)-(ii) are foreground, and thus the diagrams scale as $r^{-3}$. In contrast, although \fref{fig:persistentfigures}(iii) is of much higher order in the coupling constant $f=1.670144(95)\times 10^{-8}$, %
it  is seen in \sref{sec:pairdecay} to scale as $r^{-1}$ and is thus of greater importance in the far field regime. Parent diagrams for \fref{fig:persistentfigures}(iii) prior to mean-field expansion are shown in \freft{fig:persistentfigures}(iv)-(v), noting that in order to affect the boson pair count in interval $[r,r+\rmd r)$, only the initial decay of one member of the photon pair need take place within that interval; the rest of the diagram may be evaluated over any length scale. It is then convenient to evaluate the other photon of the pair by substituting the pseudovacuum term of the mean field theory expansion. Further, by overall neutrality of the matter source (which is made up of both electrons and positrons), the {average} onward-propagating photon field necessarily has net zero foreground component, and thus the average total photon field on the outbound leg \emph{also} evaluates to the pseudovacuum field. Nevertheless, diagrams~(iv) and~(v) are distinguishable as it is possible to spatially discriminate between the mean-field term of a photon inbound from $r$ and the nonvanishing component of a photon outbound to $r+\rmd r$.

It may at first appear problematic that \fref{fig:persistentfigures}(iii) is being evaluated in a mean-field limit in which the outbound component of the foreground fields vanishes. However, recall that an inverse-square biphoton field mediates no net force transfer beyond that accounted for by its individual constituents, and it is therefore unnecessary for an interaction which reduces the photon pair field to exhibit any net (average) transfer of momentum from this pair field to elsewhere. Further,
in the present context (with a net electrically neutral source) the single-photon exchange process likewise mediates no force. Thus the radially outbound single-photon field carries an {average} foreground momentum of zero, and reduces to the background component {on average} in the mean-field limit.
Consequently, although \fref{fig:persistentfigures}(iii) has no net outbound foreground particle species, it may nevertheless result in a reduction in the {number} of photon pairs.

Ironically, on mapping negative energy outbound pairs into positive energy inbound pairs, the resulting deviation from inverse-square decay {does} now result in an imbalance between the outward and inward photon pair pressures (with decay occurring more rapidly in the lower-$r$, higher-field regime), and %
may be anticipated to yield a net attractive force between neutral bodies not corresponding to any process in the Standard Model. This effect is small, so there is no immediate need to engage in a perturbative calculation to evaluate second- and higher-order effects, and to present accuracy it suffices to adopt a co-ordinate frame on $\Cw{18}$ in which the photon pair number density obeys the expected inverse-square profile with respect to the connection on the $\SL{2}{C}$ subframe. The coupling of the photon to $GG^\dagger$ pairs in \fref{fig:persistentfigures}(iii) is thus supplanted by an equivalent coupling to space--time curvature, which %
is by construction a {true} space--time curvature on the manifold which is the target of mapping~$\G$.

As noted above, every diagram in \freft{fig:pairinteractions}-\ref{fig:persistentfigures} also has a counterpart in which the $G^\bdag$~bosons are replaced by $W^\bdag$~bosons. However:
\begin{itemize}
\item $G^\bdag$~bosons are massless over all length and timescales, whereas $W^\bdag$ bosons are always massive over scales large compared with $\mc{L}_0$, and 
\item there is no requirement that the $G^\bdag$ or $W^\bdag$ loop in these figures should be small compared with $\mc{L}_0$. (This is discussed further in \sref{sec:curvature}.) Thus behaviour in the low-energy regime is dominated by $\mc{L}>\mc{L}_0$ in which the $W$~boson is massive but the $G$~boson is not.
\item The classical limit of the photon field around a point source may be understood as a superposition of positive and negative energy massless virtual photons. Thus massive $W$~boson loop processes in the photon pair field are heavily disfavoured relative to massless $G$~boson loops.
\end{itemize}

Taking all of this into account, %
now observe that electromagnetically neutral composite bodies are ubiquitous on classical scales, and the $K$-augmented emission process of \fref{fig:photonpairemission}(ii) is thus likewise widespread. As a result, so are the %
$G^\bdag$-mediated decay processes which take place throughout the supported photon pair fields. %
It is postulated that \fref{fig:persistentfigures}(iii) is %
the highest-weight beyond-Standard-Model diagram affecting the photon pair field in the large-$r$ regime. 

In regions where $G^\bdag$-mediated photon pair decay is the dominant beyond-Standard-Model process, it then follows that space--time curvature in the large-$r$ regime, and thus Newton's Constant, may be obtained from evaluation of \freft{fig:persistentfigures}(iv)-(v). This calculation is performed in \sref{sec:curvature}.

\subsubsection{Principle of equivalence\label{sec:principequiv}}

The above discussion has addressed photon pair emission from a prototype source comprising left- and right-handed electrons and positrons. However, for higher-generation leptons, mass is augmented by the $K$~matrices and these appear identically in the fermion mass interaction and in the $K$-augmented photon pair emission process, with
net result %
that all leptons support the beyond-Standard-Model processes in photon pair fields of \fref{fig:persistentfigures} at a rate identically proportional to their mass. Anticipating \sref{sec:elimGGpairs}, in which %
these processes are mapped to space--time curvature, this equivalence of proportion implies the weak principle of equivalence for leptons (equivalence of gravitational and inertial mass).

The mass mechanism for quarks is identical to that for leptons. It may therefore be reasonably anticipated that quarks will also respect the weak principle of equivalence.

With baryons also being composite fermions, let it be assumed that the baryon mass mechanism is the natural extension of the quark mechanism, with inertial mass being determined by coupling to background boson fields and by the nine-preon/three-quark counterpart to operator $\hat{\mc{K}}_\mu$~\Peref{IV}{eq:defMCK}. In addition to their preonic constituents, baryons also contain the gluons which bind them together---but then, so do leptons and quarks. Under the assumption that the confinement scale $\mc{L}_\preon$ is smaller than $2\mc{L}_\Omega$, the gluons are not in general able to participate in any %
mass interactions over their individual lifetimes. They therefore behave as particles without inertial mass, and consequently without gravitational mass. The sole exception to this is the unconfined ``neutral gluon'' $N_\mu$. However, this particle is not associated with the dimension-8 representation of $\SU{3}_C$ and thus is not a participant in the preon or quark confinement interactions. With the gluons being essentially massless, any transient appearances of gluons within a baryon may simply be viewed as variety in the associations of the preon lines making up the baryon (including brief time-reversals in trajectory, and the appearance of temporary loops of characteristic scale less than $\mc{L}_0$, which are thus normed to~1). The net preon composition, which interacts with the nine-preon analogue of $\hat{\mc{K}}_\mu$, remains unchanged. It is therefore reasonable to anticipate that baryonic matter {also} respects the weak principle of equivalence.

Indeed, the only exception of any significant abundance is likely to be the neutral gluon, $N_\mu$. This particle has an inertial mass of %
$80.3810(22)~\GeV/c^2$ (\tref{tab:VI:results}), but its coupling to the photon pair field (via the universal coupling mechanism of \Psref{VI}{sec:Zuniversalcplg}) is smaller by a factor of 
\begin{equation}
\frac{\frac{5}{18\left[k^{(e)}_{1}(\mc{E}_e)\,N_0\right]^4}}{1+\frac{55}{18\left[k^{(e)}_{1}(\mc{E}_e)\,N_0\right]^4}}=7.71305(59)\times 10^{-5}\label{eq:gravinertialmassratio} %
\end{equation}
resulting in a correspondingly-reduced gravitational mass-squared.
If the $N_\mu$ boson exists in any significant quantity, it has substantial inertial mass (making it a Weakly Interacting Massive Particle or WIMP) and also supports a photon pair field (which is the prerequisite for the gravity-like interaction of the $\Cw{18}$ model). It is therefore a candidate for an interesting species of dark matter which breaks the weak principle of equivalence (equality of gravitational and inertial mass).

\subsection{Evaluation of space--time curvature\label{sec:curvature}}

This Section calculates the metric in the vicinity of a nonrotating electrically neutral body, including the value of Newton's Constant. Extension to incorporate conceptually important subleading effects is discussed in \sref{sec:moreprocesses}, though these processes are numerically irrelevant in the large-$r$ regime at current numerical precision.

\subsubsection{Leading-order photon pair decay profile\label{sec:pairdecay}}

To evaluate the photon pair field decay profile as a function of radius,
first note the following:
\begin{itemize}
\item As per \sref{sec:principequiv}, the photon pair field support is expected to scale identically with inertial mass for all fermionic species, so for convenience let any mass $M$ be represented by the already-discussed prototypical chargeless, spinless neutral body made up entirely of electrons and positrons.
\item Functionally the propagator term of the Lagrangian behaves as a two-particle vertex, structurally equivalent to the mass vertex but acting with the momentum operator $k_\mu$ in lieu of the rest mass operator (which takes the form $[m,0,0,0]$ in the rest frame of the particle). 
\item On radial propagation the photon pair field supported by a mass $M$ exhibits decay as $r^{-2}$, such that on propagation from $r$ to $r+\rmd r$ the photon pair field experiences a geometrically induced decrease according to the corresponding infinitesimal relationship
\begin{equation}
A_\mu A_\nu\longrightarrow A_\mu A_\nu\left(1-\frac{2\rmd r}{r}\right).\label{eq:geometricdecay}
\end{equation}
This is associated with \fref{fig:pairfield}.
\item Alternatively, %
the photon pair may undergo decay in accordance with \freft{fig:persistentfigures}(iv)-(v). Normalising these diagrams with respect to the radial propagation of \fref{fig:pairfield} introduces the quantity $I(r)$, defined in \fref{fig:Iratio}. 
\begin{figure}
\includegraphics[width=\linewidth]{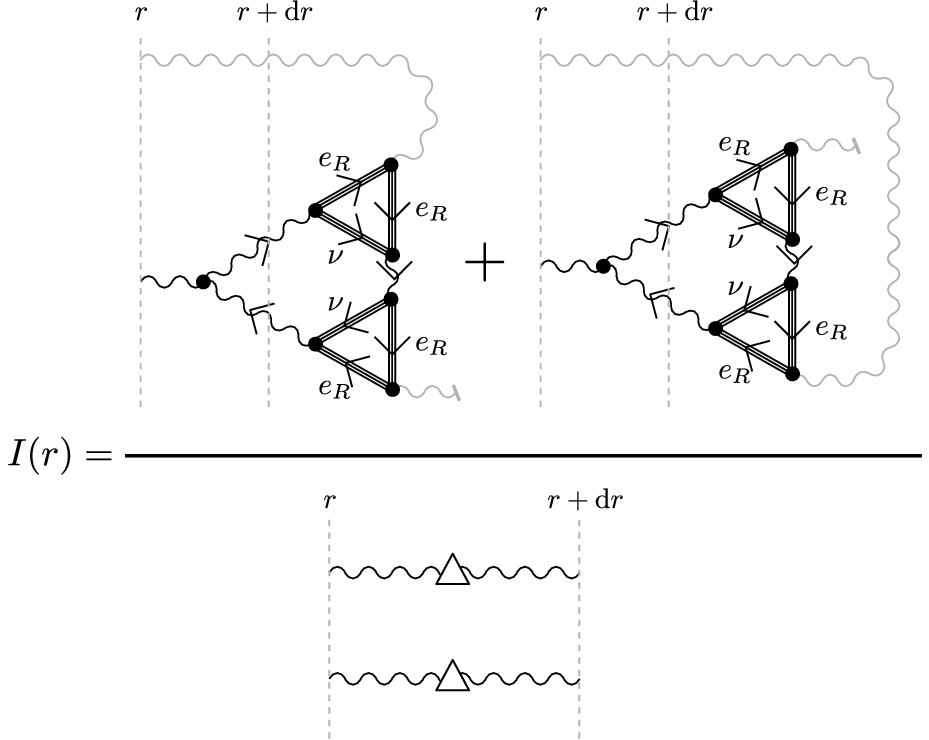}
\caption{Relative weight of diagrams associated with photon pair decay, normalised with respect to photon pair propagation. If the latter is associated with a relative decay of \prm{-{2\rmd r}/{r}}, then the former is associated with \prm{-{2I(r)\rmd r}/{r}} where \prm{I(r)} is defined as above.\label{fig:Iratio}}
\end{figure}%
The pair propagation of \fref{fig:pairfield} yields a relative decay of the pair field of $-2\rmd r/r$, as in \Eref{eq:geometricdecay} above, so that associated with \freft{fig:persistentfigures}(iv)-(v) is weighted by a further factor of $I(r)$ to yield a contribution %
of $-2I(r)\rmd r/r$.%
\item The net relative decay of the photon pair field on propagation from $r$ to $r+\rmd r$ is thus given by
\begin{equation}
A_\mu A_\nu\longrightarrow A_\mu A_\nu\left[1-\frac{2\rmd r}{r}-\frac{2I(r)\rmd r}{r}\right].\label{eq:fulldecay}
\end{equation}
\end{itemize}

Evaluation of ratio $I(r)$ may be broken down into a series of stages. First, recognise that the behaviour under study is the radial decay of foreground photon pairs. Consider the denominator, and recognise:
\begin{itemize}
\item The multiplier of $\left[k^{(e)}_1(\mc{E})\right]^4$ on \fref{fig:photonpairemission}(ii) may be interpreted as a modifier to the vertex factor which changes the rate of foreground photon pair emission. Thus each photon line inbound or outbound in the denominator of \fref{fig:Iratio} is associated with a factor of $\left[k^{(e)}_1(\mc{E})\right]^2$.
\item For a prototype body made up entirely of electrons and positrons, the characteristic particle energy $\mc{E}$ corresponds to $\mc{E}_e=m_ec^2$. %
\item Although the objects of study are foreground fields, over an interval of width $\rmd r$ the foreground and background fields are not clearly distinguishable. Although the number of photon pairs in the foreground field is clearly determined, the foreground momentum is distributed across all preons with correlators consistent with photon pairs within this region. In the large-$r$ regime the average momentum of a photon within a pair is that of the pseudovacuum, up to a small foreground perturbation. On-shell, the propagation operators thus evaluate as
\begin{equation}
\triangle^{\mu\nu}\rightarrow k^\mu k^\nu\rightarrow -\delta^{\mu\nu}{\omega_0}^2.
\end{equation}
\item In addition, as mentioned in \sref{sec:origphotpair}, foreground photon pair emission is augmented by an indirect (universality) coupling:
\begin{itemize}
\item Mean field expansion the fermions in the $Af\bar f$ vertices of \fref{fig:photonpairemission}(ii) yields a leading-order term comparable in structure to \fref{fig:Wmassbosoncorrsnew}(i).
\item In the electromagnetic sector, after integrating out the four surplus preon lines, the emitted photon pair may be expressed as continuations (perhaps up to potential charge transformations) of the background preon/antipreon pair. The associated four-fermion vertex attracts no factor save for the factor of $f^{2}$ associated with any four-boson vertex as follows from the derivative-to-field rules of \sref{sec:fgspinorscalar}~\erefr{eq:opsubvp}{eq:opsubH}.
\item In this, it has a vertex factor identical to that of the $W$~boson and thus similarly acquires an EM contribution of
\begin{equation}
\frac{1}{18\left[k^{(e)}_{1}(\mc{E}_e)\,N_0\right]^4}.
\end{equation}
\item The contribution from the scalar sector is the same as for the $W$~boson,
\begin{equation}
\frac{18}{18\left[k^{(e)}_{1}(\mc{E}_e)\,N_0\right]^4}.
\end{equation}
\item For the colour sector, recall that the photon also has coloured counterparts (the gluphotons), all massless, with unbroken $\GLTR$ symmetry. In the classical (many-particle) limit, $A$~and $C$ sector charges are separable (\aref{apdx:ACseparability}) implying that recombination of colour charges reduces the generated gluphotons to photons in the far-field limit, augmenting the $AA$ field attributable to \fref{fig:photonpairemission}. Indirect (universality) coupling augmentation of the classical biphoton field is therefore equivalent to that for the mass interaction of the $W^{\dot cc}$ boson (\sref{sec:colouredWmass}), namely a relative increase of
\begin{equation}
\frac{243}{36\left[k^{(e)}_{1}(\mc{E}_e)\,N_0\right]^4}.
\end{equation}
\item The total indirect coupling factor for beyond-Standard-Model photon pair emission is therefore
\begin{equation}
\left\{1+\frac{281}{36\left[k^{(e)}_{1}(\mc{E}_e)\,N_0\right]^4}\left[1+\OO{\alpha}\right]\right\}.\label{eq:AApairUCfactor}
\end{equation}
\end{itemize}
As with $\left[k^{(e)}_1(\mc{E})\right]^4$, under the assumption of the classical/quantum correspondence of \sref{sec:QCcorr} this factor acts to increment the number of available photon pairs, and thus a copy of factor~\eref{eq:AApairUCfactor} is associated with both the inbound pair and the outbound pair in the denominator of \fref{fig:Iratio}.
\item To evaluate FSF symmetry factors: 
\begin{itemize}
\item Recognise that the interval $[r,r+\rmd r)$ is small compared with $\mc{L}_0$  and there are four foreground preons inbound (two from $r$ and two from $r+\rmd r$) and four preons outbound, each of which may be of type $a=1$ or $a=2$. Colour is summed over and net neutral across the foreground fields, so may be ignored. There are also $N_0$ inbound and $N_0$ outbound background preons for any given charge labelling.
\item All inbound preons participate in equivalent interactions, so their associated FSFs are interchangeable. Likewise for all outbound preons.
\item First assume all inbound foreground preons have the same $A$-charge; conservation of charge requires that the outbound charges be the same. The first photon line in the diagram may incorporate any one of the $N_0+4$ inbound preons (or more strictly, incorporates one specific inbound preon but this may be associated with any of $N_0+4$ eligible FSFs), and any one of the four $N_0+4$ outbound preons (ditto). Up to a normalisation factor, the corresponding sources/sinks are associated with a FSF symmetry factor of $(N_0+4)^2$. The second photon yields $(N_0+3)^2$, and so on. There are two choices of the $A$-charge involved ($a=1$ and $a=2$). This yields two configurations each with a factor of $[(N_0+4)(N_0+3)(N_0+2)(N_0+1)]^2$.
\item Next, consider configurations where one inbound preon preon has a different charge to the others. There are eight such configurations, each yielding a factor of $[(N_0+3)(N_0+2)(N_0+1)^2]^2$.
\item Finally, there are six configurations in which two inbound preons have charge $a=1$ and two have charge $a=2$, for a factor of $[(N_0+2)(N_0+1)]^4$. 
\item Normalisation with respect to background fields as per \Psref{I}{sec:normWrtBgFields} requires that for a single background photon field source/sink the FSF symmetry factor goes to~1. That is, in the absence of foreground fields the source/sink attracts a factor of ${N_0}^2$ and this is then divided by ${N_0}^2$.
\item As all photons in the above factors are foreground, the number of background FSFs available is constant at ${N_0}$ for every term, so 
\begin{equation}
\begin{split}
(N_0+4)&\rightarrow {N_0}^{-1}(N_0+4)\\
(N_0+3)&\rightarrow {N_0}^{-1}(N_0+3)\\
&\vdots
\end{split}
\end{equation}
\item For each of these configurations, the same symmetry factors apply at the propagation operators (considered collectively) as at the sources/sinks (considered collectively), though without requiring the ${N_0}^{-1}$ normalisation factor. Instead, \Psref{I}{sec:normWrtBgFields} is taken to imply that no vertex should be drawn which involves only background fields.
\item The net FSF symmetry factor associated with the sources and sinks is thus
\begin{equation}
{N_0}^8 S_{\triangle\triangle}
\end{equation}
where $S_{\triangle\triangle}$ is a correction of $\ILO{1}$ given by
\begin{equation}
\begin{split}
S_{\triangle\triangle}:=\{&2[(N_0+4)(N_0+3)(N_0+2)(N_0+1)]^4\\
&+8[(N_0+3)(N_0+2)(N_0+1)^2]^4
\\&+6[(N_0+2)(N_0+1)]^8\}~/~({16{N_0}^{16}}).
\end{split}\label{eq:Striangletriangle}
\end{equation}
\end{itemize}
\item Following construction of the photons, their connection to the propagation vertices attracts a symmetry factor of sixteen:
\begin{itemize}
\item The first photon (let this be a photon inbound from $r$) may be connected to any of the four connection points on the two propagation vertices: Factor~4.
\item The second photon at the same radius must be connected to one of the two connection points on the other propagation vertex: Factor~2.
\item The first of the remaining two photons may be connected to the remaining connection point on either of the propagation vertices: Factor~2.
\item The final photon has only one possible connection: Factor~1.
\end{itemize}
\item The net factor associated with the denominator is thus
\begin{equation}
16\left[k^{(e)}_1(\mc{E}_e)\right]^8 {\omega_0}^4 {N_0}^{8}S_{\triangle\triangle}\left\{1+\frac{281}{36\big[k^{(e)}_{1}(\mc{E}_e)\,N_0\big]^4}\left[1+\OO{\alpha}\right]\right\}^2.
\end{equation}
\end{itemize}

Now consider the numerator:
\begin{itemize}
\item \Freft{fig:persistentfigures}(iv)-(v) both evaluate to the same value, therefore specialise to \fref{fig:persistentfigures}(iv) and multiply by two.
\item The lower photon in the pair inbound at $r$ is required to be foreground as it interacts with the $G^\bdag$~fields. The upper photon may be foreground or background, and %
the contribution of the diagram will be dominated by the background terms at large~$r$. (Note that in the numerator, these background terms of the mean field expansion truly represent the pseudovacuum. This is in contrast to the photons seen in the denominator, which carry foreground correlations superimposed upon background mean field values, and thus have instantaneous magnitudes of momemtum approximately equal to the background value while nevertheless continuing to be foreground photons.)
\item The foreground photon decay vertex within interval $[r,r+\rmd r)$ is associated with a factor of $f/\sqrt{2}$. 
\item The vertex (propagation or otherwise) within this interval which acts on the other member of the photon pair is not shown. \Fref{fig:persistentfigures}(iv) is dominated by contributions in which this member evaluates to the background term, %
and while a background photon may indeed undergo an interaction within interval $[r,r+\rmd r)$, its effect on \fref{fig:persistentfigures}(iv) is normed away as per \Psref{I}{sec:normWrtBgFields}. Alternatively it may be argued that a propagator vertex on the non-decaying photon is not a part of the minimal process for decay of a photon pair (through decay of one of its elements) and thus has no place in the limit $\Delta r\rightarrow \rmd r$. This may be contrasted with the denominator, in which the minimal process for propagation of a pair of photons across interval $\rmd r$ necessarily requires a propagator vertex on each photon.
\item Note that the vertex in \fref{fig:persistentfigures}(iv) in which this background photon does engage \emph{may} be within interval $[r,r+\rmd r)$ but is not required to be---the value of the diagram is independent of the radius at which this vertex is evaluated.
\item Since only the lower photon inbound from $r$ is foreground, only that photon attracts a factor of $\left[k^{(e)}_1(\mc{E}_e)\right]^2$. In conjunction with the quantum/classical correlation of \sref{sec:QCcorr}, the net classical factor associated with the photon field of a single pair at the lower vertex is thus $\left[k^{(e)}_1(\mc{E}_e)\right]^2/r$.
\item There are $M/m_e$ fermions capable of acting as sources of such a pair, so this factor is multiplied by $M/m_e$ to yield $\left[k^{(e)}_1(\mc{E}_e)\right]^2M/(m_e r)$.
\item An indirect (universality) coupling factor also applies to the numerator of \fref{fig:Iratio}. Although there is only a single foreground photon inbound from $r$, and none outbound to $r+\rmd r$, this photon is a part of a foreground pair augmented by factor~\eref{eq:AApairUCfactor} and thus the numerator also acquires a factor of
\begin{equation}
\left\{1+\frac{281}{36\left[k^{(e)}_{1}(\mc{E}_e)\,N_0\right]^4}\left[1+\OO{\alpha}\right]\right\}.\tagref{eq:AApairUCfactor}
\end{equation}
No factor is acquired on account of the background (mean field) photons.
\item A momentum operator $\rmi\partial_\mu$ also acts at this vertex.
\begin{itemize}
\item It may act on the photon field, the $G$~field, or the $G^\dagger$~field, with the weights of the latter two processes being half that of the former.
\item As previously discussed for the denominator term, the momentum associated with the foreground photon field is distributed across all photons within the local correlation region. On average the background contribution vanishes, but \emph{instantaneously} (averaging over magnitude but not sign), the action of the derivative operator on the $A_\mu$ field at this vertex may be represented by ${N_0}^2\rmi\omega_0$, up to an arbitrary sign and a small perturbation representing the foreground field superimposed on the pseudovacuum. The factor of ${N_0}^2$ arises from the choice of FSF fields which may be acted on by the chiral derivative operators making up the background photon whose momentum is then evaluated. 
\item There is a further factor of $\frac{1}{2}$ from the structure constants of $\SU{3}_A$.
\item The terms in which the derivative operator acts on~$G$ and~$G^\dagger$ may be ignored as there are no background $G^\bdag$~fields and the contributions which these terms yield are thus negligible at large~$r$.
\end{itemize}
\item The subsequent completion of the decay process, which may take place outside interval $[r,r+\rmd r)$, contains a further six vertices which yield a factor of $f^6/2$.
\item Two of these vertices involve pseudovacuum photons. As this portion of the diagram is not constrained to take place within interval $[r,r+\rmd r)$, in order to be non-vanishing these vertices must be within the same autocorrelation region of the pseudovacuum (i.e.~within distance and time $\mc{L}_0$ of one another in the isotropy frame). Up to FSF symmetry factors, they therefore evaluate to $-{\omega_0}^2$.
\item The FSF symmetry factors associated with the interaction vertices in \fref{fig:persistentfigures}(iv) are illustrated in \fref{fig:Nfactors}.
\begin{figure}
\includegraphics[width=\linewidth]{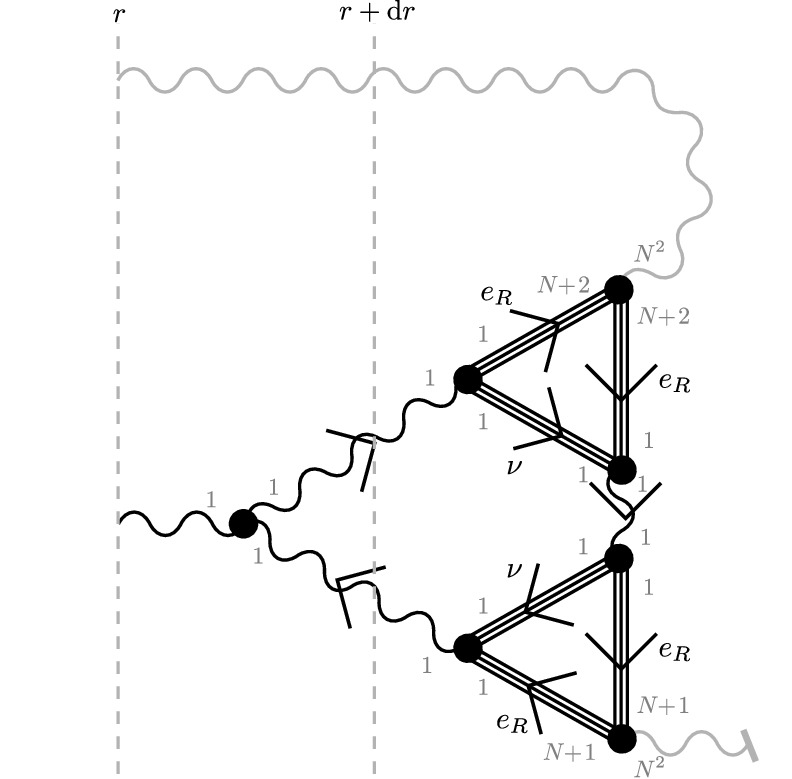} %
\caption{FSF symmetry factors associated with \pfref{fig:persistentfigures}(iv). Regarding the two vertices incorporating interactions with the background photon field, note that the fermion symmetry factors are of \prm{\ILO{N_0}} and not \prm{\ILO{{N_0}^3}} as only one preon per fermion triplet interacts with the boson.
Also note that these vertices are within \prm{\mc{L}_0} of one another and thus the symmetry factors increment, yielding \prm{(N_0+2)(N_0+1)} on the interacting preons of the inbound electrons and the same again on the interacting preons of the outbound electrons, rather than \prm{(N_0+1)(N_0+1)} apiece.\label{fig:Nfactors}}
\end{figure}%
\item There is a symmetry factor of~2 associated with the interchangeability of the two fermion triangles within the $G$~boson loop.
\item Regarding particle masses in \fref{fig:persistentfigures}(iv):
\begin{itemize}
\item On-shell photons and $G^\bdag$~bosons are massless.
\item The neutrino-family and electron-family particles may be massive or massless, with the latter being admissible over scales small compared with the characteristic mass interaction scale $\mc{L}_0$ and favoured over scales small compared with the minimum mass interaction scale $2\mc{L}_\Omega$. %
\item With the two background photon interactions being within a single autocorrelation region, the portion of the diagram connecting these two vertices may either incorporate a path which remains within the autocorrelation region and is made up of massless particles, or it may be that for any path from one vertex to the other, one or more particles %
(with the exception of the $G$~boson) may %
acquire mass (and optionally venture outside of the autocorrelation region). 
\item Considering both the mean field term illustrated in \fref{fig:persistentfigures}(iv) and small perturbations around this, the circulating foreground momentum of the photon pair decay process as a whole is within a small perturbation of zero, such that only the massless lepton solution is on-shell (or within a small perturbation of being so). (Note also that these small perturbations ensure that \fref{fig:persistentfigures}(iv) is not a tadpole diagram---this is discussed further in \sref{sec:observations}.)
\item If one of the particles connecting the two background photon interactions acquires mass, then regardless of whether it ventures out of the autocorrelation region, %
reconnection with the other background vertex involves a significantly off-shell trajectory. Thus interactions are anticipated to be dominated by conditions in which there exists at least one series of massless propagators connecting the two background photon interactions. Thus at least one of the electron lines in each triangle is massless.
\item For \fref{fig:persistentfigures}(iv) it then follows directly that the closest diagrams to on-shell are those in which either
\begin{itemize}
\item all fermions are massless, or
\item the neutrino and one electron in each triangle are massive but the location of the vertex involving both of these fermions is chosen such that the participating $G$~boson is also off-shell, and the off-shell contribution from massive fermion propagation cancels a similar off-shell contribution from the $G$~boson propagation.
\end{itemize} 
In this latter scenario the loop as a whole couples to external fields as if all participating particles were massless.
In \sref{sec:ON1} it is argued that contributions dominate in which one of the fermion vertices in each triangle is outside the correlation region, favouring the massive fermion solution. However, for purposes of the present leading-order calculation, the anticipated cancellation of off-shell contributions in this scenario permits the loop as a whole to be treated as if made up of massless species.
\end{itemize}
\item Examining the fermion triangles, at each vertex the interaction is averaged over three preons equivalent up to colour. The result is numerically unchanged by replacing each with a single interaction multiplied by a factor of three.
\item If this is done so as to bring each vertex onto the same preon within each loop, as per \fref{fig:freeloops}(i), then two of the preons within each loop are now free loops. 
\begin{figure}
\begin{center}
\includegraphics[width=\linewidth]{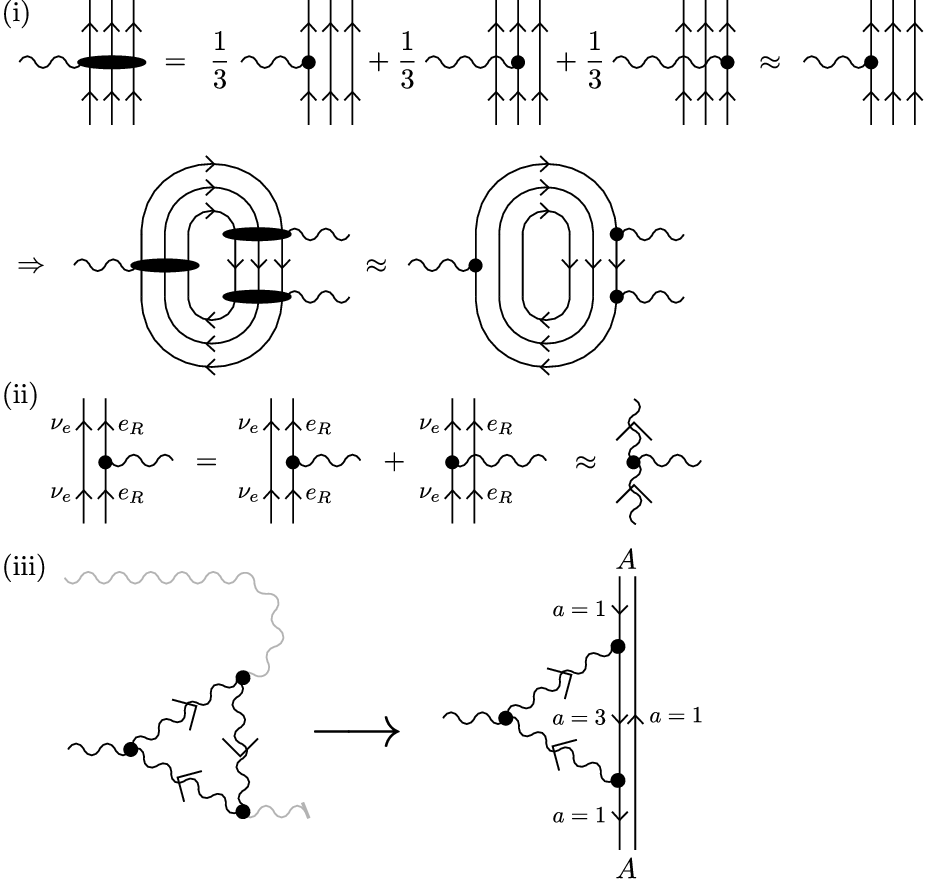}
\end{center}
\caption{(i)~Interaction vertices involving fermions are averaged over three different internal configurations. However, it is numerically equivalent to replace these interactions with a single interaction weighted by a factor of three. In the fermion loops of \pfref{fig:persistentfigures}(iv) this permits two fermion loops to be disconnected, and thus mapped to vacuum fluctuations and evaluated to yield factors of~1. (ii)~The neutrino is associated with preons having \prm{a=3}. Since these preons have no electric charge, the interaction of the photon with the \prm{a=1}~preon is equivalent to the total interaction of the photon with the entire preon pair. Up to considerations of FSF symmetry factors, this preon pair is equivalent to a \prm{G}~boson. The loop structure of \pfref{fig:persistentfigures}(iv) therefore reduces to a single \prm{G}~boson loop as shown in diagram~(iii), up to a numerical factor of \prm{f^4(N_0+2)^2(N_0+1)^2{N_0}^4} corresponding to vertices and FSF factors present in \pfref{fig:persistentfigures}(iv) but not in this diagram. If the preon decomposition of some of the bosons in diagram~(iii) is made explicit, the resulting diagram may be understood as a loop correction to photon emission/absorption by a preon with \prm{a=1}. (The photon also contains terms with \prm{a=2} but these do not couple to the \prm{G^\bdag}~fields, and this is accounted for in the factor of \prm{1/\sqrt{2}} on the \prm{A/G}~coupling coefficient.) The loop boson is massless but chiral, and the loop thus evaluates to a factor of \prm{({f^2}/{2})({1}/{4\pi})} as per Appendices~\ref{apdx:massloops}-\ref{apdx:symloops}, of which the vertex-related factor of \prm{{f^2}/{2}} has already been accounted for.\label{fig:freeloops}}
\end{figure}%
In the massless regime, diagrammatic isotopy techniques (or equivalently, vacuum normalisation) give that such loops evaluate to~1. This leaves four superfluous FSFs at each vertex, but these are cancelled by the factors of $f=\la\vp\ra^{-1}$ appearing in the definition of the fermion operators~\Peref{III}{eq:generalfermion}, so may also be removed for an average factor of~1.
\item Up to differences in FSF symmetry factors, the remaining unevaluated preons in the fermion loop are then seen to be equivalent to $G$~bosons, as per \fref{fig:freeloops}(ii), permitting reduction of the loop structure of \fref{fig:persistentfigures}(iv) to that of \fref{fig:freeloops}(iii). Note that this reduction is only valid up to a factor of $f^4(N_0+2)^2(N_0+1)^2{N_0}^4$ corresponding to vertices and FSF symmetry factors present in \fref{fig:persistentfigures}(iv) but not in \fref{fig:freeloops}(iii).
\item Selecting out a specific subdiagram as per \fref{fig:freeloops}(iv) permits evaluation of the resulting $G$~boson loop to yield a factor of $1/(4\pi)$.
\item Finally, note that \fref{fig:persistentfigures}(iv) incorporates the completed exchange of a foreground photon between a source at radius $r$ (as proxy for the original neutral source) and the $AGG^\dagger$ vertex, which attracts a factor of $S_\alpha$. This is independent of, and in addition to, the other FSF symmetry factors in \fref{fig:Nfactors}.
\end{itemize}
Putting this all together permits evaluation of the mean value of the numerator of $I(r)$, up to a sign arising from the derivative operator in the $AGG^\dagger$ vertex. The mean magnitude of the numerator is given by
\begin{equation}
\frac{Mf^7\left[k^{(e)}_1(\mc{E}_e)\right]^2\!{\omega_0}^3{N_0}^6 (N_0+2)^2(N_0+1)^2 S_\alpha }{4\sqrt{2}\pi m_e r}\left\{1+\frac{281}{36\left[k^{(e)}_{1}(\mc{E}_e)\,N_0\right]^4}\left[1+\OO{\alpha}\right]\right\}.
\end{equation}
Curiously, the sign of this term is unimportant. The interactions of \freft{fig:persistentfigures}(iv)-(v) unambiguously destroy photon pairs, whereas the ambiguity of the sign merely ensures vanishing of any emitted (net residual) species. For a neutral source this is already known. However, it is interesting to note that with a charged source, this summation over sign ensures that photon pair decay does not lead to any unexpected augmentation of the single-photon field.

The overall leading-order expression for $\la|I(r)|\ra$ is thus
\begin{align}
\la|I(r)|\ra=\,&\frac{Mf^7{N_0}^2 S_G S_\alpha}{64\sqrt{2}\pi\left[k^{(e)}_1(\mc{E}_e)\right]^6 \!{\omega_0}m_e r}\left\{1+\frac{281}{36\left[k^{(e)}_{1}(\mc{E}_e)\,N_0\right]^4}\left[1+\OO{\alpha}\right]\right\}^{-1} \label{eq:Irtree}\\
S_G:=\,&\frac{(N_0+2)^2(N_0+1)^2}{{N_0}^4S_{\triangle\triangle}},\label{eq:defSG1}
\end{align}
with the infinitesimal radial decay of the photon pair field taking the form
\begin{equation}
A_\mu A_\nu\longrightarrow A_\mu A_\nu\left(1+\frac{2\rmd r}{r}+\frac{2\la|I(r)|\ra\rmd r}{r}\right).\label{eq:|I|form}
\end{equation}

\subsubsection{Higher-order corrections\label{sec:higherorder}}

Two sources of higher-order corrections are evaluated in the present chapter. These are the $\ILO{\alpha}$ corrections to \freft{fig:persistentfigures}(iv)-(v), and the $\ILO{{N_0}^{-1}}$ corrections associated with introducing additional bosons---initially the boson from the pair field, in \sref{sec:ON1}, then also the loop boson of the $\ILO{\alpha}$ correction in \sref{sec:ON1Oalpha}. Following the approach of \Psref{VI}{sec:ONtoalpha}, these %
are evaluated as a multiplicative corrections to the fundamental and $\ILO{\alpha}$ terms respectively.

\paragraph{The \prm{\ILO{{N_0}^{-1}}} corrections\label{sec:ON1}} to the fundamental diagrams of \freft{fig:persistentfigures}(iv)-(v) are corrections which arise due to more than two vertices inhabiting the same correlation region. Regarding this:
\begin{itemize}
\item It is immediate that the two background field interactions must inhabit the same correlation region, in order that the mean field term does not vanish.
\item These vertices are connected by a series of other species, which may be massless over scales small compared with $\mc{L}_0$ and must be massive over scales large compared with $\mc{L}_0$.
\item As the separation of the background vertices is at most $\ILO{\mc{L}_0}$, the dominant contributions (being those closest to on-shell propagation) necessarily involve massless particles not leaving the correlation region.
\item There therefore exists a chain of in-zone vertices connecting the two background field interactions.
\end{itemize}
Next, posit a minimal chain comprising the two background field vertices and two additional vertices. For the moment, ignore the $AGG^\dagger$ vertex. For the remaining two vertices, being fermion/$G^\bdag$ vertices, recognise that these two vertices are free to be in or out of the correlation region. However, deleting the correlation region from the path integral approximates eliding only a very small region from the manifold. The perturbations to push the vertex away from that near-pointlike region introduce infinitesimal corrections to particle paths, say with amplitude correction $\ILO{\epsilon}$, and these corrections are symmetric in space and time and may therefore be anticipated to cancel. Overall, evaluation of the diagram is dominated by contributions where these vertices are \emph{not} in the correlation region.

Finally, consider the $AGG^\dagger$ vertex. 
\begin{itemize}
\item When the adjacent vertices are not in the correlation region, nor is this vertex by the same argument about infinitesimal perturbations.
\item When they \emph{are} in the correlation region, then they are part of the chain connecting the background field interaction vertices. By the assumption that the diagram is dominated by terms close to on-shell, species in this chain are required to be massless, and the close-to-shell requirement then ensures that the $AGG^\dagger$ vertex also lies within the correlation region.
\end{itemize}

As the diagram is symmetric under interchange of which $G$ boson couples to the photon, these two scenarios carry equal weight, and thus there are on average 4.5 vertices within the correlation region, compared with two in the mass interactions of \crefs{ch:fermion}{ch:boson}.

There are, however, no factors of $\ILO{{N_0}^{-1}}$ due to these additional vertices:
\begin{itemize}
\item The preons bringing in the foreground photon momentum must be foreground in character, and so can only arise from the scalar fields external to the region whose gradients correlate with those within the region, beinging in external momentum. At this order, this choice is unique.
\item The $e_R\nu_e G^\dagger$ and $\bar e_R\bar \nu_e G$ vertices are not background, but do not bring external foreground momentum into the correlation region.
\begin{itemize}
\item They form part of the loop circulating the introduced (net zero) foreground momentum, from the inbound preon within the inbound photon to the outbound preon within the inbound photon (up to beyond-mean-field interactions with the background field which do not need to be explicitly evaluated in the far field of the source).
\item By gauge they receive no background contribution as they involve $G^\bdag$ bosons.
\item Their momentum is thus not carried in a distributed fashion across all local FSFs, but rather the preons are in 1:1 correspondence with the FSFs from which they are constructed. If a preon which would have been inbound into the upper such vertex is interchanged with a preon inbound into the lower vertex, this results in a preon tadpole diagram on one of the two vertices.
Interchange of their FSFs is therefore not permitted within the lowest-order diagrams of \freft{fig:persistentfigures}(iv)-(v).
\item The preons in the background interaction vertices are capable of participating in FSF exchange, but have no species left to exchange FSFs with other than each other, and this is already accounted for in the mean field term. There is therefore no further factor of $\ILO{{N_0}^{-1}}$ arising from these fields.
\end{itemize}
\end{itemize}
The resulting factor of $[1+\ILO{{N_0}^{-1}}]$ due to additional vertices within the correlation region in \freft{fig:persistentfigures}(iv)-(v) therefore reduces to~1.

\begin{figure}
\includegraphics[width=\linewidth]{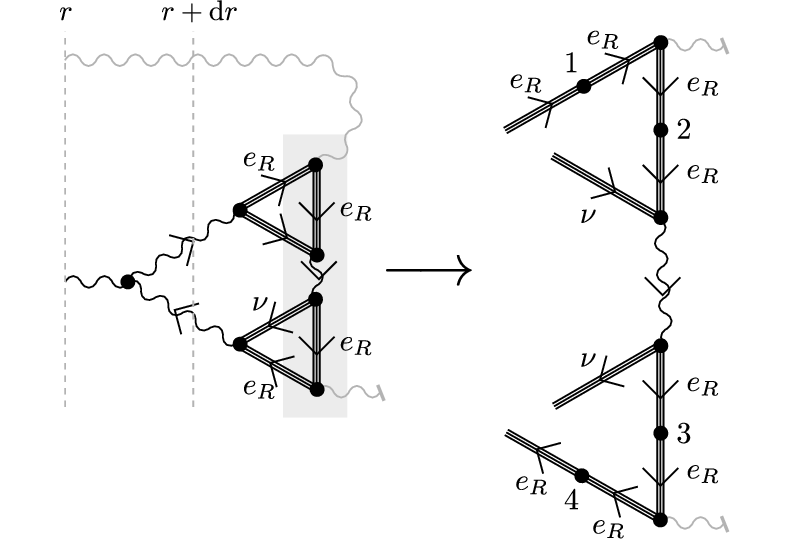}
\caption{Six diagrams give rise to \prm{\ILO{\alpha}} corrections to each of \pfreft{fig:persistentfigures}(iv)-(v). Considering \pfref{fig:persistentfigures}(iv) and magnifying the area of interest as shown, these corrections are constructed by pairwise connecting the points marked~1-4. Diagram~(i) corresponds to connection of points~1 and~2, diagram~(ii) to connection of~1 and~3, diagram~(iii) to connection of~2 and~4, diagram~(iv) to connection of~3 and~4, diagram~(v) to connection of~2 and~3, and diagram~(vi) to connection of~1 and~4. These diagrams are readily visualised from the above Figure and thus are not drawn explicitly.
An equivalent set of diagrams exists for \pfref{fig:persistentfigures}(v). Note that although the \prm{G}~bosons carry an electromagnetic charge, their coupling is heavily suppressed due to the lack of a background \prm{G}~field and thus loop corrections with vertices on the \prm{G}~bosons may be ignored.\label{fig:Oalpha}}
\end{figure}

\paragraph{The \prm{\ILO{\alpha}} corrections} to \fref{fig:persistentfigures}(iv) are given by \fref{fig:Oalpha}, where the photon loops may be viewed as generating gyromagnetic anomaly-like corrections to the interactions with background fields and/or neutrino emission processes.
Those for \fref{fig:persistentfigures}(v) are equivalent.
The first step in evaluating these diagrams is to determine which loop vertices are within the same correlation region as the two interactions with the photons from the pseudovacuum. As observed in \sref{sec:pairdecay}, at least one of the electron-family lines associated with the pseudovacuum interaction must be massless. However, the other %
is massive as per \sref{sec:ON1}. 

As also noted in \sref{sec:ON1}, there are two possible choices regarding which vertices lie within a common correlation region; reading from top to bottom in \fref{fig:Oalpha}, these are the vertices lying along the particle chains
\begin{equation}
\begin{split}
&\bgfield{A}\rightarrow e_R\rightarrow G\rightarrow e_R\rightarrow \bgfield{A}\\
&\bgfield{A}\rightarrow \bar e_R\rightarrow G^\dagger\rightarrow G^\dagger\rightarrow \bar e_R\rightarrow \bgfield{A}.
\end{split}
\end{equation}
Considering \fref{fig:Oalpha} diagram~(i), one of the two ends of the loop correction photon is therefore on a massless electron within the correlation region, and the other connects to a massive electron, doing so either within or outside the local correlation region. %
For a loop correction of energy $\mc{E}_\ell$, contributions are dominated by diagrams with loop vertex separation of $(2\mc{E}_\ell)^{-1}$ or less, as per \aref{apdx:massloops}. This may be substantially larger than $\mc{L}_0$ at low $\mc{E}_\ell$, and the second vertex of the photon thus almost universally lies outside the local correlation region.

As in \fref{fig:freeloops} it is recognised that with the neutrino being chargeless, couplings between the loop photon and an electron commute with couplings between the electron and a neutrino, up to any change in FSF symmetry factors. Thus, for example, the correction associated with \fref{fig:Oalpha} diagram~(ii) is numerically equivalent to that associated with \fref{fig:Oalpha} diagram~(i). 

With only one end of the loop photon lying within the same local correlation region as the two pseudovacuum couplings, evaluation of \freft{fig:Oalpha} diagrams~(i)-(iv) is without surprises and it follows immediately that each of is associated with a correction factor $\alpha/(2\pi)$, for a subtotal over these four diagrams of $2\alpha/\pi$. Similarly, both loop vertices in \fref{fig:Oalpha} diagram~(vi) are outside the region and thus this diagram yields a further contribution of $\alpha/(2\pi)$. In contrast, however, both vertices of the loop correction in \fref{fig:Oalpha} diagram~(v) are within the same correlation region as both pseudovacuum vertices. Following the approach described in \aref{apdx:symloops}, this diagram yields a symmetry-augmented contribution with weight $\alpha/\pi$. The total corrections due to photon loops to $\ILO{\alpha}$ thus multiply \Eref{eq:Irtree} by a factor of
\begin{equation}
\left(1+\frac{7\alpha}{2\pi}\right).
\end{equation}

\paragraph{The \prm{\ILO{{N_0}^{-1}}} corrections to the \prm{\ILO{\alpha}} corrections\label{sec:ON1Oalpha}} are conceptually similar to those of \Psref{VI}{sec:ONtoalpha}.
As already discussed, the loop corrections to \fref{fig:persistentfigures}(iv) presented in \fref{fig:Oalpha} may have between zero and two vertices within the same correlation region as the two background photons, and the same is true for the equivalent corrections to \fref{fig:persistentfigures}(v). The key question, however, is how many vertices exist within the same correlation region, and whether the preons in these vertices are capable of exchanging FSFs with the loop photon. In \sref{sec:ON1} it was established that there are on average 4.5 vertices from the original diagram within the correlation region, but for the leading-order diagram the additional 2.5 vertices were not free to participate in exchange of FSFs, so no additional factor was gained. However, this restriction no longer applies at $\ILO{\alpha{N_0}^{-1}}$:
\begin{itemize}
\item For the inbound photon from the photon pair, this is foreground in character so the FSFs from which its preons are constructed must be external to the correlation region, corresponding to arrival of an external foreground photon. The foreground loop photon extends outside the correlation region, so its associated FSFs also meet this criterion. Exchange of these FSFs is therefore permitted.
\item Note, however, that exchanging an FSF with one of the FSFs associated with the foreground loop correction also results in some of the circulating momentum entering the diagram at a different location. If this is interpreted as preon interchange, it does not yield a transformation to a different QFT diagram at the composite particle level; indeed it cannot be understood at the composite particle level unless it is reinterpreted as fluctuations in the foreground momentum. Such fluctuations are already incorporated into the emulated QFT [see \Psref{I}{sec:4momflucs}; on dropping the window approximation \Peref{I}{eq:window} this argument also extends to scales beyond $\mc{L}_0$]. Contributions arising from this interchange are therefore well-represented as corrections to \fref{fig:Oalpha}(i).
\item For the fermion vertices in the $G$~boson loop, interchange of a preon with the loop correction photon does not generate tadpole diagrams, so this also is permitted, with similar considerations regarding momentum fluctuations.
\item For the background fields, this interchange is also permitted since the momentum of a foreground {photon} (as opposed to a $G^\bdag$~boson) is carried in a distributed fashion across all FSFs in the correlation region.
\end{itemize}
Thus there are on average 4.5 vertices from the leading-order diagram which are within the local correlation region and able to participate in FSF exchange. However, two of these vertices have $G^\bdag$ bosons in place of photons. Regarding these vertices:
\begin{itemize}
\item One preon in the $G^\bdag$ boson has $A$-charge $a=3$ (or equivalently, its associated FSF has a long-range correlation on its gradient in the $a=3$ sector). This is not compatible with photon construction, so it may be ignored.
\item The other preon has a fixed $A$-charge of~2. However, the $A$-charge of a preon in the loop photon may be~1 or~2 so the odds of matching $A$-charge are $\frac{1}{2}$. This is the same as when both bosons are photons.
\item Thus in terms of symmetry factors at $\ILO{\alpha{N_0}^{-1}}$, a $G^\bdag$ vertex behaves as half a photon vertex.
\end{itemize}
This yields 3.5 as the effective number of participating photon-like vertices from the leading-order diagram, compared with two in \Psref{VI}{sec:ONtoalpha}.
It is also useful to recap that when a loop photon has a single vertex in the correlation region, there is a correction factor of
\begin{equation}
\left[1+\frac{2}{3N_0}+\OO{{N_0}^{-2}}\right]
\end{equation}
per vertex from the leading-order diagram which is within the correlation region,
which corresponds to
\begin{itemize}
\item a factor of four for the four preon lines at the vertex from the leading-order diagram,
\item weighted by the $1/6$ chance of any two preons (from photons) matching $A$-charge, $C$-charge, and species ($A/N/Z$),
\item yielding a probability of performing a single increment of an FSF count.
\item A single increment is realised through multiplication by $[1+{N_0}^{-1}+\ILO{{N_0}^{-2}}]$, and the probability weighting on the increment is realised through weighting of the ${N_0}^{-1}$ term.
\end{itemize}

Consider first \freft{fig:Oalpha}(i)-(iv), which are associated with $\ILO{\alpha}$ correction terms of weight $\alpha/(2\pi)$. The correlation region in these diagrams contains one vertex from the photon loop, and on average~3.5 vertices from the leading-order diagram. 
The resulting correction from each of these four diagrams is therefore
\begin{equation}
\frac{\alpha}{2\pi}\longrightarrow\frac{\alpha}{2\pi}\left[1+\frac{7}{2}\cdot\frac{2}{3N_0}+\OO{{N_0}^{-2}}\right].
\end{equation}

Next, consider \fref{fig:Oalpha}(v), which is associated with an $\ILO{\alpha}$ correction term of weight $\alpha/\pi$. In this diagram there are two added vertices within the correlation region, which both belong to the same photon.
Each of these vertices contains preons which may match those in the (on average) 3.5 vertices from the leading-order diagram present within the correlation region. Consider, therefore, the correction arising from matches between the vertex labelled~2 and the vertices from the leading-order diagram, ignoring the presence of vertex~3. Evaluation is as for \freft{fig:Oalpha}(i)-(iv), this time yielding a correction
\begin{equation}
\frac{\alpha}{\pi}\longrightarrow\frac{\alpha}{\pi}\left[1+\frac{7}{2}\cdot\frac{2}{3N_0}+\OO{{N_0}^{-2}}\right].
\end{equation}
An identical factor arises from matches between vertex~3 and the vertices from the leading-order diagram.

\begin{figure}
\includegraphics[width=\linewidth]{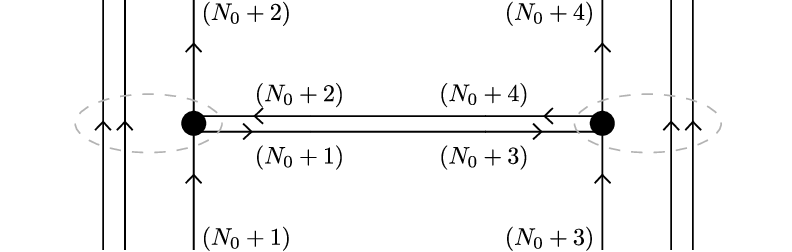}
\caption{FSF symmetry factors associated with photon emission and absorption when the entire process is contained within a single correlation region.\label{fig:fermionsymfactors_close}}
\end{figure}

Finally, consider matches between vertices~2 and~3. In treating this as a standard electromagnetic loop correction, it has implicitly been assumed that the FSF symmetry factors on the preons of these vertices are as shown in \fref{fig:fermionsymfactors},
\begin{equation}
(N_0+2)^2(N_0+1)^5(N_0+\tfrac{5}{4}). 
\end{equation}
This is further multiplied by a factor of $(N_0+1)^{-2}$ which arises from ensuring connection of the external photon lines in diagrams~(i) and~(ii), as described in \Psref{III}{sec:EWint_sym}.
However, this factor is only correct when the two vertices are separated by a distance or time large compared with $\mc{L}_0$. In the present context, the correct factors are shown in \fref{fig:fermionsymfactors_close} with the factor of $(N_0+1)^{-2}$ remaining unchanged.
The resulting FSFs differ by a factor of
\begin{equation}
\left(1+\frac{20-10\frac{1}{4}}{N_0}\right),
\end{equation}
again applied to the $\OO{\alpha}$ term of $\alpha/\pi$ arising from \fref{fig:Oalpha}(v). Putting together all effects arising from \fref{fig:Oalpha}(v) yields
\begin{equation}
\begin{split}
\frac{\alpha}{\pi}\longrightarrow\,&\frac{\alpha}{\pi}\left[1+2\cdot\frac{7}{2}\cdot\frac{2}{3N_0}+\frac{39}{4N_0}+\OO{{N_0}^{-2}}\right]\\
&=\frac{\alpha}{\pi}\left[1+\frac{173}{12N_0}+\OO{{N_0}^{-2}}\right].
\end{split}
\end{equation}

There are no $\ILO{{N_0}^{-1}}$ corrections to \fref{fig:Oalpha}(vi).

Finally, putting together all six $\ILO{\alpha}$ corrections and the $\ILO{{N_0}^{-1}}$ corrections to these corrections yields the amendment
\begin{equation}
\left(1+\frac{7\alpha}{2\pi}\right)\longrightarrow\left[1+\frac{7\alpha}{2\pi}+\frac{229\alpha}{12\pi N_0}+\OO{\frac{\alpha}{\pi{N_0}^{2}}}\right].
\end{equation}

\subsubsection{Elimination of \prm{GG^\dagger} pairs\label{sec:elimGGpairs}}

To eliminate the formation of $GG^\dagger$ pairs, it suffices to identify a choice of co-ordinate frame for which the radial profile of the photon pair field is returned to inverse-square. This replaces couplings between the $AA$ and $GG^\dagger$ fields with interactions between the photon pair field and space--time curvature. As the photon pair field has recovered an inverse-square profile, it becomes undetectable (as it no longer mediates any direct force transfer) and the resulting space--time curvature may be associated with the neutral body which acts as the source of the field. Under mapping $\G$ from $M\subset\Cw{18}$ to a manifold $\G(M)$ which is locally Minkowski, this is a true curvature on the target manifold, and the $\SLTC$ connection on $\Cw{18}$ maps to the space--time connection on $\G(M)$, appearing appropriately in the covariant derivative operator and thus acting on all (quasi)particles appearing on $\G(M)$.

Note that with the resultant $AA$ field exhibiting inverse-square propagation, although the $AA$ fields prior to gauging determine the curvature of the resulting space--time, after gauging they do not couple to it. Thus the implied graviton is massless. %

To realise the requisite co-ordinate frame, and thus identify the target space--time of mapping $\G(M)$, recognise that the total decay process of \Eref{eq:|I|form} resembles that of \Eref{eq:geometricdecay} with the radial propagation interval %
rescaled as
\begin{equation}
\rmd r\longrightarrow \rmd r\left[1+\la|I(r)|\ra\right].
\end{equation}
As this effect affects \emph{only} radial propagation, it is akin to evaluating purely geometric propagation on the space--time whose metric is obtained from that of Minkowski space--time by substituting
\begin{equation}
\rmd r^2\longrightarrow \rmd r^2\left[1+2\la|I(r)|\ra\right].
\end{equation}
Recalling that the $\mbb{R}^+$ component of the $\GL{18}{C}$ symmetry of the $\Cw{18}$ model was gauged in \Psref{VI}{sec:gss}, the volume form must be conserved, giving
\begin{equation}
\rmd t^2\longrightarrow \rmd t^2\left\{1-2\la|I(r)|\ra+\OOO{\la|I(r)|\ra^2}\right\}.
\end{equation}
and this serves to uniquely identify the target space--time as one which supports the Schwarzschild metric,
\begin{align}
\rmd s^2 =\,& -\left(1+\frac{2G_NM}{r}\right)^{-1}\rmd t^2+\left(1+\frac{2G_NM}{r}\right)\rmd r^2 %
+ r^2\rmd\Omega,\label{eq:ds2}
\end{align}
with
\begin{align}
\begin{split}
G_N%
:=\,&\frac{f^7{N_0}^2 S_G S_\alpha}{64\sqrt{2}\pi\left[k^{(e)}_1(\mc{E}_e)\right]^6 \!{\omega_0}m_e}\left(1+\frac{7\alpha}{2\pi}+\frac{229\alpha}{12\pi N_0}\right)\left\{1+\frac{281}{36\big[k^{(e)}_{1}(\mc{E}_e)\,N_0\big]^4}\right\}^{-1}\\
&\times\bm{\left[}1+\OO{\frac{\alpha}{\pi{N_0}^2}}+\OO{\frac{\alpha^2}{\pi^2}}+\OOOO{\frac{\alpha}{\pi \big[k^{(e)}_{1}(\mc{E}_e)\,N_0\big]^4}}\bm{\right]}. 
\end{split}\label{eq:fGN}
\end{align}
Taking into account the definitions of $N_0$, $f$, and $\omega_0$
as per \PEreft{VI}{eq:N04}, \Peref{VI}{eq:calcf}, and~\Peref{VI}{eq:calcomega0} 
this may be solved alongside \PErefr{VI}{eq:eieratio}{eq:Oe3} (\sref{sec:massrelationships}) 
to yield %
\begin{equation}
G_N=6.67426(230)\times 10^{-11}~\mrm{m}^3\mrm{kg}^{-1}\mrm{s}^{-2},\label{eq:ValueOfGN} %
\end{equation}
with the central value differing from observation at the level of $0.3\,\sigma_\mrm{exp}$. %

\subsubsection{Some observations on the calculation\label{sec:observations}}

It is worth commenting on a couple of unusual aspects of the above calculation.
\begin{itemize}
\item Regarding evaluation of the derivative operator at the $AGG^\dagger$ vertices in the numerator of \fref{fig:Iratio}: If the diagrams in the numerator were a constructive process, the constructed field would vanish as the value of this operator is summed randomly over sign in the classical limit. However, this process is instead \emph{destructive} of photon pairs and thus magnitude---not signed value---determines the efficacy of this process at annihilating $AA$ pairs. The efficacy of this process at constructing novanishing fields of any species in the classical limit (which requires correlated sign) is irrelevant to this effect.
\item How can \fref{fig:Iratio} yield valid expressions when the diagrams in the numerator of \fref{fig:Iratio} both appear to be tadpole diagrams in the foreground fields? In truth they are not tadpole diagrams at all---%
in each of these diagrams both the lower inbound photon and the outbound photon are summed over the total photon field (both background and foreground). By gauge choice~\eref{eq:bga45gauge} the background contribution from the lower inbound field vanishes everywhere, and hence this boson is schematically represented as foreground.\footnote{To see this, recall that by gauge choice~\peref{eq:bga45gauge} the \prm{G} and \prm{G^\dagger} bosons must be entirely foreground in character. They therefore cannot absorb or emit a photon field with correlations of background character as this would require either or both of the \prm{G} and \prm{G^\dagger} fields to likewise acquire correlations of background character in violation of the gauge condition.} Regarding the outbound photon, at the microscopic level this photon is nonvanishing and thus incorporates both background and foreground terms, such that the diagram as a whole is not a tadpole diagram. Further, the single-photon field associated with any specific pair decay process may in general be nonvanishing. However, for any neutral source composed of positive and negative charges, the net (single-photon) $A$~field necessarily vanishes in the far-field regime and this remains true even in the context of processes such as \fref{fig:Iratio}.\footnote{This is not a problem---the diagram may annihilate excitations of the photon pair field without needing to generate any net onward-propagating foreground particle field, in effect by recombining positive and negative energy/momentum components of that field to construct a net closed loop.} Given the further assumption that the foreground fields are small compared with $\mc{E}_0$, the mean field contributions illustrated in the numerator of \fref{fig:Iratio} dominate in the evaluation of the all-field contributions, and thus for numerical reasons it is convenient to represent the output photon as background in character, even though pair by pair it may retain a nonvanishing foreground component.
\item Can the upper inbound photon be foreground instead of the lower inbound photon? Yes---this is accounted for by numerical boson exchange factors. Can both be foreground? Yes, but this results in a factor scaling as $r^{-2}$ instead of $r^{-1}$ and thus is not the dominant process at large~$r$.
\item Finally, are there corresponding diagrams in which the outbound photon arises from the $AGG^\dagger$ vertex? Yes, but for these diagrams to eliminate $AA$~pairs, at least one of the inbound photons must be foreground. (Terms in which both are background correspond to interactions of background photon pairs, and are eliminated by the normalisation convention of \sref{sec:normWrtBgFields}.) As already remarked, by gauge choice~\eref{eq:bga45gauge} the photon arising from the $AGG^\dagger$ vertex must also be foreground, and thus such a diagram also scales at best as $r^{-2}$.
\end{itemize}

\subsubsection{Rotating source}

Extension to a rotating source is not considered explicitly in terms of photon pair decay in this chapter. However, noting that the argument deriving the Kerr metric is a classical one~\cite{kerr1963}, %
requiring only the Schwarzschild metric and a rotating frame of reference, its extension to the $\Cw{18}$ model is expected to be immediate within the regime of validity of \Eref{eq:ds2}, this corresponding to the classical regime and a source sufficiently close to rest in the isotropy frame of the $\Cw{18}$ model, and excluding high-curvature regimes in which it would be necessary to also take into account %
corrections of higher order in $r$ such as \freft{fig:persistentfigures}(i)-(ii), or in which the assumption that the foreground field is small is anticipated to break down. 

\subsection{Beyond the dominant process\label{sec:moreprocesses}}

\subsubsection{Other spin-2 processes\label{sec:spin2processes}}

The above discussion has centred around \fref{fig:persistentfigures}(iii) and its $\ILO{\alpha}$ corrections, this being the dominant beyond-Standard-Model spin-2 interaction. %
However, it extends directly to other beyond-Standard-Model spin-2 interactions, in which the relevant spin-2 composite boson field is mapped to a further correction to curvature such that the effect of the beyond-Standard-Model effect vanishes on the foreground field counterparts of the Standard Model and is replaced by the effect of a curved space--time. %

Understanding how all spin-2 beyond-Standard-Model processes correct photon pair propagation in the flat-space gauge thus permits more precise calculation of the geometry of space--time in the curved-space (physical) gauge, 
with photon pair propagation then being purely geometric on that curved space--time in keeping with the expectations of the Standard Model (supplemented by dark processes involving boson~$N_\mu$). %

The further extension to absorb beyond-Standard-Model spin-1 processes into the space--time curvature is discussed in \sref{sec:Rwnf}.

\subsubsection{The right weak interaction and other spin-1 effects\label{sec:Rwnf}}

While the above process may readily be extended to eliminate all spin-2 beyond-Standard-Model processes, further detail is required regarding spin-1 processes such as the right-handed weak interaction. Processes such as
\begin{equation}
\nu_e\longrightarrow \bar e_R+G
\end{equation}
may create single $G^\bdag$ bosons, which do not immediately admit a mapping to space--time curvature.

The solution is a technique previously encountered in \Psref{III}{sec:EWint_Wintdetail} and \Psref{IV}{sec:Csector} in which a process serves as a point of reference for charge on a sector. In \Psref{IV}{sec:Csector}, the identification of preon triplets in fermions as being the reference point for colour neutrality induced an active co-ordinate transformation which gave rise to the $K$~matrices. Similarly, now let it be a point of definition that all $G^\bdag$ bosons appear in $GG^\dagger$ pairs. The only other comparable imposition of an active co-ordinate transformation on the $A$~sector is the introduction of co-ordinate sleeves for $W$~boson interactions in \Psref{III}{sec:EWint_Wintdetail}, and the processes which give rise to these transformations are independent and may both be applied as required without conflict. Every $G$~boson emission is thus accompanied by a $G^\dagger$~boson emission, and the resulting pair may be immediately mapped to space--time curvature, completing the elimination of %
$G^\bdag$~bosons from the $\Cw{18}$~model.

It is tempting---but inappropriate---to attempt to also apply this approach to eliminate the neutral $N$ boson. The $N$ boson does not change the species of particles it interacts with, cannot couple to fermions, and is not a significant vector force mediator in its own right (though it may exhibit some weak couplings to the scalar boson). 
As a weakly-interacting species it is relatively stable, and thus the primary effect of its massive nature is anticipated to be as a gravitational source in its own right, not as a correction to its (negligible) effects on photon pair decay. A co-ordinate sleeve cannot be applied to such a process. For this reason, the $N$~boson is assumed not to be mapped to space--time curvature, and is predicted to be discoverable with a first generation inertial mass of $80.3810(22)~\GeV/c^2$ %
and a gravitational mass of $1381.486(37)~m_e$, equivalent to that of fermionic matter totalling $705.938(19)~\MeV/c^2$. %

\section{Results}

\subsection{Value of Newton's constant\label{sec:valueGN}}

As noted in \sref{sec:curvature}, elimination of the $G^\bdag$~bosons from the $\Cw{18}$ model induces the Schwarzschild metric
\begin{align}
\rmd s^2 =\,& -\left(1+\frac{2G_NM}{c^2 r}\right)^{-1}\rmd t^2+\left(1+\frac{2G_NM}{c^2 r}\right)\rmd r^2 %
+ r^2\rmd\Omega\tagref{eq:ds2}
\end{align}
on the target manifold of mapping $\G$ which acts on $M\subset\Cw{18}$. Further, the value of $G_N$ is fixed to be
\begin{align}
\begin{split}
G_N:=\,&\frac{f^7{N_0}^2 S_G S_\alpha}{64\sqrt{2}\pi\left[k^{(e)}_1(\mc{E}_e)\right]^6 \!{\omega_0}m_e}\left(1+\frac{7\alpha}{2\pi}+\frac{229\alpha}{12\pi N_0}\right)\left\{1+\frac{281}{36\big[k^{(e)}_{1}(\mc{E}_e)\,N_0\big]^4}\right\}^{-1}\\
&\times\bm{\left[}1+\OO{\frac{\alpha}{\pi{N_0}^2}}+\OO{\frac{\alpha^2}{\pi^2}}+\OOOO{\frac{\alpha}{\pi \big[k^{(e)}_{1}(\mc{E}_e)\,N_0\big]^4}}\bm{\right]}. 
\end{split}\tagref{eq:fGN}
\end{align}
Substituting 
as per \PEreft{VI}{eq:N04}, \Peref{VI}{eq:calcf}, and~\Peref{VI}{eq:calcomega0} 
and making factors of $c$ and $h$ explicit yields
\begin{align}
G_N=\,&\frac{\alpha^4hc\left[1+\Delta_e(m_e,\mc{E}_e)\right]^\frac{1}{2}%
{S_{18,147}}%
^\frac{1}{2}S_G}{8\pi{N_0}^{18}\left[k^{(e)}_1(\mc{E}_e)\right]^4m_e^2(1+a_e)^8{S_\alpha}^3}\nn\\
&\times\left(1+\frac{7\alpha}{2\pi}+\frac{229\alpha}{12\pi N_0}\right)\left\{1+\frac{281}{36\big[k^{(e)}_{1}(\mc{E}_e)\,N_0\big]^4}\right\}^{-1}\label{eq:alphaGN}\\
&\times\left[1+\mc{O}_b+\mc{O}_e(m_e,\mc{E}_e)+\OO{\frac{\alpha^2 m_e^2}{m_\mu^2}}\right]\nn\\
S_{18,147}:=\,&{N_0}^{-12}(N_0+2)^6(N_0+1)^6\Ptagref{VI}{eq:redefS18147}\\
S_\alpha:=\,&{N_0}^{-6}{(N_0+\tfrac{5}{4})(N_0+1)^3}{(N_0+2)^2}\Ptagref{III}{eq:defSalpha}
\end{align}
\begin{align}
\begin{split}
S_G:=16(&N_0+2)^2(N_0+1)^2{N_0}^{12}\\
\times\{&2[(N_0+4)(N_0+3)(N_0+2)(N_0+1)]^4\\
&+8[(N_0+3)(N_0+2)(N_0+1)^2]^4\\
&+6[(N_0+2)(N_0+1)]^8\}^{-1}
\end{split}
\end{align}
where $a_e$ is the gyromagnetic anomaly of the electron in the Standard Model, evaluated here to $\ILO{\alpha^2}$ \cite{petermann1958,sommerfield1958,aoyama2019}:
\begin{equation}
\begin{split}
a_e&=\frac{\alpha}{2\pi}+\left\{\frac{197}{144}+\frac{\pi^2[1-6\,\mrm{ln}(2)]}{12}+\frac{3\,\zeta(3)}{4}\right\}\frac{\alpha^2}{\pi^2}\\
&\,\p{\equiv}\,+\OO{\frac{\alpha^2m_e^2}{m_\mu^2}}\\
&\approx\frac{\alpha}{2\pi}-0.328478965579193\,\frac{\alpha^2}{\pi^2}+\OO{\frac{\alpha^2m_e^2}{m_\mu^2}}.
\end{split}\Ptagref{VI}{eq:approxae}
\end{equation}
The above equations %
fully define $G_N$ in terms of the parameterisation of \PErefr{VI}{eq:eieratio}{eq:Oe3}. Note that the error terms $\mc{O}_b$ and $\mc{O}_e(m_e,\mc{E}_e)$ incorporate error terms $\E{1}$-$\E{17}$ of \cref{ch:detail}, and also the error terms appearing in \Eref{eq:fGN} in the present chapter, which are designated $\E{22}$-$\E{24}$ respectively. The term $\ILO{\alpha^2 m_e^2/m_\mu^2}$ corresponds to error term $\E{21}$ in \aref{apdx:accessory}, and arises from truncation of $a_e$ as per \PEref{VI}{eq:approxae}. This term need not be written explicitly, as it is small compared with term $\ILO{\alpha^2/\pi^2}$ in $\mc{O}_b$, but retaining it ensures that it will not be overlooked when subsequent higher-order calculations reduce the magnitude of the error terms in $\mc{O}_b$.

Evaluating \Eref{eq:alphaGN} simultaneously with \PErefr{VI}{eq:eieratio}{eq:Oe3} yields
\begin{equation}
G_N=6.67426(230)\times 10^{-11}~\mrm{m}^3\mrm{kg}^{-1}\mrm{s}^{-2}.\tagref{eq:ValueOfGN} %
\end{equation}
The sources of uncertainty in this result include all those listed in \tref{tab:VI:errorlist}, $\E{21}$ from \aref{apdx:accessory}, plus also those listed in \tref{tab:VII:errorlist} of the present chapter. Their magnitudes are tabulated in \tref{tab:VII:erroreffects}. 
\begin{table}
\caption{\label{tab:VII:errorlist}List of supplementary sources of uncertainty in the calculation of \prm{G_N}~\peref{eq:alphaGN}.}
\begin{center}
\begin{tabular}{c|l}
Label~ & \multicolumn{1}{c}{Description}\\\hline\hline
\E{22} &~ $\ILOOOO{\alpha/\{\pi \big[k^{(e)}_{1}(\mc{E}_e)\,N_0\big]^4\}}$ term in \Eref{eq:fGN}\\
\E{23} &~ $\ILOO{\alpha/(\pi{N_0}^2)}$ term in \Eref{eq:fGN}\\
\E{24} &~ $\ILO{\alpha^2/\pi^2}$ term in \Eref{eq:fGN}
\end{tabular}
\end{center}
\end{table}
\begin{table}
\caption{\label{tab:VII:erroreffects}Contributions of different sources of error to the value of \prm{G_N}, expressed both in units of \prm{\mrm{m}^3\mrm{kg}^{-1}\mrm{s}^{-2}} and relative to the experimental error. Labels are enumerated in \ptref{tab:VII:errorlist}, and in \ptref{tab:VI:errorlist} and \paref{apdx:accessory}. For purposes of summation, sources of error are assumed independent.}
~\\
\begin{center}
\begin{tabular}{c|c|c|r}
Label&Coefficient &\multicolumn{2}{c}{Uncertainty in $G_N$}\\
\cline{3-4}
	 &&$10^{-14}~\mrm{m}^3\mrm{kg}^{-1}\mrm{s}^{-2}$		&\multicolumn{1}{c}{$\sigma_\mrm{exp}$}\\
\hline\hline %
\E{1}  & $\pm10$ & 0.19 & 1.26\\
\E{2}  & $\pm10$ & 1.58 & 10.51\\
\E{3}  & $\pm10$ & 0.00 & 0.00\\
\E{4}  & $\pm10$ & 0.00 & 0.00\\
\E{5}  & $\pm10$ & 0.00 & 0.00\\
\E{6}  & $\pm10$ & 0.00 & 0.00\\
\E{7}  & $\pm10$ & 0.19 & 1.26\\
\E{8}  & $\pm10$ & 1.58 & 10.54\\
\E{9}  & $\pm10$ & 0.02 & 0.12\\
\E{10} & $\pm10$ & 0.01 & 0.05\\
\E{11} & $\pm10$ & 0.00 & 0.00\\
\E{12} & $\pm10$ & 0.30 & 2.02\\
\E{13} & $\pm10$ & 0.02 & 0.10\\
\E{14} & $\pm10$ & 0.01 & 0.09\\
\E{15} & $\pm10$ & 0.03 & 0.22\\
\E{16} & $\pm10$ & 0.00 & 0.01\\
\E{17} & 1       & 0.00 & 0.00\\
\E{18} & $\pm1$  & 0.00 & 0.00\\
\E{19} & $\pm1$  & 0.00 & 0.00\\
\E{20} & $\pm1$  & 0.00 & 0.01\\
\E{21} & $\pm10$ & 0.00 & 0.00\\
\E{22} & $\pm10$ & 0.04 & 0.29\\
\E{23} & $\pm10$ & 0.00 & 0.03\\
\E{24} & $\pm10$ & 0.36 & 2.40\\
\hline
Total &  & 2.30 & 15.32
\end{tabular}
\end{center}
\end{table}
As in %
\sref{sec:errors}, rather than attempting to quantify the magnitudes of the higher-order corrections these have been given a generous coefficient of $\pm10$ with the intent %
of allowing for up-to-ninefold colour-related degeneracies, thus avoiding underestimation of uncertainty in the resulting figure.
The largest terms in the uncertainty results from contributions~\E{2} and~\E{8}, 
which correspond to two-loop corrections to some of the boson mass diagrams of \crefs{ch:boson}{ch:detail}.

The value of Newton's constant thus obtained is presently less precise than experimental results, but the central calculated value
is nevertheless in agreement with observation at the level of $0.3\,\sigma_\mrm{exp}$.%

\subsection{Qualitative implications\label{sec:qual}}

It is interesting to examine some key qualitative properties of the gravitational mechanism of the $\Cw{18}$ model. First, the extent to which the model truly realises a space--time curvature. The underlying $\Cw{18}$ manifold remains flat, but the emulated manifold with real co-ordinates, $\G(M)$, has a true curvature as reflected by metric~\eref{eq:ds2}. This is the same curvature as is exhibited by the $\SLTC$ subgroup of the physically motivated co-ordinate frame on $\Cw{18}$. In one sense it might therefore be argued that the underlying manifold is flat, with metric $\varepsilon^{\alpha\beta}$. In another, to the extent that there exists a space--time inhabited by the particles of the Standard Model, this space--time is curved. This is conceptually not dissimilar to the widely-used technique of treating the metric or vierbein/vielbein as a field over a Minkowski space--time, with the particles inhabiting that space--time then coupling to the indices of the metric/vielbein rather than that of the Minkowski space--time itself. The only significant difference here is in the choice of underlying manifold (and hence the construction consequently required to generate emergent normalisable quasiparticles on that manifold). 

Given this fairly robust footing for interpreting the $\SLTC$ co-ordinate frame in the physical gauge as a true curvature of the emergent space--time, at a minimum up to energy scales of $\frac{1}{2}\mc{E}_\Omega\sim 3.1~\TeV$ %
and potentially beyond, the gravitational interaction of bosons is of particular interest. All species of massive boson in the $\Cw{18}$ model break the weak principle of equivalence, which may variously be stated as the bosons having different value(s) of Newton's constant to the fermions, or the bosons having discrepancies between their gravitational and inertial masses. This is of particular significance given the existence of the neutral vector boson $N_\mu$, which is colourless, and chargeless with respect to both electromagnetic and weak interactions, but is nevertheless capable of participating in gravitational interactions, albeit with an effective gravitational mass substantially smaller than %
its inertial mass. It also interacts (weakly) with the scalar boson. With an inertial mass of $80.3810(22)~\GeV/c^2$ (see \tref{tab:VI:results}) %
and a gravitational mass only $1381.486(37)~m_e$ (where $m_e$ is the gravitational mass of the electron), equivalent to $705.938(19)~\MeV/c^2$ of fermionic matter, it is a candidate for an unusual form of WIMP dark matter. %

\section{Conclusion\label{sec:conclusions}}

This chapter has extended the $\Cw{18}$ model to support the emulation of Standard Model interactions on curved space-times which are only locally Minkowski. This extension to curved space--times is henceforth %
referred to as
the ``$\Cw{18}$ Analogue to the Standard Model In pseudo-Riemannian space--time'' (CASMIR).

CASMIR has provided a further demonstration of the predictive ability of the underlying $\Cw{18}$ model, this time permitting first-principles calculation of the value of Newton's constant. This comes in addition to the previous derivation of the masses of the tau particle and the $W$, $Z$, and Higgs bosons presented in \cref{ch:detail}, all being obtained within $0.2\,\sigma%
$ or better of observed values. In all of these calculations the only input parameters are the fine structure constant and the masses of the electron and muon, %
with no other tuning parameters.

Further testing of CASMIR may comprise further detailed numerical predictions for parameters already observed, and also for parameters and particle species not yet observed, such as the mass of the second generation $W$~boson anticipated to appear at $16.61320(46)~\TeV/c^2$ (and, indeed, its heavy $Z$ and Higgs boson companions at $18.84673(73)~\TeV/c^2$ and $25.8643(10)~\TeV/c^2$ respectively). %
The calculation of known parameters will allow further determination %
of the regime of validity of the model, while the the calculation of as-yet-unknown physical parameters (such as the mass of the $W_2$ boson) will provide for direct experimental tests of its predictive power in previously unexplored domains. 

It should be noted that the lack of any tuning parameters indicates that these predictions are already implicit in the model to arbitrarily high precision, and it is now %
simply necessary to perform the requisite calculations to obtain high-precision numerical predictions. %

\notchap{
\section*{Acknowledgements}
This research was supported in part by the Perimeter Institute for Theoretical Physics.
Research at the Perimeter Institute is supported by the Government of Canada through Industry Canada and by the Province of Ontario through the Ministry of Research and Innovation.
The author thanks the Ontario Ministry of Research and Innovation Early Researcher Awards (ER09-06-073) for financial support.
This project was supported in part through the Macquarie University Research Fellowship scheme.
This research was supported in part by the ARC Centre of Excellence in Engineered Quantum Systems (EQuS), Project No.~CE110001013.

In addition to the usual financial acknowledgements, the author would particularly like to thank the following individuals, without whom this work would have been far more difficult: Prof.~Y.~Koide for recognising a relationship between the lepton masses, and %
exploring the mechanisms which underly this remarkable property. 
Dr.~A.~N.~Chantler for the confidence to believe that a better model could be achieved.
And finally, D.~Di~Mario for a remarkable numerical relationship which is probably coincidental, but which was nevertheless %
a %
vital
beacon along the way.
}

\chapter[Further Calculations: $W$~mass in CDF~II, and Muon~$g-2$]{Calculations beyond the Standard Model: $W$~mass in CDF~II, and Muon~$g-2$\label{ch:CDF2}}

See \href{https://arxiv.org/abs/2212.01255}{arXiv:2212.01255} \cite{pfeifer2022CASM5}.

\chapter{Leptoquarks and LHCb\label{ch:LHCb}}
\begin{abstract}
The Classical Analogue to the Standard Model In pseudo-Riemannian space-time (CASMIR) is an analogue model which reproduces the quantum field theory of the Standard Model, and the curved space--time of general relativity, in appropriate limits. It has a strong beyond-Standard-Model track record, predicting the anomalous value of the $W$~boson mass measured by CDF~II, and the value of the muon gyromagnetic anomaly measured at Fermilab Muon $g-2$, with tensions $0.1\,\sigma$ and $0.2\,\sigma$ respectively\notstandalone{ (\cref{ch:CDF2})}. This \paper{} applies CASMIR to beauty quark decay in LHCb. CASMIR predicts the existence of effective leptoquarks with lifetimes small compared to $(3.09~\TeV)^{-1}$. The interactions mediated by these species do not violate lepton universality.
\end{abstract} %

\section{Introduction}

The Classical Analogue to the Standard Model In pseudo-Riemannian space-time (CASMIR)\standalone{~\cite{pfeifer2022CASM1,pfeifer2022CASM2,pfeifer2022CASM3,pfeifer2022CASM4}} is an analogue model which reproduces the quantum field theory of the Standard Model, and the curved space--time of general relativity, in appropriate limits. It correctly predicts the anomalous value of the $W$~boson mass measured by CDF~II, and the value of the muon gyromagnetic anomaly measured at Fermilab Muon $g-2$, with tensions $0.1\,\sigma$ and $0.2\,\sigma$ (upper bound $0.5\,\sigma$) respectively (compared with the Standard Model tensions of $6.9\,\sigma$ and $5.2\,\sigma$ respectively). %
With these encouraging results, %
it is natural to ask whether CASMIR may also be productively applied to the analysis of beauty quark decay.

\begin{figure}
\includegraphics[width=\linewidth]{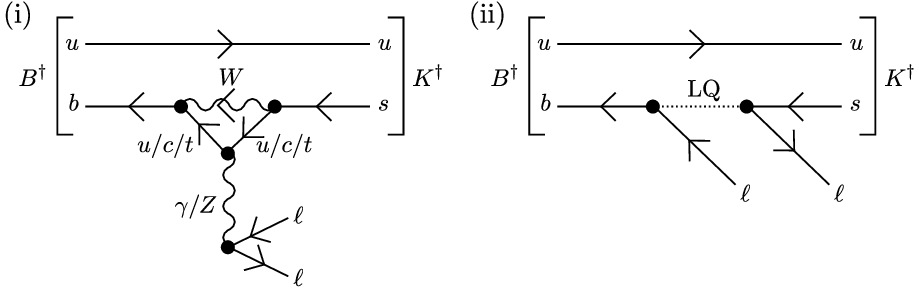}
\caption{(i)~Dominant Standard Model process for decay of \prm{B^+} mesons to \prm{K^+} mesons by lepton pair emission. The Standard Model does not predict deviation from lepton universality. (b)~Proposed tree-level leptoquark channel for \prm{\BK} decay. If different generations of leptons exhibit different coupling constants to the leptoquark, this breaks lepton universality, allowing for different rates of electron and muon pair production.\label{fig:BKdecay}}
\end{figure}

In the Standard Model the nonresonant decay of the $B^+$ meson to the $K^+$ meson (i.e.~excluding $B^+\!\rightarrow\! J/\psi \,K^+$) is prohibited at tree level by symmetry constraints, re-emerging only at the 1-loop level as shown in \fref{fig:BKdecay}(i).
This process respects lepton universality, such that electron and muon pairs from non-$J/\psi$ decay processes are produced in equal numbers (if muon and electron masses are assumed vanishingly small compared with the mass of the beauty quark).
In contrast, extensions to the Standard Model may break lepton universality and result in pair production rates with ratios other than unity. Leptoquarks are one such proposal, and introduce a tree-level diagram mediating $\BK$ decay as shown in \fref{fig:BKdecay}(ii). For appropriate coupling coefficients, this diagram may dominate over \fref{fig:BKdecay}(i) and may yield pair production ratios dependent on the ratio of the leptoquark:muon and leptoquark:electron couplings, which are not \emph{a priori} required to coincide. 

In this \paper{}, the high-energy regime of CASMIR is shown to include leptoquark-like interactions with lifetimes small compared with $(3.09~\TeV)^{-1}$. The interactions mediated by these leptoquarks are generation-independent, and thus are seen not to violate lepton universality.
This analysis provides a useful introduction to the mechanism of generation change in CASMIR.

In what follows, for simplicity all electroweak bosons are assumed achromatic, and all references to $\BK$ decay are implicitly taken to exclude the $B^+\!~\rightarrow\!J/\psi \,K^+$ channel.

\section{\prm{\BK} decay in CASMIR}

\subsection{Fermions in \prm{\BK} decay}

In CASMIR, fermions and bosons are constructed from spin-$\frac{1}{2}$ preon with dimension $L^{-\frac{1}{2}}$. Fermions comprise three such preons, and bosons two, with allowed combinations being restricted by specifying characteristics of preferred co-ordinate frames on the underlying model---these map into symmetry-breaking choices of gauge on the effective field theory of the intermediate regime, yielding the particle spectrum of the Standard Model in the low-energy regime~\notchap{\cite{pfeifer2022CASM3}}\chap{(\cref{ch:SM})}. The assumption that phenomena at higher-energy regimes may be explained in terms of the species observed at low-energy regimes then extends the emergent Standard Model particle spectrum to all length scales.

Preons carry charges with respect to two copies of $\SU{3}$ in the intermediate regime, denoted $\SU{3}_A$ and $\SU{3}_C$. Triplets of preons with identical $A$-charge and distinct $C$-charges make up the leptons, with relative phases determining the particle generation. Left-handed preon triplets $\psi^{ar}\psi^{ag}\psi^{ab}$ having a first-generation relative phase configuration are identified with the left-handed Weyl spinors $[\bar{e}_R,e_L,\nu_e]$ for $a$ taking values $[1,2,3]$ respectively, transforming in the usual way under hermitian conjugation. It is therefore convenient to denote the left-handed preons by $[\bar{e}_R^c,e_L^c,\nu_e^c]$ reflecting the lowest-generation lepton they may construct when assembled in groupings homogeneous in $A$-charge.

Quarks are similarly constructed from triplets of preons, this time inhomogeneous in $A$-charge. A quark continues to contain one preon of each colour, with its constituents being bound by the colour interaction, but the differing mass interaction of $\nu_e^c$ compared with $e_L^c$ and $\bar{e}_R^c$ results in $\nu_e^c$ preons being less tightly bound than their $e_X^c$ counterparts and exhibiting a wider distribution around their common centre of mass. In conjunction with colour shielding this results in an effective residual colour charge when the quark is observed from a distance large compared with the preon binding scale.

This paper explicitly addresses $\BK$ decay where the $B^+$ meson contains an anti-[left-handed beauty quark] denoted $\bar{b}_L$. A similar approach may be extended to $\bar{b}_R$, though with $G^\bdag$ bosons in place of $W^\bdag$ bosons in the CASMIR flat-space gauge, the coupling strength is anticipated to be much reduced. %

The preon structures associated with the anti-[left-handed beauty/strange quarks], henceforth $\bar{b}_L$ and $\bar{s}_L$, are 
\begin{equation}
\bar{\nu}_e^{c_1}\bar{\nu}_e^{c_2}\bar{e}_L^{c_3}, \quad\{c_1,c_2,c_3\}\in P\{r,g,b\}.
\end{equation}
The quarks $d_L$, $s_L$, and $b_L$ form a generation triplet, with the three preons occupying an eigenstate of the colour-mediated binding interaction. These eigenstates correpond to eigenvectors of a matrix $K_{d_L}$ which mediates the binding interaction, with eigenvalues $k_1^{(d_L)}$, $k_2^{(d_L)}$, $k_3^{(d_L)}$ in ascending order. Up to a global phase and a freedom of sign on $\rmi$ these eigenstates may be denoted
\begin{align}
\frac{1}{\sqrt{3}}\ctriplet{1}{1}{1},\quad~~\frac{1}{\sqrt{3}}\ctriplet{e^{\frac{\pi\rmi}{3}}}{e^{\frac{-\pi\rmi}{3}}}{1},\quad~~\frac{1}{\sqrt{3}}\ctriplet{1}{e^{\frac{2\pi\rmi}{3}}}{e^{\frac{-2\pi\rmi}{3}}}.
\end{align}

\subsection{Leptoquark processes in \prm{\BK} decay\label{sec:LQBK}}

\begin{figure}
\includegraphics[width=\linewidth]{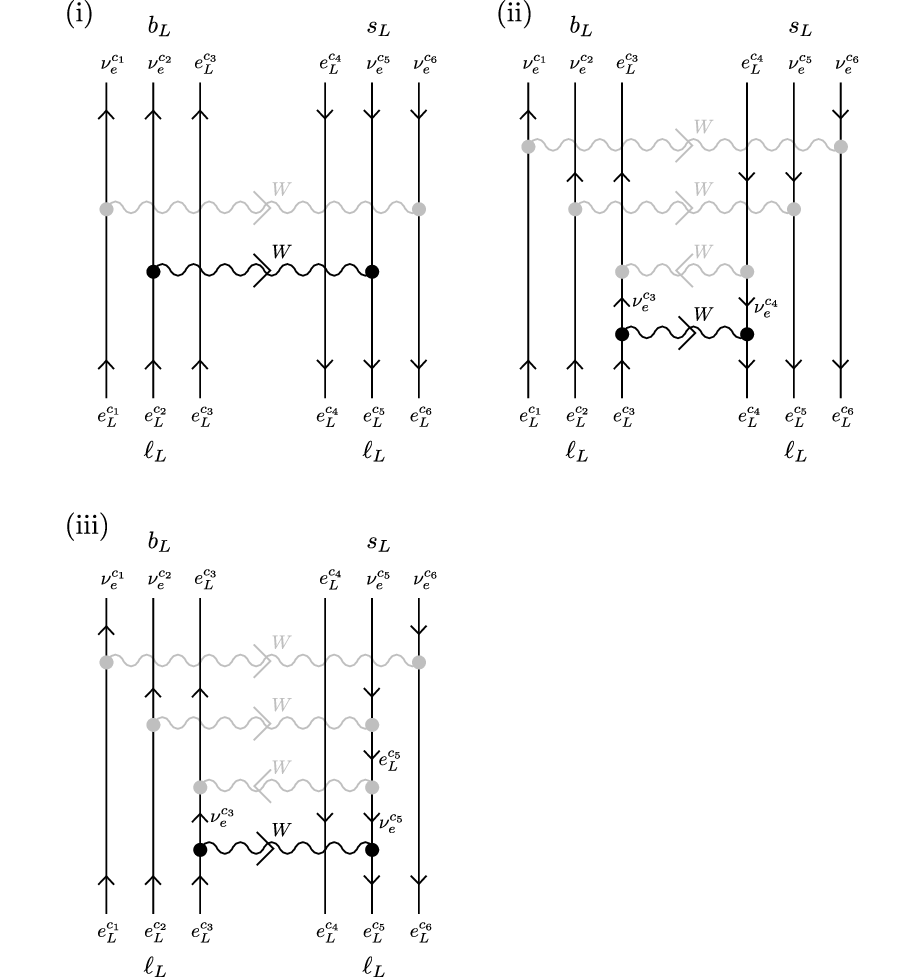}
\caption{Three example composite-leptoquark processes mediating \prm{b\!\rightarrow\!\ell\bar\ell} decay in CASMIR, being representative members of three equivalence classes. Grey lines represent local co-ordinate transforms on the gauge group, consistent with the gauge choices of \pcref{ch:SM}, which act on the preon \prm{A}-sector charges in a manner equivalent to the exchange of a massless \prm{W}~boson. In the sub-preon scalar field model these transforms correspond to induced choices of co-ordinate frame on \prm{\Cw{18}}. Labels~(i), (ii), and~(iii) permit identification of individual diagrams in the main text.\label{fig:preonBK}}
\end{figure}

In addition to the loop diagram of \fref{fig:BKdecay}(i), 
the preonic tree-level description of $\BK$ contains three distinct processes which are illustrated in \fref{fig:preonBK}.
In each, a $W$~boson is exchanged between two fermion triplets. As per \sref{sec:EWint_Wintdetail}\notchap{of \rcite{pfeifer2022CASM3}}, %
$W$~boson exchange is obligatorily accompanied by a local change in co-ordinate frame on the underlying model selected by the requirement that the charges of the $\SU{3}_A$ sector are conserved by fermion/boson interactions. Specifying the species entering and leaving the interaction suffices to uniquely constrain this transformation (up to physically meaningless perturbations). 
As discussed in \sref{sec:1storderK}\standalone{of \rcite{pfeifer2022CASM4}}, %
where this passive transformation results in a change in charge labelling across the boundary of the transformed region, it may equivalently be represented as an active transformation described by one or more bosons interacting only where species lines traverse this boundary. However, in contrast with \sref{sec:1storderK} %
this process is not energy-dependent and thus the equivalent bosons (or co-ordinate sleeves as per \sref{sec:EWint_Wintdetail}) %
are necessarily trivial aside from the net transfer of $A$-sector charges. They are represented in grey in \fref{fig:preonBK}, and interact with an effective coupling constant of~1.

It is reasonable to assign the induced (grey) vertices to occur simultaneously (in the isotropy frame of the CASMIR pseudovacuum) with the foreground (black) $W$/preon interactions which induce them.
It is then convenient %
to connect these vertices, for example as shown in \fref{fig:preonBK}, as a reflection of local conservation of $A$-sector charge.
In order that the co-ordinate transformations may be considered a part of a propagating leptoquark, it is appropriate that they be represented by bosons co-propagating with the boson which induces them. Thus in \fref{fig:preonBK} the connections of the induced boson vertices are chosen so as to always span from one fermion to the other.
Where a co-ordinate transformation admits multiple equivalent boson arrangements, these form an equivalence class from which only one need be considered (up to a numerical multiplier corresponding to the number of elements of the equivalence class). 

Next, observe that the quarks of \fref{fig:preonBK} are required to be bound within $B^+$ and $K^+$ mesons and thus the process as a whole takes place on the order of the colour confinement scale %
which %
is $\mc{L}_\preon\sim(3.09~\TeV)^{-1}$~\notchap{\cite{pfeifer2022CASM5}}\chap{(\cref{ch:CDF2})}. Over this scale, the foreground $W$ boson does not acquire mass through interactions with the CASMIR pseudovacuum.

Putting this all together, a leptoquark in CASMIR comprises an exchanged foreground boson accompanied by one or more induced co-ordinate transforms as required by the species labelling of the diagram lines, and ultimately by the far-field process (or term of said process) being described. 
Over the timescales of the leptoquark-mediated interactions, the foreground boson, and thus the leptoquark as a whole, is massless.

In evaluating the diagrams of \fref{fig:preonBK}, a further property of CASMIR must be kept in mind. 
\begin{itemize}
\item First, consider a pair of vertices which are topologically equivalent to each other but not to any other vertices in a diagram. As in QFT, the freedom to connect up either vertex in either location yields a multiplicative factor of two on the process being described. When a diagram exists as part of a set summed over colour perturbations, it suffices that the set be invariant under vertex exchange and thus that an individual diagram be invariant up to a colour perturbation mapping it onto another member of the set. An example of such a pair of vertices is the pair of preon vertices in the left fermion/leptoquark interaction of \fref{fig:preonBK}(i).
\item In CASMIR, however, a further factor arises in processes dominated by contributions arising over length scales small compared with $\mc{L}_0\approx(3.59~\GeV)^{-1}$. %
All preon fields in CASMIR correspond to gradients of Fundamental Scalar Fields (FSFs) taken with respect to directions on the underlying $\mbb{C}^{\wedge 18}$ manifold, and these fields are in general correlated over length scales of order $\mc{L}_0$ or less. Two of these correlated fields carry perturbations in their (Grassmann-valued) gradients corresponding to preons $\{\nu_e^{c_1},\nu_e^{c_2}\}$ for some colours $\{c_1,c_2\}$, and since the FSFs are common to all members of the sum over colour, the FSFs on which these derivatives act may be interchanged. The FSFs associated with the preons $\{e_L^{c_1},e_L^{c_2}\}$ may be likewise interchanged, but interchanging both sets is equivalent to interchanging the vertex to which the $W$~bosons connect, and thus is double counting. The preonic constituents of the boson are contiguous with the preons, so no further factors arise from the boson. The vertices acquire an additional factor of two from the FSFs, but only because they are within a common FSF autocorrelation region.
\item Note this exchange of the FSFs on which the anticommuting %
derivatives act is independent of an exchange of the order of the derivative operators themselves. It is thus in addition to, not instead of, fermion exchange symmetries at the level of the QFT diagram.
\end{itemize}
The same argument also applies to the right-hand fermion/leptoquark vertex of \fref{fig:preonBK}(i).

Next, consider how %
$W$~boson interactions may mediate generation changes in CASMIR. Critical to this ability is the recognition that the $W$~boson is a complex boson. When a preon couples to a $W$~boson, this coupling may impart an arbitrary phase to the preon provided the conjugate phase is applied to the $W$~boson, from where it may be factored out into a global phase acting on the diagram as a whole. Thus, in the $A$~sector, \emph{only} $W$~bosons may participate in generation-nonconserving interaction vertices.\footnote{In the $C$ sector it is anticipated that gluons are also prohibited from participating in net generation-nonconserving processes, due to the combination of unbroken $\SU{3}_C$ and phase symmetry.} Generation changes may be mediated by real $W$~bosons and also by the pale grey frame bosons when these have $W$~character in the $A$-sector.

Now consider each of the above diagrams in turn. All have the same basic weight from species and vertices. For descriptive purposes, each diagram is considered to comprise two composite fermion/leptoquark vertices regardless of the number of additional bosons induced. %

\begin{enumerate}
\item[(i)] This diagram belongs to the simplest class of processes. The conventional $W$~boson connects to two preons which are changing flavour. Four such diagrams are possible. There is also an additional factor of two at each fermion vertex arising from the FSFs as described above. With four diagrams and a factor of four from FSF interchange, this class of processes has a relative weight of~16.
\item[(ii)] This diagram belongs to a class in which the $W$~boson connects the two non-flavour-changing preons. This class contains only the diagram shown in \fref{fig:preonBK}(ii). FSF symmetry factors on the $e_L^{c}\bar{\nu}_e^{\dot c}W^\dagger\vline_{\,\dot c=c}$ vertices and their conjugates yield a factor of~4. Note that even though no numerical factor is obtained from exchange of $W$~boson connections, a factor is still acquired from interchange of the associated FSFs. The total weight of this class of processes is~4.
\item[(iii)] This diagram belongs to a class in which the $W$~boson couples a flavour-changing preon to a preon which does not change flavour. There are four configurations of the foreground boson (the conventional $W$ may couple to the non-flavour-changing preon in either the left or the right vertex, and then to either flavour-changing preon in the other vertex). The vertices on the two flavour-changing preons of the left side of \fref{fig:preonBK}(iii) are topologically equivalent for an FSF factor of two, and the net weight of this class of processes is~8.
\end{enumerate}

When one of the induced co-ordinate transformation bosons follows exactly the same trajectory as the absorbed~$W$, the two may be combined to yield a real superposition of the photon, $Z$~boson, and the dark matter $N$ boson. As this superposition is real, it cannot engage in a phase-transferring interaction at the $\mu_L\bar{b}_LW^\dagger$ vertex. This might be anticipated to suppress \fref{fig:preonBK}(ii) and \fref{fig:preonBK}(iii) except if the input and output fermion at a vertex were of the same generation (for example when the rightmost vertex is $s_L\bar\mu_LW$), breaking lepton universality. 

However, this is not in fact the case. If the $W$~boson analogues are capable of carrying phases so that the lowest grey $W$~boson in \fref{fig:preonBK}(ii) is conjugate to the black $W$~boson, rendering the $WW^\dagger$ pair real, then the other $W$~boson analogues are also capable of carrying phases and thus these induced boson-analogues are capable of supporting generation change. \Freft{fig:preonBK}(ii)-(iii) are therefore not suppressed. Alternatively, if the grey boson-analogues are chosen/defined to be incapable of imparting a phase at a vertex, then any phase arising from the black $W$~boson is unopposed and this boson continues to be capable of mediating generation change. Again \freft{fig:preonBK}(ii)-(iii) are not suppressed.

In CASMIR, the leptoquark-mediated process shown in \fref{fig:BKdecay}(ii) consequently does not break lepton universality.

\section{Conclusion\label{sec:LHCbconclusion}}

Making use of the microscopic properties of CASMIR, and calculation techniques developed in \standalone{Refs.~\cite{pfeifer2022CASM1,pfeifer2022CASM2,pfeifer2022CASM3,pfeifer2022CASM4}}\notstandalone{\crefr{ch:simplest}{ch:detail}}, it is relatively straightforward to obtain a prediction from CASMIR for leptoquark-mediated contributions to the relative rates of non-$J/\psi$-mediated muon pair and electron pair production during beauty quark decay. 
The interaction vertices obtained for the leptoquark-mediated interactions do not break lepton universality.

Deviations from $R_K=R_{K^*}=1$ consequently arise only as a result of mass-dependent effects, as in the Standard Model. Since experimental tests of $R_{K^{(*)}}$ are as yet unable to discriminate between the decay processes of \freft{fig:BKdecay}(i) and~(ii), 
tests of lepton universality in beauty quark decay %
do not currently distinguish between CASMIR and the (leptoquark-free) Standard Model.

\chapter{Additional Brief Notes\label{ch:notes}}
\section{Higher-generation weak bosons in CASMIR\label{sec:heavyWZH}}
\levelone{Brief comments}

\leveltwo{Meaning of boson generation}

In CASMIR, there exist three mass eigenstates for each of the weak bosons ($W$, $Z$, and $\bmh$, plus coloured counterparts of $W$ and $Z$). These may be identified as ``generations'' of these bosons. However, in contrast with the charged leptons where different generations correspond to different eigenstates of a $3\times 3$ mass matrix, %
for the bosons there are instead three different mass-generating processes, whose eigenstates are distinguished by rest mass alone. Thus the only prerequisite for a boson to change generations is that it should be supplied with a sufficiently large augmentation to its rest energy. (Consider the hypothetical situation of an on-shell generation-1 $W$~boson at rest which is given additional energy while remaining at rest. The most likely outcome is that this boson will become off-shell---\emph{unless} the additional energy imparted is of the correct amount to cause it to become an on-shell $W$ boson of generation~2 or~3. It is readily seen that it is meaningless to talk about the generation of an off-shell $W$~boson as the generation labels merely describe the three on-shell rest energies.)

\leveltwo{Beyond the preon scale}

The second and third generation $W$~boson states are hard to access, as their rest masses exceed the preon scale. Typically it would be expected that at energies slightly below the rest mass of an excitation, it may make meaningful contributions as a virtual particle. However, the most energy that can be borrowed towards the construction of a virtual particle is the preon scale $E_\preon=\frac{1}{2}\mc{E}_\preon=3.09~\TeV$~\notchap{\cite{pfeifer2022CASM5}}\chap{(\cref{ch:CDF2})}. Therefore, no hint whatsoever of these particles will be observed until at least within $3~\TeV$ of their rest mass (and even then, processes in which they appear may be affected by local depletion of the pseudovacuum until significantly closer than $3~\TeV$).

\levelone{Calculating the heavy weak boson masses}

The heavy weak boson masses are given directly by replacing the generation-1 vertices in the mass interactions with their generation-2 or generation-3 counterparts. Thus
\begin{align}
\frac{m_{W_i}^2}{m_W^2}&=\frac{\left[k_i^{(e)}(\mc{E}_{e_i})\right]^4\left\{1+\frac{19}{18\big[k_i^{(e)}(\mc{E}_{e_i})N_0\big]^4}\right\}}{\left[k_1^{(e)}(\mc{E}_{e})\right]^4\left\{1+\frac{19}{18\big[k_1^{(e)}(\mc{E}_{e})N_0\big]^4}\right\}}\qquad\qquad
\frac{m_{Z_i}^2}{m_Z^2}&\!\!\!\!=\frac{\left[k_i^{(e)}(\mc{E}_{e_i})\right]^4\left\{1+\frac{55}{18\big[k_i^{(e)}(\mc{E}_{e_i})N_0\big]^4}\right\}}{\left[k_1^{(e)}(\mc{E}_{e})\right]^4\left\{1+\frac{55}{18\big[k_1^{(e)}(\mc{E}_{e})N_0\big]^4}\right\}}\\
\frac{m_{c_i}^2}{m_c^2}&=\frac{\left[k_i^{(e)}(\mc{E}_{e_i})\right]^4\left\{1+\frac{131}{18\big[k_i^{(e)}(\mc{E}_{e_i})N_0\big]^4}\right\}}{\left[k_1^{(e)}(\mc{E}_{e})\right]^4\left\{1+\frac{131}{18\big[k_1^{(e)}(\mc{E}_{e})N_0\big]^4}\right\}}\qquad\qquad
\frac{m_{\bmh_i}^2}{m_\bmh^2}&\!\!\!\!=\frac{\left[k_i^{(e)}(\mc{E}_{e_i})\right]^4\left\{1+\frac{39}{18\big[k_i^{(e)}(\mc{E}_{e_i})N_0\big]^4}\right\}}{\left[k_1^{(e)}(\mc{E}_{e})\right]^4\left\{1+\frac{39}{18\big[k_1^{(e)}(\mc{E}_{e})N_0\big]^4}\right\}}.
\end{align}
Here, $m_c$ is the effective long-range gluon mass; this parameter is not normally observable except
\begin{itemize}
\item[(i)] indirectly during the propagation of gluon field holes in the pseudovacuum, which contribute a correction to particle masses as discussed in \cref{ch:detail}\standalone{ of \rcite{pfeifer2022CASM4}}, and 
\item[(ii)]~as the mass of the $80.42815(42)~\GeV/c^2$ neutral dark matter boson denoted $N_\mu$ in \standalone{\rcites{pfeifer2022CASM3}{pfeifer2022CASM4}}\notstandalone{\crefs{ch:SM}{ch:detail}}.
\end{itemize}

Hypothetically, coloured $W$ and $Z$ eigenstates (here denoted $W^{\tc}$ and $Z^{\tc}$) may also be created as discussed in \standalone{\rcites{pfeifer2022CASM4}{pfeifer2022CASM5}}\notstandalone{\crefr{ch:detail}{ch:CDF2}}. However, their rest masses are far outside the regime in which coloured weak bosons are supported ($1.79~\GeV<E<3.09~\TeV$) and thus they are not expected to be observed. If colour-rich circumstances can be contrived in which to generate these bosons, they satisfy
\begin{align}
\frac{m_{W^{\tc}_i}^2}{m_{W^\tc}^2}&=\frac{\left[k_i^{(e)}(\mc{E}_{e_i})\right]^4\left\{1+\frac{2491}{288\big[k_i^{(e)}(\mc{E}_{e_i})N_0\big]^4}\right\}}{\left[k_1^{(e)}(\mc{E}_{e})\right]^4\left\{1+\frac{2491}{288\big[k_1^{(e)}(\mc{E}_{e})N_0\big]^4}\right\}}\qquad\qquad
\frac{m_{Z^{\tc}_i}^2}{m_{Z^\tc}^2}&\!\!\!\!=\frac{\left[k_i^{(e)}(\mc{E}_{e_i})\right]^4\left\{1+\frac{3979}{1152\big[k_i^{(e)}(\mc{E}_{e_i})N_0\big]^4}\right\}}{\left[k_1^{(e)}(\mc{E}_{e})\right]^4\left\{1+\frac{3979}{1152\big[k_1^{(e)}(\mc{E}_{e})N_0\big]^4}\right\}}.
\end{align}

The calculated heavy weak boson masses (including the higher-generation dark matter boson masses $m_{N_2}$ and $m_{N_3}$) are:

~

\begin{center}
\begin{tabular}{c|r@{.}l}
Boson&\multicolumn{2}{c}{Mass ($\TeV/c^2$)}\\
\hline\hline
$m_{W_2}$&16&61320(46)\\
$m_{N_2}$&16&61320(46)\\
$m_{Z_2}$&18&84673(73)\\
$m_{\bmh_2}$&25&8643(10)\\ %
$m_{W_3}$&279&4038(77)\\
$m_{N_3}$&279&4038(77)\\
$m_{Z_3}$&316&968(12)\\
$m_{\bmh_3}$&434&991(17)
\end{tabular}$\qquad\qquad\qquad$
\begin{tabular}{c|r@{.}l}
Boson&\multicolumn{2}{c}{Mass ($\TeV/c^2$)}\\
\hline\hline
$m_{c_2}$&16&61320(10)\\
$m_{W^{\tc}_2}$&16&61320(46)\\
$m_{Z^{\tc}_2}$&18&84673(73)\\
$m_{c_3}$&279&4038(17)\\
$m_{W^{\tc}_3}$&279&4038(77)\\
$m_{Z^{\tc}_3}$&316&968(12)
\end{tabular}
\end{center}

~

~

Because CASMIR exhibits a real UV cutoff, there is a minimum energy below which these species cannot be observed, even as virtual particles. For the $W_2$ that minimum energy is predicted to be $13.5~\TeV$~\standalone{\cite{pfeifer2022CASM4}}\notstandalone{(\cref{ch:detail})}.

\section{Neutrino masses\label{sec:neutrinos}}
\levelone{No seesaw mechanism}

There is no seesaw mechanism in CASMIR. Instead, the $W$~boson coupling can be generation-nonconserving at the cost of imparting a phase to the overall wavefunction as described in \cref{ch:LHCb}\standalone{ of \rcite{pfeifer2022CASM4}}.
In contrast with the Standard Model, any $W$-mediated interaction yielding neutrinos therefore produces a superposition of all three generations, with wavefunctions differing by a phase. This provides an alternative resolution to the solar neutrino deficit, with the emitted neutrinos being divided %
between the three generations.

\levelone{Neutrino masses}

In the absence of a seesaw mechanism, there is no longer an \emph{a priori} requirement for the neutrinos to have mass. CASMIR provides a mechanism whereby neutrinos may acquire mass, and retracing the calculation of \cref{ch:detail}%
\standalone{ of \rcite{pfeifer2022CASM4}} but with neutrinos instead of electrons yields an expression
\begin{align}
\begin{split}
m_{\nu_i}^2=\,&\frac{f^2}{2}\left[k^{(\nu)}_i(\mc{E}_{\nu_i})\right]^4{\omega_0}^2 {N_0}^8 S_{18,147}\\
&\times\bm{\left(}\frac{5m_{\nu_i}^2}{\mcst^2}\left\{1+\frac{90\alpha m_{\nu_i}^2}{\pi\mcst^2}+\frac{\alpha\left(f_W m_{\nu_i}^2-48f_Zm_{\nu_i}^2+25\sum_j m_{e_j}^2\right)}{6\pi m_W^2}\right\}+\frac{40 m_{\nu_i}^2}{3m_\bmh^2\big[k_1^{(e)}(\mc{E}_e)N_0\big]^4}\bm{\right)}\\
&\times\bm{\left(}
1+\OOOO{\frac{\alpha m_{\nu_i}^2\sum_j m_{e_j}^2}{N_0\mcst^2m_W^2}}
+\OO{\frac{\alpha^2\sum_j m_{e_j}^4}{m_W^4}}+\OOOO{\frac{m_{\nu_i}^2}{m_\bmh^2\big[k_1^{(e)}(\mc{E}_e)\big]^4{N_0}^5}}\right.\\
&\left.
~~~~~~~+\OOOO{\frac{\alpha m_{\nu_i}^2\sum_j m_{e_j}^2}{m_\bmh^2\big[k^{(e)}_1(\mc{E}_e)N_0\big]^4m^2_W}}
+\OOOO{\frac{m_{\nu_i}^4}{m_\bmh^4\big[k_1^{(e)}(\mc{E}_e)\big]^4}}+\OOOO{\frac{m_{\nu_i}^6}{\mcst^6}}\bm{\right)}
\end{split}
\end{align}
where $f_W$ comes from the $m_{\nu_i}^2/m_W^2$ component of the $W$~loop correction. This term of the $W$~loop correction is usually ignored as $m_{\nu_i}^2/m_W^2\ll m_e^2/m_W^2$. While not formally evaluated here, $f_W$ is anticipated to be around~100.

Note that the neutrino mass terms all involve massive bosons; consequently the neutrino mass interaction is dependent on the neutrino having mass, and the solution $m_{\nu_i}=0$ is acceptable for any or all generations of neutrino.

Assuming at least one massive generation of neutrinos, and ignoring the higher-order corrections, for the massive neutrinos the above may be rewritten
\begin{align}
\begin{split}
1=\,&\frac{f^2}{2}\left[k^{(\nu)}_i(\mc{E}_{\nu_i})\right]^4{\omega_0}^2 {N_0}^8 S_{18,147}\\
&\times\bm{\left(}\left\{\frac{1}{\mcst^2}+\frac{25\alpha\sum_j m_{e_j}^2}{6\pi m_W^2\mcst^2}+\frac{40}{3m_\bmh^2\big[k_1^{(e)}(\mc{E}_e)N_0\big]^4}\right\}\right.\\
&~~~~~~~+\left.\left\{\frac{450\alpha m_{\nu_i}^2}{\pi\mcst^4}+\frac{5\alpha\left(f_W-48 f_Z\right)m_{\nu_i}^2}{6\pi m_W^2\mcst^2}\right\}\bm{\right)}.
\end{split}
\end{align}
Provided the neutrino mass is small compared to the electroweak scale (i.e.~$m_W^2$ etc.), the second term in curly braces is a small (perturbative) correction to the first.

By the leading order terms, all $k^{(\nu)}_i$ are (approximately) equal. However, for 
\begin{equation}
k^{(\nu)}_n(\mc{E}) = 1+\sqrt{2}\cos{\left[\theta_\nu(\mc{E})-\frac{2\pi(n-1)}{3}\right]}
\end{equation}
there are no values of $\theta_\nu$ for which all three eigenvalues even approximately coincide. Therefore at least one neutrino must be massless.

Further, restricting to the leading order expression
\begin{align}
\begin{split}
1=\,&\frac{f^2}{2}\left[k^{(\nu)}_i(\mc{E}_{\nu_i})\right]^4{\omega_0}^2 {N_0}^8 S_{18,147}\left\{\frac{1}{\mcst^2}+\frac{25\alpha\sum_j m_{e_j}^2}{6\pi m_W^2\mcst^2}+\frac{40}{3m_\bmh^2\big[k_1^{(e)}(\mc{E}_e)N_0\big]^4}\right\}
\end{split}
\end{align}
and solving for $k^{(\nu)}_i$ yields $k^{(\nu)}_i\approx 16$. %
However, the maximum value of $k^{(\nu)}_i$ for any $\theta_\nu$ is $\sqrt{2}$. Hence there are no nonzero solutions  for neutrino masses in the regime $m_{\nu_i}\ll m_e\ll m_W$. All light neutrinos are therefore massless:
\begin{equation}
m_{\nu_i}=0\quad\forall~i.
\end{equation}

\tbf{NOTE:} This argument does not exclude the possibility of additional superheavy neutrino mass eigenstates $(m_{\nu_i}\geq m_W)$, which would be beyond the scope of the present perturbative approach.

\chapter{Higgs Physics\label{ch:Higgs}}

\section{Higgs production cross-sections\label{sec:HiggssxBR}}

The most pressing question regarding Higgs physics in CASMIR is whether it is capable of reproducing production cross-sections consistent with those observed at the LHC, with \srefs{sec:addconstrs}{sec:scalbosint} suggesting that Higgs production as described by CASMIR might be suppressed by a factor of $\big[k^{(e)}_1(\mc{E}_e)N_0\big]^{-2}\approx 1.7\times 10^{-2}$ relative to the Standard Model at collision energies small compared with $E_\Omega=3.09~\TeV$. However, this does not in fact occur.

In order for the additional factor~$\big[k^{(e)}_1(\mc{E}_e)N_0\big]^{-2}$ to be associated with a Higgs vertex, that vertex must satisfy the following criteria:
\begin{enumerate}
\item the vertex energy must be small compared with $E_\Omega=3.09~\TeV$, %
\item the Higgs boson must be removed to the far field (in practice, propagation beyond $\mc{L}_\Omega$), and
\item {there must be a summation, explicit or implicit, over the exchange of multiple Higgs bosons}, which leads to a cancellation over phase. This cancellation is described in \sref{sec:scalbosint} below \Eref{eq:bghh*}, beginning ``Averaging over multiple such interactions involves a sum over phase\ldots''.
\end{enumerate}
While generation of Higgs at LHC satisfies the first two requirements, each detected foreground Higgs emission in the LHC is a single event, and the third requirement consequently does not apply. A single foreground Higgs emission therefore does not attract factors of~$\big[k^{(e)}_1(\mc{E}_e)N_0\big]^{-2}$. 

In contrast, Higgs loop corrections to particle mass as described in multiple contexts in \cref{ch:detail}, and to the muon gyromagnetic anomaly as described in \sref{sec:muongyrocalc}, involve the ongoing continual transmission of Higgs bosons along loop trajectories, with these ongoing processes only being represented schematically by the single figure, much as the EM interaction diagram of \fref{fig:EM} does not represent exchange of just one photon. Thus these processes {do} attract the reduction factors when the energy and far field requirements are met, assuming the diagram meets any other requirements as per \sref{sec:scalbosint}.

With the additional suppressing factor being conditional on the interaction being described and the environment in which it takes place, the Higgs component of the Lagrangian is also found to be more complex than originally indicated in \Eref{eq:Lscalarprop}, in which this factor was assumed to be unconditional. Nevertheless, this simplification typically suffices for Higgs loops and sustained Higgs fields in the low-energy regime, although it is unsuited to single-Higgs events.

In brief, the summation over phase of \sref{sec:scalbosint} does not apply to the emission of single Higgs bosons, and the Higgs emission vertices in CASMIR are therefore not suppressed by factors of $\big[k^{(e)}_1(\mc{E}_e)N_0\big]^{-2}\approx 1.7\times 10^{-2}$ per vertex compared with the Standard Model.

Still yet to be determined is whether the emission of a single Higgs boson is best described in CASMIR by the scalar field $\bmh$ or by the rescaled scalar field $\tilde{\bmh}$ (\sref{sec:scalbosint}), and whether vertices attract additional factors of $9$ due to degeneracy of contributing diagrams, and/or~$\frac{1}{9}$ per scalar boson (real or virtual) due to a given preon coupling to only one term in the nine-part sum making up the composite field $\bmh$ (See, for example, \sref{sec:Wmass_scalbosloop}.) %

\section{Parity conservation}

As another point of interest, parity conservation in CASMIR requires that Higgs bosons always be emitted in $\bmh\bmh^*$ pairs, for a total of two holomorphic and two antiholomorphic derivative operators acting on the FSFs. This holds regardless of whether the Higgs bosons are emitted by bosons or by fermions. When a single real Higgs $\bmh$ is produced at LHC, it is therefore necessarily accompanied by a virtual conjugate $\bmh^*$ which (typically) shares both vertices with the real Higgs.

\bibliographystyle{apsrev4-2}
\bibliography{Paper.bib}

\begin{thebibliography}{43}%
\makeatletter
\providecommand \@ifxundefined [1]{%
 \@ifx{#1\undefined}
}%
\providecommand \@ifnum [1]{%
 \ifnum #1\expandafter \@firstoftwo
 \else \expandafter \@secondoftwo
 \fi
}%
\providecommand \@ifx [1]{%
 \ifx #1\expandafter \@firstoftwo
 \else \expandafter \@secondoftwo
 \fi
}%
\providecommand \natexlab [1]{#1}%
\providecommand \enquote  [1]{``#1''}%
\providecommand \bibnamefont  [1]{#1}%
\providecommand \bibfnamefont [1]{#1}%
\providecommand \citenamefont [1]{#1}%
\providecommand \href@noop [0]{\@secondoftwo}%
\providecommand \href [0]{\begingroup \@sanitize@url \@href}%
\providecommand \@href[1]{\@@startlink{#1}\@@href}%
\providecommand \@@href[1]{\endgroup#1\@@endlink}%
\providecommand \@sanitize@url [0]{\catcode `\\12\catcode `\$12\catcode
  `\&12\catcode `\#12\catcode `\^12\catcode `\_12\catcode `\%12\relax}%
\providecommand \@@startlink[1]{}%
\providecommand \@@endlink[0]{}%
\providecommand \url  [0]{\begingroup\@sanitize@url \@url }%
\providecommand \@url [1]{\endgroup\@href {#1}{\urlprefix }}%
\providecommand \urlprefix  [0]{URL }%
\providecommand \Eprint [0]{\href }%
\providecommand \doibase [0]{https://doi.org/}%
\providecommand \selectlanguage [0]{\@gobble}%
\providecommand \bibinfo  [0]{\@secondoftwo}%
\providecommand \bibfield  [0]{\@secondoftwo}%
\providecommand \translation [1]{[#1]}%
\providecommand \BibitemOpen [0]{}%
\providecommand \bibitemStop [0]{}%
\providecommand \bibitemNoStop [0]{.\EOS\space}%
\providecommand \EOS [0]{\spacefactor3000\relax}%
\providecommand \BibitemShut  [1]{\csname bibitem#1\endcsname}%
\let\auto@bib@innerbib\@empty
\bibitem [{\citenamefont {Onsager}(1944)}]{onsager1944}%
  \BibitemOpen
  \bibfield  {author} {\bibinfo {author} {\bibfnamefont {L.}~\bibnamefont
  {Onsager}},\ }\href {https://doi.org/10.1103/PhysRev.65.117} {\bibfield
  {journal} {\bibinfo  {journal} {Phys. Rev.}\ }\textbf {\bibinfo {volume}
  {65}},\ \bibinfo {pages} {117} (\bibinfo {year} {1944})}\BibitemShut
  {NoStop}%
\bibitem [{\citenamefont {Suzuki}(1976)}]{suzuki1976}%
  \BibitemOpen
  \bibfield  {author} {\bibinfo {author} {\bibfnamefont {M.}~\bibnamefont
  {Suzuki}},\ }\href {https://doi.org/10.1143/PTP.56.1454} {\bibfield
  {journal} {\bibinfo  {journal} {Prog. Theor. Phys.}\ }\textbf {\bibinfo
  {volume} {56}},\ \bibinfo {pages} {1454} (\bibinfo {year}
  {1976})}\BibitemShut {NoStop}%
\bibitem [{\citenamefont {Srivastava}(1990)}]{srivastava1990}%
  \BibitemOpen
  \bibfield  {author} {\bibinfo {author} {\bibfnamefont {G.~P.}\ \bibnamefont
  {Srivastava}},\ }\href@noop {} {\emph {\bibinfo {title} {The Physics of
  Phonons}}}\ (\bibinfo  {publisher} {Taylor \& Francis},\ \bibinfo {year}
  {1990})\BibitemShut {NoStop}%
\bibitem [{\citenamefont {Drazin}\ and\ \citenamefont
  {Johnson}(1989)}]{drazin1989}%
  \BibitemOpen
  \bibfield  {author} {\bibinfo {author} {\bibfnamefont {P.~G.}\ \bibnamefont
  {Drazin}}\ and\ \bibinfo {author} {\bibfnamefont {R.~S.}\ \bibnamefont
  {Johnson}},\ }\href@noop {} {\emph {\bibinfo {title} {Solitons: An
  Introduction}}},\ \bibinfo {edition} {2nd}\ ed.\ (\bibinfo  {publisher}
  {University Press},\ \bibinfo {address} {Cambridge},\ \bibinfo {year}
  {1989})\BibitemShut {NoStop}%
\bibitem [{\citenamefont {Cooper}(1956)}]{cooper1956}%
  \BibitemOpen
  \bibfield  {author} {\bibinfo {author} {\bibfnamefont {L.~N.}\ \bibnamefont
  {Cooper}},\ }\href {https://doi.org/10.1103/PhysRev.104.1189} {\bibfield
  {journal} {\bibinfo  {journal} {Phys. Rev.}\ }\textbf {\bibinfo {volume}
  {104}},\ \bibinfo {pages} {1189} (\bibinfo {year} {1956})}\BibitemShut
  {NoStop}%
\bibitem [{\citenamefont {Bardeen}\ \emph {et~al.}(1957)\citenamefont
  {Bardeen}, \citenamefont {Cooper},\ and\ \citenamefont
  {Schrieffer}}]{bardeen1957}%
  \BibitemOpen
  \bibfield  {author} {\bibinfo {author} {\bibfnamefont {J.}~\bibnamefont
  {Bardeen}}, \bibinfo {author} {\bibfnamefont {L.~N.}\ \bibnamefont
  {Cooper}},\ and\ \bibinfo {author} {\bibfnamefont {J.~R.}\ \bibnamefont
  {Schrieffer}},\ }\href {https://doi.org/10.1103/PhysRev.108.1175} {\bibfield
  {journal} {\bibinfo  {journal} {Phys. Rev.}\ }\textbf {\bibinfo {volume}
  {108}},\ \bibinfo {pages} {1175} (\bibinfo {year} {1957})}\BibitemShut
  {NoStop}%
\bibitem [{\citenamefont {Maynard}(2001)}]{maynard2001}%
  \BibitemOpen
  \bibfield  {author} {\bibinfo {author} {\bibfnamefont {J.~D.}\ \bibnamefont
  {Maynard}},\ }\href {https://doi.org/10.1103/RevModPhys.73.401} {\bibfield
  {journal} {\bibinfo  {journal} {Rev. Mod. Phys.}\ }\textbf {\bibinfo {volume}
  {73}},\ \bibinfo {pages} {401} (\bibinfo {year} {2001})}\BibitemShut
  {NoStop}%
\bibitem [{\citenamefont {Dragoman}\ and\ \citenamefont
  {Dragoman}(2004)}]{dragoman2004}%
  \BibitemOpen
  \bibfield  {author} {\bibinfo {author} {\bibfnamefont {D.}~\bibnamefont
  {Dragoman}}\ and\ \bibinfo {author} {\bibfnamefont {M.}~\bibnamefont
  {Dragoman}},\ }\href {https://www.springer.com/gp/book/9783540201472} {\emph
  {\bibinfo {title} {Quantum--Classical Analogies}}},\ The Frontiers
  Collection\ (\bibinfo  {publisher} {Springer-Verlag},\ \bibinfo {address}
  {Berlin Heidelberg},\ \bibinfo {year} {2004})\BibitemShut {NoStop}%
\bibitem [{\citenamefont {Lewenstein}\ \emph {et~al.}(2007)\citenamefont
  {Lewenstein}, \citenamefont {Sanpera}, \citenamefont {Ahufinger},
  \citenamefont {Damski}, \citenamefont {Sen(De)},\ and\ \citenamefont
  {Sen}}]{lewenstein2007}%
  \BibitemOpen
  \bibfield  {author} {\bibinfo {author} {\bibfnamefont {M.}~\bibnamefont
  {Lewenstein}}, \bibinfo {author} {\bibfnamefont {A.}~\bibnamefont {Sanpera}},
  \bibinfo {author} {\bibfnamefont {V.}~\bibnamefont {Ahufinger}}, \bibinfo
  {author} {\bibfnamefont {B.}~\bibnamefont {Damski}}, \bibinfo {author}
  {\bibfnamefont {A.}~\bibnamefont {Sen(De)}},\ and\ \bibinfo {author}
  {\bibfnamefont {U.}~\bibnamefont {Sen}},\ }\\\href
  {https://doi.org/10.1080/00018730701223200} {\bibfield  {journal} {\bibinfo
  {journal} {Adv. Phys.}\ }\textbf {\bibinfo {volume} {56}},\ \bibinfo {pages}
  {243} (\bibinfo {year} {2007})}\BibitemShut {NoStop}%
\bibitem [{\citenamefont {Visser}\ \emph {et~al.}(2002)\citenamefont {Visser},
  \citenamefont {Barcel{\'o}},\ and\ \citenamefont {Liberati}}]{visser2002}%
  \BibitemOpen
  \bibfield  {author} {\bibinfo {author} {\bibfnamefont {M.}~\bibnamefont
  {Visser}}, \bibinfo {author} {\bibfnamefont {C.}~\bibnamefont
  {Barcel{\'o}}},\ and\ \bibinfo {author} {\bibfnamefont {S.}~\bibnamefont
  {Liberati}},\ }\href {https://doi.org/10.1023/A:1020180409214} {\bibfield
  {journal} {\bibinfo  {journal} {Gen. Rel. and Grav.}\ }\textbf {\bibinfo
  {volume} {34}},\ \bibinfo {pages} {1719} (\bibinfo {year}
  {2002})}\BibitemShut {NoStop}%
\bibitem [{\citenamefont {Liberati}\ \emph {et~al.}(2009)\citenamefont
  {Liberati}, \citenamefont {Girelli},\ and\ \citenamefont
  {Sindoni}}]{liberati2009}%
  \BibitemOpen
  \bibfield  {author} {\bibinfo {author} {\bibfnamefont {S.}~\bibnamefont
  {Liberati}}, \bibinfo {author} {\bibfnamefont {F.}~\bibnamefont {Girelli}},\
  and\ \bibinfo {author} {\bibfnamefont {L.}~\bibnamefont {Sindoni}},\
  }\href@noop {} {\bibinfo {title} {Analogue models for emergent gravity}},\
  \bibinfo {howpublished}
  {\href{https://arxiv.org/abs/0909.3834v1}{arXiv:0909.3834v1 [gr-qc]}}
  (\bibinfo {year} {2009})\BibitemShut {NoStop}%
\bibitem [{\citenamefont {Barcel{\'{o}}}\ \emph {et~al.}(2011)\citenamefont
  {Barcel{\'{o}}}, \citenamefont {Liberati},\ and\ \citenamefont
  {Visser}}]{barcelo2011}%
  \BibitemOpen
  \bibfield  {author} {\bibinfo {author} {\bibfnamefont {C.}~\bibnamefont
  {Barcel{\'{o}}}}, \bibinfo {author} {\bibfnamefont {S.}~\bibnamefont
  {Liberati}},\ and\ \bibinfo {author} {\bibfnamefont {M.}~\bibnamefont
  {Visser}},\ }\href {https://doi.org/10.12942/lrr-2011-3} {\bibfield
  {journal} {\bibinfo  {journal} {Living Reviews in Relativity}\ }\textbf
  {\bibinfo {volume} {14}},\ \bibinfo {pages} {3} (\bibinfo {year}
  {2011})}\BibitemShut {NoStop}%
\bibitem [{\citenamefont {Unruh}(1981)}]{unruh1981}%
  \BibitemOpen
  \bibfield  {author} {\bibinfo {author} {\bibfnamefont {W.~G.}\ \bibnamefont
  {Unruh}},\ }\href {https://doi.org/10.1103/PhysRevLett.46.1351} {\bibfield
  {journal} {\bibinfo  {journal} {Phys. Rev. Lett.}\ }\textbf {\bibinfo
  {volume} {46}},\ \bibinfo {pages} {1351} (\bibinfo {year}
  {1981})}\BibitemShut {NoStop}%
\bibitem [{\citenamefont {Garay}\ \emph {et~al.}(2000)\citenamefont {Garay},
  \citenamefont {Anglin}, \citenamefont {Cirac},\ and\ \citenamefont
  {Zoller}}]{garay2000}%
  \BibitemOpen
  \bibfield  {author} {\bibinfo {author} {\bibfnamefont {L.~J.}\ \bibnamefont
  {Garay}}, \bibinfo {author} {\bibfnamefont {J.~R.}\ \bibnamefont {Anglin}},
  \bibinfo {author} {\bibfnamefont {J.~I.}\ \bibnamefont {Cirac}},\ and\
  \bibinfo {author} {\bibfnamefont {P.}~\bibnamefont {Zoller}},\ }\href
  {https://doi.org/10.1103/PhysRevLett.85.4643} {\bibfield  {journal} {\bibinfo
   {journal} {Phys. Rev. Lett.}\ }\textbf {\bibinfo {volume} {85}},\ \bibinfo
  {pages} {4643} (\bibinfo {year} {2000})}\BibitemShut {NoStop}%
\bibitem [{\citenamefont {Garay}\ \emph {et~al.}(2001)\citenamefont {Garay},
  \citenamefont {Anglin}, \citenamefont {Cirac},\ and\ \citenamefont
  {Zoller}}]{garay2001}%
  \BibitemOpen
  \bibfield  {author} {\bibinfo {author} {\bibfnamefont {L.~J.}\ \bibnamefont
  {Garay}}, \bibinfo {author} {\bibfnamefont {J.~R.}\ \bibnamefont {Anglin}},
  \bibinfo {author} {\bibfnamefont {J.~I.}\ \bibnamefont {Cirac}},\ and\
  \bibinfo {author} {\bibfnamefont {P.}~\bibnamefont {Zoller}},\ }\href
  {https://doi.org/10.1103/PhysRevA.63.023611} {\bibfield  {journal} {\bibinfo
  {journal} {Phys. Rev. A}\ }\textbf {\bibinfo {volume} {63}},\ \bibinfo
  {pages} {023611} (\bibinfo {year} {2001})}\BibitemShut {NoStop}%
\bibitem [{\citenamefont {Lahav}\ \emph {et~al.}(2010)\citenamefont {Lahav},
  \citenamefont {Itah}, \citenamefont {Blumkin}, \citenamefont {Gordon},
  \citenamefont {Rinott}, \citenamefont {Zayats},\ and\ \citenamefont
  {Steinhauer}}]{lahav2010}%
  \BibitemOpen
  \bibfield  {author} {\bibinfo {author} {\bibfnamefont {O.}~\bibnamefont
  {Lahav}}, \bibinfo {author} {\bibfnamefont {A.}~\bibnamefont {Itah}},
  \bibinfo {author} {\bibfnamefont {A.}~\bibnamefont {Blumkin}}, \bibinfo
  {author} {\bibfnamefont {C.}~\bibnamefont {Gordon}}, \bibinfo {author}
  {\bibfnamefont {S.}~\bibnamefont {Rinott}}, \bibinfo {author} {\bibfnamefont
  {A.}~\bibnamefont {Zayats}},\ and\ \bibinfo {author} {\bibfnamefont
  {J.}~\bibnamefont {Steinhauer}},\ }\\\href
  {http://dx.doi.org/10.1103/PhysRevLett.105.240401} {\bibfield  {journal}
  {\bibinfo  {journal} {Phys. Rev. Lett.}\ }\textbf {\bibinfo {volume} {105}},\
  \bibinfo {pages} {240401} (\bibinfo {year} {2010})}\BibitemShut {NoStop}%
\bibitem [{\citenamefont {Gordon}(1923)}]{gordon1923}%
  \BibitemOpen
  \bibfield  {author} {\bibinfo {author} {\bibfnamefont {W.}~\bibnamefont
  {Gordon}},\ }\href {https://doi.org/10.1002/andp.19233772202} {\bibfield
  {journal} {\bibinfo  {journal} {Ann. Phys.}\ }\textbf {\bibinfo {volume}
  {377}},\ \bibinfo {pages} {421} (\bibinfo {year} {1923})}\BibitemShut
  {NoStop}%
\bibitem [{\citenamefont {Leonhardt}\ and\ \citenamefont
  {Piwnicki}(1999)}]{leonhardt1999}%
  \BibitemOpen
  \bibfield  {author} {\bibinfo {author} {\bibfnamefont {U.}~\bibnamefont
  {Leonhardt}}\ and\ \bibinfo {author} {\bibfnamefont {P.}~\bibnamefont
  {Piwnicki}},\ }\href {https://doi.org/10.1103/PhysRevA.60.4301} {\bibfield
  {journal} {\bibinfo  {journal} {Phys. Rev. A}\ }\textbf {\bibinfo {volume}
  {60}},\ \bibinfo {pages} {4301} (\bibinfo {year} {1999})}\BibitemShut
  {NoStop}%
\bibitem [{\citenamefont {Leonhardt}\ and\ \citenamefont
  {Piwnicki}(2000)}]{leonhardt2000}%
  \BibitemOpen
  \bibfield  {author} {\bibinfo {author} {\bibfnamefont {U.}~\bibnamefont
  {Leonhardt}}\ and\ \bibinfo {author} {\bibfnamefont {P.}~\bibnamefont
  {Piwnicki}},\ }\href {https://doi.org/10.1103/PhysRevLett.84.822} {\bibfield
  {journal} {\bibinfo  {journal} {Phys. Rev. Lett.}\ }\textbf {\bibinfo
  {volume} {84}},\ \bibinfo {pages} {822} (\bibinfo {year} {2000})}\BibitemShut
  {NoStop}%
\bibitem [{\citenamefont {Jacobson}\ and\ \citenamefont
  {Volovik}(1998)}]{jacobson1998}%
  \BibitemOpen
  \bibfield  {author} {\bibinfo {author} {\bibfnamefont {T.~A.}\ \bibnamefont
  {Jacobson}}\ and\ \bibinfo {author} {\bibfnamefont {G.~E.}\ \bibnamefont
  {Volovik}},\ }\href {https://doi.org/10.1103/PhysRevD.58.064021} {\bibfield
  {journal} {\bibinfo  {journal} {Phys. Rev. D}\ }\textbf {\bibinfo {volume}
  {58}},\ \bibinfo {pages} {064021} (\bibinfo {year} {1998})}\BibitemShut
  {NoStop}%
\bibitem [{\citenamefont {Reznik}(2000)}]{reznik2000}%
  \BibitemOpen
  \bibfield  {author} {\bibinfo {author} {\bibfnamefont {B.}~\bibnamefont
  {Reznik}},\ }\href {https://doi.org/10.1103/PhysRevD.62.044044} {\bibfield
  {journal} {\bibinfo  {journal} {Phys. Rev. D}\ }\textbf {\bibinfo {volume}
  {62}},\ \bibinfo {pages} {044044} (\bibinfo {year} {2000})}\BibitemShut
  {NoStop}%
\bibitem [{\citenamefont {Sch\"utzhold}\ and\ \citenamefont
  {Unruh}(2005)}]{schutzhold2005}%
  \BibitemOpen
  \bibfield  {author} {\bibinfo {author} {\bibfnamefont {R.}~\bibnamefont
  {Sch\"utzhold}}\ and\ \bibinfo {author} {\bibfnamefont {W.~G.}\ \bibnamefont
  {Unruh}},\ }\href {https://doi.org/10.1103/PhysRevLett.95.031301} {\bibfield
  {journal} {\bibinfo  {journal} {Phys. Rev. Lett.}\ }\textbf {\bibinfo
  {volume} {95}},\ \bibinfo {pages} {031301} (\bibinfo {year}
  {2005})}\BibitemShut {NoStop}%
\bibitem [{\citenamefont {Sch\"utzhold}\ and\ \citenamefont
  {Unruh}(2002)}]{schutzhold2002}%
  \BibitemOpen
  \bibfield  {author} {\bibinfo {author} {\bibfnamefont {R.}~\bibnamefont
  {Sch\"utzhold}}\ and\ \bibinfo {author} {\bibfnamefont {W.~G.}\ \bibnamefont
  {Unruh}},\ }\href {https://doi.org/10.1103/PhysRevD.66.044019} {\bibfield
  {journal} {\bibinfo  {journal} {Phys. Rev. D}\ }\textbf {\bibinfo {volume}
  {66}},\ \bibinfo {pages} {044019} (\bibinfo {year} {2002})}\BibitemShut
  {NoStop}%
\bibitem [{\citenamefont {Koide}(2001)}]{koide2000}%
  \BibitemOpen
  \bibfield  {author} {\bibinfo {author} {\bibfnamefont {Y.}~\bibnamefont
  {Koide}},\ }in\ \href@noop {} {\emph {\bibinfo {booktitle} {Proc. 30th Int.
  Conf. High-energy Phys., Osaka, Japan}}},\ \bibinfo {editor} {edited by\
  \bibinfo {editor} {\bibfnamefont {C.~S.}\ \bibnamefont {Lim}}\ and\ \bibinfo
  {editor} {\bibfnamefont {T.}~\bibnamefont {Yamanaka}}}\ (\bibinfo
  {publisher} {World Scientific},\ \bibinfo {address} {Singapore},\ \bibinfo
  {year} {2001})\ \Eprint {https://arxiv.org/abs/arXiv:hep-ph/0005137v1}
  {arXiv:hep-ph/0005137v1} \BibitemShut {NoStop}%
\bibitem [{\citenamefont {Koide}(1983)}]{koide1983}%
  \BibitemOpen
  \bibfield  {author} {\bibinfo {author} {\bibfnamefont {Y.}~\bibnamefont
  {Koide}},\ }\href {https://doi.org/10.1016/0370-2693(83)90644-5} {\bibfield
  {journal} {\bibinfo  {journal} {Phys. Lett. B}\ }\textbf {\bibinfo {volume}
  {120}},\ \bibinfo {pages} {161} (\bibinfo {year} {1983})}\BibitemShut
  {NoStop}%
\bibitem [{\citenamefont {Peskin}\ and\ \citenamefont
  {Schroeder}(1995)}]{peskin1995}%
  \BibitemOpen
  \bibfield  {author} {\bibinfo {author} {\bibfnamefont {M.~E.}\ \bibnamefont
  {Peskin}}\ and\ \bibinfo {author} {\bibfnamefont {D.~V.}\ \bibnamefont
  {Schroeder}},\ }\href@noop {} {\emph {\bibinfo {title} {An Introduction to
  Quantum Field Theory}}}\ (\bibinfo  {publisher} {Westview Press},\ \bibinfo
  {year} {1995})\BibitemShut {NoStop}%
\bibitem [{\citenamefont {{The ATLAS
  Collaboration}}(2023)}]{the-ATLAS-collaboration2023}%
  \BibitemOpen
  \bibfield  {author} {\bibinfo {author} {\bibnamefont {{The ATLAS
  Collaboration}}},\ } {\emph
  {\bibinfo {title} {Improved {$W$} boson mass measurement using
  $\sqrt{s}=7~\mathrm{TeV}$ proton-proton collisions with the {ATLAS}
  detector}}},\ \bibinfo {type} {Tech. Rep.}\ \bibinfo {number}
  {\href {https://cds.cern.ch/record/2853290}{ATLAS-CONF-2023-004}}\ (\bibinfo  {institution} {CERN},\ \bibinfo {year}
  {2023})\BibitemShut {NoStop}%
\bibitem [{\citenamefont {Kitaev}(2006)}]{kitaev2006}%
  \BibitemOpen
  \bibfield  {author} {\bibinfo {author} {\bibfnamefont {A.}~\bibnamefont
  {Kitaev}},\ }\href {https://doi.org/10.1016/j.aop.2005.10.005} {\bibfield
  {journal} {\bibinfo  {journal} {Ann. Phys.}\ }\textbf {\bibinfo {volume}
  {321}},\ \bibinfo {pages} {2} (\bibinfo {year} {2006})}\BibitemShut {NoStop}%
\bibitem [{\citenamefont {Bonderson}(2007)}]{bonderson2007}%
  \BibitemOpen
  \bibfield  {author} {\bibinfo {author} {\bibfnamefont {P.~H.}\ \bibnamefont
  {Bonderson}},\ }\emph {\bibinfo {title} {Non-Abelian Anyons and
  Interferometry}},\ \href
  {http://resolver.caltech.edu/CaltechETD:etd-06042007-101617} {Ph.D. thesis},\
  \bibinfo  {school} {California Institute of Technology} (\bibinfo {year}
  {2007})\BibitemShut {NoStop}%
\bibitem [{\citenamefont {Tiesinga}\ \emph {et~al.}(2018)\citenamefont
  {Tiesinga}, \citenamefont {Mohr}, \citenamefont {Newell},\ and\ \citenamefont
  {Taylor}}]{tiesinga2018}%
  \BibitemOpen
  \bibfield  {author} {\bibinfo {author} {\bibfnamefont {E.}~\bibnamefont
  {Tiesinga}}, \bibinfo {author} {\bibfnamefont {P.~J.}\ \bibnamefont {Mohr}},
  \bibinfo {author} {\bibfnamefont {D.~B.}\ \bibnamefont {Newell}},\ and\
  \bibinfo {author} {\bibfnamefont {B.~N.}\ \bibnamefont {Taylor}}} (\bibinfo
  {year} {2018}),\ \bibinfo {note} {({Web} {Version} 8.1). Database developed
  by J.~Baker, M.~Douma, and S.~Kotochigova. Available at
  \href{http://physics.nist.gov/constants}{http://physics.nist.gov/constants},
  National Institute of Standards and Technology, Gaithersburg, MD
  20899}\BibitemShut {NoStop}%
\bibitem [{\citenamefont {Penrose}(1971)}]{penrose1971}%
  \BibitemOpen
  \bibfield  {author} {\bibinfo {author} {\bibfnamefont {R.}~\bibnamefont
  {Penrose}},\ }in\ \href@noop {} {\emph {\bibinfo {booktitle} {Combinatorial
  Mathematics and its Applications}}},\ \bibinfo {editor} {edited by\ \bibinfo
  {editor} {\bibfnamefont {D.~J.~A.}\ \bibnamefont {Welsh}}}\ (\bibinfo
  {publisher} {Academic Press},\ \bibinfo {year} {1971})\ pp.\ \bibinfo {pages}
  {221--244}\BibitemShut {NoStop}%
\bibitem [{\citenamefont {Pfeifer}\ \emph {et~al.}(2010)\citenamefont
  {Pfeifer}, \citenamefont {Corboz}, \citenamefont {Buerschaper}, \citenamefont
  {Aguado}, \citenamefont {Troyer},\ and\ \citenamefont {Vidal}}]{pfeifer2010}%
  \BibitemOpen
  \bibfield  {author} {\bibinfo {author} {\bibfnamefont {R.~N.~C.}\
  \bibnamefont {Pfeifer}}, \bibinfo {author} {\bibfnamefont {P.}~\bibnamefont
  {Corboz}}, \bibinfo {author} {\bibfnamefont {O.}~\bibnamefont {Buerschaper}},
  \bibinfo {author} {\bibfnamefont {M.}~\bibnamefont {Aguado}}, \bibinfo
  {author} {\bibfnamefont {M.}~\bibnamefont {Troyer}},\ and\ \bibinfo {author}
  {\bibfnamefont {G.}~\bibnamefont {Vidal}},\ }\\\href
  {https://doi.org/10.1103/PhysRevB.82.115126} {\bibfield  {journal} {\bibinfo
  {journal} {Phys. Rev. B}\ }\textbf {\bibinfo {volume} {82}},\ \bibinfo
  {pages} {115126} (\bibinfo {year} {2010})}\BibitemShut {NoStop}%
\bibitem [{\citenamefont {Pfeifer}(2011)}]{pfeifer2011}%
  \BibitemOpen
  \bibfield  {author} {\bibinfo {author} {\bibfnamefont {R.~N.~C.}\
  \bibnamefont {Pfeifer}},\ }\emph {\bibinfo {title} {Simulation of {Anyons}
  {Using} {Symmetric} {Tensor} {Network} {Algorithms}}},\ \href@noop {} {Ph.D.
  thesis},\ \bibinfo  {school} {The University of Queensland} (\bibinfo {year}
  {2011}),\ \bibinfo {note}
  {\href{https://arxiv.org/abs/1202.1522v2}{arXiv:1202.1522v2}}\BibitemShut
  {NoStop}%
\bibitem [{\citenamefont {Aoyama}\ \emph {et~al.}(2019)\citenamefont {Aoyama},
  \citenamefont {Kinoshita},\ and\ \citenamefont {Nio}}]{aoyama2019}%
  \BibitemOpen
  \bibfield  {author} {\bibinfo {author} {\bibfnamefont {T.}~\bibnamefont
  {Aoyama}}, \bibinfo {author} {\bibfnamefont {T.}~\bibnamefont {Kinoshita}},\
  and\ \bibinfo {author} {\bibfnamefont {M.}~\bibnamefont {Nio}},\ }\href
  {https://doi.org/10.3390/atoms7010028} {\bibfield  {journal} {\bibinfo
  {journal} {Atoms}\ }\textbf {\bibinfo {volume} {7}},\ \bibinfo {pages} {28}
  (\bibinfo {year} {2019})}\BibitemShut {NoStop}%
\bibitem [{\citenamefont {Workman}\ \emph {et~al.}()\citenamefont {Workman}
  \emph {et~al.}}]{workman2022}%
  \BibitemOpen
  \bibfield  {author} {\bibinfo {author} {\bibfnamefont {R.~L.}\ \bibnamefont
  {Workman}} \emph {et~al.},\ }\href {https://doi.org/10.1093/ptep/ptac097}
  {\bibfield  {journal} {\bibinfo  {journal} {Prog. Theor. Exp. Phys.}\
  }\textbf {\bibinfo {volume} {2022}},\ \bibinfo {pages} {083C01 (2022) and
  2023 update}}\BibitemShut {NoStop}%
\bibitem [{\citenamefont {Aad}\ \emph {et~al.}(2023)\citenamefont {Aad} \emph
  {et~al.}}]{aad2023}%
  \BibitemOpen
  \bibfield  {author} {\bibinfo {author} {\bibfnamefont {G.}~\bibnamefont
  {Aad}} \emph {et~al.} (\bibinfo {collaboration} {ATLAS Collaboration}),\
  }\href {https://doi.org/10.1103/PhysRevLett.131.251802} {\bibfield  {journal}
  {\bibinfo  {journal} {Phys. Rev. Lett.}\ }\textbf {\bibinfo {volume} {131}},\
  \bibinfo {pages} {251802} (\bibinfo {year} {2023})}\BibitemShut {NoStop}%
\bibitem [{\citenamefont {Pfeifer}(2022)}]{code2022}%
  \BibitemOpen
  \bibfield  {author} {\bibinfo {author} {\bibfnamefont {R.~N.~C.}\
  \bibnamefont {Pfeifer}},\ }\href@noop {} {\bibinfo {title} {Software for the
  {Classical} {Analogue} to the {Standard} {Model} {In} pseudo-{Riemannian}
  space time {(CASMIR)}}},\ \bibinfo {howpublished}
  {\href{https://www.academia.edu/65931512}{https://www.academia.edu/65931512}}
  (\bibinfo {year} {2022})\BibitemShut {NoStop}%
\bibitem [{\citenamefont {Petermann}(1958)}]{petermann1958}%
  \BibitemOpen
  \bibfield  {author} {\bibinfo {author} {\bibfnamefont {A.}~\bibnamefont
  {Petermann}},\ }\href {https://doi.org/10.1016/0029-5582(58)90065-8}
  {\bibfield  {journal} {\bibinfo  {journal} {Nuclear Physics}\ }\textbf
  {\bibinfo {volume} {5}},\ \bibinfo {pages} {677} (\bibinfo {year}
  {1958})}\BibitemShut {NoStop}%
\bibitem [{\citenamefont {Sommerfield}(1958)}]{sommerfield1958}%
  \BibitemOpen
  \bibfield  {author} {\bibinfo {author} {\bibfnamefont {C.~M.}\ \bibnamefont
  {Sommerfield}},\ }\href {https://doi.org/10.1016/0003-4916(58)90003-4}
  {\bibfield  {journal} {\bibinfo  {journal} {Ann. Phys.}\ }\textbf {\bibinfo
  {volume} {5}},\ \bibinfo {pages} {26} (\bibinfo {year} {1958})}\BibitemShut
  {NoStop}%
\bibitem [{\citenamefont {Kerr}(1963)}]{kerr1963}%
  \BibitemOpen
  \bibfield  {author} {\bibinfo {author} {\bibfnamefont {R.~P.}\ \bibnamefont
  {Kerr}},\ }\href {https://doi.org/10.1103/PhysRevLett.11.237} {\bibfield
  {journal} {\bibinfo  {journal} {Phys. Rev. Lett.}\ }\textbf {\bibinfo
  {volume} {11}},\ \bibinfo {pages} {237} (\bibinfo {year} {1963})}\BibitemShut
  {NoStop}%
\bibitem [{\citenamefont {Di~Mario}()}]{di-mario2003}%
  \BibitemOpen
  \bibfield  {author} {\bibinfo {author} {\bibfnamefont {D.}~\bibnamefont
  {Di~Mario}},\ }\href@noop {} {}\bibinfo {howpublished}
  {\href{http://digilander.libero.it/bubblegate/ephys.html}{http://digilander.libero.it/bubblegate/ephys.html},
  \emph{The Physics Deep}, retrieved 31st October 2007},\ \bibinfo {note}
  {hosting articles \emph{Magnetic Anomaly in Black Hole Electrons} (2003),
  \emph{Planck Permittivity and Electron Force} (2003), \emph{Reality of the
  Planck Mass} (2003), \emph{Electric Field as Variation of Gravity}
  (2005)}\BibitemShut {NoStop}%
\bibitem [{\citenamefont {Pfeifer}(2023{\natexlab{a}})}]{pfeifer2022CASM5}%
  \BibitemOpen
  \bibfield  {author} {\bibinfo {author} {\bibfnamefont {R.~N.~C.}\
  \bibnamefont {Pfeifer}},\ }\href {https://doi.org/10.48550/arXiv.2212.01255}
  {\bibinfo {title} {Falsifiable analog model predictions of {$W$~mass} in
  {CDF~II} and {ATLAS}}},\ \bibinfo {howpublished} {arXiv:2212.01255 with
  updates at
  \href{https://www.academia.edu/65931513}{https://www.academia.edu/65931513}}
  (\bibinfo {year} {2023}{\natexlab{a}})\BibitemShut {NoStop}%
\bibitem [{\citenamefont {Pfeifer}(2023{\natexlab{b}})}]{pfeifer2022CASM3}%
  \BibitemOpen
  \bibfield  {author} {\bibinfo {author} {\bibfnamefont {R.~N.~C.}\
  \bibnamefont {Pfeifer}},\ }
  {\bibinfo {title} {A classical analogue to the {Standard} {Model}, chapter 3:
  {Standard} {Model} particle spectrum from scalar fields on
  $\mathbb{C}^{\wedge 18}$}},\ \bibinfo {howpublished} {\href {https://doi.org/10.48550/arXiv.0805.3819}{arXiv:0805.3819} with
  updates at
  \href{https://www.academia.edu/65931513}{https://www.academia.edu/65931513}}
  (\bibinfo {year} {2023}{\natexlab{b}})\BibitemShut {NoStop}%
\end{thebibliography}%

\end{document}